\documentclass[10pt]{article}
\usepackage{natbib,setspace,lscape,longtable}
\usepackage{natbib,epsfig,graphicx}
\usepackage{mathrsfs,amsmath,amsthm,amssymb,color, verbatim}
\allowdisplaybreaks
\newcommand{\keywords}[1]{\vspace{1ex}\noindent\textbf{Keywords:} #1}
\usepackage{multirow}
\usepackage{bm}
\usepackage{hyperref}
\usepackage{bbm}

\usepackage{dsfont}
\usepackage{tabularx}

\usepackage{dsfont}
\usepackage{IEEEtrantools}
\usepackage{xr} 

\usepackage{geometry}
\usepackage{float}
\usepackage{array}
\usepackage{enumerate}
\usepackage{comment}
\usepackage{setspace}
\usepackage{amsfonts}
\usepackage{indentfirst}
\usepackage{booktabs}
\usepackage{appendix}
\usepackage{caption}
\usepackage{graphicx, subfig}

\numberwithin{equation}{section}
\newtheorem{theorem}{Theorem}
\newtheorem{lemma}{Lemma}
\newtheorem{proposition}{Proposition}
\newtheorem{cor}{Corollary}

\theoremstyle{definition}
\newtheorem{definition}{Definition}

\theoremstyle{remark}
\newtheorem{example}{Example}
\newtheorem{remark}{Remark}
\newtheorem{asm}{Assumption}

\theoremstyle{plain}

\theoremstyle{definition}

\newtheorem{definitionD}{Definition}

\theoremstyle{remark}

\newtheorem{asmD}{Assumption}

\renewcommand{\thedefinitionD}{D.\arabic{definitionD}}
\renewcommand{\theasmD}{D.\arabic{asmD}}

\def\1{\mathbf{1}}

\def\hat{\widehat}

\def\beq{\begin{equation}}
	\def\eeq{\end{equation}}
\def\beqs{\begin{equation*}}
	\def\eeqs{\end{equation*}}
\def\beqr{\begin{eqnarray}}
	\def\eeqr{\end{eqnarray}}
\def\beqrs{\begin{eqnarray*}}
	\def\eeqrs{\end{eqnarray*}}
\def\bet{\begin{theorem}}
	\def\eet{\end{theorem}}
\def\bel{\begin{lemma}}
	\def\eel{\end{lemma}}
\def\bep{\begin{proposition}}
	\def\eep{\end{proposition}}
\def\bg{\begin{figure}[tbph]\begin{center}}
		\def\eg{\end{center}\end{figure}}

\def\bc{\begin{center}}
	\def\ec{\end{center}}

\hypersetup{
  colorlinks=true,
  linkcolor={blue},
  filecolor={Maroon},
  citecolor={blue},
  urlcolor={blue}}

\title{Penalized Network Cross-Validation for Nested Models by Edge-Sampling}
\author{Bokai Yang\thanks{
    The first two authors contributed equally to this work. }\hspace{.2cm}\\
    Qiuzhen College, Tsinghua University\\
    and \\
    Yuanxing Chen \\
    Yau Mathematical Sciences Center, Tsinghua University\\
    and\\
    Yuhong Yang\thanks{
    Corresponding author. Email: yyangsc@mail.tsinghua.edu.cn.
    }\\
    Yau Mathematical Sciences Center, Tsinghua University}

\begin{document}

\maketitle
\begin{abstract}
    In the network literature, a wide range of statistical models has been proposed to exploit structural patterns in the data. Therefore, model selection between different models is a
    fundamental problem. However, systematic theoretical understanding remains limited when comparisons involve different model classes. To address this challenging issue, we propose a penalized edge-sampling cross-validation framework for nested network model selection. By incorporating a model complexity penalty into the evaluation process, our method effectively mitigates the overfitting tendency of cross-validation and adapts to varying model structures. This framework supports comparisons among widely used models, including stochastic block models (SBMs), degree-corrected SBMs (DCBMs), and graphon models, providing the first consistency guarantees for model selection across these settings to the best of our knowledge. Empirical evaluations, including both simulated data and the ``Political Books'' network, demonstrate that our method yields stable and accurate performance across various scenarios.
\end{abstract}

\newpage

\tableofcontents

\newpage

\section{Introduction}\label{sec-intro}

The exponential growth of network data in contemporary scientific research has profoundly advanced our understanding of complex network systems.  
Network modeling frameworks have evolved into indispensable analytical tools, finding widespread applications in diverse domains including online social networks \citep{jin2013understanding}, biological regulation networks \citep{friston2011functional,emmert2014gene}, and information-technological networks \citep{adamic2005political,pagani2013power}.
Since the seminal work of Erd\"os-R\'enyi's random graph model \citep{erdos1960evolution}, the field has witnessed substantial theoretical advancements through the development of various statistical models for network characterization.
Among these, the stochastic block model (SBM) \citep{holland1983stochastic} has emerged as a fundamental framework for community detection research over the past two decades. Its notable extensions, such as the affiliation stochastic block model \citep{frank1982cluster} and the degree-corrected block model (DCBM) \citep{karrer2011stochastic}, further enhance its flexibility in real-world applications.
Furthermore, the graphon model \citep{Aldous1981rep} provides a powerful nonparametric generalization for modeling the statistical behaviors of edges between nodes.

Given the diversity of network model classes, selecting an appropriate network model and its associated community structure remains a challenging problem.
Existing work has mainly focused on determining the number of communities within a given model class, for example through likelihood-based \citep{wang2017likelihood,hu2020corrected}, spectral \citep{le2022estimating,hwang2024estimation,wu2024eigengap}, and testing-based approaches \citep{jin2023optimal,ma2021determining}. 
After completing the initial draft of this article, we became aware of two related studies that investigate model comparison across related block models, for example via goodness-of-fit statistics \citep{Jin2025goodness} and test-based objective functions under a fixed number of communities $K$ \citep{Bhadra2026unified}. 
However, these approaches are typically restricted to a fixed number of communities, a single prespecified model class, or pairwise hypothesis-testing formulations, and therefore do not provide a unified solution to simultaneous model class and community structure selection.
Moreover, a principled framework that accommodates the consideration of both parametric and nonparametric network models, such as stochastic block models and graphons, is still lacking. 
In this work, we address this gap by proposing a penalized network cross-validation framework that enables simultaneous within-class and cross-class model selection, supported by explicit theoretical guarantees on model selection consistency.

We adopt cross-validation (CV), a widely recognized model selection framework with well-established theoretical foundations dating back to \cite{stone1974cv} and \cite{geisser1975pred}, and further developed in a substantial body of subsequent work (e.g., \citealp{shao1993cv,van2003unified,yang2007consistency,arlot2010survey,lei2025moderncv}). 
The network literature has also incorporated CV, with adaptations such as node-splitting \citep{chen2018network} and edge-sampling \citep{li2020network} designed to accommodate the dependence structure inherent in network data. 
More recently, \cite{Chakrabarty2025subsampling} proposed a subsampling-based CV framework (NETCROP) for comparing several parametric network models. These pioneering contributions demonstrate CV's strong empirical performance and establish consistency results for model selection within fixed model classes, such as the SBM and the DCBM.
Despite these advances, the theoretical underpinnings of CV for network models remain underdeveloped. Current theoretical analyses primarily focus on within-class model selection problems -- for instance, determining the number of communities in the SBM or DCBM -- with an emphasis on preventing underfitting.
Moreover, even in such within-class comparisons, existing methods usually rely on a prespecified upper bound on the number of communities, which may affect selection accuracy and restrict adaptability. 
Crucially, rigorous theoretical guarantees for CV-based selection across fundamentally different model classes, particularly between parametric block models and nonparametric graphon models, are still absent.

In this paper, we bridge the aforementioned gaps by establishing theoretical guarantees for cross-validation-based model selection across multiple network modeling scenarios.
Specifically, we introduce a novel penalized nested network cross-validation framework for model selection on undirected nested networks.
Building upon the edge-sampling scheme introduced by \cite{li2020network}, we generate training and evaluation sets, while employing model-specific matrix completion techniques to reconstruct connectivity patterns from partially observed networks. The key theoretical innovation of our approach lies in the introduction of a complexity-penalized loss function during model evaluation.
Unlike previous CV analyses that struggle to prevent overfitting -- with several studies providing only partial solutions -- our penalty term explicitly scales with model complexity to effectively regularize the selection process. 
This represents a significant advancement, as our penalization scheme induces necessary regularization across the model complexity space that cannot be achieved through standard CV alone. 
As a result, our method can avoid the need for prespecifying an upper bound on the number of communities, a requirement commonly imposed in existing approaches, thereby offering greater flexibility in practice. The optimal model is then selected through minimization of this penalized loss.

Our experimental results on the ``Political Books'' network dataset from \cite{krebs2004} demonstrate the practical advantages of our proposed method. 
While \cite{chen2018network} exclusively employed the SBM for this dataset (citing concerns about overfitting with its 105 nodes only), our analysis reveals that the DCBM class achieves significantly better fit than the SBM across various data splitting ratios (detailed in Section \ref{real_data}). 
These findings not only validate our method's ability to flexibly identify the most suitable model class in practical settings but also provide a more nuanced understanding of the network's underlying structure.

This work makes two contributions to address current methodological limitations.
Firstly, we develop a comprehensive framework for nested model selection, introducing a new penalized cross-validation methodology applicable to diverse network model classes, and the penalized CV idea may be applicable in other important learning problems.
Secondly, we establish rigorous theoretical guarantees for consistent model selection both within and across model classes (e.g., between the SBM, DCBM, and nonparametric graphon models). To the best of our knowledge, this work provides the first unified CV-based framework that delivers consistency guarantees for both within-class and cross-class network model selection.

The rest of the article is organized as follows. Section \ref{methodology} introduces the model specifications and the proposed penalized nested network cross-validation procedure, along with the key theoretical insights underlying our methodology.
Sections \ref{chooseK}--\ref{sbm_graphon} establish the theoretical foundations of our approach across various scenarios. In Section \ref{chooseK} we address intra-model selection for SBMs and DCBMs. 
In Section \ref{sbm_dcbm}, we analyze inter-model comparisons between different block model variants, including: the affiliation SBM versus general SBM, and general SBM versus DCBM.
Section \ref{sbm_graphon} advances the theoretical framework further by examining selection between constrained block models and more flexible graphon models, bridging classical and flexible network modeling paradigms.
Section \ref{simulation} presents comprehensive simulation studies that demonstrate the superior performance of our penalized approach compared to existing cross-validation methods in key scenarios.
Section \ref{real_data} provides a data example to demonstrate the utility of our methods. Finally, Section \ref{sec-diss} concludes with a discussion of remaining challenges and promising directions for future research.

\section{Methodology} 
\label{methodology}

\subsection{Notation}
For any positive integer $n$, we denote $[n]=\{1,\dots, n\}$.
For any matrix $M=[M_{ij}]$, we denote $\|M\|_F=\sqrt{\sum_{ij}M_{ij}^2}$ as its Frobenius norm and $\|M\|_{\infty}:=\max_{ij}|M_{ij}|$. We denote $\mathbf{1}_r$ as the vector of all 1s with length $r$, and $I_r$ as the identity matrix of order $r$. For a finite set $\mathcal{A}$, we denote $|\mathcal{A}|$ as its cardinality. Besides, let $\mathds{1}$ represent the indicator function.

\subsection{Nested Relationship of Network Models} \label{algorithm}

Throughout the paper, we consider an undirected network of $n$ nodes and its symmetric adjacency matrix $A$. 
Assume that all upper-triangular entries $A_{ij}\,(i<j)$ are independently drawn from Bernoulli distributions with $\mathbb{P}(A_{ij}=1)=P_{ij}$, while diagonal entries are zero. Typically, the symmetric probability matrix $P$ is assumed to follow certain structures, such as low-rank structures (e.g., Stochastic Block Model (SBM) or Degree-Corrected Block Model (DCBM)) or graphon models with some prespecified structure. 
Several nested relationships exist among these network models.
Leveraging this characteristic, this study employs a penalized nested network cross-validation (PNN-CV) procedure to select the optimal model from nested candidate models, where the optimal model refers to the best model within the best model class, representing the overall most accurate description of the observed network's underlying structure.

We first provide a formal definition of nested network models. For a network model $\delta$, let $\mathcal{P}$ be the set of all probability matrices allowed under the model. 
\begin{definition}
	
	Let $\delta^{(\ell)}$, $1\leq \ell\leq q$, denote a collection of network models for some $q\geq2$. 
	We say that the collection $\{\delta^{(\ell)}\}_{\ell=1}^{q}$ is nested if, for all $1\le \ell<\ell'\le q$, model $\delta^{(\ell)}$ is nested within model $\delta^{(\ell')}$ in the sense that $\mathcal{P}^{(\ell)}\subset \mathcal{P}^{(\ell')}$, where the inclusion is taken with respect to the space of admissible probability matrices.
\end{definition}

\begin{example}[Determination of $K$ within SBMs]
	Under the SBM framework, consider a network with $n$ nodes and $K\,(K\geq1)$ communities. Let  $c=(c_1,\ldots,c_n)^\top$ be a node membership vector, where $c_i\in[K]$ indicates the community membership of node $i$, such that the range of $c$ covers $[K]$. We denote by $\mathcal{C}_K$ the set of all membership vectors that satisfy these conditions.
	The adjacency matrix satisfies
	$\mathbb{P}(A_{ij}=1)=B_{c_ic_j}$ for $i\ne j\in[n],$
	where $B\in[0,1]^{K\times K}$ is a symmetric matrix specifying the block-wise connection probabilities. Denote by $\mathcal{B}_K$ the set of all such block-wise connection probability matrices. Denote the model associated with $K$ communities by $\delta^{(K)}$, and thus the corresponding $\mathcal{P}^{(K)}$ can be written as 
	\begin{equation}
		\mathcal{P}^{(K)}=\left\{P:P_{ij}=B_{c_ic_j}(1-\delta_{ij}),\ c\in\mathcal{C}_K\text{ and }B\in\mathcal{B}_K\right\},\label{P_in_SBM}
	\end{equation}
	where $\delta_{ij}$ is the Kronecker delta.
	Let $\mathcal{K}$ be the candidate set of the specified community numbers. Then the sequence $\{\delta^{(K)}\}_{K\in\mathcal{K}}$ forms a nested sequence of network models, as shown in Supplementary Material (Section B). 
\end{example}

\begin{example}[Nested relationships among affiliation SBMs, general SBMs, DCBMs, and graphon models]
	
	The stochastic block model admits a special subclass known as the affiliation stochastic block model \citep{frank1982cluster,Allman2011identi,matias2017statistical}.
	In this model, all within-community connection probabilities are identical and denoted by $p_{\mathrm{wc}}$, while all between-community connection probabilities are identical and denoted by $p_{\mathrm{bc}}$. 
	Denoting the affiliation SBM with $K$ communities by $\delta^{(1,K)}$, its corresponding parameter space can be written as
	$$\mathcal{P}^{(1,K)}=\left\{P:P_{ij}=B_{c_ic_j}(1-\delta_{ij}),\text{ where }B=p_{bc}\mathbf{1}_K\mathbf{1}_K^{\top}+(p_{wc}-p_{bc})I_{K}\text{ and }c\in\mathcal{C}_K,\,p_{wc},p_{bc}\in[0,1]\right\}.$$
	The general SBM with $K$ communities, denoted by $\delta^{(2,K)}$, allows for arbitrary block-wise connection probabilities.
	Its parameter space $\mathcal{P}^{(2,K)}$ coincides with $\mathcal{P}^{(K)}$ defined in \eqref{P_in_SBM}.    
	The degree-corrected block model (DCBM) further extends the general SBM by incorporating node-specific degree parameters $\theta_i$, $i\in[n]$, to capture degree heterogeneity. 
	Formally, given a community membership vector $c\in\mathcal C_K$, a block-wise probability matrix $B\in\mathcal B_K$, and a degree  parameter vector $\theta=(\theta_1,\ldots,\theta_n)^\top$, the connection probability between nodes $i$ and $j$ is defined as
	$
	\mathbb{P}(A_{ij}=1)=\theta_i\theta_jB_{c_ic_j}$ for $i\ne j\in[n].
	$
	This representation characterizes the DCBM through the triplet $(c,B,\theta)$. 
	To ensure identifiability, we impose the constraint $\sum_{i: c_i=k}\theta^2_i=|\{i:c_i=k\}|$ for each $k \in [K]$, which will be formally stated and discussed in Section \ref{dcbm_K}. 
	For a given $c$, let $\Theta_c$ denote the corresponding admissible parameter space for $\theta$.
	Denoting the DCBM with $K$ communities by $\delta^{(3,K)}$, the associated parameter space is defined by
	$$\mathcal{P}^{(3,K)}=\left\{P:P_{ij}=\theta_i\theta_jB_{c_ic_j}(1-\delta_{ij}),\ c\in\mathcal{C}_K,B\in\mathcal{B}_K\text{ and }\theta\in\Theta_c\right\}.$$
	
	The graphon model serves as a nonparametric generalization capable of representing substantially richer network structures and traces back to the foundational work of \cite{Aldous1981rep}. 
	Formally, a graphon model is defined by a symmetric graphon function $f:[0,1]\times[0,1]\to[0,1]$.
	Given latent variables $\{\xi_i\}_{i=1}^n$ independently drawn from the uniform distribution on $[0,1]$, the adjacency matrix $A$ satisfies
	$\mathbb{P}(A_{ij}=1)=f(\xi_i,\xi_j)$ for $i<j\in[n]$.
	Denoting the graphon model by $\delta^{(4)}$, its induced parameter space is 
	$$\mathcal{P}^{(4)}=\left\{P: P_{ij}=f(\xi_i,\xi_j)(1-\delta_{ij})\right\},$$
	where $\xi_i$ are realizations of $\text{i.i.d.}\,\mathrm{Uniform}(0,1)$ variables and $f$ is an arbitrary graphon function. 
	For a fixed number of communities $K$, it can be verified that $\mathcal{P}^{(1,K)}\subset \mathcal{P}^{(2,K)}\subset \mathcal{P}^{(3,K)}\subset \mathcal{P}^{(4)}$, establishing a nested hierarchy among affiliation SBMs, general SBMs, DCBMs, and graphon models. Detailed derivations are provided in Supplementary Materials (Section B).
\end{example}

\subsection{Penalized Nested Network Cross-Validation Procedure}

Let $0<w_n<1$ be a training proportion, that is, for each edge pair $(i,j)$ with $i<j$, it is assigned to the training set with probability $w_n$ and to the evaluation set with probability $1-w_n$. The lower triangular edge pairs are assigned by symmetry. 
Let $\mathcal{E}\subset [n]\times[n]$ be the training edge set  and $\mathcal E^c$ be the evaluation edge set corresponding to the complement of set $\mathcal E$. 
Given an adjacency matrix $A$ and a training edge set $\mathcal E$, let $Y$ be a partially observed adjacency matrix such that $Y_{ij}=A_{ij}$ if $(i,j)\in\mathcal E$ and $Y_{ij}=0$ otherwise.
Suppose that all candidate models can be applied either to the rescaled matrix $Y/w_n$ or to an adjusted version of $Y$ accounting sampling probability $w_n$.
For example, \cite{li2020network} developed a CV method to determine the optimal community number
within SBM/DCBM based on the adjusted $Y$, obtained by applying a low-rank truncation algorithm to $Y$. The Neighborhood Smoothing method (discussed in Section \ref{sbm_graphon}) proposed by \cite{zhang2017estimating} can be directly applied to $Y/w_n$.


For any nested network model sequence $\mathcal{C}:=\{\delta^{(m)},m\in[M]\}$, let $\hat{\delta}^{(m)}$ denote the estimation procedure corresponding to the $m$-th model $\delta^{(m)}$ and $\hat{P}^{(m)}$ denote the estimator of $P$ obtained by $\hat{\delta}^{(m)}$.
Let $\mathcal C$ be the candidate set comprising all models in the sequence. The prediction loss for this estimator is defined as 
\[
\ell_{m}(A,\mathcal{E}^c)=\sum_{(i,j)\in\mathcal{E}^c}(A_{ij}-\hat{P}^{(m)}_{ij})^2.
\]
Then the optimal model is given by 
\[
\hat{m}=\arg\min_{m\in[M]}\Big\{L_{m}(A,\mathcal{E}^c):=\frac{1}{|\mathcal E^c|}\ell_{m}(A,\mathcal{E}^c)+d_{m}\lambda_n\Big\},
\]
where $d_{m}$ is a certain complexity measure of ${\delta}^{(m)}$ that can be flexibly specified on a case-by-case basis. The penalty term $\lambda_n$ may be deterministic or a random variable in our context.

\begin{table}[htbp]
	\centering\renewcommand\arraystretch{1}{
		\begin{tabular}{p{15cm}}
			\hline
			\textbf{Algorithm 1} Penalized Nested Network Cross-Validation (PNN-CV) \\
			\hline
			\textbf{Input:} an adjacency matrix $A$, the training proportion $w$, a candidate set $\mathcal{C}$, the model complexity parameter $d_{m},\ m\in[M]$, the penalty order $\lambda_n$, and the number of replications $S$.\\
			\textbf{Output:} the best model with $\hat m$.\\
			\textbf{Step 1:} For $s=1,\dots,S.$
			\begin{enumerate}[(a)]
				\item Randomly choose a subset of node pairs $\mathcal E_s$ with probability $w$ as the training set.
				\item For each $m$, fit the model $\delta^{(m)}$ on adjusted (or unadjusted) $Y$ to obtain a corresponding estimator $\hat{P}^{(m)}_s$. 
				\item Evaluate the penalized loss of each $m$ by
				\[
				L_{m}(A,\mathcal{E}_s^c)=\frac{1}{|\mathcal E_s^c|}\sum_{(i,j)\in\mathcal{E}_s^c}(A_{ij}-\hat{P}^{(m)}_{s,ij})^2+d_{m}\lambda_n.
				\]
			\end{enumerate}
			\textbf{Step 2}: Determine the best model by
			\[
			\hat{m}=\arg\min_{m}\ S^{-1}\sum_{s=1}^S L_{m}(A,\mathcal{E}_s^c).
			\]\\
			\hline
		\end{tabular}
	}
\end{table}

A general algorithm for nested-model-based model selection is summarized in Algorithm 1. We introduce a penalty term, $\lambda_n$, to address the challenge of preventing overfitting, where cross-validation (CV) alone may only prevent underfitting. While the algorithm utilizes multiple random data splits for robustness, V-fold cross-validation can be adopted for practical implementation.
The success of Algorithm 1 hinges on two parameters: the training proportion $w_n$ and the penalty term $\lambda_n$. 
The subsequent theoretical analysis suggests that taking $w_n$ as a fixed constant is typically appropriate.
A feasible choice of $\lambda_n$ is determined by the candidate set $\mathcal{C}$.
For example, the recommended $\lambda_n$ for selecting the community number $K$ within SBMs differs from that used when choosing between SBM and graphon.

To investigate the asymptotic properties of this model selection procedure, we establish a general theoretical framework and derive key insights into the behavior of nested model sequences.
Our analysis begins by introducing some necessary definitions to characterize the convergence rates governing estimator behavior, and separation bounds essential for model distinguishability.

\begin{definition}\label{general_upper_bound}
	An estimation procedure $\hat{\delta}^{(m)}$ is said to attain a convergence rate $a_{n,w,m}=o(1)$ if the corresponding estimator $\hat P^{(m)}$ satisfies
	
	$$
	\frac{1}{n^2}\big\|\hat P^{(m)}-P\big\|_F^2=O_{\mathbb P}(a_{n,w,m}).
	$$
\end{definition}

\begin{definition}\label{general_lower_bound}
	An estimation procedure $\hat{\delta}^{(m)}$ based on $\mathcal{E}$ is said to be separated from the true probability matrix $P$ by at least an order $b_{n,w,m}>0$ if the estimator $\hat P^{(m)}$ satisfies
	\[
	\frac{1}{|\mathcal{E}^c|}\sum_{(i,j)\in\mathcal{E}^c}\big(\hat P_{ij}^{(m)}-P_{ij}\big)^2=\Omega_{\mathbb P}(b_{n,w,m}).
	\]
\end{definition}

Definition \ref{general_upper_bound} characterizes the upper bound on the estimation error for an estimation procedure $\hat{\delta}^{(m)}$, while Definition \ref{general_lower_bound} provides a lower bound on the estimation error for $\hat{\delta}^{(m)}$. Model $\delta^{(m^*)}(m^*\in[M])$ is assumed to be the true model in the sense that the true probability matrix $P^*$ satisfies $P^*\in\mathcal{P}^{(m^*)}$, but for any $m'<m^*$, $P^*\notin\mathcal{P}^{(m')}$.

\bigskip
\begin{asm}\label{a1}
	For $m'>m^*$, i.e., for all larger alternative models $\delta^{(m')}$ in the nested sequence, the estimator $\hat P^{(m')}$ of the estimation procedure $\hat{\delta}^{(m')}$ converges to the true probability matrix $P$ at a rate $a_{n,w,m'}$.
\end{asm}

\begin{asm}\label{a2}
	For $m'<m^*$, i.e., for all smaller alternative models $\delta^{(m')}$ in the nested sequence, they are asymptotically separated from the true probability matrix $P$ by at least an order $b_{n,w,m'}$.
\end{asm}

\begin{asm}\label{a3}
	(1) $(1-w_n)b_{n,w}/\max\left\{\sqrt{a_{n,w}(1-w_n)\|P\|_{\infty}},a_{n,w}\right\}=\omega(1)$, where $b_{n,w}:=\min_{m'<m^*}b_{n,w,m'}$ and $a_{n,w}:=\max_{m'\ge m^*}a_{n,w,m'}$; (2) $n^2(1-w_n)\min\{b_{n,w},1\}=\omega(1)$.
\end{asm}

Assumption \ref{a1} characterizes the asymptotic convergence rate of the estimator under the true model specification. Assumption \ref{a2} establishes the critical requirement that all competing smaller models necessarily exhibit strictly inferior performance compared to the true model.
The model selection consistency is guaranteed by Assumption \ref{a3}, which serves as the cornerstone of our theoretical framework, and it shall be verified in various practical cases. Note that we allow $\|P\|_{\infty}\rightarrow 0$ to accommodate the general possibility of sparse network modeling.



\begin{proposition}\label{general_prop}
	Under Assumptions 1 -- 3, let $d_{m'}$ be an increasing sequence representing the complexities of the nested model sequence. If $\max\left\{\sqrt{a_{n,w}(1-w_n)\|P\|_{\infty}},a_{n,w}\right\}\ll (1-w_n)\lambda_n\ll (1-w_n)b_{n,w}$, then we have
	$
	\mathbb P(\hat m=m^*)\rightarrow 1.
	$
\end{proposition}

Proposition \ref{general_prop} establishes a general theoretical framework for our penalized nested network cross-validation approach. 
It shows that model selection consistency can be achieved in nested models by verifying Assumptions \ref{a1} -- \ref{a3}.
Given a true model class with its associated model parameters, explicit characterizations of the orders $a_{n,w}$ and $b_{n,w}$ are derived in the following sections.

\section{Selection within the Same Block Model Class} \label{chooseK}

\subsection{Determination of K within the SBM Models}\label{sbm_K}
In this subsection, we demonstrate the application of our algorithm for determining the number of communities $K$ in the context of the stochastic block model (SBM). 
Two well-known cross-validation methods \citep{chen2018network,li2020network} were developed to prevent underfitting, achieving consistent selection of the true number of communities $K^*$ as the network size grows.
While these methods successfully address the underfitting problem, they remain vulnerable to overfitting -- a critical limitation that may lead to systematic overestimation of the number of communities. 
In contrast, our proposed penalized nested network cross-validation method offers a unified solution that simultaneously addresses both underfitting and overfitting issues.

Before applying our method, we first need to recover the adjacency matrix from the partially observed matrix $Y$.
Leveraging the low-rank structure inherent in the SBM, where the rank corresponds to $K^*$, we employ the low-rank matrix approximation method proposed in \cite{li2020network}.
Specifically, for a prespecified rank $k$, the complete adjacency matrix can be estimated via a singular value decomposition (SVD) with thresholding: $\hat{A}^{(k)}=S_H(Y, k)/w,$
where the operator $S_H(Y, k)$ performs a rank-$k$ truncated SVD on matrix $Y$. To elaborate, if the SVD of $Y$ is written as
$
Y= UDV^{\top},
$
where $ D = \operatorname{diag}(\sigma_1, \ldots, \sigma_n) $ with  
$\sigma_1 \geq \sigma_2 \geq \cdots \geq \sigma_n \geq 0,$ then  
$
S_H(Y, k) = UD_{k}V^{\top},
$
where  
$D_{k} = \operatorname{diag}(\sigma_1, \ldots, \sigma_{k}, 0, \ldots, 0)$ is the diagonal matrix retaining only the top $k$ singular values while setting the remaining singular values to zero.

After obtaining the recovered adjacency matrix $\hat{A}^{(k)}$, we employ the spectral clustering algorithm to estimate community labels $\hat{c}^{(k)}$. 
The choice of spectral clustering is motivated by its computational efficiency and favorable consistency properties, provided that the expected degree grows as $\Omega(\log n)$ and the spectral norm deviation between the observed and true probability matrices remains sufficiently small \citep{lei2015consistency}.

Once the estimated community labels $\hat{c}^{(k)}$ are obtained, we estimate $P$ through the following steps. 
First, we estimate the block-wise probabilities as
\begin{equation} \label{SBMest}
	\hat{B}_{k_1k_2}^{(k)}=\frac{\sum_{(i,j)\in \mathcal{E}}A_{ij}\mathds{1}\{\hat{c}_i^{(k)}=k_1,\hat{c}_j^{(k)}=k_2\}}{\sum_{(i,j)\in \mathcal{E}}\mathds{1}\{\hat{c}_i^{(k)}=k_1,\hat{c}_j^{(k)}=k_2\}},\quad k_1,k_2\in[k].
\end{equation}
The estimated probability matrix is constructed as $\hat{P}_{ij}^{(k)}=\hat{B}_{\hat{c}_i^{(k)}\hat{c}_j^{(k)}}^{(k)}$ for any $i,j\in[n]$.
For the model selection task, we aim to identify the true number of communities $K^*$. In contrast to prior works \citep{chen2018network,li2020network}, which restrict the candidate set $\mathcal{K}$ to a bounded range, we allow $\mathcal{K}$ to be unbounded, defaulting to the largest possible choice, $[n]$.
Finally, following Algorithm 1, we determine the optimal model by minimizing the penalized loss.
We formalize this complete procedure as Algorithm 2 (see Supplementary Materials, Section C for complete pseudo-code).  Similarly, we provide analogous implementations for Algorithms 3--6 in the supplementary materials.

To establish the asymptotic properties of Algorithm 2, we consider the following assumptions.
\begin{asm}[Balanced community structure]\label{a4}
	There exists a constant $0<\pi_0\le ({K^*})^{-1}$ such that $\min_k n_k\ge\pi_0 n$, where $n_k=|\{i\in[n]:c_i=k\}|$ is the community size for the $k$-th community.
\end{asm}

\begin{asm}[Degree condition of the stochastic block model]\label{a5}
	The true block-wise probability matrix is given by $B_n=\rho_nB_0$, where $B_0$ is a fixed $K^*\times K^*$ nonsingular symmetric matrix with entries in $[0,1]$. We assume that $K^*$ is fixed and that $B_0$ has distinct rows. The network sparsity parameter $\rho_n$ satisfies $n\rho_nw_n^2=\Omega(\log n)$.
\end{asm}

The stated condition slightly strengthens the standard requirements for stochastic block models \citep{chen2018network,li2020network} by additionally accounting for the training proportion $w_n$. Since $0<w_n<1$, it implies the usual sparsity condition $\rho_n=\Omega(\log n/n)$ and ensures that the training network contains sufficient information for low-rank recovery. This condition is mild when $w_n$ is bounded away from zero. If the training proportion $w_n$ is too small, the training edge set may fail to preserve the original block structure, thereby losing critical information about the rank constraint. Consequently, inadequate training performance may compromise the model selection procedure. Now we state the selection consistency result for $K^*$ in the stochastic block model.

\begin{theorem}[Model selection consistency within the Stochastic Block Model]
	\label{them2} 
	Under Assumptions \ref{a4} and \ref{a5}, let $d_k=k(k+1)/2$. Suppose the penalty term $\lambda_n$ satisfies 
	\begin{enumerate}
		\item $\lambda_n=o_{\mathbb{P}}(\rho_n^2).$
		\item For any constant $C>0$, $\mathbb{P}\left[\lambda_n<{C\rho_n}/{(nw_n)}\right]\ll1/n$.
		\item For any constant $C>0$, $\mathbb{P}\left[\lambda_n<{C\log n}/({n^2(1-w_n)})\right]\ll1/n$.
	\end{enumerate}  
	Then $\mathbb{P}(\hat{K}=K^*)\to1$ as $n\to\infty$.
	Here, the rates in Assumptions \ref{a1}--\ref{a3} are $a_{n,w}={\rho_n}/{(nw_n)}$ and $b_{n,w}=\rho_n^2$.
\end{theorem}

\begin{remark}
	Notice that here we do not assume that the candidate set $\mathcal{K}$ is bounded, as the proposed penalty term can help deal with an unbounded candidate set $\mathcal{K}$, even if it is the whole set $[n]$. Therefore, in contrast to Proposition~\ref{general_prop}, which establishes the theory under a finite model comparison regime, the current setting demands a more delicate control of the penalty term. Furthermore, we allow the penalty term $\lambda_n$ here to be a random variable. In the three assumptions for $\lambda_n$, when $w_n$ is a constant, the third assumption is a special case of the second assumption, as $\rho_n=\Omega(\log n/n)\Rightarrow{\rho_n}/{n}=\Omega({\log n}/{n^2})$.
\end{remark}

\subsection{Determination of K within the DCBM Models}\label{dcbm_K}
This subsection develops our framework for model selection under the degree–corrected stochastic block model (DCBM). The DCBM has been extensively studied in the literature, with a large body of work establishing consistent community detection procedures when the number of communities is known \citep{zhao2012consistency,lei2015consistency,jin2015fast}. 
Several methods have also been proposed for consistently estimating the true number of communities $K^*$ within the DCBM framework \citep{hu2020corrected,jin2023optimal,le2022estimating}. 
More recently, \cite{Chakrabarty2025subsampling} established selection consistency results for the DCBM under a subsampling-based cross–validation scheme, with an emphasis on preventing underfitting. 
In contrast, our proposed PNN-CV procedure attains full consistency by simultaneously guarding against both underfitting and overfitting.

The DCBM is characterized by the parameter triplet $(c,B,\theta)$. To ensure model identifiability, we assume that $\sum_{i\in \mathcal G_{\ell}}\theta_i^2=n_{\ell}$ for $\ell\in[K^*]$, where $\mathcal G_{\ell}=\{i\in[n]:c_i=l\}$ denotes the $\ell$-th community and $n_{\ell}=|\mathcal G_{\ell}|$ is its community size. Building upon the procedure described in Section \ref{sbm_K}, for each candidate $k$ and each data split, we first recover an estimate $\hat{A}^{(k)}$ of the adjacency matrix from the partially observed matrix $Y$. Unlike the SBM case, we then apply a spherical spectral clustering algorithm to estimate the community labels $\hat{c}^{(k)}$, in order to accommodate degree heterogeneity. To account for degree heterogeneity, we introduce an additional normalization step as follows. Specifically, we define a scaled version of the degree heterogeneity parameters by
\begin{equation}\label{scaletheta}
	\theta_i'=\theta_i/\sqrt{n_{c_i}},\quad i\in[n],
\end{equation} 
which satisfies $\sum_{i\in \mathcal{G}_{\ell}}(\theta_i')^2=1$ for each $\ell\in[K^*]$. The corresponding scaled block probability matrix $B'$ is defined accordingly. To estimate $\theta_i'$, we first compute $\hat{U}^{(k)}$, the matrix formed by the top $k$ eigenvectors of $\hat{A}^{(k)}$. The $i$th estimated scaled degree parameter $\hat{\theta}_i'^{(k)}$ is then defined as the $\ell_2$ norm of the $i$-th row of $\hat{U}^{(k)}$, namely, $\hat{\theta}_i'^{(k)}=\left(\sum_{j=1}^{k}(\hat{U}^{(k)}_{ij})^2\right)^{1/2}$.
Given the estimated labels $\hat{c}^{(k)}$ and scaled degree parameters $\hat{\theta}'^{(k)}$, we estimate the scaled block probability matrix $B'$ as follows,
\begin{equation}\label{DCBMB}
	\hat{B}'^{(k)}_{k_1k_2}=\frac{\sum_{(i,j)\in \mathcal{E}}A_{ij}\mathds{1}\{\hat{c}^{(k)}_i=k_1,\hat{c}^{(k)}_j=k_2\}}{\sum_{(i,j)\in \mathcal{E}}\hat{\theta}'^{(k)}_i\hat{\theta}'^{(k)}_j\mathds{1}\{\hat{c}^{(k)}_i=k_1,\hat{c}^{(k)}_j=k_2\}},\quad k_1,k_2\in[k].
\end{equation}
The estimated probability matrix under the DCBM is then constructed as $\hat{P}^{(k)}_{ij}=\hat{\theta}'^{(k)}_i\hat{\theta}'^{(k)}_j\hat{B}'^{(k)}_{\hat{c}^{(k)}_i\hat{c}^{(k)}_j}$. Finally, we compute the penalized loss defined in Algorithm 1 and select the optimal model by minimizing this criterion. The complete procedure is summarized in Algorithm 3.

To establish the theoretical properties of Algorithm 3, we first introduce a set of assumptions on the degree parameters $\theta$.
\begin{asm}\label{a7}
	There exist two constants $\theta_{\max}>\theta_{\min}>0$ such that $\theta_{\min}\leq\theta_i\leq\theta_{\max}$ for all $i\in[n]$.
\end{asm}
Assumption \ref{a7} is standard in the literature on degree-corrected models and has been widely used in prior work \citep{jin2015fast,lei2015consistency,chen2018network}.

\begin{asm}[Degree condition of the degree-corrected block model]\label{a5rev}
	The true block-wise probability matrix is given by $B_n=\rho_nB_0$, where $B_0$ is a fixed $K^*\times K^*$ positive-definite symmetric matrix with entries in $(0,1]$. We assume that $K^*$ is fixed and that $B_0$ has distinct rows. The network sparsity parameter $\rho_n$ satisfies the following conditions: (i) $\rho_n=\Omega(\frac{\log{n}}{n})$ and (ii) $\rho_n\gg\max\{1/(\sqrt{n}w_n),(\log n)^{1/4}/(\sqrt{n}(1-w_n)^{1/4})\}$.
\end{asm}

Compared with the conditions imposed under the SBM, we additionally require $B_0$ to be positive definite and its entries lie in $(0,1]$ so as to guarantee stronger spectral separation, which is essential for the analysis of the DCBM. We also strengthen the sparsity requirement. In particular, when the training proportion $w_n$ is fixed, condition (ii) reduces to $\rho_n \gg (\log n)^{1/4}/\sqrt{n}$, which relaxes the corresponding assumption in \cite{Chakrabarty2025subsampling}. We are now ready to state the selection consistency result for $K^*$ under the DCBM.

\begin{theorem}[Model selection consistency under the degree-corrected block model]
	\label{them_dcbm} 
	Under Assumptions \ref{a4} and \ref{a7}--\ref{a5rev}, let $d_k=k(k+3)/2$. Suppose the penalty term $\lambda_n$ satisfies: 
	\begin{enumerate}
		\item $\lambda_n=o_{\mathbb{P}}(\rho_n^2).$
		\item For any constant $C>0$, $\mathbb{P}\left[\lambda_n<C\cdot\max\left\{{\rho_n}/{(w_n\sqrt{n})},\sqrt{{\rho_n^3\log n}/{(w_nn)}}\right\}\right]\ll1/n$.
		\item For any constant $C>0$, $\mathbb{P}\left[\lambda_n<C\sqrt{{\log n}/{(n^2(1-w_n))}}\right]\ll1/n$.
	\end{enumerate}  
	Then $\mathbb{P}(\hat{K}=K^*)\to1\quad\mathrm{as}\quad n\to\infty.$ Here, the rates in Assumptions \ref{a1}--\ref{a3} are given by $a_{n,w}=\max\left\{{\rho_n}/{(\sqrt{n}w_n)},\sqrt{{\log n\rho_n^3}/{(nw_n)}}\right\}$ and $b_{n,w}=\rho_n^2$.
\end{theorem}
\begin{remark}
	Among the three conditions imposed on the penalty term, when $\rho_n\gg \sqrt{\log n/n}$ -- a regime covering most commonly studied sparsity rates -- the third condition follows from the second one.
\end{remark}

\section{Selection between Different Block Models}\label{sbm_dcbm}


\subsection{Selection between the Affiliation and General SBM Models} \label{selectSBM}


This subsection presents the details for model selection between the affiliation model and the general SBM. 
The key distinction between them lies in the parameter constraints that all diagonal blocks share the uniform within-community connection probability $p_{\mathrm{wc}}$, and all off-diagonal blocks share the uniform between-community connection probability $p_{\mathrm{bc}}$.
While the affiliation model's parameter parsimony offers computational advantages and structural consistency, its restrictive assumptions may introduce substantial bias when the true network structure violates its uniformity constraints. Our methodology simultaneously addresses two critical tasks: model specification identification (affiliation vs. general SBM) and true community number ($K^*$) determination.

For ease of reference, throughout this subsection, we denote the affiliation model with $k_1$ communities as Model $\delta^{(1,k_1)}$ and the general SBM with $k_2$ communities as Model $\delta^{(2,k_2)}$.
Building on Theorem \ref{them2}, Algorithm 2 yields an estimator $\hat K$ that converges to $K^*$ under mild regularity conditions. This consistent community number estimate enables rigorous model comparison between the two specifications through our proposed evaluation framework.

Given the estimated community number $\hat{K}$, we denote the estimated community labels under models $\delta^{(1,\hat K)}$ and $\delta^{(2,\hat K)}$ as $\hat{c}$.
For the general SBM model, the estimated probability matrix $\hat{P}^{(2)}$ is constructed in the same manner as described in Section \ref{chooseK}, that is,
$
\hat{B}^{(2)}_{k_1k_2}=\left({\sum_{(i,j)\in \mathcal{E}}A_{ij}\mathds{1}\{\hat{c}_i=k_1,\hat{c}_j=k_2\}}\right)/{\hat{n}^{\mathcal{E}}_{k_1k_2}}\text{ for}\ k_1,k_2\in[\hat{K}]$ and $\hat{P}^{(2)}_{ij}=\hat{B}^{(2)}_{\hat{c}_i\hat{c}_j}\text{ for}\ i,j\in[n].
$

For the affiliation model, the estimated probability matrix $\hat P^{(1)}$ is constructed as follows. Let
$
\hat{n}^{\mathcal{E}}_1=\left|\{(i,j) \in \mathcal{E} : \hat{c}_i=\hat{c}_j\}\right|\text{ and } \hat{n}^{\mathcal{E}}_2=\left|\{(i,j) \in \mathcal{E} : \hat{c}_i \neq\hat{c}_j\}\right|,
$
and define
$$
\hat{p}_{\mathrm{wc}}=\frac{\sum_{(i,j)\in\mathcal{E}}A_{ij}\mathds{1}\{\hat{c}_i=\hat{c}_j\}}{\hat{n}^{\mathcal{E}}_1},\quad\quad\hat{p}_{\mathrm{bc}}=\frac{\sum_{(i,j)\in\mathcal{E}}A_{ij}\mathds{1}\{\hat{c}_i\neq\hat{c}_j\}}{\hat{n}^{\mathcal{E}}_2}.
$$
Finally, the estimated probability matrix $\hat P^{(1)}$ is constructed as $\hat{P}^{(1)}_{ij}=\hat{p}_{\mathrm{wc}}$ if $\hat{c}_i=\hat{c}_j$ and $\hat{P}^{(1)}_{ij}=\hat{p}_{\mathrm{bc}}$ otherwise. Given the model class complexity parameter $d_{1\hat K},d_{2\hat K}$ for two model classes, the penalized loss is computed as 
$L_{m}(A,\mathcal{E}_s^c)=\left(\sum_{(i,j)\in\mathcal{E}_s^c}(A_{ij}-\hat{P}^{(m)}_{s,ij})^2\right)/|\mathcal E_s^c|+d_{m\hat{K}}\lambda_n,$
with the optimal model selected accordingly. We formalize this procedure as Algorithm 4 and establish the following selection consistency results.

\begin{theorem}[Model selection consistency between the affiliation model and the general SBM] \label{them3}
	Assume the conditions in Theorem \ref{them2} hold. Let $d_{1k}\equiv2$ and $d_{2k}=k(k+1)/2$ be the model complexities of each model. For some $m^*\in\{1,2\}$ and $K^*\in\mathcal K$ with $K^*\geq 2$, if the true model is $\delta^{(m^*,K^*)}$, then $\mathbb P\left[(\hat m,\hat K)=(m^*,K^*)\right]\rightarrow 1.$
	Here, the rates in Assumptions \ref{a1}--\ref{a3} corresponding to sequence $\{\delta^{(1,K^*)},\delta^{(2,K^*)}\}$ are $a_{n,w}={\rho_n}/{(nw_n)}$ and $b_{n,w}=\rho_n^2$.
\end{theorem}


\subsection{Selection between the SBM Models and DCBM Models} \label{selectdcBM}
In this subsection, we study model selection between the SBM and the DCBM, addressing the practically important question of whether degree heterogeneity needs to be incorporated. Existing contributions, such as \cite{Bhadra2026unified} and \cite{Jin2025goodness}, approach this problem primarily from a hypothesis-testing perspective. The method in \cite{Bhadra2026unified} requires the true number of communities to be known in advance (or to be accurately estimated by external procedures), while \cite{Jin2025goodness} focuses on relatively simple network settings and does not fully develop theoretical guarantees for more general DCBM structures. 
In contrast, our proposed PNN-CV framework provides a unified model selection mechanism that simultaneously determines the number of communities and distinguishes between the SBM and the DCBM, with formal consistency guarantees. For notational clarity, throughout this subsection, we denote the SBM with $k_1$ communities as Model $\delta^{(1,k_1)}$ and the DCBM with $k_2$ communities as Model $\delta^{(2,k_2)}$.

Building upon the procedure described in Section \ref{dcbm_K}, we apply Algorithm 3 under the DCBM framework to obtain a consistent estimator $\hat{K}$ of the true number of communities $K^*$. 
Several alternative methods are also available for estimating $K^*$, including the stGoF procedure proposed by \cite{jin2023optimal} and the spectral estimator (BHMC) introduced by \cite{le2022estimating}. 
Notably, when the true underlying model is the SBM (a special case of the DCBM with $\theta_i\equiv1$ for all nodes), Algorithm 3 remains valid and continues to provide a consistent estimator of $K^*$.

Following the procedure outlined earlier, our algorithm proceeds as follows. For each sample splitting, 
we estimate the adjacency matrix $\hat{A}^{(\widehat K)}$ via rank-$\hat{K}$ truncated singular value thresholding based on the training edge set. 
We then apply a spherical spectral clustering algorithm to $\hat{A}^{(\widehat K)}$ to obtain the estimated community labels $\hat{c}$ under both candidate models, namely the SBM $\delta^{(1,\hat K)}$ and the DCBM $\delta^{(2,\hat K)}$. 
Given the estimated labels $\hat{c}$, we proceed to estimate the corresponding probability matrices for both models.
For the SBM, the estimator $\hat{P}^{(1)}$ is constructed according to  \eqref{SBMest},
while for the DCBM, the estimator $\hat{P}^{(2)}$ follows \eqref{DCBMB}. 
Then, we compute the penalized loss defined in Algorithm 4 and select the optimal model by minimizing this criterion. The complete procedure is summarized in Algorithm 5.

\begin{definition} \label{d4}
	For a DCBM model $(c,B,\theta)$, we say it has degree heterogeneity of order at least $a_{n,K^*}$ if 
	\begin{equation}\label{degree_heterogeneity}
		\frac{1}{n^2}\sum_{k_1,k_2\in[K^*]}\sum_{(i_1,j_1),(i_2,j_2)\in \mathcal G_{k_1}\times \mathcal G_{k_2}}\frac{(\theta_{i_1}\theta_{j_1}-\theta_{i_2}\theta_{j_2})^2}{|\mathcal G_{k_1}\times \mathcal G_{k_2}|}=\Omega(a_{n,K^*}).
	\end{equation}
\end{definition}
Definition \ref{d4} is essential for distinguishing between the SBM and DCBM models. 
If all $\theta_i$ are identical, the left-hand side of \eqref{degree_heterogeneity} equals 0. However, given $i\in\mathcal G_k$, if $\theta_i$ varies within a community $\mathcal G_k\subset[n]$, then the left-hand side of \eqref{degree_heterogeneity} becomes large, reflecting significant degree heterogeneity. Specifically, for any pair of communities $(k_1,k_2)$,
$$
\sum_{(i_1,j_1),(i_2,j_2)\in \mathcal G_{k_1}\times \mathcal G_{k_2}}\frac{(\theta_{i_1}\theta_{j_1}-\theta_{i_2}\theta_{j_2})^2}{|\mathcal G_{k_1}\times \mathcal G_{k_2}|}\geq\sum_{j\in \mathcal G_{k_2}}\frac{\theta_j^2}{|\mathcal G_{k_2}|}\sum_{i_1,i_2\in \mathcal G_{k_1}}\frac{(\theta_{i_1}-\theta_{i_2})^2}{|\mathcal G_{k_1}|},
$$
which demonstrates that the degree heterogeneity is proportional to the variance of $\theta$. 
Consequently, a larger variance in $\theta_i$ indicates greater deviation from the SBM model and a larger order $a_{n,K^*}$.
\begin{asm} \label{a8}
	The true DCBM model satisfies that either it is SBM, or it has degree heterogeneity of order at least $a_{n,K^*}$ that satisfies $a_{n,K^*}\gg \max\{1/(\sqrt{n}\rho_nw_n),\sqrt{\log n/(n\rho_nw_n)}\}$.
\end{asm}

\begin{remark}
	(i) Assumption \ref{a8} requires that the model is either an SBM or is sufficiently distinct from an SBM. A similar assumption is made in \cite{Bhadra2026unified}. This is reasonable because if $\theta$ is merely a small perturbation of a constant vector, the model becomes difficult to distinguish from an SBM model. 
	(ii) The order provided in Assumption \ref{a8} is mild. Notice that
	\begin{align*}
		&\frac{1}{n^2}\sum_{k_1,k_2\in[K^*]}\sum_{(i_1,j_1),(i_2,j_2)\in\mathcal G_{k_1}\times \mathcal G_{k_2}}\frac{1}{|\mathcal G_{k_1}\times \mathcal G_{k_2}|}(\theta_{i_1}\theta_{j_1}-\theta_{i_2}\theta_{j_2})^2\\=&\frac{2}{n^2}\sum_{k_1,k_2\in[K^*]}\sum_{(i,j)\in\mathcal G_{k_1}\times \mathcal G_{k_2}}(\theta_{i}\theta_j-\bar{\theta}_{k_1}\bar{\theta}_{k_2})^2,
	\end{align*}
	where $\bar{\theta}_k$ is the average of $\theta_i$ for all $i\in\mathcal G_k$. If the variance of $\theta$ within each community is of constant order, this sum is of constant order under the assumption of balanced community structure.
\end{remark}

We now present our consistency results for model selection between SBM and DCBM.

\begin{theorem}[Model selection consistency between the SBM and the DCBM]\label{them4}
	Assume Assumptions \ref{a4}, \ref{a7}--\ref{a8} hold. Let $d_{1k}=k(k+1)/2$ and $d_{2k}=k(k+3)/2$ represent the model complexities of each model, and let $\lambda_n$ satisfy
	$\lambda_n=o_{\mathbb{P}}(\rho_n^2a_{n,K^*})$ and conditions 2--3 in Theorem \ref{them_dcbm}. For some $m^*\in\{1,2\}$ and $K^*\in\mathcal K$, if the true model is $\delta^{(m^*,K^*)}$, then $\mathbb P\left[(\hat m,\hat K)=(m^*,K^*)\right]\rightarrow 1.$
	Here, the rates in Assumptions \ref{a1}--\ref{a3} corresponding to sequence $\{\delta^{(1,K^*)},\delta^{(2,K^*)}\}$ are $a_{n,w}=\max\left\{{\rho_n}/{(\sqrt{n}w_n)},\sqrt{{\log n\rho_n^3}/{(nw_n)}}\right\}$ and $b_{n,w}=\rho_n^2a_{n,K^*}$.
\end{theorem}
\begin{remark}
	The separation condition in Assumption~\ref{a8} ensures that the lower-order requirement on $\lambda_n$ is compatible with its upper bound, so the admissible range for $\lambda_n$ is nonempty.
\end{remark}

\section{Selection between Block Models and Graphon Models} \label{sbm_graphon}

This section addresses the model selection problem between the SBM and the graphon model.
This framework extends the classical parametric versus nonparametric model selection paradigm to network analysis.
The literature offers several established approaches for graphon estimation. Notable examples include the Universal Singular Value Thresholding (USVT) algorithm \citep{chatterjee2015usvt} and the Neighborhood Smoothing (NS) algorithm \citep{zhang2017estimating}.
Given an estimator for the probability matrix $P$ generated by the graphon function $f$, Algorithm 6 provides a systematic procedure for model selection between these two models.

For ease of reference, throughout this subsection, we denote the SBM model with $k_1$ communities as Model $\delta^{(1,k_1)}$ and the graphon model as Model $\delta^{(2)}$.
Drawing from Definition \ref{general_upper_bound}, which outlines the order of an algorithm, we give some assumptions regarding the true graphon model and the chosen estimation algorithm as follows.
\begin{asm} \label{a9}
	The chosen graphon estimation procedure achieves an order $a_{n,w}$ (under suitable conditions for the true graphon function). Specifically, for the estimator $\hat P^{(2)}$ obtained from the estimation procedure fitted on the partially observed adjacency matrix $Y/w_n$, the following holds:
	$$\frac{1}{n^2}\big\|\hat{P}^{(2)}-P\big\|_F^2=O_{\mathbb P}(a_{n,w}).$$
\end{asm}

Assumption \ref{a9} provides an upper bound for the estimation error when the true model is indeed a graphon model satisfying certain mild conditions. Building upon Corollary 1 in \cite{zhang2017estimating}, we have the following result for the NS estimator:
\begin{proposition}
	Suppose the underlying model is a graphon with $f(x,y)=\rho_n f_0(x,y)$, where $f_0$, $\rho_n$, and $w_n$ satisfy the regularity and sparsity conditions detailed in the Supplementary Material (Section A.7), and the NS algorithm is employed as the estimation method. Then, the estimator $\hat{P}^{(2)}$ fitted on the partially observed matrix $Y/w_n$ satisfies for two positive constants $C_1$ and $C_2$,
	$$
	\mathbb{P}\left\{\frac{1}{n^2}\big\|\hat{P}^{(2)}-P\big\|_F^2\geq C_1\frac{\rho_n}{w_n}\left(\frac{\log n}{n}\right)^{\frac{1}{2}}\right\}\leq n^{-C_2}.
	$$
	Consequently, the estimation error of $\hat{P}^{(2)}$ achieves an order of ${\rho_n}/{w_n}\cdot\left({\log n}/{n}\right)^{1/2}$.
\end{proposition}

Now given the estimators $\hat{P}^{(1,k)}$ for the SBM model as in Section \ref{chooseK} and $\hat{P}^{(2)}$ for the chosen graphon estimation algorithm, as well as the model complexity parameter $d_{1k}:k\in\mathcal{K}$ for each rank $k$ in the SBM and $d_2$ for the graphon model, we evaluate the penalized loss for each model class $m$ by
$L_{m}(A,\mathcal{E}_s^c)=\left(\sum_{(i,j)\in\mathcal{E}_s^c}(A_{ij}-\hat{P}^{(m)}_{s,ij})^2\right)/|\mathcal{E}_s^c|+d_{m}'\lambda_n,$
where $d_2'=d_2$ and $d_1'=d_{1\hat{K}}$ and choose the best model accordingly. We refer to this procedure as Algorithm 6.

The next definition is similar to Definition \ref{d4}, which characterizes how far the true graphon model is distinct from the SBM model.
\begin{definition}\label{def5}
	We say that the $n\times n$ probability matrix $P$ generated by graphon $f=\rho_nf_0$ is at least $b_{n,\mathcal{K}}$ -varying if, uniformly for any $k\in\mathcal{K}$ and any balanced labeling $c$ with $k$ communities (i.e. every community has cardinality exceeding ${\gamma_0n}/{k}$ for some constant $\gamma_0$), the following holds:
	\begin{equation} \label{vargraphon}
		\frac{1}{n^2}\sum_{k_1,k_2\in[k]}\sum_{(i,j)\in \mathcal G_{k_1}\times\mathcal G_{k_2}}\left(P_{ij}-\frac{\sum_{(i',j')\in \mathcal G_{k_1}\times\mathcal G_{k_2}}P_{i'j'}}{|\mathcal G_{k_1}\times\mathcal G_{k_2}|}\right)^2=\Omega(b_{n,\mathcal{K}}),
	\end{equation}
	where $\mathcal G_{k}=\{i\in[n]:c_i=k\}$ denotes the $k$-th community under labeling $c$.
\end{definition}
Definition \ref{def5} requires that the left-hand side of \eqref{vargraphon} exceeds $b_{n,\mathcal{K}}$ for all such partitions. 
Such an assumption is sensible or even necessary because if, for some labeling, the left-hand side of \eqref{vargraphon} is small, the true probability matrix may closely resemble an SBM with community structure defined by that labeling, making the two models difficult to distinguish.
Consequently, we adopt the following assumption.
\begin{asm}\label{a10}
	The true model is either an SBM satisfying Assumptions \ref{a4}--\ref{a5}, or a graphon model with the true probability matrix being at least $b_{n,\mathcal{K}}$-varying, where 
	\begin{equation}
		b_{n,\mathcal{K}}\gg \max\left\{\sqrt{\frac{a_{n,w}\rho_n}{(1-w_n)}},\frac{a_{n,w}}{1-w_n},\sqrt{\frac{\rho_n^3\log n}{w_n\sqrt{n}}},\frac{\max_{K\in\mathcal{K}}\log K}{n(1-w_n)^2}\right\}. \label{eqn_varying}
	\end{equation}
\end{asm}
\begin{remark}
	Assumption \ref{a10} is essential to ensure the distinguishability of the two models. For the right-hand side, for most cases, the dominating term is the first term. That is, \eqref{eqn_varying} is equivalent to $b_{n,\mathcal{K}}\gg \sqrt{{a_{n,w}\rho_n}/{(1-w_n)}}$ for most cases. This assumption is stronger than the assumption for $\rho_n$ proposed in Theorems \ref{them2} and \ref{them4}, but it is not unattainable. A detailed analysis is as follows.
	
	If the graphon function $f_0$ exhibits a constant-order standard deviation within blocks for any balanced block structure, where a simple example is $f_0(x,y)=(x+y)/2$, then the left-hand side of \eqref{vargraphon} is of order $O_{\mathbb{P}}(\rho_n^2)$. In our case, if the NS algorithm is employed, Assumption \ref{a10} becomes
	$$
	b_{n,\mathcal{K}}\gg\max\left\{\frac{\rho_n}{\sqrt{w_n(1-w_n)}}\left(\frac{\log n}{n}\right)^{1/4},\frac{\rho_n}{w_n(1-w_n)}\sqrt{\frac{\log n}{n}},\sqrt{\frac{\rho_n^3\log n}{w_n\sqrt{n}}},\frac{\max_{K\in\mathcal{K}}\log K}{n(1-w_n)^2}\right\}.
	$$
	Plugging in $b_{n,\mathcal{K}}=\rho_n^2$, with $w_n$ being of constant order and $\rho_n\gg \left({\log n}/{n}\right)^{1/4}$, Assumption \ref{a10} holds.
\end{remark}

\begin{asm}[balanced estimated label]\label{a12}
	Let $\hat{c} \in [k]^n$ be the estimated cluster labels obtained from a spectral clustering or spherical spectral clustering algorithm constrained to produce at most $k$ clusters. We assume that the clustering output is balanced in the following sense: there exists a constant $\pi_0' \in (0,1)$ such that
	$
	\min_{1 \leq \ell \leq k} \hat{n}_\ell \geq {\pi_0' n}/{k},\text{ where } \hat{n}_\ell := \left|\left\{ i \in [n] : \hat{c}_i = \ell \right\}\right|.
	$
\end{asm}
\begin{remark}
	Assumption \ref{a12} is imposed as a technical regularization on the estimated labels, rather than as a requirement on the intrinsic clustering performance under model misspecification. 
	This condition can always be enforced through a simple post-processing step, such as merging or trimming clusters whose sizes fall below a prespecified threshold. 
	When the true underlying model admits a block structure, standard clustering algorithms are known to satisfy this assumption automatically. In contrast, when the true model is a graphon, Assumption \ref{a12} rules out vanishing clusters in the misspecified SBM approximation, thereby ensuring well-controlled cluster sizes and enabling a meaningful lower bound on the approximation error in this underfitting regime.
\end{remark}

We now state our consistency result for model selection between the SBM and the graphon model.
\begin{theorem}[SBM versus graphon model]\label{them5}
	Assume that Assumptions \ref{a9}--\ref{a12} hold. 
	Let $d_{1k}=k(k+1)/2$ and $d_{2}=3\sqrt{n/\log n}$ denote the model complexities of the SBM and the graphon model, respectively. 
	Suppose further that $w_n^{-1}\max_{K\in\mathcal{K}} K=o(n^{1/4}/\log^{1/4} n)$. Assume that the penalty parameter $\lambda_n$ satisfies the following conditions:
	(1) $\lambda_n\cdot \sqrt{n/\log n}=o_{\mathbb{P}}(b_{n,\mathcal{K}})$; (2) $\lambda_n=\omega_{\mathbb{P}}\left({\max\left\{\sqrt{a_{n,w}(1-w_n)\rho_n},a_{n,w}\right\}}/\left({(1-w_n)\cdot \sqrt{n/\log n}}\right)\right)$; (3) $\lambda_n$ satisfies the conditions imposed in Theorem \ref{them2}. Then, the true model $\delta^*$ will be chosen with probability tending to 1.
	Here, the rates in Assumptions \ref{a1}--\ref{a3} corresponding to sequence $\{\delta^{(1,K^*)},\delta^{(2)}\}$ are $a_{n,w}=a_{n,w}$ (from Assumption \ref{a9}) and $b_{n,w}=b_{n,\mathcal{K}}$.
\end{theorem}

\begin{remark}
	A portion of Assumption \ref{a10} is necessary to ensure the existence of $\lambda_n$, while the remaining conditions are employed in the technical proof of Theorem \ref{them5}.
\end{remark}

\begin{remark}
	We restrict the maximum number of candidate communities to be of order $o(n^{1/4}/\log^{1/4} n)$ to maintain a sufficiently large separation between the complexity terms of the two competing models. In particular, the graphon complexity $d_2=3\sqrt{n/\log n}$ reflects the effective degrees of freedom of the smoothing-based estimator, since it can be written in the form $\hat{P}=SA$, where $S$ has nonnegative entries and its diagonal entries are of order $1/\sqrt{n\log n}$. The multiplicative constant serves as a calibration factor to align the penalty scale of graphon estimators with that of parametric block models, and does not affect the
	asymptotic rate. If $K$ grows faster, the resulting complexity gap would no longer be adequate to support the consistency analysis. Independently, the rate $\max K=\Omega(n^{1/2})$ arises as a technical barrier in the proof. Beyond this regime, the current analytical techniques fail to provide sufficient control and different arguments would be needed.
\end{remark}

To illustrate the generality of our framework, we also consider the comparison between DCBM and graphon models under the assumption that the degree heterogeneity parameters ${\theta_i}$ are known, an assumption also adopted in previous works such as \citet{hu2020corrected,lei2016goodness,gao2018community}. This comparison follows the same logic as our SBM vs. graphon case with the same constraints on the penalty term $\lambda_n$. Due to space constraints, we provide the formal procedure, definitions, assumptions, and theoretical guarantees in the Supplementary Materials (Section D).

\begin{remark}
	To ensure consistent model selection across different network models (affiliation model, SBM, DCBM, and graphon models), we need to carefully choose the scaling of $\lambda_n$. When comparing within block models, we recommend the choice of penalty to be $\lambda_n=0.001\hat{\rho}_n^2/\sqrt{\log n}$, where $\hat{\rho}_n:=\sum_{ij}A_{ij}/n(n-1)$ is a plug-in estimator for the sparsity parameter $\rho_n$, satisfying the conditions required by Theorems \ref{them2} -- \ref{them4}. For the specific comparison between SBM/DCBM and graphon models (using NS), we recommend the choice of penalty to be $\lambda_n=0.1\hat{\rho}_n^2/\sqrt{n}$, where $\hat{\rho}_n$ is the same plug-in estimator as before, such that conditions in Theorem \ref{them5} are satisfied, as long as $\rho_n\gg\log^{3/4}n/n^{1/4}$. Sensitivity analysis for the penalty choice is provided in the Supplementary Materials (Section E).
\end{remark}


\section{Simulation Studies} \label{simulation}

In this section, we conduct comprehensive numerical simulations to evaluate the finite-sample performance of the proposed PNN-CV approach. Throughout the simulations, we adopt a splitting ratio of $w=0.9$, under which PNN-CV can be implemented using a single 10-fold CV procedure (corresponding to $S=10$ replications). 
Following \cite{yang2007consistency}, we further considered voting- and averaging-based variants to stabilize the selection via repeated 10-fold CV. However, as they bring only marginal performance gains, we omit their results and use voting only later in real data analyses for improved stability.

To assess the effectiveness and flexibility of our framework, we consider two types of comparisons.
First, we compare PNN-CV with its unpenalized counterpart (PNN-CV0) to illustrate the role of penalization in model selection.
Second, in scenarios where existing methods are applicable, we benchmark PNN-CV and its variants against representative cross-validation-based competitors, including ECV \citep{li2020network}, NCV \citep{chen2018network}, and NETCROP \citep{Chakrabarty2025subsampling}.

\subsection{Simulation Settings}
In this subsection, two main simulation studies are presented.
Simulation 1 examines the ability of the proposed method to distinguish affiliation SBMs from general SBMs while simultaneously selecting the number of communities.
Simulation 2 investigates model selection between SBMs and DCBMs under both balanced and imbalanced community structures, and includes comparisons with existing methods.
Additional simulation results are reported in the Supplementary Materials (Section E).

\textbf{Simulation 1: Affiliation SBMs and general SBMs.} We fix $K^*=3$ with a regularization parameter $\lambda_n=0.001\hat{\rho}_n^2/\sqrt{\log n}$ where we plug in an estimator $\sum_{ij} A_{ij}/n(n-1)$ for $\rho_n$. We adopt a balanced community structure, with label $c$ generated from a multinomial distribution $\mathcal M(n, \Pi)$ with equal community probabilities $\Pi=(1/K^*,\dots,1/K^*)$. When determining the number of communities, we employ an adaptive search procedure initialized at $K=1$, and at each iteration, a new candidate value of $K$ is evaluated. The current best value of $K$ is updated by comparing the mean penalized squared error, and the procedure terminates when no improvement is observed after five consecutive iterations, and the current best $K$ is selected as the final estimate $\widehat K$.
\begin{itemize}
	\item When the true model is the affiliation SBM, we set $B=rB_0$ with 
	$B_0=(1-\beta)I_{K^*}+\beta\mathbf{1}_{K^*}\mathbf{1}_{K^*}^\top$, where $\beta$ controls the ratio of between- to within-community edge probabilities. 
	We take $r=0.1$, $\beta=0.4$, and $n\in\{300,600,1200,2000\}$.
	\item When the true model is the general SBM, we set $B=rB_0$, where the diagonal entries of $B_0$ are drawn independently from $\mathcal U(0.6,1)$ and the off-diagonal entries from $\mathcal U(0.1,0.3)$. 
	We take $r=0.1$ and $n\in\{300,600,1200\}$.
\end{itemize}

\textbf{Simulation 2: General SBMs and DCBMs.} 
In this simulation, we compare PNN-CV with existing CV-based methods, and additionally report results for a hybrid PNN-CV+BHMC approach, where BHMC \citep{le2022estimating} is first used to estimate the number of communities.
To accommodate space constraints while covering a broader range of scenarios, 
we consider both balanced and imbalanced settings. 
Specifically, the case $K^*=3$ corresponds to a balanced configuration, whereas the case $K^*=5$ represents an imbalanced setting.
For $K^*=3$, we set the community proportion vector to $\Pi=\{1/3,1/3,1/3\}$ with $n\in\{300,600,900\}$. 
For $K^*=5$, we consider an imbalanced configuration 
$\Pi=\{1/10,1/10,1/5,3/10,3/10\}$ with 
$n\in\{1000,2000,3000,6000\}$, which represents a substantially more challenging regime. 
As in Simulation~1, we set $\lambda_n=0.001\hat{\rho}_n^2/\sqrt{\log n}$.
\begin{itemize}
	\item When the true model is an SBM, the probability matrix is set the same as in Simulation 1.
	
	\item When the true model is a DCBM, we set $B=rB_0$ with $r=0.1$, where the diagonal and off-diagonal entries of $B_0$ are drawn from $\mathcal U(0.8,1)$ and $\mathcal U(0.2,0.4)$, respectively. 
	The degree parameters $\theta_i$ are sampled from $\mathcal U(0.1,1)$ and normalized for identifiability, yielding $P_{0,ij}=\theta_i\theta_j B_{c_i c_j}$. 
\end{itemize}

\subsection{Simulation Results}
\begin{figure}[htbp]
	\centering
	\subfloat[$\beta=0.4$ and $n=300$]{
		\includegraphics[width =4cm]{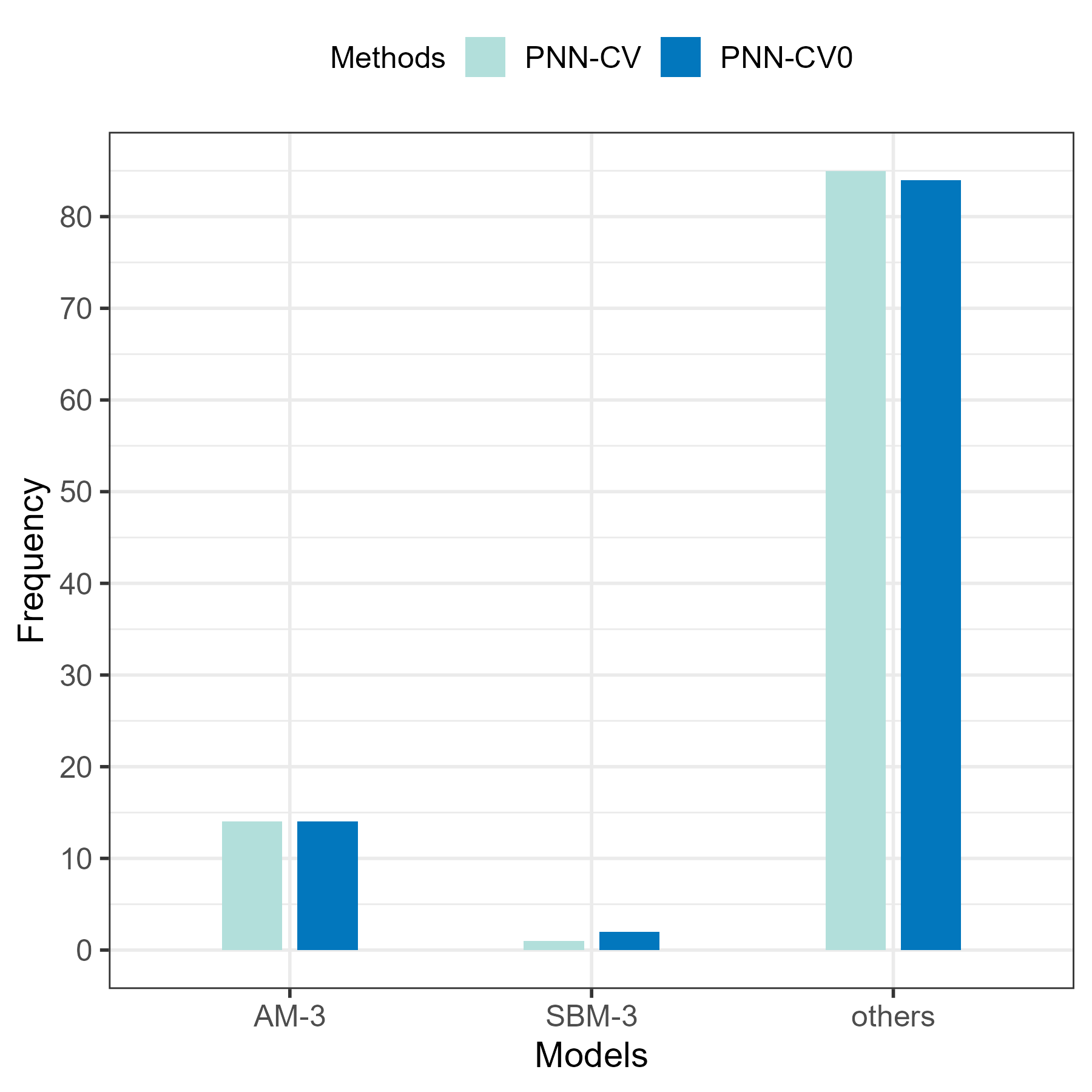}}
	\subfloat[$\beta=0.4$ and $n=600$]{
		\includegraphics[width =4cm]{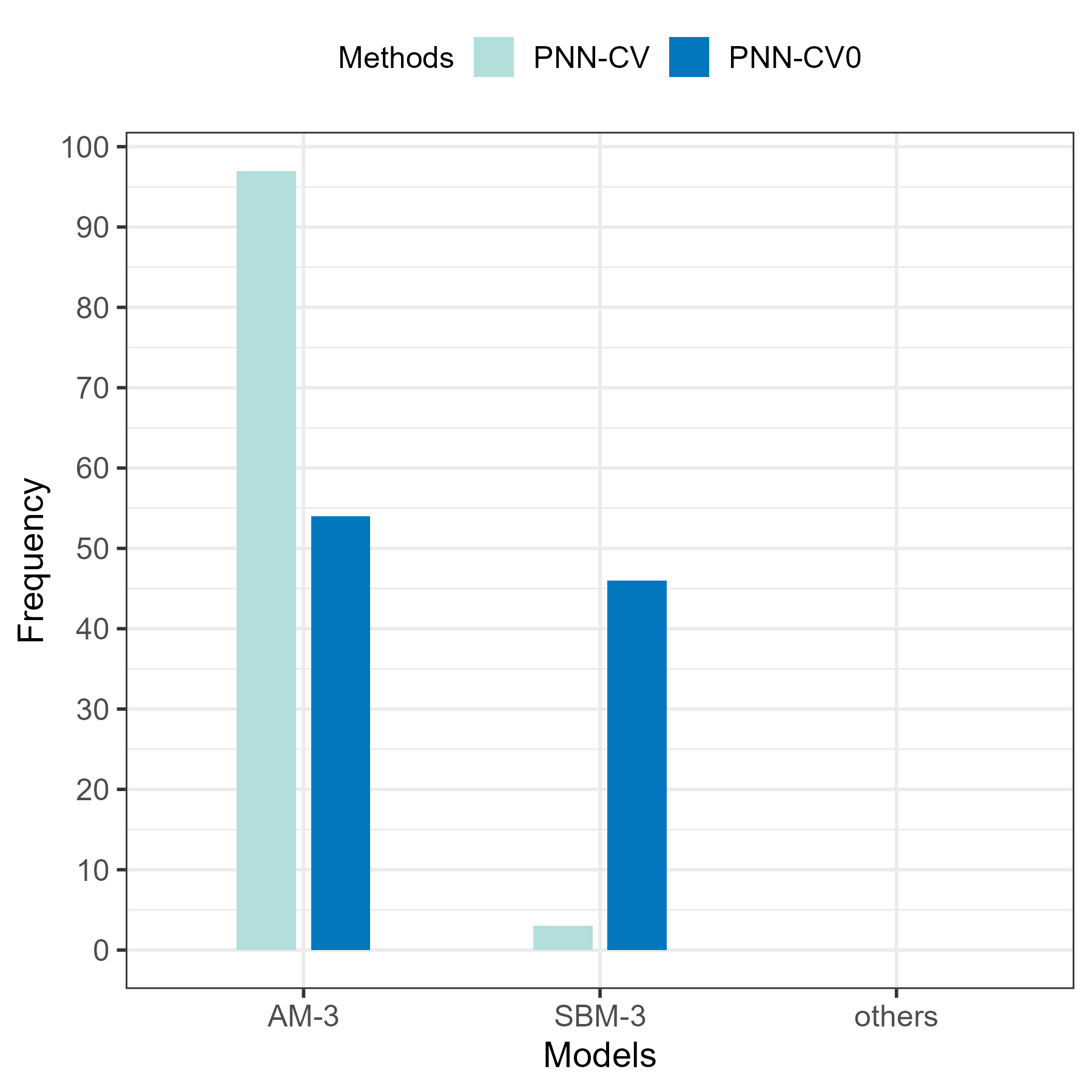}}
	\subfloat[$\beta=0.4$ and $n=1200$]{
		\includegraphics[width =4cm]{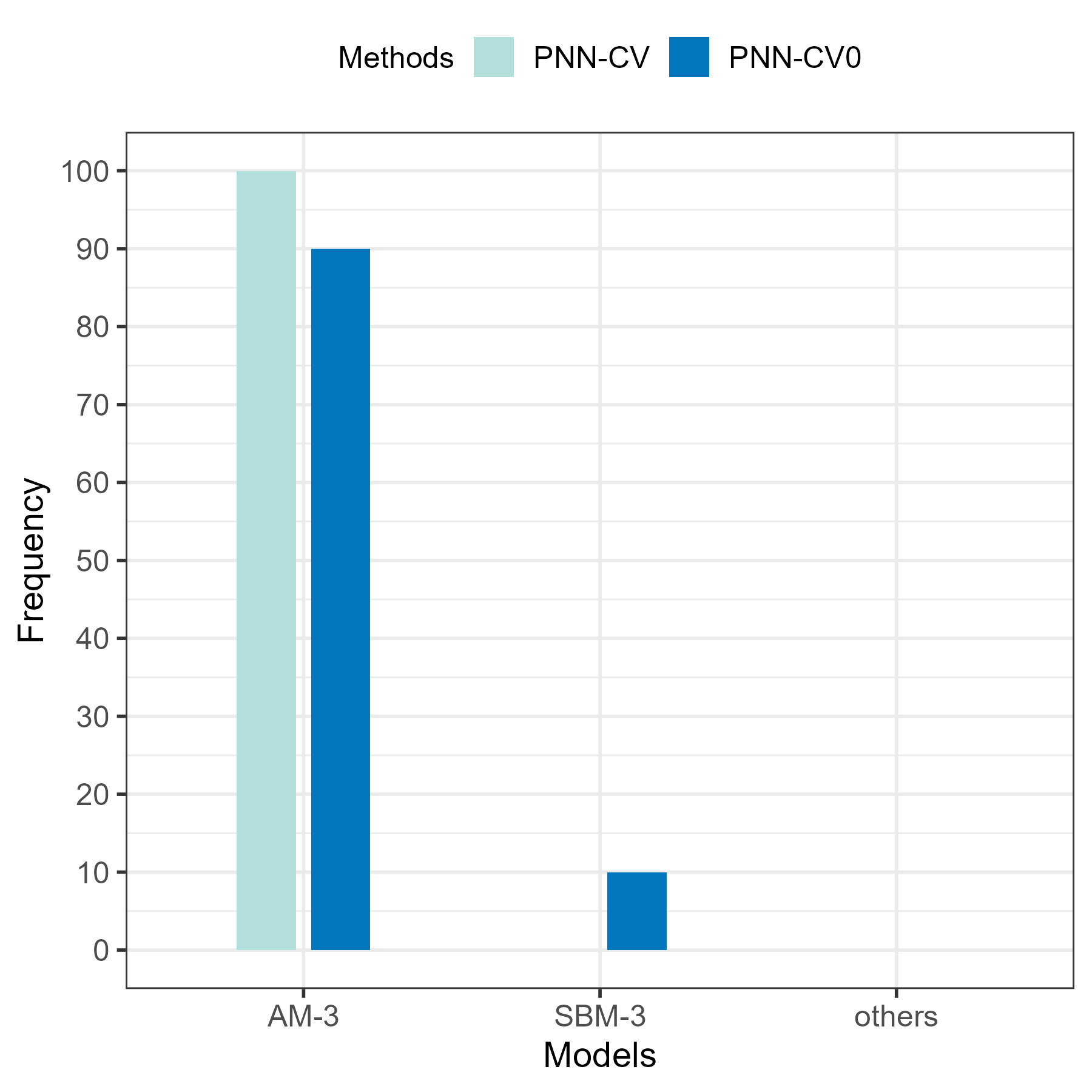}}
	\subfloat[$\beta=0.4$ and $n=2000$]{
		\includegraphics[width =4cm]{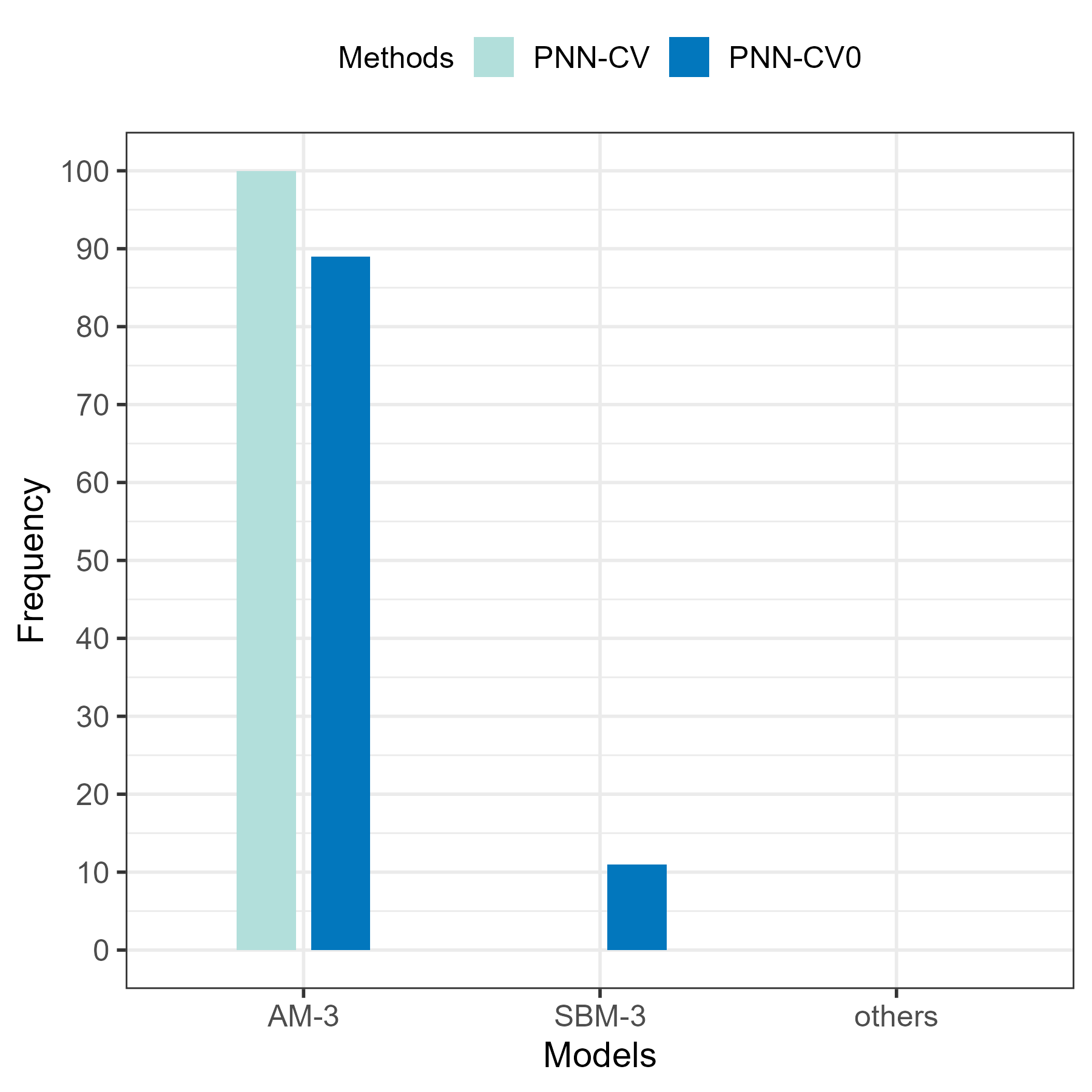}}\\
	\centering
	\caption{Barplots for affiliation SBM data in Simulation 1.}\label{barplot_simulation22}
\end{figure}

\begin{figure}[htbp]
	\centering
	\subfloat[$n=300$ and $K=3$]{
		\includegraphics[width =4cm]{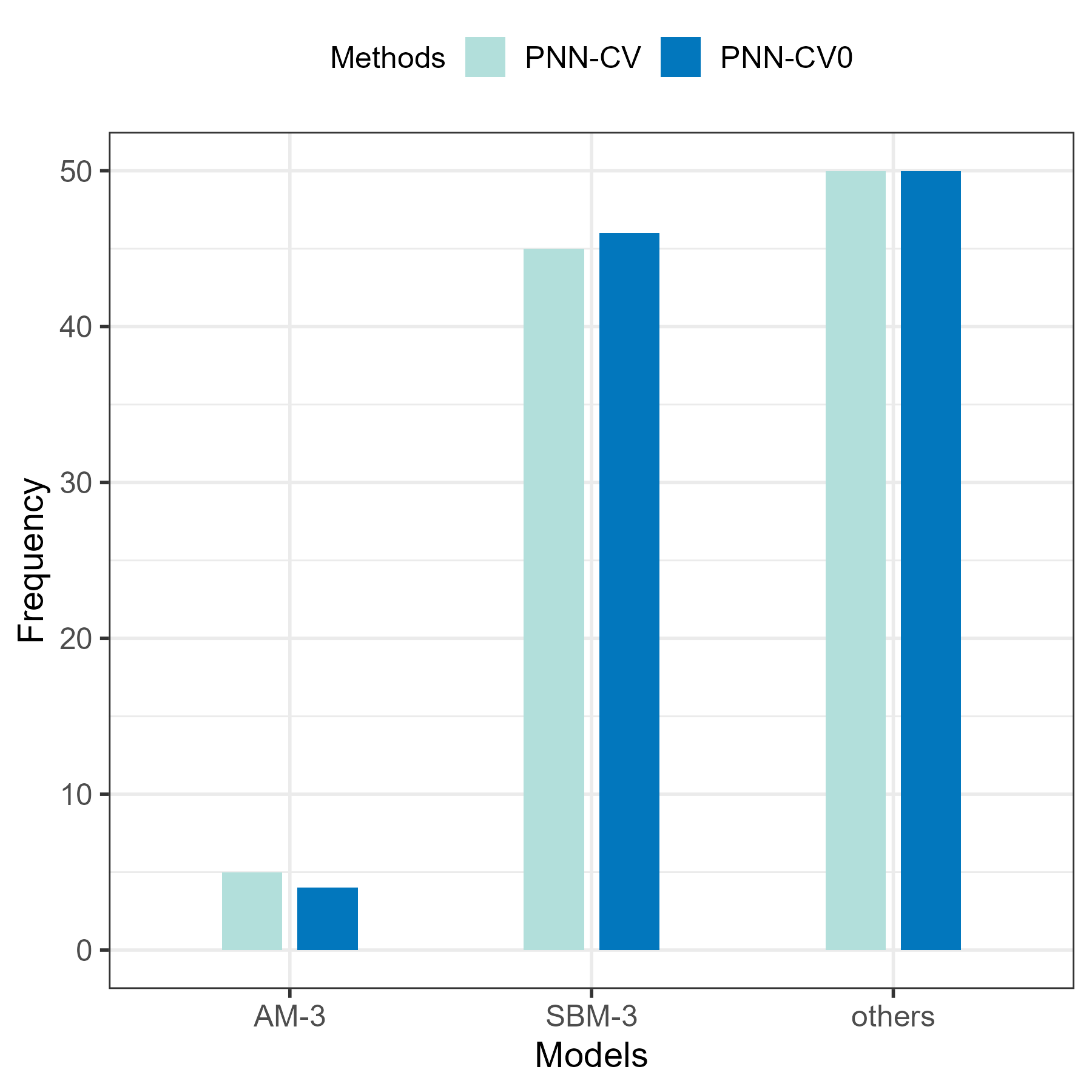}}
	\subfloat[$n=600$ and $K=3$]{
		\includegraphics[width =4cm]{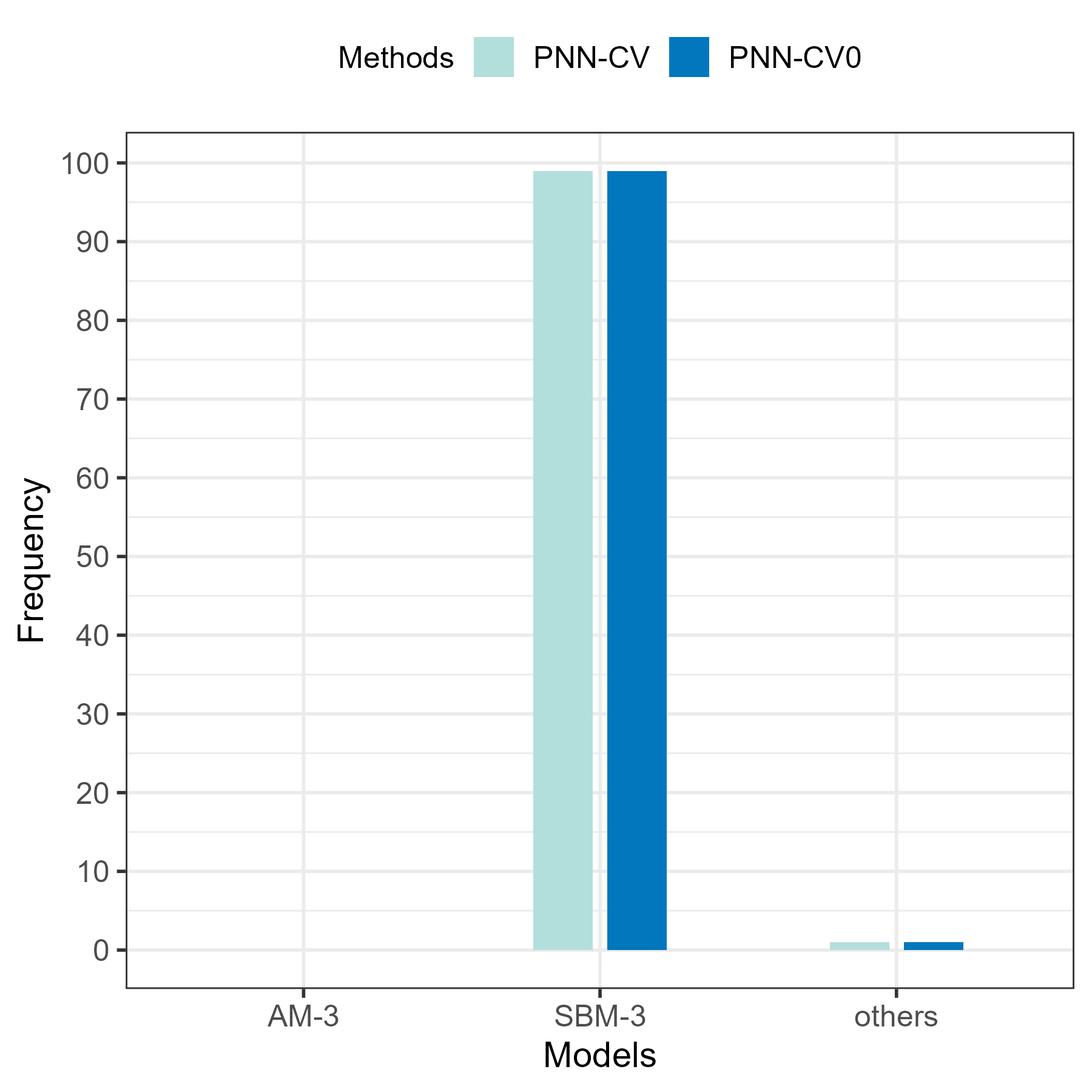}}
	\subfloat[$n=1200$ and $K=3$]{
		\includegraphics[width =4cm]{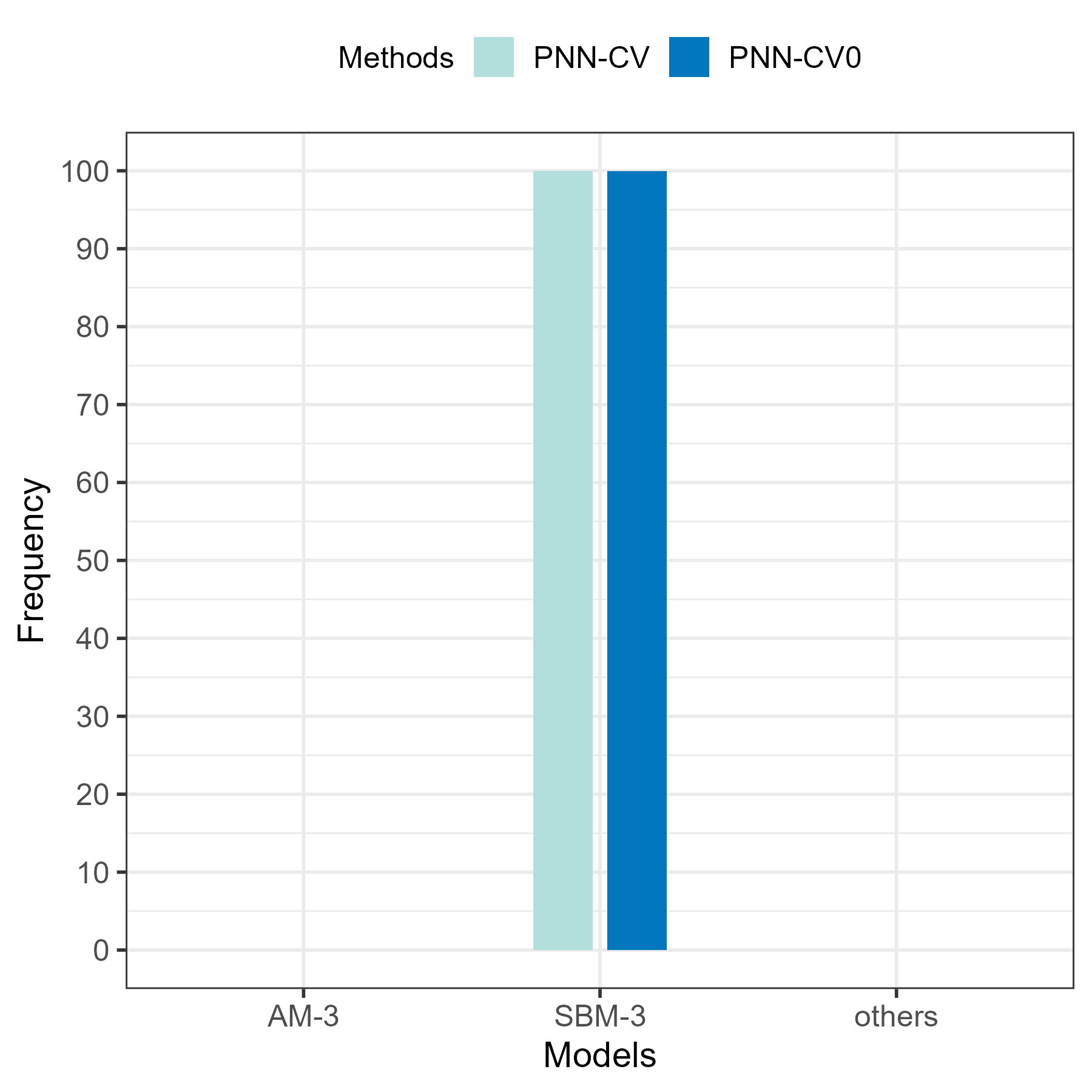}}\\
	\centering
	\caption{Barplots for general SBM data in Simulation 1.}\label{barplot_simulation23}
\end{figure}

Figures \ref{barplot_simulation22} and \ref{barplot_simulation23} show the selection frequencies across candidate models, with the first two columns corresponding to the affiliation model with 3 communities and the SBM model with 3 communities. Furthermore, we compare the performance of our proposed PNN-CV method with that of its counterpart without the penalty term, referred to as the PNN-CV0 method. As in Figure \ref{barplot_simulation22} (the true model is the affiliation model with 3 communities), for PNN-CV, the height of column ``AM-3'' asymptotically approaches 100 with increasing sample sizes, indicating simultaneous consistency in both community number selection and model identification. For the alternative PNN-CV0, the aggregated height concentrates at columns ``AM-3'' and ``SBM-3'', indicating consistency in community number selection. However, even at the maximum sample size ($n=2,000$), PNN-CV0 fails to achieve perfect model selection accuracy (peaking around 90 in column ``AM-3''). This observed ceiling effect strongly suggests that the absence of penalty terms in PNN-CV0 is likely to induce irreducible model mis-specification error. 
In Figure \ref{barplot_simulation23}, it is observed that when the true data-generating process follows a general SBM specification, both our methods and the alternative approaches achieve consistent joint selection with larger sample sizes.

Table~\ref{simulation3} reports the proportion of correct model selections (out of 100 replications) for distinguishing between the SBM and the DCBM. When $K^*=3$, PNN-CV exhibits rapid convergence under both SBM and DCBM data-generating processes. For the more challenging imbalanced case with $K^*=5$, the success rates of PNN-CV display a clear upward trend as the sample size $n$ increases.
Compared to existing cross-validation approaches (ECV, NCV, NETCROP), the proposed PNN-CV framework consistently delivers top-tier performance across all considered scenarios. 
Notably, the hybrid PNN-CV+BHMC approach demonstrates superior performance in small to moderate sample sizes, highlighting the complementary benefits of accurate community number estimation and principled model-class selection. 
These results indicate that the proposed framework can be naturally combined with existing methods to further enhance finite-sample performance, underscoring its adaptivity.

\begin{table}[htbp]
	\centering\renewcommand\arraystretch{0.9}{\scriptsize
		\caption{Selection between SBM and DCBM.
		}\label{simulation3}
		\setlength{\tabcolsep}{2mm}{
			\begin{tabular}{ccccccccc}
				
				\hline
				\cline{1-9}
				&&\multicolumn{3}{c}{$K=3$}&\multicolumn{4}{c}{$K=5$}\\
				\cmidrule(lr){3-5}
				\cmidrule(lr){6-9}
				True model	 &  Method&$n=300$&$n=600$&$n=900$&$n=1000$&$n=2000$&$n=3000$&$n=6000$\\
				\hline
				
				SBM    & PNN-CV+BHMC    &0.87   &1.00   &1.00  &0.12  &0.78    &0.98   &1.00\\
				& PNN-CV	        &0.50   &0.99   &1.00  &0.01  &0.19    &0.55   &0.98\\
				& ECV	        &0.46   &0.98   &1.00  &0.02  &0.20    &0.56   &0.99\\
				& NCV            &0.49   &0.93   &1.00  &0.03  &0.13    &0.31   &0.87\\
				& NETCROP        &0.00   &0.28   &0.40  &0.00  &0.14    &0.50   &0.91\\
				\hline
				
				DCBM   & PNN-CV+BHMC    &0.47   &1.00   &1.00  &0.00  &0.52    &0.99   &1.00\\
				& PNN-CV	        &0.21   &0.96   &1.00  &0.00  &0.02    &0.35   &0.98\\
				& ECV	        &0.35   &1.00   &1.00  &0.00  &0.04    &0.33   &1.00\\
				& NCV            &0.00   &0.34   &0.90  &0.00  &0.01    &0.15   &0.91\\
				& NETCROP        &0.00   &0.68   &0.92  &0.00  &0.00    &0.02   &0.40\\
				\hline
				\cline{1-9}	
			\end{tabular}
	}}
\end{table}

To summarize, Simulations 1 and 2 clearly demonstrate the advantages of our proposed PNN-CV framework across a range of model selection tasks. 
In the first setting, which aims to distinguish affiliation SBMs from general SBMs, the inclusion of a penalty term in PNN-CV leads to significantly improved model selection accuracy, particularly when the true underlying model is relatively simple. 
In the second setting, focusing on discrimination between SBMs and DCBMs, PNN-CV achieves competitive performance relative to existing CV-based methods, while its hybrid variant consistently outperforms the benchmarks.
Together, these results illustrate both the effectiveness and the adaptability of our approach.
Additional simulations we presented in supplementary materials further support the theoretical advantages of our framework, showing that PNN-CV remains competitive or superior across diverse settings.

\section{Real Data Application} \label{real_data}

In this section, we apply the proposed PNN-CV+BHMC method with voting to the well-known network dataset ``Political Books'' from \cite{krebs2004}, also used in \cite{chen2018network}, \cite{le2022estimating} and \cite{hwang2024estimation}, with the number of replications in voting set to 20. It has $n=105$ nodes, each representing a political book. An edge between a pair of nodes
indicates frequent co-purchasing of these two books. The books were manually labeled as one of three categories: ``neutral'' (13 nodes), ``liberal'' (43 nodes), and ``conservative'' (49 nodes). 

\cite{chen2018network} focused on fitting this dataset using the SBM model, arguing that the sample size is not sufficiently large to support fitting a DCBM. In contrast, our goal is to apply the proposed method to select the most suitable model and to investigate whether a DCBM provides a better fit despite the limited sample size.

Since the dataset has been manually labeled into three categories, we first consider an experimental setting where we treat $K = 3$ as the ground truth and compare candidate models under this constraint. We also conduct a second setting where we do not assume the correctness of the manual labeling and instead apply our full model selection procedure to compare SBMs and DCBMs. The selection result is stable across a wide range of splitting ratios (following the approach in \cite{zhan202224}), while the result produced by ECV varies across different folds. Detailed sensitivity results are reported in the Supplement Material (Section F).

\begin{figure}[htbp]
	\centering
	\subfloat[DCBM-3]{
		\includegraphics[width =7.5cm]{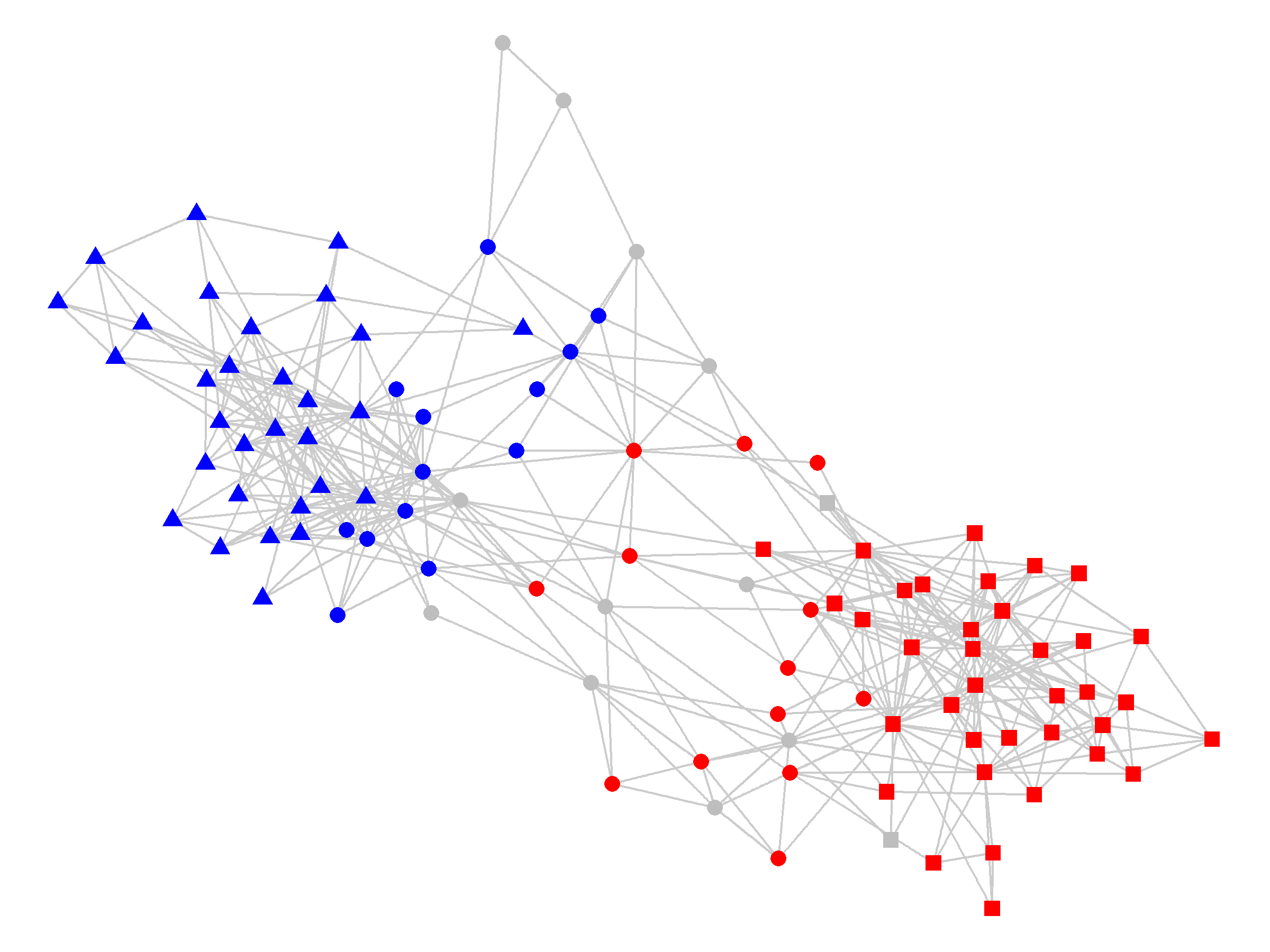}}
	\subfloat[DCBM-4]{
		\includegraphics[width =7.5cm]{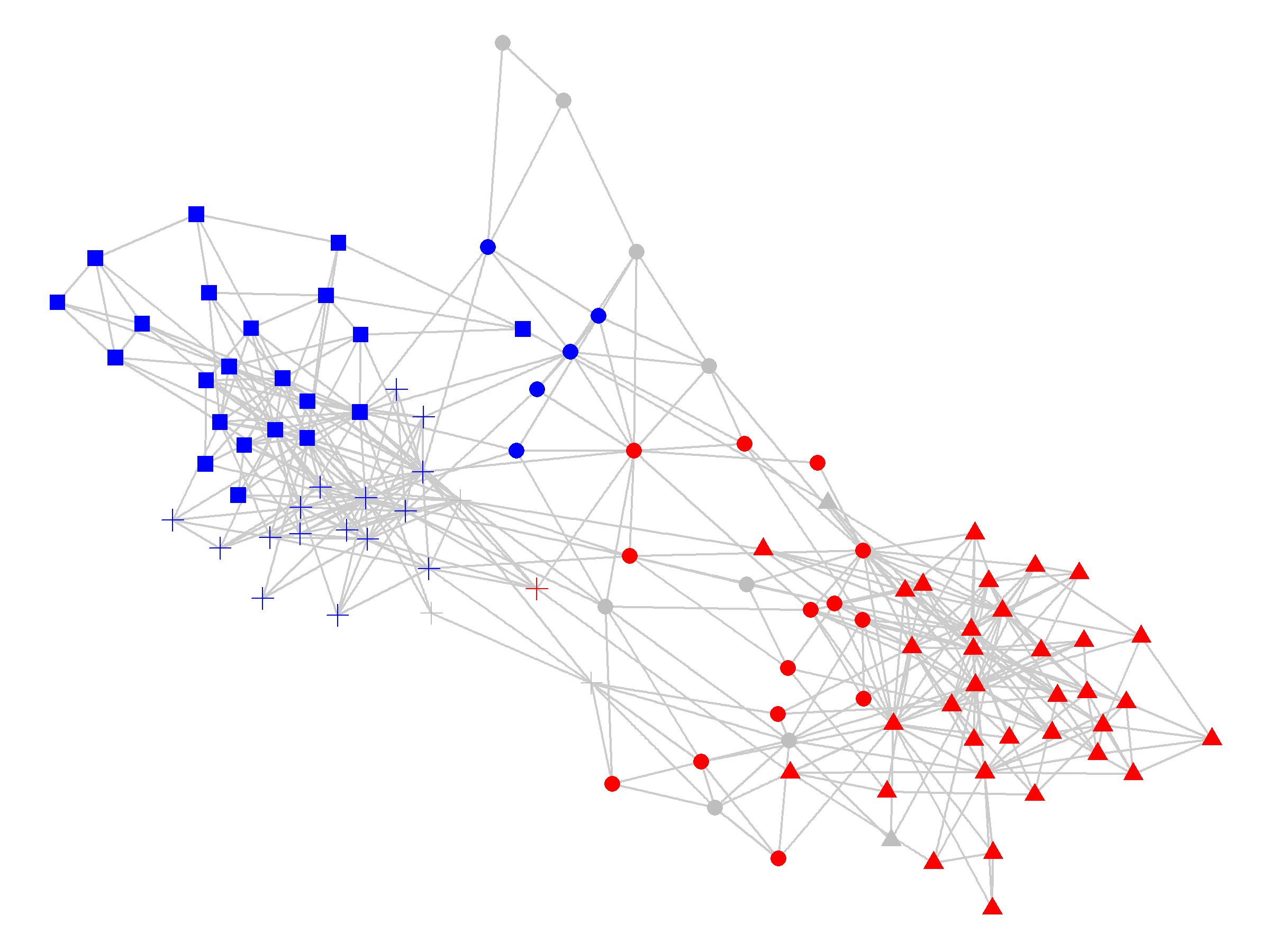}}\\
	\centering
	\caption{Estimated community structure for DCBM models}\label{polbooks_visualize}
\end{figure}

Our method consistently favors DCBM over SBM under both settings, and selects DCBM-4 consistently if no known ground truth on $K$ is assumed. Figure \ref{polbooks_visualize} visualizes the community structures estimated by spherical spectral clustering. Blue nodes represent liberal books, red nodes represent conservative books, and gray nodes denote neutral books. Under DCBM-3, communities roughly correspond to ideological categories, while under DCBM-4, the liberal group is further split, revealing finer heterogeneity.

We note that both DCBM-3 and DCBM-4 are reasonable under different interpretations of the data. However, regardless of the choice of $K$, DCBM clearly provides a better fit than SBM. In fact, within each estimated community, the degrees of the nodes vary substantially from as low as 3 to as high as 25, despite the total number of nodes being only 105. Such large inside-community degree variation makes the SBM assumption of degree homogeneity somehow implausible for this dataset.

Moreover, we remark that even when fitting the model using SBM-3 as in prior studies \citep{chen2018network}, the observed degree variation within certain communities still ranges from 3 to 25, suggesting that the degree heterogeneity is inherent in the data rather than an artifact of fitting with DCBM. Consistent with this observation, the goodness-of-fit statistic of \cite{Jin2025goodness} is notably smaller for DCBM than for SBM, further supporting the suitability of DCBM for this dataset.

\section{Discussion}\label{sec-diss}

In this article, we proposed PNN-CV, a penalized cross-validation framework for model selection in nested network models. Our primary contribution lies in establishing theoretical guarantees for consistent model selection across different model classes, including the first (to our knowledge) results for distinguishing among SBM, DCBM, and graphon models. While simulation studies and real data analysis further support the practical effectiveness of PNN-CV, the core value of our work is providing a unified, theoretically justified solution to cross-validated model selection in networks.

Looking forward, several promising directions remain open. First, the current work focuses on the symmetric network setting. Extending penalized cross-validation to more general structures such as bipartite or directed networks presents new challenges and would significantly broaden the applicability of the framework. Second, in many real-world applications, networks are often accompanied by covariate information. How to effectively incorporate such covariates into the model selection process, and how to select between models that involve covariates, remains an exciting avenue for future research. We expect that our general penalized CV strategy could be adapted and extended to handle such enriched modeling settings.

\newpage
\appendix
\renewcommand{\thetheorem}{A.\arabic{theorem}} 
\renewcommand{\thelemma}{A.\arabic{lemma}}     
\renewcommand{\theproposition}{A.\arabic{proposition}}
\renewcommand{\thedefinition}{A.\arabic{definition}}
\renewcommand{\theasm}{A.\arabic{asm}}
\renewcommand{\theremark}{A.\arabic{remark}}
\renewcommand{\thedefinitionD}{D.\arabic{definitionD}}
\renewcommand{\theasmD}{D.\arabic{asmD}}
\renewcommand{\thecor}{D.\arabic{cor}}
\renewcommand{\thefigure}{E.\arabic{figure}}
\renewcommand{\thetable}{E.\arabic{table}}
\setcounter{figure}{0}
\setcounter{table}{0}
\setcounter{theorem}{0} 
\setcounter{lemma}{0}   
\setcounter{proposition}{0}
\setcounter{definition}{0}
\setcounter{asm}{0}
\begin{center}
    \Large \textbf{Supplementary Materials for ``Penalized Network Cross-Validation for Nested Models by Edge-Sampling''}
\end{center}

In this supplementary file, we provide the proofs of the theoretical results and additional materials including discussions, supplementary numerical results, and pseudo-codes. In Section A, we present detailed proofs for the propositions and theorems stated in the main article. In Section B, we provide derivations for the illustrative examples in Section 2.2. In Section C, we include the complete pseudo-code for the algorithms discussed in Sections 3 -- 5. In Section D, we formally discuss the DCBM versus graphon comparison, introducing the full procedure, definitions, assumptions, and theoretical guarantees. In Section E, we conduct additional simulations to further validate our framework under various settings and provide a comprehensive sensitivity analysis for the tuning parameter. In Section F, we display supplementary analysis and illustration results for the ``Political Books'' dataset, and implement our method on four additional real datasets across various sample sizes.

\section{Proofs}
We focus on a single split in the penalized edge-split cross-validation procedure, since consistency for a single split immediately extends to any fixed number S of splits. For notational simplicity, we will write $w_n$ simply as $w$ where there is no conflict. 

\subsection{Notations and Supporting Lemmas}
We start with additional notations. For any vector $\mathbf{x}=(x_1,\ldots,x_d)$, we denote $\|\mathbf{x}\|=\|\mathbf{x}\|_2=\sqrt{\sum_{i}x_i^2}$ as its Euclidean norm, $\|\mathbf{x}\|_1=\sum_{i}|x_i|$ and $\|\mathbf{x}\|_{\infty}=\max_{i}|x_i|$. For two vectors $\mathbf{x}=(x_1,\ldots,x_d)$ and $\mathbf{y}=(y_1,\ldots,y_d)$, we denote $\langle\mathbf{x},\mathbf{y}\rangle=\sum_ix_iy_i$ as the inner product of two vectors. For any matrix $M=[M_{ij}]$, we denote $\|M\|=\|M\|_2=\max_{\|\mathbf{x}\|=1}\|M\mathbf{x}\|$ as its spectral norm. We also denote $M_{i*}$ as the $i$-th row of the matrix $M$.

Throughout the proofs, for any fitted block whose denominator is zero, we set the corresponding block estimate equal to zero. All concentration inequalities are applied to the upper-triangular dyads $\{(i,j):i<j\}$. Whenever a sum is written over a symmetric edge set, each undirected dyad is counted twice; this affects only universal constants.

We list some supporting lemmas that will be used in our proofs.
\begin{lemma}[Hoeffding's inequality]\label{lemma_hoeffding}
	Let $X_1, X_2, \ldots, X_n$ be independent random variables such that $X_i \in [a_i, b_i]$ almost surely. Let $S = X_1 + \cdots + X_n$. Then for all $t > 0$:
	\[
	\mathbb{P}(S - \mathbb{E}S \geq t) \leq \exp\left( -\frac{2t^2}{\sum_{i=1}^n (b_i - a_i)^2} \right).
	\]
\end{lemma}

\begin{lemma}[Bernstein's inequality]\label{lemma_bernstein}
	Let $X_1, X_2, \ldots, X_n$ be independent random variables such that $\mathbb{E}[X_i] = 0$ for all $i$, and assume that $|X_i| \leq M$ almost surely. Then for all $t > 0$,
	\[
	\mathbb{P}\left[ \sum_{i=1}^n X_i \geq t \right] \leq \exp\left( -\frac{t^2/2}{\sum_{i=1}^n \mathbb{E}[X_i^2] + \frac{1}{3}Mt} \right).
	\]
\end{lemma}

We first give a concentration lemma of the partially observed matrix, which is Theorem 1 in \cite{li2020network}.
\begin{lemma}[Concentration of $\hat{A}$]\label{lem1}
	Let $P$ be a probability matrix of rank $K$ and let $d$ be an upper bound of the expected node degree, that is, defined to be any value such that $\max_{ij}P_{ij}\leq d/n$. Let $A$ be an adjacency matrix with edges sampled independently and $\mathbb{E}(A) = P$. Let $\mathcal{E}$ be the training edge set with proportion $w\geq C_1(\log n)/n$ for some absolute constant $C_1$, and let $Y$ be the corresponding partially observed matrix. If $d \geq C_2 \log n$ for some absolute constant $C_2$, then with probability at least $1 - 3n^{-\delta}$ for some $\delta > 0$, the completed matrix $\hat{A}$ obtained by performing rank-$K$ truncated SVD on matrix $Y$ with $\hat A=S_H(Y,K)/w$ satisfies
	\[
	\| \hat{A} - P \| \leq \tilde{C} \max \left\{ \left( \frac{K d^2}{n w} \right)^{1/2}, \left( \frac{d}{w} \right)^{1/2}, \frac{\sqrt{\log n}}{w} \right\},
	\]
	where $\tilde{C} = \tilde{C}(\delta, C_1, C_2)$ is a constant that depends only on $C_1$, $C_2$ and $\delta$. This also implies
	\[
	\frac{\| \hat{A} - P \|_F^2}{n^2} \leq \frac{\tilde{C}^2}{2} \max \left( \frac{K^2 d^2}{n^3 w}, \frac{K d}{n^2 w}, \frac{K \log n}{n^2 w^2} \right).
	\]
\end{lemma}

In their work, they only give the proof of Lemma \ref{lem1} in the directed network scenario and use it directly in the undirected scenario, so we may follow them and use this lemma in the undirected scenario directly.

\subsection{Proof of Proposition 1}

\begin{proof}[Proof of Proposition 1]
	Assume the true data-generating model is $\delta^{(m^*)}$. For any $m$, we denote
	$$\ell_{m}(A,\mathcal{E}^c):=\sum_{(i,j)\in\mathcal{E}^c}\left(A_{ij}-\hat{P}^{(m)}_{ij}\right)^2,$$
	and for any $m'\ne m^*$, we denote
	$$\ell_{m',m^*}(A,\mathcal{E}^c)=\ell_{m'}(A,\mathcal{E}^c)-\ell_{m^*}(A,\mathcal{E}^c).$$ 
	Denote the oracle loss by 
	\begin{equation}\label{oracle_loss}
		\ell_0(A,\mathcal{E}^c):=\sum_{(i,j)\in\mathcal{E}^c}(A_{ij}-P_{ij})^2.
	\end{equation}
	Denote the event
	$$\Omega:=\left\{\frac{n(n-1)(1-w)}{2}\leq|\mathcal{E}^c|\leq2n(n-1)(1-w)\right\}.$$
	It is easy to show that cardinality $|\mathcal{E}^c|$ follows twice a Binomial distribution $\mathcal B(\frac{n(n-1)}{2},1-w)$. Let $\mu:=\mathbb E |\mathcal{E}^c|=n(n-1)(1-w).$
	Then by the multiplicative Chernoff bounds,
	$$\mathbb P\left(|\mathcal{E}^c|\leq\frac{\mu}{2}\right)\leq\exp\left(-\frac{\mu}{16}\right)=\exp\left\{-\frac{n(n-1)(1-w)}{16}\right\},$$
	and
	$$\mathbb P(|\mathcal{E}^c|\geq2\mu)\leq\exp\left(-\frac{\mu}{6}\right)=\exp\left\{-\frac{n(n-1)(1-w)}{6}\right\}.$$
	Consequently,
	$$
	\mathbb P(\Omega^c)
	\leq
	2\exp\left\{-\frac{n(n-1)(1-w)}{16}\right\},
	$$
	which tends to zero provided that $n^2(1-w)\to\infty$.
	Now, we separately analyze the behavior of $\ell_{m',m}(A,\mathcal{E}^c)$ for cases where $m'<m^*$ and $m'>m^*$ under the condition that event $\Omega$ holds.
	
	\textbf{Case 1.} $m'>m^*$.
	
	It is noteworthy that all estimators $\hat{P}^{(m')}$ are constructed solely based on the training edge set $\mathcal{E}$, and thus maintain statistical independence from the evaluation edges $A_{ij}:(i,j)\in\mathcal{E}^c$. This structural independence allows us to derive that
	\begin{align*}
		\left|\ell_{m'}(A,\mathcal{E}^c)-\ell_{0}(A,\mathcal{E}^c)\right|=&\left|\sum_{(i,j)\in\mathcal{E}^c}\left[(A_{ij}-\hat{P}^{(m')}_{ij})^2-\left(A_{ij}-P_{ij}\right)^2\right]\right|\\
		\le&\sum_{(i,j)\in\mathcal{E}^c}\left|\hat{P}_{ij}^{(m')}-P_{ij}\right|(2A_{ij}+P_{ij}+\hat{P}_{ij}^{(m')})\\
		\leq&2\sum_{(i,j)\in\mathcal{E}^c}\left|\hat{P}_{ij}^{(m')}-P_{ij}\right|\left(A_{ij}+P_{ij}+\left|\hat{P}_{ij}^{(m')}-P_{ij}\right|\right)\\
		\leq&2\left[\sum_{(i,j)\in\mathcal{E}^c}(\hat{P}_{ij}^{(m')}-P_{ij})^2\right]^{\frac{1}{2}}\left(\sum_{(i,j)\in\mathcal{E}^c}A_{ij}\right)^{\frac{1}{2}}+2\sum_{(i,j)\in\mathcal{E}^c}\left|\hat{P}_{ij}^{(m')}-P_{ij}\right|P_{ij}\\
		&+2\sum_{(i,j)\in\mathcal{E}^c}\left|\hat{P}_{ij}^{(m')}-P_{ij}\right|^2\\
		\leq& O_{\mathbb{P}}\left(\|\hat{P}^{(m')}-P\|_F\sqrt{\|P\|_{\infty}|\mathcal{E}^c|}\right)+2\|\hat{P}^{(m')}-P\|_F^2,
	\end{align*}
	where the last inequality follows from Bernstein's inequality and the Cauchy–Schwarz inequality.
	By Assumption 1, $\|\hat{P}^{(m')}-P\|_F^2=O_\mathbb{P}(n^2a_{n,w})$ for all $m'\geq m^*$ by the nested model assumption. Conditioned on event $\Omega$, we can obtain
	$$ \left|\ell_{m'}(A,\mathcal{E}^c)-\ell_{0}(A,\mathcal{E}^c)\right|= n^2O_{\mathbb{P}}\left(\sqrt{a_{n,w}(1-w)\|P\|_{\infty}}+a_{n,w}\right)=n^2O_{\mathbb{P}}\left(\max\left\{\sqrt{a_{n,w}(1-w)\|P\|_{\infty}},a_{n,w}\right\}\right)$$
	for all $m'\geq m^*$. Therefore, 
	\begin{align}\label{prop1_over}
		\ell_{m',m^*}(A,\mathcal{E}^c)&=\left(\ell_{m'}(A,\mathcal{E}^c)-\ell_{0}(A,\mathcal{E}^c)\right)-\left(\ell_{m^*}(A,\mathcal{E}^c)-\ell_{0}(A,\mathcal{E}^c)\right)\nonumber\\
		&\geq-n^2O_{\mathbb{P}}\left(\max\left\{\sqrt{a_{n,w}(1-w)\|P\|_{\infty}},a_{n,w}\right\}\right).
	\end{align}
	
	\textbf{Case 2.} $m'<m^*$.
	
	Notice that 
	$$\mathbb{E}[\ell_{m'}(A,\mathcal{E}^c)-\ell_{0}(A,\mathcal{E}^c)]=\sum_{(i,j)\in\mathcal{E}^c}(P_{ij}-\hat{P}^{(m')}_{ij})^2,
	$$
	where the expectation is taken over the evaluation entries. 
	Therefore, by Hoeffding's inequality, we have 
	$$\mathbb{P}\left\{\ell_{m'}(A,\mathcal{E}^c)-\ell_{0}(A,\mathcal{E}^c)\leq\frac{1}{2}\mathbb{E}[\ell_{m'}(A,\mathcal{E}^c)-\ell_{0}(A,\mathcal{E}^c)]\right\}\leq\exp\left(-\frac{\sum_{(i,j)\in\mathcal{E}^c}(P_{ij}-\hat{P}^{(m')}_{ij})^2}{8}\right).$$
	Under Assumptions 2 and 3, for $m'<m^*$, $\sum_{(i,j)\in\mathcal{E}^c}\big(\hat P_{ij}^{(m')}-P_{ij}\big)^2=\Omega_{\mathbb P}(|\mathcal{E}^c|b_{n,w})=\Omega_{\mathbb P}(n^2(1-w)b_{n,w})=\omega_{\mathbb P}(1)$. Therefore, with probability tending to 1,
	$$\ell_{m'}(A,\mathcal{E}^c)-\ell_{0}(A,\mathcal{E}^c)\geq\frac{1}{2}\mathbb{E}[\ell_{m'}(A,\mathcal{E}^c)-\ell_{0}(A,\mathcal{E}^c)]=\frac{1}{2}\sum_{(i,j)\in\mathcal{E}^c}(P_{ij}-\hat{P}^{(m')}_{ij})^2=\Omega_{\mathbb{P}}(n^2(1-w)b_{n,w}).$$ 
	Thus by Assumption 3 again, 
	\begin{align}\label{prop1_under}
		\ell_{m',m^*}(A,\mathcal{E}^c)&=\left(\ell_{m'}(A,\mathcal{E}^c)-\ell_{0}(A,\mathcal{E}^c)\right)-\left(\ell_{m^*}(A,\mathcal{E}^c)-\ell_{0}(A,\mathcal{E}^c)\right)\\
		&\geq\Omega_{\mathbb{P}}(n^2(1-w)b_{n,w})-O_{\mathbb{P}}\left(n^2\max\left\{\sqrt{a_{n,w}(1-w)\|P\|_{\infty}},a_{n,w}\right\}\right)\nonumber \\
		&=\Omega_{\mathbb{P}}(n^2(1-w)b_{n,w}).
	\end{align}
	For the consistency, we just need to prove
	\begin{equation}
		\mathbb{P}\left[L_{m^*}(A,\mathcal{E}^c)>L_{m'}(A,\mathcal{E}^c)\right]\to0,\quad\forall\ m'\neq m^*. \label{prop1_cons}
	\end{equation}
	When $m'>m^*$, under the condition for $\lambda_n$, increasing sequence $d_{m'}$, and \eqref{prop1_over},
	\begin{align*}
		\mathbb{P}\left[L_{m^*}(A,\mathcal{E}^c)>L_{m'}(A,\mathcal{E}^c)\right]&=\mathbb{P}[-\ell_{m',m^*}(A,\mathcal{E}^c)>(d_{m'}-d_{m^*})|\mathcal{E}^c|\lambda_n]\\
		&\leq\mathbb{P}[-\ell_{m',m^*}(A,\mathcal{E}^c)>1/2(d_{m'}-d_{m^*})n(n-1)(1-w)\lambda_n]\to 0.
	\end{align*}
	Similarly, \eqref{prop1_cons} holds by \eqref{prop1_under}, when $m'<m^*$.
\end{proof}

\subsection{Proof of Theorem 1}
Our proof in this section will mainly follow the framework in \cite{li2020network}. We denote $P$ here by $P_{ij}=B_{c_ic_j}$, instead of $P_{ij}=B_{c_ic_j}(1-\delta_{ij})$, as is commonly done in previous works such as \cite{lei2015consistency} and \cite{chen2018network} for simplicity. Notice that the norm of the difference of two $P$'s are of constant order, and thus won't affect the concentration result we used in this section.



Under Assumption 5, in the SBM framework, $d$ in Lemma \ref{lem1} can be chosen as $n\rho_n$. Furthermore, we do not assume that the true $K^*$ changes with $n$, thus the concentration result in Lemma \ref{lem1} becomes 
$$\|\hat{A} - P\|\leq\tilde{C}\sqrt{\frac{n\rho_n}{w}}.$$

Now we state the consistency result of spectral clustering on the estimated $\hat{A}$.

\begin{lemma}[Community recovery under Stochastic Block Model]\label{lem2}
	Under Assumptions 4 and 5, let $\hat{A}$ be the recovered adjacency matrix in Algorithm 2 with $k$ being the true $K^*$. Assume that $\rho_n\geq C \frac{\log n}{n}$.  Let label estimator $\hat{c}$ be the output of fitting the spectral clustering algorithm (with $(1+\epsilon)$-approximate $k$-means) on $\hat{A}$. Then $\hat{c}$ coincides with the true label $c$ on all but $O(\frac{1}{\rho_nw})$ nodes within each of the $K^*$ communities (up to a permutation of block labels), with probability tending to one. \label{labelesti}
\end{lemma}

\begin{proof}[Proof of Lemma \ref{lem2}]
	We exactly follow the proof of Theorem 3.1 in \cite{lei2015consistency}, while changing some key formulas as follows. Let $\hat{U},U\in\mathbb{R}^{n\times K^*}$ be the $K^*$ leading eigenvectors of $\hat{A}$ and $P$. For some $K^*$-dimensional orthogonal matrix $Q$, we have
	$$\|\hat{U}-UQ\|_F\leq\frac{2\sqrt{2K^*}}{n_{\min}\rho_n\lambda_{\min}}\|\hat{A}-P\|\leq\frac{2\sqrt{2K^*}}{n_{\min}\rho_n\lambda_{\min}}\tilde{C}\sqrt{\frac{n\rho_n}{w}},$$
	where $\lambda_{\min}$ is the smallest absolute eigenvalue of $B_0$, and $n_{\min}$ is the cardinality of the smallest community. Let $S_k$ be the node subset in the $k$-th community $\mathcal{G}_k$ that may be misclassified to other communities under the label estimator $\hat{c}$. Following \cite{lei2015consistency}, denote $n_k=|\mathcal{G}_k|$, we have
	$$
	\sum_{k=1}^{K^*}|S_k|\left(\frac{1}{n_k}+\frac{1}{\max\{n_l:l\neq k\}}\right)\leq8(2+\epsilon)\|\hat{U}-UQ\|_F^2\leq64(2+\epsilon)\tilde{C}^2\frac{K^*n}{n_{\min}^2\lambda_{\min}^2\rho_nw}.
	$$
	Then by Assumption 4, $n_{\min}\geq\pi_0 n$, so
	$$
	\sum_{k=1}^{K^*}\frac{|S_k|}{n_k}\leq64\tilde{C}^2(2+\epsilon)\frac{K^*}{\pi_0^2\lambda_{\min}^2n\rho_nw}=O\left(\frac{1}{n\rho_nw}\right).
	$$
	Thus within each community (and in the whole network), $\hat{c}$ coincides with the true label $c$ on all but $O(\frac{1}{\rho_nw})$ nodes.
\end{proof}

We also need the following lemma, which is Lemma 4 in \cite{li2020network}, to characterize the behavior of the estimated labels when we use a model with $K'(<K^*)$ communities.

\begin{lemma}\label{lem3}
	Assume the network is drawn from the stochastic block model with $K^*$ communities satisfying Assumptions 4 -- 5. Suppose we cluster the nodes into $K'(< K^*)$ communities. Define $\mathcal I_{k_1 k_2} = (\mathcal G_{k_1} \times \mathcal G_{k_2})\, \cap\, \mathcal{E}^c$ and $\hat{\mathcal I}_{k_1 k_2} = (\hat{\mathcal G}_{k_1} \times \hat{\mathcal G}_{k_2})\, \cap\, \mathcal{E}^c$. Then, with probability tending to 1, there must exist $l_1, l_2, l_3 \in [K^*]$ and $k_1, k_2 \in [K']$ such that
	\begin{enumerate}
		\setlength{\itemsep}{0.05cm}
		\item $|\hat{\mathcal I}_{k_1 k_2} \cap \mathcal I_{l_1 l_2}| \geq \tilde{c}n^2(1-w)$
		\item $|\hat{\mathcal I}_{k_1 k_2} \cap \mathcal I_{l_1 l_3}| \geq \tilde{c}n^2(1-w)$
		\item $B_{0,l_1 l_2} \neq B_{0,l_1 l_3}$ where $B_{0,ij}$ denotes the $(i,j)$-th element of $B_0$.
	\end{enumerate}
\end{lemma}

Denote 
$$\ell_{k}(A,\mathcal{E}^c):=\sum_{(i,j)\in\mathcal{E}^c}\left(A_{ij}-\hat{P}^{(k)}_{ij}\right)^2$$
and 
$$\ell_{K',K^*}(A,\mathcal{E}^c):=\ell_{K'}(A,\mathcal{E}^c)-\ell_{K^*}(A,\mathcal{E}^c).$$

\begin{proposition}\label{propa1}
	When $K'<K^*$, we have $$\ell_{K',K^*}(A,\mathcal{E}^c)\geq\Omega_{\mathbb{P}}\left(n^2\rho_n^2(1-w)\right)-O_{\mathbb{P}}\left(\frac{n\rho_n(1-w)}{w}+\frac{1-w}{w^2}\right).$$
	Therefore as long as $n\rho_nw\rightarrow\infty$, we have $\ell_{K',K^*}(A,\mathcal{E}^c)=\Omega_{\mathbb{P}}\left(n^2\rho_n^2(1-w)\right)$.
\end{proposition}

\begin{proof}[Proof of Proposition \ref{propa1}]
	We still follow the proof of Theorem 3 in \cite{li2020network}. We will present the crucial modifications of their proof and further details can be found in \cite{li2020network}. Denote the oracle loss $\ell_0(A,\mathcal{E}^c)$ exactly as \eqref{oracle_loss}. Then 
	$$\ell_{K',K^*}(A,\mathcal{E}^c)=\left(\ell_{K'}(A,\mathcal{E}^c)-\ell_{0}(A,\mathcal{E}^c)\right)-\left(\ell_{K^*}(A,\mathcal{E}^c)-\ell_{0}(A,\mathcal{E}^c)\right).$$ 
	
	\textbf{Step 1.} Upper bound for $\ell_{K^*}(A,\mathcal{E}^c)-\ell_{0}(A,\mathcal{E}^c)$.
	
	We define several sets of entries:
	$$\mathcal{Q}_{k_1,k_2,l_1,l_2}=\{(i,j):c_i=l_1,c_j=l_2,\hat{c}_i=k_1,\hat{c}_j=k_2\}$$
	$$\mathcal{U}_{k_1,k_2,l_1,l_2}=\{(i,j)\in\mathcal{E}:c_i=l_1,c_j=l_2,\hat{c}_i=k_1,\hat{c}_j=k_2\}$$
	$$\mathcal{T}_{k_1,k_2,l_1,l_2}=\{(i,j)\in\mathcal{E}^c:c_i=l_1,c_j=l_2,\hat{c}_i=k_1,\hat{c}_j=k_2\}.$$
	Let $\mathcal{T}_{\cdot,\cdot,l_1,l_2}:=\cup_{k_1,k_2}\mathcal{T}_{k_1,k_2,l_1,l_2}$ be the union taken over the first two entries, and similarly define $\mathcal{T}_{k_1,k_2,\cdot,\cdot}$, $\mathcal{U}_{\cdot,\cdot,l_1,l_2}$, $\mathcal{U}_{k_1,k_2,\cdot,\cdot}$, $\mathcal{Q}_{\cdot,\cdot,l_1,l_2}$ and $\mathcal{Q}_{k_1,k_2,\cdot,\cdot}$. By Lemma \ref{lem3}, we have $|\hat{\mathcal{G}}_k\Delta\mathcal{G}_k|=O_{\mathbb{P}}(\frac{1}{\rho_nw})$ and  $|\cup_k(\hat{\mathcal{G}}_k\Delta\mathcal{G}_k)|=O_{\mathbb{P}}(\frac{1}{\rho_nw})$, thus we have
	$$\left|\mathcal{Q}_{k_1,k_2,\cdot,\cdot}\Delta\mathcal{Q}_{\cdot,\cdot,k_1,k_2}\right|=O_{\mathbb{P}}\left(\frac{n}{\rho_nw}\right),\ \text{ and }\left|\cup_{k_1,k_2}(\mathcal{Q}_{k_1,k_2,\cdot,\cdot}\Delta\mathcal{Q}_{\cdot,\cdot,k_1,k_2})\right|=O_{\mathbb{P}}\left(\frac{n}{\rho_nw}\right).$$
	Consequently,
	$$|\mathcal{U}_{k_1,k_2,\cdot,\cdot}|\geq cn^2w$$
	and
	\begin{equation}
		\left|\cup_{k_1,k_2}(\mathcal{U}_{k_1,k_2,\cdot,\cdot}\Delta\mathcal{U}_{\cdot,\cdot,k_1,k_2})\right|=O_{\mathbb{P}}\left(\frac{n}{\rho_n}\right),\quad\left|\cup_{k_1,k_2}(\mathcal{T}_{k_1,k_2,\cdot,\cdot}\Delta\mathcal{T}_{\cdot,\cdot,k_1,k_2})\right|=O_{\mathbb{P}}\left(\frac{n(1-w)}{\rho_nw}\right).\label{propa1UT}
	\end{equation}
	Denote $\hat{B}^{(K)}$ simply as $\hat{B}$. By Bernstein's inequality, we have
	\begin{align*}\label{Bbern}
		|\hat{B}_{kl}-B_{kl}|=&|\frac{\sum_{\mathcal{U}_{k,l,\cdot,\cdot}}A_{ij}}{|\mathcal{U}_{k,l,\cdot,\cdot}|}-B_{kl}|\nonumber\\
		\leq&\frac{|\mathcal{U}_{\cdot,\cdot,k,l}|}{|\mathcal{U}_{k,l,\cdot,\cdot}|}\left|\frac{\sum_{\mathcal{U}_{\cdot,\cdot,k,l}}A_{ij}}{|\mathcal{U}_{\cdot,\cdot,k,l}|}-B_{kl}\right|+\left|1-\frac{|\mathcal{U}_{\cdot,\cdot,k,l}|}{|\mathcal{U}_{k,l,\cdot,\cdot}|}\right|B_{kl}\nonumber\\
		&+\frac{|\mathcal{U}_{\cdot,\cdot,k,l}\Delta \mathcal{U}_{k,l,\cdot,\cdot}|}{|\mathcal{U}_{k,l,\cdot,\cdot}|}\frac{\sum_{\mathcal{U}_{\cdot,\cdot,k,l}\Delta \mathcal{U}_{k,l,\cdot,\cdot}}A_{ij}}{|\mathcal{U}_{\cdot,\cdot,k,l}\Delta \mathcal{U}_{k,l,\cdot,\cdot}|}\nonumber\\
		\leq& O_{\mathbb{P}}\left(\sqrt{\frac{\rho_n}{n^2w}}\right)+O_{\mathbb{P}}\left(\frac{1}{n\rho_nw}\right)\rho_n+O_{\mathbb{P}}\left(\frac{1}{n\rho_nw}\right)\rho_n=O_{\mathbb{P}}\left(\frac{1}{nw}\right).
	\end{align*}
	Here the last bound follows by the same argument as in the proof of Theorem 3 of \citet{li2020network}; alternatively, it can be obtained by bounding the total number of edges incident to the misclassified nodes through the maximum training degree. A similar argument will be applied in the proofs of Theorem 2 \& 4.
	
	Now comparing $\ell_{K^*}$ and $\ell_0$, we have
	\begin{align*}
		\ell_{K^*}(A, \mathcal{E}^c) - \ell_0(A, \mathcal{E}^c) 
		=& \sum_{k_1, k_2, l_1, l_2} \sum_{(i,j) \in \mathcal{T}_{k_1, k_2, l_1, l_2}} 
		\left[ (A_{ij}-\hat{B}_{k_1 k_2})^2 -(A_{ij}-B_{l_1 l_2})^2\right] \\
		=& \sum_{k_1, k_2} \sum_{(i,j) \in \mathcal{T}_{k_1, k_2, k_1, k_2}} 
		\left[ (A_{ij}-\hat{B}_{k_1 k_2})^2 -(A_{ij}-B_{k_1 k_2})^2 \right] \\
		&+ \sum_{(k_1, k_2) \ne (l_1, l_2)} \sum_{(i,j) \in \mathcal{T}_{k_1, k_2, l_1, l_2}} 
		\left[(A_{ij}-\hat{B}_{k_1 k_2})^2 -(A_{ij}-B_{l_1 l_2})^2\right] \\
		:=& \mathcal{I} + \mathcal{II}.
	\end{align*}
	Thus we have
	\begin{align*}
		|\mathcal{I}|\leq& \sum_{k_1, k_2} \sum_{(i,j) \in \mathcal{T}_{k_1, k_2, k_1, k_2}} 
		2|\hat{B}_{k_1 k_2} - B_{k_1 k_2}| \left(A_{ij} + B_{k_1 k_2} + |\hat{B}_{k_1 k_2} - B_{k_1 k_2}| \right) \notag \\
		\leq& O_{\mathbb{P}}\left( \frac{1}{nw} n^2(1-w) \rho_n \right) 
		+ O_{\mathbb{P}}\left( \frac{1}{nw} n^2(1-w) \rho_n \right) 
		+ O_{\mathbb{P}}\left( \left(\frac{1}{nw} \right)^2 n^2(1-w) \right)\\
		=& O_{\mathbb{P}}\left(\frac{n\rho_n(1-w)}{w}+\frac{1-w}{w^2}\right);
	\end{align*}
	\begin{align*}
		|\mathcal{II}| 
		\leq& \sum_{(k_1, k_2) \ne (l_1, l_2)} \sum_{(i,j) \in \mathcal{T}_{k_1, k_2, l_1, l_2}} 
		(2A_{ij} + \hat{B}_{k_1 k_2} + B_{l_1 l_2})(\hat{B}_{k_1 k_2} + B_{l_1 l_2})\\
		=& \sum_{(k_1, k_2) \ne (l_1, l_2)} \sum_{(i,j) \in \mathcal{T}_{k_1, k_2, l_1, l_2}} 
		(2A_{ij} + B_{k_1 k_2} + B_{l_1 l_2} + (\hat{B}_{k_1 k_2} - B_{k_1 k_2}))\\
		&(B_{k_1 k_2} + B_{l_1 l_2} + (\hat{B}_{k_1 k_2} - B_{k_1 k_2}))\\
		=& O_{\mathbb{P}}\left( \frac{n(1-w)}{\rho_nw} \rho_n^2 \right) = O_{\mathbb{P}}\left(\frac{n\rho_n(1-w)}{w}\right).
	\end{align*}
	Combining the above two formulas, we have
	$$\ell_{K^*}(A, \mathcal{E}^c) - \ell_0(A, \mathcal{E}^c) =O_{\mathbb{P}}\left(\frac{n\rho_n(1-w)}{w}+\frac{1-w}{w^2}\right).$$
	
	\textbf{Case 2.} Lower bound for $\ell_{K'}(A,\mathcal{E}^c)-\ell_{0}(A,\mathcal{E}^c).$
	
	We still follow the lines of the proof of Theorem 3 in \cite{li2020network}. Without loss of generality, assume the $k_1 ,k_2$ and $l_1 ,l_2 ,l_3$ in Lemma \ref{lem3} are 1,2 and 1,2,3 respectively. Then
	
	\begin{align*}
		\ell_{K'}(A,\mathcal{E}^c)-\ell_{0}(A,\mathcal{E}^c) 
		\geq& \sum_{(i,j) \in \mathcal{T}_{1,2,1,2}} \left[(A_{ij}-\hat{p})^2- (A_{ij}-B_{12})^2 \right]+\sum_{(i,j) \in \mathcal{T}_{1,2,1,3}} \left[ (A_{ij}-\hat{p})^2-(A_{ij}-B_{13})^2 \right]\\
		&+ \sum_{{(k_1, k_2, l_1, l_2) \notin \{(1,2,1,2), (1,2,1,3)\}}} 
		\sum_{(i,j) \in \mathcal{T}_{k_1, k_2, l_1, l_2}} 
		\left[(A_{ij}-\hat{p}_{k_1,k_2,l_1,l_2})^2-(A_{ij}-B_{l_1 l_2})^2\right]\\
		:=& \mathcal{III} + \mathcal{IV} + \mathcal{V},
	\end{align*}
	where $\hat{p}$ is the average of $A_{ij}$ over $\mathcal{T}_{1,2,1,2} \cup \mathcal{T}_{1,2,1,3}$ and $\hat{p}_{k_1,k_2,l_1,l_2}$ is the average of $A_{ij}$ over $\mathcal{T}_{k_1,k_2,l_1,l_2}$. Here $\hat{p}=t\hat{p}_1+(1-t)\hat{p}_2$, where $\hat{p}_1=\hat{p}_{1,2,1,2}$ and $\hat{p}_2=\hat{p}_{1,2,1,3}$, with $t=\frac{|\mathcal{T}_{1,2,1,2}|}{|\mathcal{T}_{1,2,1,2}|+|\mathcal{T}_{1,2,1,3}|}.$ We have $|\mathcal{T}_{1,2,1,2}|\sim|\mathcal{T}_{1,2,1,3}|$, so $t$ and $1-t$ are of constant order. Denote correspondingly $p_1=B_{12}$, $p_2=B_{13}$, and denote $f(p):=\sum_{(i,j)\in\mathcal{T}_{1,2,1,2}}(A_{ij}-p)^2$. We have
	$$\mathcal{III}=f(t\hat{p}_1+(1-t)\hat{p}_2)-f(p_1)\geq|\mathcal{T}_{1,2,1,2}|(1-t)^2|\hat{p}_1-\hat{p}_2|^2+f(\hat{p}_1)-f(p_1).$$
	By Lemma \ref{lem3}, $|\mathcal{T}_{1,2,1,2}|\geq cn^2(1-w)$, so by Bernstein inequality, 
	$$|\hat{p}_1-\hat{p}_2|\geq|p_1-p_2|-|\hat{p}_1-p_1|-|\hat{p}_2-p_2|\geq|p_1-p_2|-O_{\mathbb{P}}\left(\sqrt{\frac{\rho_n}{n^2(1-w)}}\right)\geq c_{K'}\rho_n$$
	for some constant $c_{K'}$. Similarly as Step 1 but with the bound on $|\hat{p}_1-p_1|$ replaced by $O_{\mathbb{P}}\left(\sqrt{\frac{\rho_n}{n^2(1-w)}}\right)$, we get
	$$|f(\hat{p}_1)-f(p_1)|=O_{\mathbb{P}}\left(n\rho_n^{3/2}(1-w)^{1/2}\right).$$
	Thus combining the above formulas, we have
	$$\mathcal{III}\geq \Omega\left(n^2\rho_n^2(1-w)\right).$$
	Similarly $\mathcal{IV}\geq \Omega\left(n^2\rho_n^2(1-w)\right)$, and 
	$$\mathcal{V}=-\sum_{{(k_1, k_2, l_1, l_2) \notin \{(1,2,1,2), (1,2,1,3)\}}}|\mathcal{T}_{k_1, k_2, l_1, l_2}|(\hat{p}_{k_1, k_2, l_1, l_2}-B_{l_1l_2})^2\geq-O_{\mathbb{P}}(\rho_n).$$
	Combining all the results, we have $\ell_{K'}(A,\mathcal{E}^c)-\ell_{K^*}(A,\mathcal{E}^c)\geq\Omega_{\mathbb{P}}\left(n^2\rho_n^2(1-w)\right).$
\end{proof}

\begin{proposition}\label{propa2}
	When $K^*<K'\leq 2K^*$, we have $$\ell_{K'}(A, \mathcal{E}^c) - \ell_0(A, \mathcal{E}^c) \geq-O_{\mathbb{P}}\left(\rho_n\right).$$
\end{proposition}
\begin{proof}[Proof of Proposition \ref{propa2}]
	We use the same notation $\mathcal{T}_{k_1,k_2,l_1,l_2}$ as before. Denote $\hat{p}_{k_1,k_2,l_1,l_2}$ as the average of $A_{ij}$ over $\mathcal{T}_{k_1,k_2,l_1,l_2}$. Then since $\hat{P}^{(K')}_{ij}$ is constant on $\mathcal{T}_{k_1,k_2,l_1,l_2}$, and $\sum_{\mathcal{T}}(A_{ij}-p)^2$ is minimized at the average of $A_{ij}$ over $\mathcal{T}$, we have
	\begin{align*}
		\ell_{K'}(A,\mathcal{E}^c)-\ell_{0}(A,\mathcal{E}^c)=&\sum_{k_1,k_2,l_1,l_2}\sum_{(i,j)\in\mathcal{T}_{k_1,k_2,l_1,l_2}}\left[\left(A_{ij}-\hat{P}^{(K')}_{ij}\right)^2-\left(A_{ij}-B_{l_1l_2}\right)^2\right]\\
		\geq&\sum_{k_1,k_2,l_1,l_2}\sum_{(i,j)\in\mathcal{T}_{k_1,k_2,l_1,l_2}}\left[\left(A_{ij}-\hat{p}_{k_1k_2l_1l_2}\right)^2-\left(A_{ij}-B_{l_1l_2}\right)^2\right]\\
		=&-\sum_{k_1,k_2,l_1,l_2}|\mathcal{T}_{k_1,k_2,l_1,l_2}|\left(\hat{p}_{k_1k_2l_1l_2}-B_{l_1l_2}\right)^2\geq-O_{\mathbb{P}}(\rho_n).
	\end{align*}
	Thus the conclusion holds.
\end{proof}

\begin{proposition}\label{propaplus}
	When $K'>2K^*$, we have $$\mathbb{P}\left[\ell_{K'}(A,\mathcal{E}^c)-\ell_{0}(A,\mathcal{E}^c)\leq-2(K^*)^2K'^2\log n \right]\leq\frac{2(K^*)^2}{n^2}.$$
\end{proposition}
\begin{proof}[Proof of Proposition \ref{propaplus}]
	We use the same notation as in the proof of Proposition \ref{propa2}, and we also have
	\begin{align*}
		\ell_{K'}(A,\mathcal{E}^c)-\ell_{0}(A,\mathcal{E}^c)=&\sum_{k_1,k_2,l_1,l_2}\sum_{(i,j)\in\mathcal{T}_{k_1,k_2,l_1,l_2}}\left[\left(A_{ij}-\hat{P}^{(K')}_{ij}\right)^2-\left(A_{ij}-B_{l_1l_2}\right)^2\right]\\
		\geq&\sum_{k_1,k_2,l_1,l_2}\sum_{(i,j)\in\mathcal{T}_{k_1,k_2,l_1,l_2}}\left[\left(A_{ij}-\hat{p}_{k_1k_2l_1l_2}\right)^2-\left(A_{ij}-B_{l_1l_2}\right)^2\right]\\
		=&-\sum_{k_1,k_2,l_1,l_2}|\mathcal{T}_{k_1,k_2,l_1,l_2}|\left(\hat{p}_{k_1k_2l_1l_2}-B_{l_1l_2}\right)^2.
	\end{align*}
	
	Here the $K'$ may diverge to $n$, so we need a more careful control of the probability. Denote the estimated label under $K'$ communities as $\hat{c}$. Therefore, 
	\begin{align*}
		&\mathbb{P}\left[\ell_{K'}(A,\mathcal{E}^c)-\ell_{0}(A,\mathcal{E}^c)\leq-2(K^*)^2K'^2\log n\right]\\
		\leq&\mathbb{P}\left[\sum_{k_1,k_2,l_1,l_2}|\mathcal{T}_{k_1,k_2,l_1,l_2}|\left(\hat{p}_{k_1k_2l_1l_2}-B_{l_1l_2}\right)^2\geq 2(K^*)^2K'^2\log n\right].
	\end{align*}
	Now we shall estimate the bound of $\sum_{k_1,k_2,l_1,l_2}|\mathcal{T}_{k_1,k_2,l_1,l_2}|\left(\hat{p}_{k_1k_2l_1l_2}-B_{l_1l_2}\right)^2$. For each $\{k_1,k_2,l_1,l_2\}$, by Hoeffding's inequality, we have
	$$\mathbb{P}\left[|\hat{p}_{k_1k_2l_1l_2}-B_{l_1l_2}|\geq \sqrt{\frac{2\log n}{|\mathcal{T}_{k_1,k_2,l_1,l_2}|}}\right]
	\leq 2\exp\left(-\frac{4\log n|\mathcal{T}_{k_1,k_2,l_1,l_2}|}{|\mathcal{T}_{k_1,k_2,l_1,l_2}|}\right)\leq 2n^{-4},$$
	i.e.,
	$$\mathbb{P}\left[|\mathcal{T}_{k_1,k_2,l_1,l_2}|\left(\hat{p}_{k_1k_2l_1l_2}-B_{l_1l_2}\right)^2\geq 2\log n\right]\leq 2n^{-4}.$$
	Therefore, by pigeonhole principle, if $\sum_{k_1,k_2,l_1,l_2}|\mathcal{T}_{k_1,k_2,l_1,l_2}|\left(\hat{p}_{k_1k_2l_1l_2}-B_{l_1l_2}\right)^2\geq 2(K^*)^2K'^2\log n$, then there must exist some $\{k_1,k_2,l_1,l_2\}$, such that $|\mathcal{T}_{k_1,k_2,l_1,l_2}|\left(\hat{p}_{k_1k_2l_1l_2}-B_{l_1l_2}\right)^2\geq 2\log n$. Thus, 
	\begin{align*}
		&\mathbb{P}\left[\sum_{k_1,k_2,l_1,l_2}|\mathcal{T}_{k_1,k_2,l_1,l_2}|\left(\hat{p}_{k_1k_2l_1l_2}-B_{l_1l_2}\right)^2\geq2(K^*)^2K'^2\log n\right]\\
		\le & \sum_{k_1,k_2,l_1,l_2}\mathbb{P}\left[|\mathcal{T}_{k_1,k_2,l_1,l_2}|\left(\hat{p}_{k_1k_2l_1l_2}-B_{l_1l_2}\right)^2\geq2\log n\right]\\
		\leq & 2(K^*)^2K'^2n^{-4}\leq 2(K^*)^2n^{-2}.
	\end{align*}
\end{proof}

\begin{proof}[Proof of Theorem 1]
	Recall the event 
	$$\Omega=\left\{\frac{n(n-1)(1-w)}{2}\leq|\mathcal{E}^c|\leq2n(n-1)(1-w)\right\}$$
	defined in the proof of Proposition 1, where we obtain that $\mathbb{P}(\Omega)\to1$. In Proposition \ref{propa1} and \ref{propa2}, we proved that 
	$$\ell_{K^*}(A, \mathcal{E}^c) - \ell_0(A, \mathcal{E}^c) =O_{\mathbb{P}}\left(\frac{n\rho_n(1-w)}{w}\right),$$
	for all $K'<K^*$,
	$$\ell_{K'}(A, \mathcal{E}^c) - \ell_0(A, \mathcal{E}^c) =\Omega_{\mathbb{P}}\left(n^2\rho_n^2(1-w)\right),$$
	and for all $K^*<K'\leq 2K^*$,
	$$\ell_{K'}(A, \mathcal{E}^c) - \ell_0(A, \mathcal{E}^c) \geq-O_{\mathbb{P}}\left(\rho_n\right).$$
	Since the true $K^*$ is fixed, for any $\delta>0$, there exists an event $\Omega_0(\delta)$, such that with $n$ sufficiently large, this event has probability larger than $1-\delta$, and there exist constants $C_1(\delta)$, $C_2(\delta)$ and $C_3(\delta)$ such that on this event, 
	\begin{equation}
		\ell_{K^*}(A, \mathcal{E}^c) - \ell_0(A, \mathcal{E}^c) \leq C_1(\delta)\frac{n\rho_n(1-w)}{w},\label{bound_1}
	\end{equation}
	for all $K'<K^*$,
	\begin{equation}
		\ell_{K'}(A, \mathcal{E}^c) - \ell_0(A, \mathcal{E}^c) \geq C_2(\delta)n^2\rho_n^2(1-w),\label{bound_2}
	\end{equation}
	and for all $K^*<K'\leq 2K^*$,
	\begin{equation}
		\ell_{K'}(A, \mathcal{E}^c) - \ell_0(A, \mathcal{E}^c) \geq-C_3(\delta)\rho_n.\label{bound_3}
	\end{equation}
	We first prove the consistency conditioned on $\Omega\cap\Omega_0(\delta)$, for any $\delta>0$. Since the true $K^*$ is fixed, we just need to prove
	\begin{equation}
		\mathbb{P}\left[L_{K^*}(A,\mathcal{E}^c)>L_{K'}(A,\mathcal{E}^c)\right]\to0,\quad\forall\ K'\leq 2K^*\text{ and }K'\neq K^*, \label{theo1_cons_bounded}
	\end{equation}
	and
	\begin{equation}
		\sum_{K'>2K^*}\mathbb{P}\left[L_{K^*}(A,\mathcal{E}^c)>L_{K'}(A,\mathcal{E}^c)\right]\to0. \label{theo1_cons_unbounded}
	\end{equation}
	We first consider \eqref{theo1_cons_bounded}. When $K^*<K'\leq 2K^*$, notice that
	\begin{align*}
		&\mathbb{P}\left[L_{K^*}(A,\mathcal{E}^c)>L_{K'}(A,\mathcal{E}^c)\right]=\mathbb{P}[-\ell_{K',K^*}(A,\mathcal{E}^c)>(d_{K'}-d_{K^*})|\mathcal{E}^c|\lambda_n]\\
		\leq&\mathbb{P}\left[\left(\ell_{K^*}(A,\mathcal{E}^c)-\ell_{0}(A,\mathcal{E}^c)\right)-\left(\ell_{K'}(A,\mathcal{E}^c)-\ell_{0}(A,\mathcal{E}^c)\right)>1/4(K'(K'+1)-K^*(K^*+1))n(n-1)(1-w)\lambda_n\right]\to0
	\end{align*}
	by \eqref{bound_1}, \eqref{bound_3} and the assumption $\lambda_n=\omega_{\mathbb{P}}\left(\frac{\rho_n}{nw}\right)$.
	Similarly \eqref{theo1_cons_bounded} holds when $K'<K^*$, by \eqref{bound_1}, \eqref{bound_2} and $\lambda_n=o_{\mathbb{P}}(\rho_n^2)$.
	
	Now we prove \eqref{theo1_cons_unbounded}. When $K'>2K^*$, we have
	\begin{align*}
		&\mathbb{P}\left[L_{K^*}(A,\mathcal{E}^c)>L_{K'}(A,\mathcal{E}^c)\right]\\
		\leq&\mathbb{P}\left[\left(\ell_{K^*}(A,\mathcal{E}^c)-\ell_{0}(A,\mathcal{E}^c)\right)-\left(\ell_{K'}(A,\mathcal{E}^c)-\ell_{0}(A,\mathcal{E}^c)\right)>1/4(K'(K'+1)-K^*(K^*+1))n(n-1)(1-w)\lambda_n\right]\\
		\leq&\mathbb{P}\left[\left(\ell_{K^*}(A,\mathcal{E}^c)-\ell_{0}(A,\mathcal{E}^c)\right)-\left(\ell_{K'}(A,\mathcal{E}^c)-\ell_{0}(A,\mathcal{E}^c)\right)>1/8K'^2n(n-1)(1-w)\lambda_n\right]\\
		\overset{(*)}{\leq}&\mathbb{P}\left[C_1(\delta)\frac{n\rho_n(1-w)}{w}+2(K^*)^2K'^2\log n>1/8K'^2n(n-1)(1-w)\lambda_n\right]+\frac{2(K^*)^2}{n^2}\\
		\leq&\mathbb{P}\left[C_1(\delta)\frac{n\rho_n(1-w)}{w}>1/16(K^*)^2n(n-1)(1-w)\lambda_n\right]+\mathbb{P}\left[2(K^*)^2K'^2\log n>1/16K'^2n(n-1)(1-w)\lambda_n\right]+\frac{2(K^*)^2}{n^2},
	\end{align*}
	where (*) comes from \eqref{bound_1} and Proposition \ref{propaplus}. Notice that for any $C>0$, we have $\mathbb{P}\left[\lambda_n<\frac{C\rho_n}{nw}\right]\ll1/n$, then as $n\to\infty$, $$\mathbb{P}\left[C_1(\delta)\frac{n\rho_n(1-w)}{w}>1/16(K^*)^2n(n-1)(1-w)\lambda_n\right]\ll\frac{1}{n},$$
	thus
	$$\sum_{K'>2K}\mathbb{P}\left[C_1(\delta)\frac{n\rho_n(1-w)}{w}>1/16(K^*)^2n(n-1)(1-w)\lambda_n\right]=o(1).$$
	Similarly, since for any $C>0$, we have $\mathbb{P}\left[\lambda_n<\frac{C\log n}{n^2(1-w)}\right]\ll1/n$, then as $n\to\infty$,
	$$\sum_{K'>2K}\mathbb{P}\left[2(K^*)^2K'^2\log n>1/16K'^2n(n-1)(1-w)\lambda_n\right]=o(1).$$
	Therefore, as $n\to\infty$,
	$$\sum_{K'>2K^*}\mathbb{P}\left[L_{K^*}(A,\mathcal{E}^c)>L_{K'}(A,\mathcal{E}^c)\right]\leq o(1)+o(1)+\sum_{K'>2K^*}\frac{2(K^*)^2}{n^2}\leq o(1)+\frac{2(K^*)^2}{n}\to0,$$
	and \eqref{theo1_cons_unbounded} is proved. Since we are conditioning on $\Omega\cap\Omega_0(\delta)$, we have for all $\delta>0$, when n is sufficiently large,
	$$\mathbb{P}[\hat{K}=K^*]>1-\delta.$$
	Let $\delta\to0$, and thus Theorem 1 holds.
\end{proof}

\subsection{Proof of Theorem 2}
Similarly as in the proof of Theorem 1 and for the same reason, we can denote $P$ here by $P_{ij}=\theta_i\theta_jB_{c_ic_j}$, instead of $P_{ij}=\theta_i\theta_jB_{c_ic_j}(1-\delta_{ij})$.

We first give the consistency result of spherical spectral clustering on $\hat{A}$ when the number of clusters is correctly specified, which is similar to Lemma \ref{lem2}.

\begin{lemma}\label{lem_ssc}
	Let $\hat{A}$ be the recovered adjacency matrix as above. Assume that $\rho_n\geq C \frac{\log n}{n}$.  Let $\hat{\mathbf{c}}$ be the output of spherical spectral clustering with $k$-median on $\hat{A}$. Then $\hat{\mathbf{c}}$ coincides with the true $\mathbf{c}$ on all but $O(\frac{\sqrt{n}}{\sqrt{\rho_nw}})$ nodes within each of the K communities (up to a permutation of block labels), with probability tending to one.
\end{lemma}
\begin{proof}[Proof of Lemma \ref{lem_ssc}]
	Recall that $n_k = |\{i : c_i = k\}|$. Following \cite{lei2015consistency}, define $\bm{\theta}_k = \{\theta_i\}_{c_i = k}$ and
	$$\nu_k = \frac{1}{n_k^2} \sum_{i: c_i = k} \frac{\|\bm{\theta}_k\|^2}{\theta_i^2}.
	$$
	
	Under Assumption 6 and the idenfication condition, we have
	
	\[
	\nu_k \leq \frac{1}{n_k^2} \sum_{i:c_i=k} \frac{n_k}{\theta_{\min}^2} = \frac{1}{\theta_{\min}^2}.
	\]
	
	Furthermore, when Assumption 4 and 6 hold, we have
	
	\[
	\frac{\sum_k n_k^2 \nu_k^2}{\min_k n_k^2}
	\leq \frac{\sum_k n_k^2}{n^2\pi_0^2 \theta_{\min}^4}
	\leq \frac{K^*}{\pi_0^2 \theta_{\min}^4} = O(1).
	\]
	Therefore follow the proof of Corollary 4.3 in \cite{lei2015consistency} and substituting the bound of $\|\hat{U}-UQ\|_F$ by 
	\begin{equation}\label{eqn:spectral}
		\|\hat{U}-UQ\|_F\leq\frac{2\sqrt{2K^*}}{n_{\min}\rho_n\lambda}\tilde{C}\sqrt{\frac{n\rho_n}{w}}
	\end{equation}
	for orthogonal matrix $Q$ that minimizes $\|\hat{U}-UQ\|_F$, the result of the lemma can be proved.
\end{proof}

The next lemma is the Lemma 4.1 in \cite{lei2015consistency}, which states the spectral structure of the DCBM.

\begin{lemma}\label{lem_dcbmspec}
	Let $U\Lambda U^{\top}$ be the eigen-decomposition of the true probability matrix $P$ of the DCBM parametrized by $(c,B,\theta)$. Then there exists a $K^*\times K^*$ orthonormal matrix $V$ such that 
	$$U_{i*}=\theta_i'V_{k*},\quad\forall1\leq k\leq {K^*},i\in\mathcal{G}_k,$$
	here $\theta_i'$ is the scaled version of $\theta_i$ introduced in (3.1) in Section 3.
\end{lemma}

Now we shall give a control of the maximum row distance (or, the $(2,\infty)$ norm) between the estimated eigenvectors and the true eigenvectors.

For two column orthogonal matrices $U_1,U_2\in\mathbb{O}^{n\times k}$, we define the canonical angles as the main diagonal elements of the $k\times k$ diagonal matrix
$$\Theta(U_1,U_2):=\mathrm{diag}(\cos^{-1}(\sigma_1(U_1^{\top}U_2)),\ldots,\cos^{-1}(\sigma_k(U_1^{\top}U_2))).$$ 
This matrix can be seen as a measurement of the distance of the two subspaces generated by the columns of $U_1$ and $U_2$. We have the following Davis-Kahan $\sin \Theta$ Theorem. 

\begin{lemma}[Davis-Kahan $\sin \Theta$ Theorem]\label{lemma_davis}
	Let $U$ and $\hat{U}$ be the top $K^*$ eigenvector matrices of $P$ and $\hat{A}$ respectively. Let $\Lambda=\mathrm{diag}(\lambda_1,\ldots,\lambda_{K^*})>0$ be the nonzero eigenvalues of $P$. Then we have 
	$$\|\sin\Theta(\hat{U},U)\|_2\leq\frac{2\|\hat{A}-P\|_2}{\lambda_{K^*}}.$$
\end{lemma}

We also have the following well-known relationship between $\|\hat{U}-UQ\|_2$ and $\|\sin\Theta(\hat{U},U)\|_2$ where $Q$ minimizes $\|\hat{U}-UQ\|_2$ as in \eqref{eqn:spectral}, see e.g. \cite{Cai2018pertubation} and \cite{Cape2019twoinfty}.

\begin{lemma}\label{lemma_sinmatrix}
	We have
	$$\|\sin\Theta(\hat{U},U)\|_2\leq\|\hat{U}-UQ\|_2\leq\sqrt{2}\|\sin\Theta(\hat{U},U)\|_2.$$
	Moreover, we also have
	$$\frac{1}{2}\|\sin\Theta(\hat{U},U)\|^2_2\leq\|U^{\top}\hat{U}-Q\|_2\leq\|\sin\Theta(\hat{U},U)\|^2_2.$$
\end{lemma}

Notice that the top $K^*$ eigenvectors of $\hat{A}$ are exactly same as the top $K^*$ eigenvectors of $Y/w$ by the construction. Write $\hat{A}=\hat{U}\hat{\Lambda}\hat{U}^{\top}$ as the top $K^*$ rank eigenvector decomposition of $Y/w$. Then we have the following decomposition of $\hat{U}-UQ$.

\begin{lemma}[Theorem 3.1 of \cite{Cape2019twoinfty}]\label{lemma_decomposition}
	Denote $H=Y/w-P$. If $Y/w$ has rank no less than $K^*$, then $\hat{U} - UQ \in \mathbb{R}^{n \times K^*}$ admits the decomposition
	\begin{align}
		\hat{U} - U Q 
		&= (I -U U^\top) H UQ\hat{\Lambda}^{-1}\nonumber \\
		&\quad + (I - UU^\top) H(\hat{U} - UQ) \hat{\Lambda}^{-1}\nonumber \\
		&\quad + (I -UU^\top) P(\hat{U} - UU^\top \hat{U}) \hat{\Lambda}^{-1}\nonumber  \\
		&\quad + U (U^\top \hat{U} - Q).\label{decomposition}
	\end{align}
	
\end{lemma}
Notice that by Lemma \ref{lem_dcbmspec}, we can write $U=\Theta'V$ for some $K^*\times K^*$ orthonormal matrix $V$ and 
$\Theta'\in\mathbb{O}^{n\times K^*}$ is a column orthogonal matrix with
$\Theta'_{ic_i}=\theta'_i$ and other entries being 0.

We also state some technical properties of the $(2,\infty)$ matrix norm.

\begin{lemma}[Proposition 6.5 and 6.6 in \cite{Cape2019twoinfty}]\label{lemma_2inftynorm}
	For $A \in \mathbb{R}^{p_1 \times p_2}$, $B \in \mathbb{R}^{p_2 \times p_3}$ and $C \in \mathbb{R}^{p_4 \times p_1}$, then
	\begin{align*}
		\| AB \|_{2,\infty} &\leq \| A \|_{2 ,\infty} \| B \|_2;  \\
		\| CA \|_{2 ,\infty} &\leq \| C \|_{\infty} \| A \|_{2,\infty}. 
	\end{align*}
	For $A \in \mathbb{R}^{r \times s}$, $U \in \mathbb{O}_{p_1, r}$ and $V \in \mathbb{O}_{p_2, s}$,
	\begin{align*}
		\| A \|_2 &= \| UA \|_2 = \| AV^\top \|_2 = \| UAV^\top \|_2;  \\
		\| A \|_{2 ,\infty} &= \| AV^\top \|_{2,\infty}.
	\end{align*}
	However, $\| UA \|_{2 ,\infty}$ need not be equal to $\| A \|_{2,\infty}$.
\end{lemma}

Now we have the following result.
\begin{proposition}\label{prop_rowwise}
	Assume that $n\rho_nw\gg\log n$. Then we have $$\|\hat{U}-UQ\|_{2,\infty}=O_{\mathbb{P}}\left(\max\left\{\frac{1}{n\rho_n w},\sqrt{\frac{\log n}{n^2\rho_n w}}\right\}\right).$$
\end{proposition}

In the proof of this proposition, all $C$ represents some constant $C$.
\begin{proof}[Proof of Proposition \ref{prop_rowwise}]
	Following the proof of Theorem 1 of \cite{li2020network} (which is just Lemma \ref{lem1}), we can get
	$$\|H\|=\|\frac{Y}{w}-P\|\leq\tilde{C} \max \left\{ \left( \frac{{K^*} d^2}{n w} \right)^{1/2}, \left( \frac{d}{w} \right)^{1/2}, \frac{\sqrt{\log n}}{w}  \right\}\leq\tilde{C}\sqrt{\frac{n\rho_n}{w}}$$
	holds with probability larger than $1-3n^{-\delta}$. Together with $\lambda_K^*\geq Cn\rho_n$, we know that with probability tending to 1, we have $\lambda_{K^*}\geq 2\|H\|_{2}$.
	By Weyl's theorem, we have 
	$$|\lambda_{K^*}-\hat{\lambda}_{K^*}|\leq\|H\|_2\Rightarrow\hat{\lambda}_{K^*}\geq\frac{1}{2}\lambda_{K^*},$$
	thus $Y/w$ has rank no less than $K^*$ and Lemma \ref{lemma_decomposition} is available. Since $P$ has rank $K^*$ and $U=\Theta'V$ is the top $K^*$ eigenvectors of $P$, we have the third term in \eqref{decomposition} is 0. Denote the orthogonal complement of $U$ as $U_{\perp}$. Therefore, $I-UU^\top=U_{\perp}U_{\perp}^{\top}$. Apply Lemma \ref{lemma_2inftynorm} to \eqref{decomposition}, we have
	\begin{align}
		\|U_{\perp}U_{\perp}^{\top} H(\hat{U} - \Theta'V Q) \hat{\Lambda}^{-1}\|_{2,\infty}&\leq\|U_{\perp}U_{\perp}^{\top}\|_{2,\infty}\|H\|_2\|\hat{U} - \Theta'V Q\|_{2}|\hat{\lambda}_{K^*}|^{-1}\nonumber\\
		\|U_{\perp}U_{\perp}^{\top} H\Theta' V Q \hat{\Lambda}^{-1}\|_{2,\infty}&\leq\|U_{\perp}U_{\perp}^{\top}\|_{\infty}\|H\Theta'\|_{2,\infty}\| V Q\|_{2}|\hat{\lambda}_{K^*}|^{-1}\nonumber\\
		\|\Theta' V ((\Theta' V)^\top \hat{U} - Q)\|_{2,\infty}&\leq\|\Theta' V\|_{2,\infty}\|(\Theta' V)^\top \hat{U} - Q\|_2.\label{2inftyentrywise1}
	\end{align}
	
	We have $\|U_{\perp}U_{\perp}^{\top}\|_{2,\infty}=\|U_{\perp}\|_{2,\infty}\leq1$. By Lemma \ref{lemma_davis}, Lemma \ref{lemma_sinmatrix} , we have
	$$\|\hat{U} - \Theta' V Q\|_{2}\leq2\sqrt{2}\|H\|_2/\lambda_{K^*}$$
	and
	$$\|(\Theta'V)^\top \hat{U} - Q\|_2\leq4\|H\|_2^2/\lambda_{K^*}^2.$$
	Moreover, by the relationship between norms and by Assumption 4 and 6, we have 
	$$\|U_{\perp}U_{\perp}^{\top}\|_{\infty}=\|I-\Theta' V (\Theta'V)^\top\|_{\infty}\leq1+\sqrt{n}\|\Theta'V (\Theta'V)^\top\|_{2,\infty}\leq1+\sqrt{n}\|\Theta'\|_{2,\infty}=1+C_1$$
	for some constant $C_1$. We also have
	$$\|\Theta'V\|_{2,\infty}\leq\|\Theta'\|_{2,\infty}\|V\|_2\leq C_1n^{-1/2}$$
	for the same constant $C_1$.
	
	By a rearrangement of the nodes, without loss of generality, we can assume that the set of nodes in the $j$-th cluster is $I_{j}:=\{\sum_{k=1}^{j-1}n_{k}+1,\ldots,\sum_{k=1}^{j}n_{k}\}$. The $i$-th row of $H\Theta'$ has form 
	$$(H\Theta')_{i*}=\left(\sum_{j\in I_{1}}\theta_{j}'H_{ij},\ldots,\sum_{j\in I_{K^*}}\theta_{j}'H_{ij}\right).$$
	By Bernstein's inequality, for all $i\in[n]$ and $l\in[K^*]$,
	$$\mathbb{P}\left(\left|\sum_{j\in I_{l}}\theta_{j}'H_{ij}\right|\geq t\right)\leq2\exp\left(-\frac{t^2}{2\sum_{j
			\in I_{l}}\theta_j'^2P_{ij}/w+\frac{2t}{3w}\max_{j\in I_l}\theta_j'}\right)\leq2\exp\left(-\frac{t^2w}{2\rho +\frac{2C_1t}{3}n^{-1/2}}\right).$$
	Take $t=2\sqrt{\rho \log n/w}$, then as long as $n\rho_nw\gg\log n$, we have
	$$\mathbb{P}\left(\left|\sum_{j\in I_{l}}\theta_j'H_{ij}\right|\geq 2\sqrt{\frac{\rho\log n}{w}}\right)\leq2n^{-2}.$$
	Use a union bound on all $i$ and $l$, we have
	$$\mathbb{P}\left(\left|\sum_{j\in I_{l}}\theta_j'H_{ij}\right|\leq 2\sqrt{\frac{\rho\log n}{w}},\ \text{for all }i,l\right)\geq1-2K^*n^{-1},$$
	which tends to 1 as $n\to\infty$. As a result, with probability tending to 1, we have
	$$\|H\Theta'\|_{2,\infty}\leq2\sqrt{\rho K^*\log n/w}=:C_2\sqrt{\rho\log n/w}.$$
	Now the inequalities in \eqref{2inftyentrywise1} becomes
	\begin{align}
		\|U_{\perp}U_{\perp}^{\top} H(\hat{U} - \Theta'V Q) \hat{\Lambda}^{-1}\|_{2,\infty}&\leq C\|H\|^2_2|\lambda_{k^*}|^{-2}\leq C/(n\rho_n w)\nonumber\\
		\|U_{\perp}U_{\perp}^{\top} H\Theta'V Q \hat{\Lambda}^{-1}\|_{2,\infty}&\leq C\sqrt{\rho\log n/w}|\lambda_{K^*}|^{-1}\leq C\sqrt{\log n/(n^2\rho_n w)}\nonumber\\
		\|\Theta' V ((\Theta' V)^\top \hat{U} - Q)\|_{2,\infty}&\leq Cn^{-1/2}\|H\|^2_2|\lambda_{K^*}|^{-2}\leq C/(n\rho_n w).\label{2inftyentrywise2}
	\end{align}
	Then the desired result follows.
\end{proof}

From this control of the $(2,\infty)$ norm, we can get the following bounded result for the estimated degree heterogeneity parameter.
\begin{lemma}\label{lem_thetacontrol}
	Assume that the sparsity parameter satisfies $\sqrt{n}\rho_nw\to\infty$. Then there exists constant $\tilde{\theta}_{\min}$ and $\tilde{\theta}_{\max}$, such that the estimator $\hat{\theta}'$ of $\theta'$ obtained with true number of communities $K^*$ satisfies with probability tending to 1,
	$$\tilde{\theta}_{\min}\leq\sqrt{n}\hat{\theta}'_i\leq\tilde{\theta}_{\max},\quad\text{for all }i\in[n].$$
\end{lemma}

\begin{proof}[Proof of Lemma \ref{lem_thetacontrol}]
	By definition of $\theta_i'$ and $\hat{\theta}_i'$, we have
	\begin{align*}
		|\hat{\theta}_i'-\theta_i'|=&\left|\|\hat{U}_{i*}\|_2-\|U_{i*}\|_2\right|=\left|\|\hat{U}_{i*}\|_2-\|U_{i*}Q\|_2\right|=\left|\|\hat{U}_{i*}\|_2-\|(UQ)_{i*}\|_2\right|\\
		\leq&\|\hat{U}_{i*}-(UQ)_{i*}\|_2\leq\|\hat{U}-UQ\|_{2,\infty}.
	\end{align*}
	Thus by Proposition \ref{prop_rowwise}, we have
	$$|\sqrt{n}\theta_i'-\sqrt{n}\hat{\theta}_i'|=O_{\mathbb{P}}\left(\max\left\{\frac{1}{\sqrt{n}\rho_nw},\sqrt{\frac{\log n}{n\rho_nw}}\right\}\right)\to0$$
	as $\sqrt{n}\rho_nw\to\infty$.
\end{proof}

\begin{proposition}\label{prop_DCBMupper}
	If the true model is DCBM with $K^*$ communities, then under Algorithm 3, $$|\ell_{K^*}(A,\mathcal{E}^c)-\ell_0(A,\mathcal{E}^c)|=O_{\mathbb{P}}\left(\max\left\{\frac{n^{3/2}\rho_n}{w},\sqrt{\frac{n^3\log n\rho_n^3}{w}}\right\}(1-w)\right),$$ as long as the assumptions of Theorem 2 hold.
\end{proposition}

\begin{proof}[Proof of Proposition \ref{prop_DCBMupper}]
	By Proposition \ref{prop_rowwise}, with appropriate choice of $Q$, we have $$||\hat{U}-UQ||_{2,\infty}\leq C\left\{\frac{1}{n\rho_nw},\sqrt{\frac{\log n}{n^2\rho_nw}}\right\}$$
	with probability tending to 1. Therefore similar as the derivation in Lemma \ref{lem_thetacontrol}, we have for every $i\in[n]$,
	\begin{equation}\label{eqn_theta}
		|\sqrt{n}\hat{\theta}_i'-\sqrt{n}\theta_i'|\leq C\left\{\frac{1}{\sqrt{n}\rho_nw},\sqrt{\frac{\log n}{n\rho_nw}}\right\}.
	\end{equation}
	
	For $1\leq k<l\leq K^*$, we have
	$$n^{-1}\hat{B}_{kl}'=\frac{\sum_{\mathcal{U}_{k,l,\cdot,\cdot}}A_{ij}}{\sum_{\mathcal{U}_{k,l,\cdot,\cdot}}(\sqrt{n}\hat{\theta}_i')(\sqrt{n}\hat{\theta}_j')}.$$
	We consider a pseudo-oracle estimator of $n^{-1}B'$:
	$$n^{-1}\hat{B}_{kl}'^*=\frac{\sum_{\mathcal{U}_{k,l,\cdot,\cdot}}A_{ij}}{\sum_{\mathcal{U}_{k,l,\cdot,\cdot}}(\sqrt{n}\theta_i')(\sqrt{n}\theta_j')}.$$
	Then we have
	
	\begin{align*}
		|n^{-1}\hat{B}_{kl}'^*-n^{-1}B_{kl}'|=&|\frac{\sum_{\mathcal{U}_{k,l,\cdot,\cdot}}A_{ij}}{\sum_{\mathcal{U}_{k,l,\cdot,\cdot}}(\sqrt{n}\theta_i')(\sqrt{n}\theta_j')}-n^{-1}B_{kl}'|\\
		\leq&\frac{\sum_{\mathcal{U}_{\cdot,\cdot,k,l}}(\sqrt{n}\theta_i')(\sqrt{n}\theta_j')}{\sum_{\mathcal{U}_{k,l,\cdot,\cdot}}(\sqrt{n}\theta_i')(\sqrt{n}\theta_j')}\left|\frac{\sum_{\mathcal{U}_{\cdot,\cdot,k,l}}A_{ij}}{\sum_{\mathcal{U}_{\cdot,\cdot,k,l}}(\sqrt{n}\theta_i')(\sqrt{n}\theta_j')}-n^{-1}B_{kl}'\right|\\
		&+\left|(1-\frac{\sum_{U_{\cdot,\cdot,k,l}}(\sqrt{n}\theta_i')(\sqrt{n}\theta_j')}{\sum_{\mathcal{U}_{k,l,\cdot,\cdot}}(\sqrt{n}\theta_i')(\sqrt{n}\theta_j')})n^{-1}B_{kl}'\right|\\
		&+\frac{\sum_{\mathcal{U}_{\cdot,\cdot,k,l}\Delta U_{k,l,\cdot,\cdot}}(\sqrt{n}\theta_i')(\sqrt{n}\theta_j')}{\sum_{\mathcal{U}_{k,l,\cdot,\cdot}}(\sqrt{n}\theta_i')(\sqrt{n}\theta_j')}\frac{\sum_{\mathcal{U}_{\cdot,\cdot,k,l}\Delta U_{k,l,\cdot,\cdot}}A_{ij}}{\sum_{\mathcal{U}_{\cdot,\cdot,k,l}\Delta \mathcal{U}_{k,l,\cdot,\cdot}}(\sqrt{n}\theta_i')(\sqrt{n}\theta_j')}\\
		=&:\mathcal{I}+\mathcal{II}+\mathcal{III}.
	\end{align*}
	Notice by Assumption 6 and balanced community structure, there exists constant $\theta_{\max}'$, $\theta_{\min}'$ such that $\theta_{\min}'\leq\sqrt{n}\theta_i'\leq\theta_{\max}'$ holds for all $i$. By Bernstein's inequality, 
	\begin{align*}
		&\mathbb{P}\left\{\left|\sum_{\mathcal{U}_{\cdot,\cdot,k,l}}A_{ij}-n^{-1}\sum_{\mathcal{U}_{\cdot,\cdot,k,l}}(\sqrt{n}\theta_i')(\sqrt{n}\theta_j')B'_{kl}\right|\geq t\right\}\\&\leq \exp\left(-\frac{\frac{1}{2}t^2}{\sum_{\mathcal{U}_{\cdot,\cdot,k,l}}(\sqrt{n}\theta_i')(\sqrt{n}\theta_j')(n^{-1}B'_{kl})(1-n^{-1}B'_{kl})+\frac{1}{3}t}\right).
	\end{align*}
	Therefore, 
	$$\left|\sum_{\mathcal{U}_{\cdot,\cdot,k,l}}A_{ij}-n^{-1}\sum_{\mathcal{U}_{\cdot,\cdot,k,l}}(\sqrt{n}\theta_i')(\sqrt{n}\theta_j')B'_{kl}\right|=O_{\mathbb{P}}\left(\sqrt{\rho_n\sum_{\mathcal{U}_{\cdot,\cdot,k,l}}(\sqrt{n}\theta_i')(\sqrt{n}\theta_j')}\right),$$
	since $n^{-1}B'$ is of order $\rho_n$. Notice $\sqrt{n}\theta_i'$ is of constant order, so 
	\begin{equation}
		\sum_{T}(\sqrt{n}\theta_i')(\sqrt{n}\theta_j')=\Theta(|T|),\quad\text{for every index set T}. \label{T}
	\end{equation}
	Therefore, since both $|\mathcal{U}_{\cdot,\cdot,k,l}|$ and $|\mathcal{U}_{k,l,\cdot,\cdot}|$ are of order $n^2w$, we have
	$$\mathcal{I}\leq O_{\mathbb{P}}\left(\frac{\sqrt{\rho_n\sum_{\mathcal{U}_{\cdot,\cdot,k,l}}(\sqrt{n}\theta_i')(\sqrt{n}\theta_j')}}{\sum_{\mathcal{U}_{k,l,\cdot,\cdot}}(\sqrt{n}\theta_i')(\sqrt{n}\theta_j')}\right)=O_{\mathbb{P}}(\sqrt{\frac{\rho_n}{n^2w}}).$$
	
	For $\mathcal{II}$, $n^{-1}B'_{kl}=O_{\mathbb{P}}(\rho_n)$, and by \eqref{T} and Lemma \ref{lem_ssc},
	\begin{align*}
		&\left|1-\frac{\sum_{\mathcal{U}_{\cdot,\cdot,k,l}}(\sqrt{n}\theta_i')(\sqrt{n}\theta_j')}{\sum_{\mathcal{U}_{k,l,\cdot,\cdot}}(\sqrt{n}\theta_i')(\sqrt{n}\theta_j')}\right|\leq\frac{\sum_{\mathcal{U}_{\cdot,\cdot,k,l}\Delta \mathcal{U}_{k,l,\cdot,\cdot}}(\sqrt{n}\theta_i')(\sqrt{n}\theta_j')}{\sum_{\mathcal{U}_{k,l,\cdot,\cdot}}(\sqrt{n}\theta_i')(\sqrt{n}\theta_j')}=\Theta_{\mathbb{P}}(\frac{|\mathcal{U}_{\cdot,\cdot,k,l}\Delta \mathcal{U}_{k,l,\cdot,\cdot}|}{|\mathcal{U}_{k,l,\cdot,\cdot}|})\\
		&=O_{\mathbb{P}}(\frac{1}{\sqrt{n\rho_nw}}).
	\end{align*}
	Therefore
	$$\mathcal{II}\leq O_{\mathbb{P}}(\rho_n)O_{\mathbb{P}}(\frac{1}{\sqrt{n\rho_nw}})=O_{\mathbb{P}}(\sqrt{\frac{\rho_n}{nw}}).$$
	
	For $\mathcal{III}$, by Hoeffding inequality, $\sum_{\mathcal{U}_{\cdot,\cdot,k,l}\Delta \mathcal{U}_{k,l,\cdot,\cdot}}A_{ij}/\sum_{\mathcal{U}_{\cdot,\cdot,k,l}\Delta \mathcal{U}_{k,l,\cdot,\cdot}}(\sqrt{n}\theta_i')(\sqrt{n}\theta_j')$ is of the same order as $n^{-1}B_{kl}'$, thus $O_{\mathbb{P}}(\rho_n).$ Therefore, we also have
	$$\mathcal{III}\leq O_{\mathbb{P}}\left(\sqrt{\frac{\rho_n}{nw}}\right).$$
	In conclusion, we have 
	\begin{equation}
		|n^{-1}\hat{B}_{kl}'^*-n^{-1}B_{kl}'|=O_{\mathbb{P}}(\sqrt{\frac{\rho_n}{nw}}). \label{estB1}
	\end{equation}
	
	Now we estimate the distance between our estimator $n^{-1}\hat{B}_{kl}'$ and the pseudo-oracle estimator. By \eqref{eqn_theta}, we have
	\begin{align*}
		&|(\sqrt{n}\hat{\theta}_i')(\sqrt{n}\hat{\theta}_j')-(\sqrt{n}\theta_i')(\sqrt{n}\theta_j')|\leq\theta_{\max}'|(\sqrt{n}\hat{\theta}_i')-(\sqrt{n}\theta_i')|+\tilde{\theta}_{\max}|(\sqrt{n}\hat{\theta}_j')-(\sqrt{n}\theta_j')|\\
		&=O_{\mathbb{P}}\left(\max\left\{\frac{1}{\sqrt{n}\rho_nw},\sqrt{\frac{\log n}{n\rho_nw}}\right\}\right).
	\end{align*}
	Therefore, the difference of the denominators of two estimators satisfies
	\begin{align}
		&|\sum_{\mathcal{U}_{k,l,\cdot,\cdot}}(\sqrt{n}\hat{\theta}_i')(\sqrt{n}\hat{\theta}_j')-\sum_{\mathcal{U}_{k,l,\cdot,\cdot}}(\sqrt{n}\theta_i')(\sqrt{n}\theta_j')|
		\leq\sum_{\mathcal{U}_{k,l,\cdot,\cdot}}|(\sqrt{n}\hat{\theta}_i')(\sqrt{n}\hat{\theta}_j')-(\sqrt{n}\theta_i')(\sqrt{n}\theta_j')|\nonumber\\
		\leq&|\mathcal{U}_{k,l,\cdot,\cdot}|O_{\mathbb{P}}\left(\max\left\{\frac{1}{\sqrt{n}\rho_nw},\sqrt{\frac{\log n}{n\rho_nw}}\right\}\right)=O_{\mathbb{P}}\left(\max\left\{\frac{n^{3/2}}{\rho_n},\sqrt{\frac{n^3\log nw}{\rho_n }}\right\}\right), \label{eqn_diff1}
	\end{align}
	where the last equality follows from $|U_{k,l,\cdot,\cdot}|=O_{\mathbb{P}}(n^2w)$. Denote $\Gamma_1:=\sum_{\mathcal{U}_{k,l,\cdot,\cdot}}(\sqrt{n}\theta_i')(\sqrt{n}\theta_j')$ ,$\Gamma_2:=\sum_{\mathcal{U}_{k,l,\cdot,\cdot}}A_{ij}$, and denote the difference in \eqref{eqn_diff1} as $\Delta_1$. Then we have $\Gamma_1$ is of order $n^2w$, and $\Gamma_2$ is of order $n^2\rho_nw$. Therefore
	\begin{align}
		|n^{-1}\hat{B}_{kl}'-n^{-1}\hat{B}_{kl}'^*|&=|\frac{\Gamma_2}{\Gamma_1+\Delta_1}-\frac{\Gamma_2}{\Gamma_1}|=\frac{\Gamma_2|\Delta_1|}{\Gamma_1(\Gamma_1+\Delta_1)}
		\leq\frac{O_{\mathbb{P}}\left(\max\left\{{n^{7/2}w},\sqrt{n^7\log nw^3\rho_n }\right\}\right)}{\Theta_{\mathbb{P}}(n^4w^2)}\nonumber\\
		&=O_{\mathbb{P}}\left(\max\left\{\frac{1}{\sqrt{n}w},\sqrt{\frac{\log n\rho_n}{nw}}\right\}\right). \label{estB2}
	\end{align}
	
	Combining \eqref{estB1} and \eqref{estB2}, we can get
	$$|n^{-1}\hat{B}_{kl}'-n^{-1}B'_{kl}|\leq O_{\mathbb{P}}\left(\max\left\{\frac{1}{\sqrt{n}w},\sqrt{\frac{\log n\rho_n}{nw}}\right\}\right),$$
	since $\sqrt{n}\rho_nw\to\infty$. Therefore, since $n^{-1}B_{kl}'$ is of order $\rho_n$, we have for all $1\leq k<l\leq K^*$, $$n^{-1}\hat{B}_{kl}'=n^{-1}B'_{kl}\left(1+O_{\mathbb{P}}\left(\max\left\{\frac{1}{\sqrt{n}\rho_nw},\sqrt{\frac{\log n}{n\rho_nw}}\right\}\right)\right).$$ 
	That is, 
	\begin{equation}
		\hat{B}_{kl}'=B'_{kl}\left(1+O_{\mathbb{P}}\left(\max\left\{\frac{1}{\sqrt{n}\rho_nw},\sqrt{\frac{\log n}{n\rho_nw}}\right\}\right)\right). \label{estBres}
	\end{equation}
	For $k=l$, \eqref{estBres} also holds with a similar proof. 
	
	Therefore, for $(i,j)\in \mathcal{T}_{k,l,k,l}$, we have
	$$\hat{P}_{ij}=\hat{\theta}_i'\hat{\theta}_j'\hat{B}_{kl}'=\left(1+O_{\mathbb{P}}\left(\max\left\{\frac{1}{\sqrt{n}\rho_nw},\sqrt{\frac{\log n}{n\rho_nw}}\right\}\right)\right)^3\theta_i\theta_jB_{kl}'=\left(1+O_{\mathbb{P}}\left(\max\left\{\frac{1}{\sqrt{n}\rho_nw},\sqrt{\frac{\log n}{n\rho_nw}}\right\}\right)\right)P_{ij}.$$
	Since $P_{ij}$ is of order $\rho_n$, we have 
	\begin{equation}
		|\hat{P}_{ij}-P_{ij}|=O_{\mathbb{P}}\left(\max\left\{\frac{1}{\sqrt{n}w},\sqrt{\frac{\log n\rho_n}{nw}}\right\}\right),\quad\text{for }(i,j)\in(\cup_{k,l}\mathcal{T}_{k,l,k,l}). \label{Mhat1}
	\end{equation}
	For $(i,j)\in(\cup_{k,l}\mathcal{T}_{k,l,k,l})^c$, by Lemma \ref{lem_thetacontrol}, we have $\hat{\theta}_i'=O_{\mathbb{P}}(\theta_i')$ and $\hat{\theta}_j'=O_{\mathbb{P}}(\theta_j')$, since they are both of $n^{-1/2}$ order. Furthermore, for $(k_1,k_2)\neq(l_1,l_2)$, we have
	$$|\hat{B}_{k_1k_2}-B_{l_1l_2}|\leq|\hat{B}_{k_1k_2}-B_{k_1k_2}|+B_{k_1k_2}+B_{l_1l_2}=O_{\mathbb{P}}(\rho_n),$$
	therefore $\hat{B}_{k_1k_2}=O_{\mathbb{P}}(B_{l_1l_2})$. As a result, for $(i,j)\in\mathcal{E}^c$ but $\notin (\cup_{k,l}\mathcal{T}_{k,l,k,l})$, assume $(i,j)\in \mathcal{T}_{k_1,k_2,l_1,l_2}$, then
	$$\hat{P}_{ij}=\hat{\theta}_i'\hat{\theta}_j'\hat{B}_{k_1k_2}'=O_{\mathbb{P}}(\theta_i'\theta_j'B_{l_1l_2}')=O_{\mathbb{P}}(P_{ij}).$$
	
	Therefore, 
	\begin{equation}
		|\hat{P}_{ij}-P_{ij}|=O_{\mathbb{P}}(\rho_n),\quad{\text{for }(i,j)\in\mathcal{E}^c\text{ but}\notin(\cup_{k,l}\mathcal{T}_{k,l,k,l}).} \label{Mhat2}
	\end{equation}
	
	Now we can estimate $\ell_{K^*}(A,\mathcal{E}^c)-\ell_0(A,\mathcal{E}^c)$. We have
	\begin{align*}
		\ell_{K^*}(A,\mathcal{E}^c)-\ell_0(A,\mathcal{E}^c) =& \sum_{k_1,k_2,l_1,l_2} \sum_{(i,j) \in \mathcal{T}_{k_1,k_2,l_1,l_2}} \left[ (A_{ij}- \hat{P}_{ij})^2-(A_{ij}-P_{ij})^2 \right] \\
		=& \sum_{k_1,k_2} \sum_{(i,j) \in \mathcal{T}_{k_1,k_2,k_1,k_2}} \left[ (A_{ij}-\hat{P}_{ij})^2-(A_{ij}-P_{ij})^2 \right] \\
		&+ \sum_{(k_1,k_2) \neq (l_1,l_2)} \sum_{(i,j) \in \mathcal{T}_{k_1,k_2,l_1,l_2}} \left[(A_{ij}-\hat{P}_{ij})^2 - (A_{ij}-P_{ij})^2 \right] \\
		=:& \mathcal{IV} + \mathcal{V}.
	\end{align*}
	
	By \eqref{Mhat1}, we have as $\sqrt{n}\rho_nw\to\infty$,
	\begin{align}
		|\mathcal{IV}|\leq&2\sum_{k_1,k_2} \sum_{(i,j) \in \mathcal{T}_{k_1,k_2,k_1,k_2}}|\hat{P}_{ij}-P_{ij}|(A_{ij}+P_{ij}+|\hat{P}_{ij}-P_{ij}|)\nonumber\\
		\leq&O_{\mathbb{P}}\left(\max\left\{\frac{1}{\sqrt{n}w},\sqrt{\frac{\log n\rho_n}{nw}}\right\}\rho_nn^2(1-w)\right)+O_{\mathbb{P}}\left(\max\left\{\frac{1}{nw^2},\frac{\log n\rho_n}{nw}\right\}n^2(1-w)\right)\nonumber\\
		=&O_{\mathbb{P}}\left(\max\left\{\frac{n^{3/2}\rho_n}{w},\sqrt{\frac{n^3\log n\rho_n^3}{w}}\right\}(1-w)\right).\label{loss1}
	\end{align}
	
	By \eqref{Mhat2}, we have
	\begin{align}
		|\mathcal{V}|\leq&2\sum_{(k_1,k_2) \neq (l_1,l_2)} \sum_{(i,j) \in \mathcal{T}_{k_1,k_2,l_1,l_2}}|\hat{P}_{ij}-P_{ij}|(A_{ij}+P_{ij}+|\hat{P}_{ij}-P_{ij}|)\nonumber\\
		=&O_{\mathbb{P}}(n\sqrt{\frac{n}{\rho_nw}}\rho_n^2)=O_{\mathbb{P}}(n^{\frac{3}{2}}{\rho_n}^{\frac{3}{2}}w^{-\frac{1}{2}}(1-w)).\label{loss3}
	\end{align}
	
	Combining \eqref{loss1}, \eqref{loss3}, we get
	$$|\ell_{K^*}(A,\mathcal{E}^c)-\ell_0(A,\mathcal{E}^c)|=O_{\mathbb{P}}\left(\max\left\{\frac{n^{3/2}\rho_n}{w},\sqrt{\frac{n^3\log n\rho_n^3}{w}}\right\}(1-w)\right).$$
\end{proof}

Now we consider the underfitting case. 
\begin{lemma}\label{lem_nonvanishing_truncated}
	Suppose that Assumptions~4, 6, and 7 hold. Let $U=\Theta'V$ be the population eigenvector
	representation in Lemma~A.7, and let $Q$ be the orthogonal Procrustes
	matrix used in Proposition~A.4. For $1\leq K'<K^*$ and $k\in[K^*]$,
	define
	\[
	s_{k,K'}:=\left\|(VQ)_{k}^{(K')}\right\|_2,
	\]
	where $(VQ)^{(K')}$ represents the first $K'$ columns of $VQ$. Then there exists a constant $c_s>0$ such that
	\[
	\mathbb P\left(
	\min_{1\leq K'<K^*}\min_{1\leq k\leq K^*}s_{k,K'}
	\geq c_s
	\right)\longrightarrow1.
	\]
\end{lemma}

\begin{proof}[Proof of Lemma \ref{lem_nonvanishing_truncated}]
	Let
	\[
	D_n=\operatorname{diag}\left(
	\sqrt{\frac{n_1}{n}},\ldots,\sqrt{\frac{n_{K^*}}{n}}
	\right)
	\]
	and define
	\[
	M_n=D_nB_0D_n.
	\]
	Under the identifiability condition
	$\sum_{i\in G_k}\theta_i^2=n_k$, the nonzero eigenvectors of $P$ are
	given by $U=\Theta'V$, where the columns of $V$ are the eigenvectors of
	$M_n$.
	
	Let $\eta_1=Ve_1$ be the leading unit eigenvector of $M_n$. Set
	\[
	b_{\min}=\min_{k,l}B_{0,kl}>0,
	\qquad
	b_{\max}=\max_{k,l}B_{0,kl}.
	\]
	Since $n_k/n\geq\pi_0$, every entry of $M_n$ lies between
	$\pi_0b_{\min}$ and $b_{\max}$. By the Perron--Frobenius theorem, sign of
	$\eta_1$ can be chosen such that it is entrywise positive. Moreover, for any $k,r\in[K^*]$,
	\[
	\frac{\eta_1(k)}{\eta_1(r)}
	=
	\frac{\sum_l(M_n)_{kl}\eta_1(l)}
	{\sum_l(M_n)_{rl}\eta_1(l)}
	\leq
	\frac{b_{\max}}{\pi_0b_{\min}}.
	\]
	Since $\|\eta_1\|_2=1$ and $K^*$ is fixed, there exists a constant
	$c_\eta>0$ such that
	\[
	\min_{1\leq k\leq K^*}\eta_1(k)\geq c_\eta.
	\]
	
	Furthermore, the family
	\[
	\left\{
	D B_0 D:
	D=\operatorname{diag}(\sqrt{\pi_1},\ldots,\sqrt{\pi_{K^*}}),
	\ \pi_k\geq\pi_0,\ \sum_k\pi_k=1
	\right\}
	\]
	is compact and consists of entrywise positive matrices. Hence its
	Perron eigengap is uniformly positive. It follows from the
	Davis--Kahan theorem that, for some
	$\varepsilon_n\in\{-1,1\}$,
	\[
	\|\widehat u_1-\varepsilon_nu_1\|_2=o_{\mathbb P}(1),
	\]
	where $\widehat u_1$ and $u_1$ denote the leading eigenvectors of the
	sample and population matrices, respectively.
	
	Let $q_1=Qe_1$. Since $U$ has orthonormal columns,
	\[
	\begin{aligned}
		\|q_1-\varepsilon_ne_1\|_2
		&=
		\|U(q_1-\varepsilon_ne_1)\|_2\\
		&\leq
		\|Uq_1-\widehat u_1\|_2
		+
		\|\widehat u_1-\varepsilon_nu_1\|_2\\
		&\leq
		\|UQ-\widehat U\|_2
		+
		\|\widehat u_1-\varepsilon_nu_1\|_2
		=o_{\mathbb P}(1).
	\end{aligned}
	\]
	Therefore, uniformly over $k\in[K^*]$,
	\[
	\begin{aligned}
		|(VQ)_{k,1}|
		&=
		|e_k^\top Vq_1|\\
		&\geq
		|e_k^\top V(\varepsilon_ne_1)|
		-\|q_1-\varepsilon_ne_1\|_2\\
		&\geq
		c_\eta-o_{\mathbb P}(1).
	\end{aligned}
	\]
	Since
	\[
	s_{k,K'}
	=
	\left\|(VQ)_{k}^{(K')}\right\|_2
	\geq |(VQ)_{k,1}|,
	\]
	the conclusion follows by taking, for example, $c_s=c_\eta/2$.
\end{proof}
For a specific rank $K'<K^*$, we first control the estimated degree heterogeneity parameter $\{\hat{\theta'_i}^{(K')}\}_{i=1}^n$, which is the $L_2$ norm of $\hat{U}^{(K')}$, the top $K'$ eigenvectors of $Y/w$, or, the first $k'$ columns of $\hat{U}$. Therefore, we define
$$\theta'^{(K')}_i=\left\|\left(\Theta'(VQ)^{(K')}\right)_{i*}\right\|_2,$$
where $(VQ)^{(K')}$ represents the first $K'$ columns of $VQ$. Therefore, $\theta'^{(K')}_i$ is exactly the norm of the $i$-th row of the matrix consists of the first $k$ columns of $UQ$. For
$i\in G_k$, we have
\[
\theta_i'^{(K')}=\theta_i'\left\|(VQ)_{k}^{(K')}\right\|_2=s_{k,K'}\theta_i'.
\]
Consequently, if $i\in G_k$ and $j\in G_l$, then
$\theta_i'^{(K')}\theta_j'^{(K')}=s_{k,K'}s_{l,K'}\theta_i'\theta_j'.$
Define
\[
B_{kl}'^{(K')}=\frac{B_{kl}'}{s_{k,K'}s_{l,K'}}.
\]
It follows that
\[
P_{ij}=\theta_i'^{(K')}\theta_j'^{(K')}B_{c_ic_j}'^{(K')}.
\]

By the preceding lemma, with probability tending to one, $c_s\leq s_{k,K'}\leq1$ uniformly over $k\in[K^*]$ and $1\leq K'<K^*$. Moreover, Assumptions~4
and~6 imply
\[
\theta_{\min}
\leq
\sqrt n\,\theta_i'
\leq
\frac{\theta_{\max}}{\sqrt{\pi_0}}.
\]
Therefore,
\[
c_s\theta_{\min}
\leq
\sqrt n\,\theta_i'^{(K')}
\leq
\frac{\theta_{\max}}{\sqrt{\pi_0}}
\]
uniformly over $i\in[n]$ and $1\leq K'<K^*$, with probability tending to one. We then have the following control.

\begin{lemma}\label{lem_underfitthetacontrol}
	Assume that $\sqrt{n}\rho_nw\to\infty$. Then there exist constants $\tilde{\theta}_{\min}^{(K')}$ and $\tilde{\theta}_{\max}^{(K')}$, such that the estimator $\hat{\theta}'^{(K')}$ of $\theta'^{(K')}$ obtained with $K'<K^*$ communities satisfies with probability tending to 1,
	$$\tilde{\theta}_{\min}^{(K')}\leq\sqrt{n}\hat{\theta'_i}^{(K')}\leq\tilde{\theta}_{\max}^{(K')},\quad\text{for all }i\in[n].$$
	Moreover, we have
	$$\left|\sqrt{n}\hat{\theta'_i}^{(K')}-\sqrt{n}{\theta'_i}^{(K')}\right|=O_{\mathbb{P}}\left(\max\left\{\frac{1}{\sqrt{n}\rho_n w},\sqrt{\frac{\log n}{n\rho_n w}}\right\}\right).$$
\end{lemma}
\begin{proof}[Proof of lemma \ref{lem_underfitthetacontrol}]
	Notice that 
	$$\left\|\hat{U}^{(K')}-\Theta'(VQ)^{(K')}\right\|_{2,\infty}=\left\|\left(\hat{U}-UQ\right)^{(K')}\right\|_{2,\infty}\leq\|\hat{U}-UQ\|_{2,\infty}=O_{\mathbb{P}}\left(\max\left\{\frac{1}{n\rho_n w},\sqrt{\frac{\log n}{n^2\rho_n w}}\right\}\right).$$
	Thus, using the same argument as in Lemma \ref{lem_thetacontrol}, we have 
	$$\left|\sqrt{n}\hat{\theta'_i}^{(K')}-\sqrt{n}{\theta'_i}^{(K')}\right|\leq\sqrt{n}\left\|\hat{U}^{(K')}-\Theta'(VQ)^{(K')}\right\|_{2,\infty}=O_{\mathbb{P}}\left(\max\left\{\frac{1}{\sqrt{n}\rho_n w},\sqrt{\frac{\log n}{n\rho_n w}}\right\}\right)\to0$$
	as $\sqrt{n}\rho_nw\to\infty$. 
\end{proof}
\begin{lemma}[Extension of Lemma B6 in \cite{Chakrabarty2025subsampling}]\label{lem_cha}
	For any $(a_i,p_i,w_i)\in\mathbb{R}^2\times[0,\infty)$, $i\in[n]$,
	$$\sum_{i=1}^n(a_i-p_i)^2w_i\geq\sum_{i=1}^n(a_i-\bar{a}_w)^2w_i-2\sum_{i=1}^n(a_i-\bar{a}_w)p_iw_i,$$
	where $\bar{a}_w=\sum_{i=1}^{n}a_iw_i/\sum_{i=1}^nw_i$. Furthermore, when all the $p_i$ are the same, we have
	$$\sum_{i=1}^n(a_i-p)^2w_i=\sum_{i=1}^n(a_i-\bar{a}_w)^2w_i+\sum_{i=1}^nw_i(\bar{a}_w-p)^2.$$
\end{lemma}

Now we can give a lower bound for the difference between the loss and the oracle loss in the underfitting case. Let
$D_{s,K'}=\operatorname{diag}
\left(s_{1,K'},\ldots,s_{K^*,K'}\right).$
Then the rescaled block matrix defined above can be written as
\[
B'^{(K')}
=
n\rho_n\widetilde B_{0,n}^{(K')},
\qquad
\widetilde B_{0,n}^{(K')}
=
D_{s,K'}^{-1}D_nB_0D_nD_{s,K'}^{-1}.
\]
On the high-probability event established above, the matrices
$\widetilde B_{0,n}^{(K')}$ have uniformly separated rows. Indeed, for
any $k\ne l$, the difference between its $k$-th and $l$-th rows can be written as
\[
\widetilde B_{0,n,k*}^{(K')}
-
\widetilde B_{0,n,l*}^{(K')}
=
\left(
\frac{\sqrt{\pi_{k,n}}}{s_{k,K'}}e_k
-
\frac{\sqrt{\pi_{l,n}}}{s_{l,K'}}e_l
\right)^\top
B_0D_nD_{s,K'}^{-1}.
\]
Hence,
\[
\begin{aligned}
	\left\|
	\widetilde B_{0,n,k*}^{(K')}
	-
	\widetilde B_{0,n,l*}^{(K')}
	\right\|_2
	&=
	\left\|
	\left(B_0D_nD_{s,K'}^{-1}\right)^\top
	\left(
	\frac{\sqrt{\pi_{k,n}}}{s_{k,K'}}e_k
	-
	\frac{\sqrt{\pi_{l,n}}}{s_{l,K'}}e_l
	\right)
	\right\|_2\\
	&\geq
	\sigma_{\min}\left(B_0D_nD_{s,K'}^{-1}\right)
	\left\|
	\frac{\sqrt{\pi_{k,n}}}{s_{k,K'}}e_k
	-
	\frac{\sqrt{\pi_{l,n}}}{s_{l,K'}}e_l
	\right\|_2.
\end{aligned}
\]
We next bound the two factors on the right-hand side separately. Since $B_0$ is positive definite,
$\sigma_{\min}(B_0)=\lambda_{\min}(B_0)>0.
$
Moreover,
$\sigma_{\min}(D_n)=\min_{1\leq r\leq K^*}\sqrt{\pi_{r,n}}\geq
\sqrt{\pi_0}.$
Because $VQ$ is an orthogonal matrix, the Euclidean norm of each of its rows is one. Therefore,
\[
s_{r,K'}
=
\left\|(VQ)_{r^{(K')}}\right\|_2
\leq
\left\|(VQ)_{r*}\right\|_2
=
1,
\]
and consequently
\[
\sigma_{\min}\left(D_{s,K'}^{-1}\right)
=
\min_{1\leq r\leq K^*}\frac{1}{s_{r,K'}}
=
\frac{1}{\max_{1\leq r\leq K^*} s_{r,K'}}
\geq 1.
\]
Using the inequality
$\sigma_{\min}(ABC)\geq\sigma_{\min}(A)\sigma_{\min}(B)\sigma_{\min}(C),$
we obtain
\[
\sigma_{\min}\left(B_0D_nD_{s,K'}^{-1}\right)\geq\sigma_{\min}(B_0)\sigma_{\min}(D_n)\sigma_{\min}\left(D_{s,K'}^{-1}\right)\geq\lambda_{\min}(B_0)\sqrt{\pi_0}.
\]
On the other hand, since $e_k$ and $e_l$ are orthogonal,
\[
\left\|
\frac{\sqrt{\pi_{k,n}}}{s_{k,K'}}e_k-\frac{\sqrt{\pi_{l,n}}}{s_{l,K'}}e_l\right\|_2^2=\frac{\pi_{k,n}}{s_{k,K'}^2}+\frac{\pi_{l,n}}{s_{l,K'}^2}\geq\pi_{k,n}+\pi_{l,n}\geq2\pi_0,
\]
where the first inequality again follows from $s_{k,K'},s_{l,K'}\leq1$. 
Combining the above bounds yields
\[
\left\|
\widetilde B_{0,n,k*}^{(K')}
-
\widetilde B_{0,n,l*}^{(K')}
\right\|_2
\geq
\sqrt{2}\,\pi_0\lambda_{\min}(B_0).
\]
Thus, the rows of $\widetilde B_{0,n}^{(K')}$ are uniformly separated over
$k\neq l$ and $1\leq K'<K^*$.
Therefore, the separation constant required in Lemma \ref{lem3} may be chosen
uniformly over $K'<K^*$ and over the realization of the training data. 

\begin{proposition}\label{prop_dcbmunder}
	Assume that $\sqrt{n}\rho_nw\to\infty$. Then we have
	$$\ell_{K'}(A,\mathcal{E}^c)-\ell_{0}(A,\mathcal{E}^c)=\Omega_{\mathbb{P}}(n^2\rho_n^2(1-w)).$$
\end{proposition}
\begin{proof}
	Note that Lemma \ref{lem3} still holds in the DCBM case, where $B_0$ is replaced by $\widetilde B^{(K')}_0$. Without loss of generality, assume the $k_1 ,k_2$ and $l_1 ,l_2 ,l_3$ in Lemma \ref{lem3} are 1,2 and 1,2,3 respectively. Denote
	$$\hat{p}_{k_1,k_2,l_1,l_2}=\frac{\sum_{(i,j)\in\mathcal{T}_{k_1,k_2,l_1,l_2}}A_{ij}\theta_i\theta_j}{\sum_{(i,j)\in\mathcal{T}_{k_1,k_2,l_1,l_2}}\theta_i^2\theta_j^2},\qquad \hat{p}=\frac{\sum_{(i,j)\in\mathcal{T}_{1,2,1,2}\cup\mathcal{T}_{1,2,1,3}}A_{ij}\hat{\theta'_i}^{(K')}\hat{\theta'_j}^{(K')}}{\sum_{(i,j)\in\mathcal{T}_{1,2,1,2}\cup\mathcal{T}_{1,2,1,3}}\left(\hat{\theta'_i}^{(K')}\hat{\theta'_j}^{(K')}\right)^2}=\hat{\lambda}\hat{p}_1+(1-\hat{\lambda})\hat{p}_2,$$
	
	Then by the two scenarios in Lemma \ref{lem_cha},
	\begin{align*}
		&\ell_{K'}(A,\mathcal{E}^c)-\ell_{0}(A,\mathcal{E}^c) \\
		=& \sum_{(i,j) \in \mathcal{T}_{1,2,1,2}} \left[(A_{ij}-\hat{P}_{ij}^{(K')})^2- (A_{ij}-P_{ij})^2 \right]+\sum_{(i,j) \in \mathcal{T}_{1,2,1,3}} \left[(A_{ij}-\hat{P}_{ij}^{(K')})^2- (A_{ij}-P_{ij})^2 \right]\\
		&+ \sum_{{(k_1, k_2, l_1, l_2) \notin \{(1,2,1,2), (1,2,1,3)\}}} 
		\sum_{(i,j) \in \mathcal{T}_{k_1, k_2, l_1, l_2}} 
		\left[(A_{ij}-\hat{P}_{ij}^{(K')})^2- (A_{ij}-P_{ij})^2 \right]\\
		=&\sum_{(i,j) \in \mathcal{T}_{1,2,1,2}} \left[\left(\frac{A_{ij}}{\hat{\theta'_i}^{(K')}\hat{\theta'_j}^{(K')}}-\frac{\hat{P}_{ij}^{(K')}}{\hat{\theta'_i}^{(K')}\hat{\theta'_j}^{(K')}}\right)^2\left(\hat{\theta'_i}^{(K')}\hat{\theta'_j}^{(K')}\right)^2- (A_{ij}-P_{ij})^2 \right]\\	
		&+\sum_{(i,j) \in \mathcal{T}_{1,2,1,3}} \left[\left(\frac{A_{ij}}{\hat{\theta'_i}^{(K')}\hat{\theta'_j}^{(K')}}-\frac{\hat{P}_{ij}^{(K')}}{\hat{\theta'_i}^{(K')}\hat{\theta'_j}^{(K')}}\right)^2\left(\hat{\theta'_i}^{(K')}\hat{\theta'_j}^{(K')}\right)^2- (A_{ij}-P_{ij})^2 \right]\\
		&+ \sum_{{(k_1, k_2, l_1, l_2) \notin \{(1,2,1,2), (1,2,1,3)\}}} 
		\sum_{(i,j) \in \mathcal{T}_{k_1, k_2, l_1, l_2}} 
		\left[\left(\frac{A_{ij}}{\theta_i\theta_j}-\frac{\hat{P}_{ij}^{(K')}}{\theta_i\theta_j}\right)^2\theta_i^2\theta_j^2-\left(\frac{A_{ij}}{\theta_i\theta_j}-B_{l_1l_2}\right)^2\theta_i^2\theta_j^2 \right]\\
		\geq&\sum_{(i,j) \in \mathcal{T}_{1,2,1,2}} \left[\left(\frac{A_{ij}}{\hat{\theta'_i}^{(K')}\hat{\theta'_j}^{(K')}}-\hat{p}\right)^2\left(\hat{\theta'_i}^{(K')}\hat{\theta'_j}^{(K')}\right)^2- (A_{ij}-P_{ij})^2 \right]\\
		&+\sum_{(i,j) \in \mathcal{T}_{1,2,1,3}} \left[\left(\frac{A_{ij}}{\hat{\theta'_i}^{(K')}\hat{\theta'_j}^{(K')}}-\hat{p}\right)^2\left(\hat{\theta'_i}^{(K')}\hat{\theta'_j}^{(K')}\right)^2- (A_{ij}-P_{ij})^2 \right]\\
		&+ \sum_{{(k_1, k_2, l_1, l_2) \notin \{(1,2,1,2), (1,2,1,3)\}}} 
		\sum_{(i,j) \in \mathcal{T}_{k_1, k_2, l_1, l_2}} 
		\left[\left(\frac{A_{ij}}{\theta_i\theta_j}-\hat{p}_{k_1,k_2,l_1,l_2}\right)^2\theta_i^2\theta_j^2-\left(\frac{A_{ij}}{\theta_i\theta_j}-B_{l_1l_2}\right)^2\theta_i^2\theta_j^2 \right]\\
		&-2\sum_{{(k_1, k_2, l_1, l_2) \notin \{(1,2,1,2), (1,2,1,3)\}}} 
		\sum_{(i,j) \in \mathcal{T}_{k_1, k_2, l_1, l_2}}\left(\frac{A_{ij}}{\theta_i\theta_j}-\hat{p}_{k_1,k_2,l_1,l_2}\right)\hat{P}_{ij}^{(K')}\theta_i\theta_j\\
		:=& \mathcal{I} + \mathcal{II} + \mathcal{III}-\mathcal{IV},
	\end{align*}
	For notation simplicity, from now on we will write $\hat{\theta'_i}^{(K')}$ and $\theta_i'^{(K')}$ simply as $\hat{\check{\theta}}_i$ and $\check{\theta}_i$. We first analyze $\mathcal{I}$. Notice that
	\begin{align*}
		\mathcal{I}=&\sum_{(i,j) \in \mathcal{T}_{1,2,1,2}} \left[\left(\frac{A_{ij}}{\left(\sqrt{n}\hat{\check{\theta}}_i\right)\left(\sqrt{n}\hat{\check{\theta}}_j\right)}-n^{-1}\hat{p}\right)^2\left(\sqrt{n}\hat{\check{\theta}}_i\right)^2\left(\sqrt{n}\hat{\check{\theta}}_j\right)^2- (A_{ij}-P_{ij})^2 \right]\\
		=&\sum_{(i,j) \in \mathcal{T}_{1,2,1,2}} \left[\left(A_{ij}-n^{-1}\hat{p}\left(\sqrt{n}\hat{\check{\theta}}_i\right)\left(\sqrt{n}\hat{\check{\theta}}_j\right)\right)^2- (A_{ij}-P_{ij})^2 \right]\\
		=&\sum_{(i,j) \in \mathcal{T}_{1,2,1,2}} \left[\left(A_{ij}-n^{-1}\hat{p}\left(\sqrt{n}\check{\theta}_i\right)\left(\sqrt{n}\check{\theta}_j\right)\right)^2- (A_{ij}-P_{ij})^2 \right]\\
		&+\sum_{(i,j) \in \mathcal{T}_{1,2,1,2}} \left[\left(A_{ij}-n^{-1}\hat{p}\left(\sqrt{n}\hat{\check{\theta}}_i\right)\left(\sqrt{n}\hat{\check{\theta}}_j\right)\right)^2- \left(A_{ij}-n^{-1}\hat{p}\left(\sqrt{n}\check{\theta}_i\right)\left(\sqrt{n}\check{\theta}_j\right)\right)^2\right]\\
		:=&\mathcal{I}'+E_1,
	\end{align*}
	where by Lemma \ref{lem_underfitthetacontrol}
	\begin{align}
		|E_1|\leq&\sum_{(i,j)\in\mathcal{T}_{1,2,1,2}}\left|2A_{ij}-n^{-1}\hat{p}\left(\sqrt{n}\check{\theta}_i\right)\left(\sqrt{n}\check{\theta}_j\right)-n^{-1}\hat{p}\left(\sqrt{n}\hat{\check{\theta}}_i\right)\left(\sqrt{n}\hat{\check{\theta}}_j\right)\right|\nonumber\\
		&\cdot\left|n^{-1}\hat{p}\left(\left(\sqrt{n}\hat{\check{\theta}}_i\right)\left(\sqrt{n}\hat{\check{\theta}}_j\right)-\left(\sqrt{n}\check{\theta}_i\right)\left(\sqrt{n}\check{\theta}_j\right)\right)\right|\nonumber\\
		\leq&|\mathcal{T}_{1,2,1,2}|O_{\mathbb{P}}(\rho_n)\cdot O_{\mathbb{P}}(\rho_n)\cdot\left(\tilde{\theta}_{\max}^{(K')}\left|\sqrt{n}\hat{\check{\theta}}_i-\sqrt{n}\check{\theta}_i\right|+\theta_{\max}^{(K')}\left|\sqrt{n}\hat{\check{\theta}}_j-\sqrt{n}\check{\theta}_j\right|\right)\nonumber\\
		=&O_{\mathbb{P}}\left(n^2\rho_n^2(1-w)\max\left\{\frac{1}{\sqrt{n}\rho_n w},\sqrt{\frac{\log n}{n\rho_n w}}\right\}\right)=o_{\mathbb{P}}\left(n^2\rho_n^2(1-w)\right),\label{eqn_E1}
	\end{align}
	as $\sqrt{n}\rho_nw\to\infty$. Next, we define $p'$ as the oracle form of $\hat{p}$ as 
	$$p'=\frac{\sum_{(i,j)\in\mathcal{T}_{1,2,1,2}\cup\mathcal{T}_{1,2,1,3}}A_{ij}\check{\theta}_i\check{\theta}_j}{\sum_{(i,j)\in\mathcal{T}_{1,2,1,2}\cup\mathcal{T}_{1,2,1,3}}\check{\theta}_i^2\check{\theta}_j^2}=\lambda p_1'+(1-\lambda)p_2'$$
	where
	$$p_1'=\frac{\sum_{(i,j)\in\mathcal{T}_{1,2,1,2}}A_{ij}\check{\theta}_i\check{\theta}_j}{\sum_{(i,j)\in\mathcal{T}_{1,2,1,2}}\check{\theta}_i^2\check{\theta}_j^2},\qquad p_2'=\frac{\sum_{(i,j)\in\mathcal{T}_{1,2,1,3}}A_{ij}\check{\theta}_i\check{\theta}_j}{\sum_{(i,j)\in\mathcal{T}_{1,2,1,3}}\check{\theta}_i^2\check{\theta}_j^2},$$
	$$\lambda=\frac{\sum_{(i,j)\in\mathcal{T}_{1,2,1,2}}\check{\theta}_i^2\check{\theta}_j^2}{\sum_{(i,j)\in\mathcal{T}_{1,2,1,2}\cup\mathcal{T}_{1,2,1,3}}\check{\theta}_i^2\check{\theta}_j^2}.$$
	Furthermore, we denote $B'^{(K')}$ as $\check{B}$, and denote $p_1=\check{B}_{12}$, $p_2=\check{B}_{13}$.
	Thus 
	\begin{align*}
		\mathcal{I}'=&\sum_{(i,j) \in \mathcal{T}_{1,2,1,2}} \left[\left(A_{ij}-n^{-1}p'\left(\sqrt{n}\check{\theta}_i\right)\left(\sqrt{n}\check{\theta}_j\right)\right)^2- (A_{ij}-P_{ij})^2 \right]\\
		&+\sum_{(i,j) \in \mathcal{T}_{1,2,1,2}} \left[\left(A_{ij}-n^{-1}\hat{p}\left(\sqrt{n}\check{\theta}_i\right)\left(\sqrt{n}\check{\theta}_j\right)\right)^2-\left(A_{ij}-n^{-1}p'\left(\sqrt{n}\check{\theta}_i\right)\left(\sqrt{n}\check{\theta}_j\right)\right)^2 \right]\\
		:=&\mathcal{I}''+E_2,
	\end{align*}
	where
	\begin{align*}
		|E_2|\leq&\sum_{(i,j)\in\mathcal{T}_{1,2,1,2}}\left|2A_{ij}-n^{-1}\hat{p}\left(\sqrt{n}\check{\theta}_i\right)\left(\sqrt{n}\check{\theta}_j\right)-n^{-1}p'\left(\sqrt{n}\check{\theta}_i\right)\left(\sqrt{n}\check{\theta}_j\right)\right|\left|n^{-1}\hat{p}-n^{-1}p'\right|\left(\sqrt{n}\check{\theta}_i\right)\left(\sqrt{n}\check{\theta}_j\right)\\
		\leq&O_{\mathbb{P}}(n^2\rho_n(1-w))\left|n^{-1}\hat{p}-n^{-1}p'\right|.
	\end{align*}
	Denote $$T_1=\sum_{(i,j)\in\mathcal{T}_{1,2,1,2}\cup\mathcal{T}_{1,2,1,3}}\left(\sqrt{n}\check{\theta}_i\right)^2\left(\sqrt{n}\check{\theta}_j\right)^2,\qquad \Delta_1=\sum_{(i,j)\in\mathcal{T}_{1,2,1,2}\cup\mathcal{T}_{1,2,1,3}}\left(\left(\sqrt{n}\hat{\check{\theta}}_i\right)^2\left(\sqrt{n}\hat{\check{\theta}}_j\right)^2-\left(\sqrt{n}\check{\theta}_i\right)^2\left(\sqrt{n}\check{\theta}_j\right)^2\right),$$
	$$T_2=\sum_{(i,j)\in\mathcal{T}_{1,2,1,2}\cup\mathcal{T}_{1,2,1,3}}A_{ij}\left(\sqrt{n}\check{\theta}_i\right)\left(\sqrt{n}\check{\theta}_j\right),\qquad \Delta_2=\sum_{(i,j)\in\mathcal{T}_{1,2,1,2}\cup\mathcal{T}_{1,2,1,3}}A_{ij}\left(\left(\sqrt{n}\hat{\check{\theta}}_i\right)\left(\sqrt{n}\hat{\check{\theta}}_j\right)-\left(\sqrt{n}\check{\theta}_i\right)\left(\sqrt{n}\check{\theta}_j\right)\right).$$
	Thus, $T_1=\Theta_{\mathbb{P}}(n^2(1-w))$, $T_2=\Theta_{\mathbb{P}}(n^2\rho_n(1-w))$, and we have
	\begin{align*}
		|\Delta_1|\leq&\sum_{(i,j)\in\mathcal{T}_{1,2,1,2}\cup\mathcal{T}_{1,2,1,3}}\left(\tilde{\theta}_{\max}^{(K')}\right)^2\left|\left(\sqrt{n}\hat{\check{\theta}}_i\right)^2-\left(\sqrt{n}\check{\theta}_i\right)^2\right|+\left(\theta_{\max}^{(K')}\right)^2\left|\left(\sqrt{n}\hat{\check{\theta}}_j\right)^2-\left(\sqrt{n}\check{\theta}_j\right)^2\right|\\
		\leq&\left(\tilde{\theta}_{\max}^{(K')}+\theta_{\max}^{(K')}\right)\sum_{(i,j)\in\mathcal{T}_{1,2,1,2}\cup\mathcal{T}_{1,2,1,3}}\left(\tilde{\theta}_{\max}^{(K')}\right)^2\left|\sqrt{n}\hat{\check{\theta}}_i-\sqrt{n}\check{\theta}_i\right|+\left(\theta_{\max}^{(K')}\right)^2\left|\sqrt{n}\hat{\check{\theta}}_j-\sqrt{n}\check{\theta}_j\right|\\
		=&O_{\mathbb{P}}\left(n^2(1-w)\max\left\{\frac{1}{\sqrt{n}\rho_n w},\sqrt{\frac{\log n}{n\rho_n w}}\right\}\right).
	\end{align*}
	Furthermore,
	\begin{align*}
		|\Delta_2|\leq&\sum_{(i,j)\in\mathcal{T}_{1,2,1,2}\cup\mathcal{T}_{1,2,1,3}}A_{ij}\left(\tilde{\theta}_{\max}^{(K')}\left|\sqrt{n}\hat{\check{\theta}}_i-\sqrt{n}\check{\theta}_i\right|+\theta_{\max}^{(K')}\left|\sqrt{n}\hat{\check{\theta}}_j-\sqrt{n}\check{\theta}_j\right|\right)\\
		=&O_{\mathbb{P}}\left(n^2\rho_n(1-w)\max\left\{\frac{1}{\sqrt{n}\rho_n w},\sqrt{\frac{\log n}{n\rho_n w}}\right\}\right).
	\end{align*}
	Thus
	$$\left|n^{-1}\hat{p}-n^{-1}p'\right|=\left|\frac{T_2+\Delta_2}{T_1+\Delta_1}-\frac{T_2}{T_1}\right|\leq\frac{|\Delta_1T_2|+|\Delta_2T_1|}{|T_1+\Delta_1||T_1|}=O_{\mathbb{P}}\left(\rho_n\max\left\{\frac{1}{\sqrt{n}\rho_n w},\sqrt{\frac{\log n}{n\rho_n w}}\right\}\right),$$
	which results in
	\begin{equation}\label{eqn_E2}
		|E_2|=O_{\mathbb{P}}\left(n^2\rho_n^2(1-w)\max\left\{\frac{1}{\sqrt{n}\rho_n w},\sqrt{\frac{\log n}{n\rho_n w}}\right\}\right)=o_{\mathbb{P}}\left(n^2\rho_n^2(1-w)\right).
	\end{equation}
	We finally bound $\mathcal{I}''$. Let
	$$f(p):=\sum_{(i,j) \in \mathcal{T}_{1,2,1,2}}\left(\frac{A_{ij}}{\left(\sqrt{n}\check{\theta}_i\right)\left(\sqrt{n}\check{\theta}_j\right)}-n^{-1}p\right)^2\left(\sqrt{n}\check{\theta}_i\right)^2\left(\sqrt{n}\check{\theta}_j\right)^2.$$
	Then by Lemma \ref{lem_cha}, $f(p)$ is minimized at $p_1'$, and in addition we have
	\begin{align*}
		\mathcal{I}''=&f(p')-f(p_1')+f(p_1')-f(p_1)\\
		=&\sum_{(i,j) \in \mathcal{T}_{1,2,1,2}}\left(\sqrt{n}\check{\theta}_i\right)^2\left(\sqrt{n}\check{\theta}_j\right)^2(1-\lambda)^2\left|n^{-1}p_1'-n^{-1}p_2'\right|^2+f(p_1')-f(p_1).
	\end{align*}
	Here, $\sum_{(i,j) \in \mathcal{T}_{1,2,1,2}}\left(\sqrt{n}\check{\theta}_i\right)^2\left(\sqrt{n}\check{\theta}_j\right)^2$ is of order $n^2(1-w)$, $\lambda$ is of constant order, and 
	$$|n^{-1}p_1'-n^{-1}p_2'|\geq|n^{-1}p_1-n^{-1}p_2|-|n^{-1}p_1'-n^{-1}p_1|-|n^{-1}p_2'-n^{-1}p_2|.$$
	By Bernstein's inequality, 
	\begin{align*}
		&\mathbb{P}\left\{\left|\sum_{\mathcal{T}_{1,2,1,2}}\left(\sqrt{n}\check{\theta}_i\right)\left(\sqrt{n}\check{\theta}_j\right)A_{ij}-\sum_{\mathcal{T}_{1,2,1,2}}\left(\sqrt{n}\check{\theta}_i\right)^2\left(\sqrt{n}\check{\theta}_j\right)^2n^{-1}\check{B}_{12}\right|\geq t\right\}\\&\leq \exp\left(-\frac{\frac{1}{2}t^2}{\sum_{\mathcal{T}_{1,2,1,2}}\left(\sqrt{n}\check{\theta}_i\right)^2\left(\sqrt{n}\check{\theta}_j\right)^2(n^{-1}\check{B}_{kl})(1-n^{-1}\check{B}_{kl})+\frac{\left(\theta_{\max}^{(K')}\right)^2}{3}t}\right).
	\end{align*}
	Therefore, 
	$$\left|\sum_{\mathcal{T}_{1,2,1,2}}\left(\sqrt{n}\check{\theta}_i\right)\left(\sqrt{n}\check{\theta}_j\right)A_{ij}-\sum_{\mathcal{T}_{1,2,1,2}}\left(\sqrt{n}\check{\theta}_i\right)^2\left(\sqrt{n}\check{\theta}_j\right)^2n^{-1}\check{B}_{12}\right|=O_{\mathbb{P}}\left(\sqrt{\rho_n\sum_{\mathcal{T}_{1,2,1,2}}\left(\sqrt{n}\check{\theta}_i\right)^2\left(\sqrt{n}\check{\theta}_j\right)^2}\right),$$
	since $n^{-1}\check{B}$ is of order $\rho_n$. Notice $\sqrt{n}\check{\theta}_i$ is of constant order, so 
	\begin{equation*}
		\sum_{T}\left(\sqrt{n}\check{\theta}_i\right)^2\left(\sqrt{n}\check{\theta}_j\right)^2=\Theta(|T|),\quad\text{for all index set T},
	\end{equation*}
	thus we obtain $$|n^{-1}p_1'-n^{-1}p_1|=O_{\mathbb{P}}\left(\sqrt{\frac{\rho_n}{n^2(1-w)}}\right).$$
	Similarly, $|n^{-1}p_2'-n^{-1}p_2|$ has the same order, and thus we get
	\begin{equation}\label{eqn_I1}
		|n^{-1}p_1'-n^{-1}p_2'|=\Omega_{\mathbb{P}}(\rho_n)\Rightarrow\sum_{(i,j) \in \mathcal{T}_{1,2,1,2}}\left(\sqrt{n}\check{\theta}_i\right)^2\left(\sqrt{n}\check{\theta}_j\right)^2(1-\lambda)^2\left|n^{-1}p_1'-n^{-1}p_2'\right|^2=\Omega(n^2\rho_n^2(1-w)).
	\end{equation}
	For the remainder term, notice
	\begin{align*}
		f(p_1')-f(p_1)=&\sum_{(i,j) \in \mathcal{T}_{1,2,1,2}}\left|2A_{ij}-n^{-1}p_1'\left(\sqrt{n}\check{\theta}_i\right)\left(\sqrt{n}\check{\theta}_j\right)\right|\left|n^{-1}p_1'-n^{-1}p_1\right|\left(\sqrt{n}\check{\theta}_i\right)\left(\sqrt{n}\check{\theta}_j\right)\\
		=&O_{\mathbb{P}}(n(1-w)^{1/2}\rho_n^{3/2})=o_{\mathbb{P}}(n^2\rho_n^2(1-w)).
	\end{align*}
	Combining this with Equations \eqref{eqn_E1}, \eqref{eqn_E2}, \eqref{eqn_I1}, we arrived at the conclusion that 
	\begin{equation}\label{eqn_I}
		\mathcal{I}=\Omega_{\mathbb{P}}(n^2\rho_n^2(1-w)).
	\end{equation}
	Using a similar argument, one can also show that
	\begin{equation}\label{eqn_II}
		\mathcal{II}=\Omega_{\mathbb{P}}(n^2\rho_n^2(1-w)).
	\end{equation}
	
	For $\mathcal{III}$, we define
	$$\Theta_{k_1,k_2,l_1,l_2}=\sum_{(i,j) \in \mathcal{T}_{k_1, k_2, l_1, l_2}}\theta_i^2\theta_j^2,$$
	which is the same order as $|\mathcal{T}_{k_1,k_2,l_1,l_2}|$. By the second equation in Lemma \ref{lem_cha}, we have
	\begin{equation}\label{eqn_III}
		\mathcal{III}=-\sum_{{(k_1, k_2, l_1, l_2) \notin \{(1,2,1,2), (1,2,1,3)\}}}\Theta_{k_1,k_2,l_1,l_2}(\hat{p}_{k_1,k_2,l_1,l_2}-B_{l_1l_2})^2=-O_{\mathbb{P}}(\rho_n)
	\end{equation}
	by Bernstein's inequality.
	
	Finally, for $\mathcal{IV}$, we have for each $(k_1,k_2,l_1,l_2)$, we condition on the training set. Therefore, we may treat $\hat{P}_{ij}^{(K')}$ as fixed, since it only depends on the training set. By independence, the entries in the test set is still independent. Denote
	$$\mu_{k_1,k_2,l_1,l_2}:=B_{l_1l_2}\sum_{(i,j) \in \mathcal{T}_{k_1, k_2, l_1, l_2}}\hat{P}^{(K')}_{ij}\theta_i\theta_j.$$
	Then, we have
	\begin{align*}
		&\left|\sum_{(i,j) \in \mathcal{T}_{k_1, k_2, l_1, l_2}}\left(\frac{A_{ij}}{\theta_i\theta_j}-\hat{p}_{k_1,k_2,l_1,l_2}\right)\hat{P}^{(K')}_{ij}\theta_i\theta_j\right|\\
		\leq&\left|\sum_{(i,j) \in \mathcal{T}_{k_1, k_2, l_1, l_2}}A_{ij}\hat{P}^{(K')}_{ij}-\mu_{k_1,k_2,l_1,l_2}\right|+\sum_{(i,j) \in \mathcal{T}_{k_1, k_2, l_1, l_2}}\hat{P}^{(K')}_{ij}\theta_i\theta_j|\hat{p}_{k_1,k_2,l_1,l_2}-B_{l_1l_2}|.
	\end{align*}
	By Bernstein's inequality, 
	\begin{align*}
		\mathbb{P}\left\{\left.\left|\sum_{(i,j)\in\mathcal{T}_{k_1,k_2,l_1,l_2}}A_{ij}\hat{P}^{(K')}_{ij}-\mu_{k_1,k_2,l_1,l_2}\right|\geq t\Large \right| \mathfrak{F}_{\mathcal{E}}\right\}\leq \exp\left(-\frac{\frac{1}{2}t^2}{\sum_{\mathcal{T}_{k_1,k_2,l_1,l_2}}\left(\hat{P}^{(K')}_{ij}\right)^2\theta_i\theta_jB_{l_1l_2}+\frac{1}{3}t}\right).
	\end{align*}
	Thus, since $\hat{P}^{(K')}_{ij}\leq 1$ and taking $t=C\sqrt{\sum_{\mathcal{T}_{k_1,k_2,l_1,l_2}}\theta_i\theta_jB_{l_1l_2}}$, we have
	$$\left|\sum_{(i,j)\in\mathcal{T}_{k_1,k_2,l_1,l_2}}A_{ij}\hat{P}^{(K')}_{ij}-\mu_{k_1,k_2,l_1,l_2}\right|=O_{\mathbb{P}}\left(\sqrt{\rho_n|T_{k_1,k_2,l_1,l_2}|}\right).$$
	Similarly bounding the second term, we can get
	$$\left|\sum_{(i,j) \in \mathcal{T}_{k_1, k_2, l_1, l_2}}\left(\frac{A_{ij}}{\theta_i\theta_j}-\hat{p}_{k_1,k_2,l_1,l_2}\right)\hat{P}^{(K')}_{ij}\theta_i\theta_j\right|=O_{\mathbb{P}}\left(\sqrt{\rho_n|T_{k_1,k_2,l_1,l_2}|}\right),$$
	and thus
	$$\mathcal{IV}=-2\sum_{{(k_1, k_2, l_1, l_2) \notin \{(1,2,1,2), (1,2,1,3)\}}}\sum_{(i,j) \in \mathcal{T}_{k_1, k_2, l_1, l_2}}\left(\frac{A_{ij}}{\theta_i\theta_j}-\hat{p}_{k_1,k_2,l_1,l_2}\right)\hat{P}^{(K')}_{ij}\theta_i\theta_j\geq-O_{\mathbb{P}}\left(\sqrt{n^2\rho_n(1-w)}\right).$$
	Combining with Equations \eqref{eqn_I}, \eqref{eqn_II} and \eqref{eqn_III}, the desired result follows.
\end{proof}

Finally we deal with the overfitting case.
\begin{proposition}\label{prop_dcbmover1}
	When $K^*<K'\leq 2K^*$, we have $$\ell_{K'}(A,\mathcal{E}^c)-\ell_0(A,\mathcal{E}^c)\geq-O_{\mathbb{P}}\left(\sqrt{n^2\rho_n(1-w)}\right).$$
\end{proposition}
\begin{proof}[Proof of Proposition \ref{prop_dcbmover1}]
	We use the same notation as in the proof of Proposition \ref{prop_dcbmunder}. Then, by Lemma \ref{lem_cha}, we have
	\begin{align*}
		\ell_{K'}(A,\mathcal{E}^c)-\ell_{0}(A,\mathcal{E}^c)=&\sum_{k_1,k_2,l_1,l_2}\sum_{(i,j)\in\mathcal{T}_{k_1,k_2,l_1,l_2}}\left[\left(\frac{A_{ij}}{\theta_i\theta_j}-\frac{\hat{P}^{(K')}_{ij}}{\theta_i\theta_j}\right)^2\theta_i^2\theta_j^2-\left(\frac{A_{ij}}{\theta_i\theta_j}-B_{l_1l_2}\right)^2\theta_i^2\theta_j^2\right]\\
		\geq&\sum_{k_1,k_2,l_1,l_2}\sum_{(i,j)\in\mathcal{T}_{k_1,k_2,l_1,l_2}}\left[\left(\frac{A_{ij}}{\theta_i\theta_j}-\hat{p}_{k_1k_2l_1l_2}\right)^2\theta_i^2\theta_j^2-\left(\frac{A_{ij}}{\theta_i\theta_j}-B_{l_1l_2}\right)^2\theta_i^2\theta_j^2\right]\\
		&-2\sum_{k_1, k_2, l_1, l_2}\sum_{(i,j) \in \mathcal{T}_{k_1, k_2, l_1, l_2}}\left(\frac{A_{ij}}{\theta_i\theta_j}-\hat{p}_{k_1,k_2,l_1,l_2}\right)\hat{P}^{(K')}_{ij}\theta_i\theta_j\\
		=&-\sum_{k_1,k_2,l_1,l_2}\Theta_{k_1,k_2,l_1,l_2}\left(\hat{p}_{k_1k_2l_1l_2}-B_{l_1l_2}\right)^2\\
		&-2\sum_{k_1, k_2, l_1, l_2}\sum_{(i,j) \in \mathcal{T}_{k_1, k_2, l_1, l_2}}\left(\frac{A_{ij}}{\theta_i\theta_j}-\hat{p}_{k_1,k_2,l_1,l_2}\right)\hat{P}^{(K')}_{ij}\theta_i\theta_j\\
		:=&-\mathcal{I}-\mathcal{II}.
	\end{align*}
	Using the same arguments as those used to control of term $\mathcal{III}$ and $\mathcal{IV}$ in the proof of Proposition \ref{prop_dcbmunder}, we have $\mathcal{I}=O_{\mathbb{P}}(\rho_n)$ and $\mathcal{II}=O_{\mathbb{P}}\left(\sqrt{n^2\rho_n(1-w)}\right)$, thus the conclusion holds.
\end{proof}

\begin{proposition}\label{prop_dcbmover2}
	When $K'>2K^*$, we have $$\mathbb{P}\left[\ell_{K'}(A,\mathcal{E}^c)-\ell_{0}(A,\mathcal{E}^c)\leq-2(K^*)^2K'^2\log n-\frac{8\theta_{\max}^2}{\theta_{\min}^2}nK^*K'\sqrt{\log n(1-w)} \right]\leq\frac{8(K^*)^2}{n^2}.$$
\end{proposition}
\begin{proof}[Proof of Proposition \ref{prop_dcbmover2}]
	Use a similar argument as above, we have
	\begin{align*}
		&\ell_{K'}(A,\mathcal{E}^c)-\ell_{0}(A,\mathcal{E}^c)\\
		\geq&-\sum_{k_1,k_2,l_1,l_2}\Theta_{k_1,k_2,l_1,l_2}\left(\hat{p}_{k_1k_2l_1l_2}-B_{l_1l_2}\right)^2
		-2\sum_{k_1, k_2, l_1, l_2}\sum_{(i,j) \in \mathcal{T}_{k_1, k_2, l_1, l_2}}\left(A_{ij}\hat{P}^{(K')}_{ij}-\mu_{k_1,k_2,l_1,l_2}\right)\\
		&-2\sum_{k_1, k_2, l_1, l_2}\sum_{(i,j) \in \mathcal{T}_{k_1, k_2, l_1, l_2}}\hat{P}^{(K')}_{ij}\theta_i\theta_j\left(\hat{p}_{k_1,k_2,l_1,l_2}-B_{l_1l_2}\right)\\
		:=&-\mathcal{I}-\mathcal{II}-\mathcal{III}.
	\end{align*}
	Similarly we shall condition on the training entries and we need a more precise bound of tail probability since $K'$ may tend to $n$. For each $(k_1,k_2,l_1,l_2)$, by Hoeffding inequality, we have
	$$\mathbb{P}\left[\left.\left|\sum_{(i,j) \in \mathcal{T}_{k_1, k_2, l_1, l_2}}A_{ij}\theta_i\theta_j-B_{kl}\Theta_{k_1,k_2,l_1,l_2}\right|\geq\sqrt{2\Theta_{k_1,k_2,l_1,l_2}\log n}\right|\mathfrak{F}_{\mathcal{E}}\right]\leq2\exp\left(-\frac{4\Theta_{k_1,k_2,l_1,l_2}\log n}{\sum_{(i,j) \in \mathcal{T}_{k_1, k_2, l_1, l_2}}\theta_i^2\theta_j^2}\right)=2n^{-4},$$
	which results in
	\begin{equation}\label{eqn_hoef1}
		\mathbb{P}\left[\left.\left|\hat{p}_{k_1,k_2,l_1,l_2}-B_{kl}\right|\geq\sqrt{\frac{2\log n}{\Theta_{k_1,k_2,l_1,l_2}}}\right|\mathfrak{F}_{\mathcal{E}}\right]\leq2n^{-4},\quad\text{ for all }(k_1,k_2,l_1,l_2).
	\end{equation}
	Therefore, by pigeonhole principle and a union bound, we have
	\begin{equation}\label{eqn_probI}
		\mathbb{P}\left[\mathcal{I}\geq2(K^*)^2K'^2\log n\right]\leq\sum_{k_1,k_2,l_1,l_2}\mathbb{P}\left[\Theta_{k_1,k_2,l_1,l_2}\left(\hat{p}_{k_1k_2l_1l_2}-B_{l_1l_2}\right)^2\geq2\log n\right]
		\leq 2(K^*)^2K'^2n^{-4}\leq 2(K^*)^2n^{-2}.
	\end{equation}
	Furthermore, we have
	\begin{align*}
		&\mathbb{P}\left[\mathcal{III}\geq\frac{4\theta_{\max}^2}{\theta_{\min}^2}nK^*K'\sqrt{\log n(1-w)}\right]\\
		\leq&\mathbb{P}\left[\sum_{k_1, k_2, l_1, l_2}\sum_{(i,j) \in \mathcal{T}_{k_1,k_2,l_1,l_2}}\theta_i\theta_j\left|\hat{p}_{k_1,k_2,l_1,l_2}-B_{kl}\right|\geq\frac{2\theta_{\max}^2}{\theta_{\min}^2}nK^*K'\sqrt{\log n(1-w)}\right].
	\end{align*}
	Define $\Theta_{k_1,k_2,l_1,l_2}'=\sum_{(i,j) \in \mathcal{T}_{k_1,k_2,l_1,l_2}}\theta_i\theta_j$. Notice that 
	\begin{align*}
		\sum_{k_1,k_2,l_1,l_2}\Theta_{k_1,k_2,l_1,l_2}'\sqrt{\frac{2\log n}{\Theta_{k_1,k_2,l_1,l_2}}}\leq&\sum_{k_1,k_2,l_1,l_2} \theta_{\max}^2|\mathcal{T}_{k_1,k_2,l_1,l_2}|\sqrt{\frac{2\log n}{\theta_{\min}^4|\mathcal{T}_{k_1,k_2,l_1,l_2}|}}\\
		=&\frac{\sqrt{2\log n}\theta_{\max}^2}{\theta_{\min}^2}\sum_{k_1,k_2,l_1,l_2}\sqrt{|\mathcal{T}_{k_1,k_2,l_1,l_2}|}\\
		(\text{Cauchy-Schwarz})\leq&\frac{\sqrt{2\log n}\theta_{\max}^2}{\theta_{\min}^2}\sqrt{\sum_{k_1,k_2,l_1,l_2}|\mathcal{T}_{k_1,k_2,l_1,l_2}|}\sqrt{\sum_{k_1,k_2,l_1,l_2}1}\\
		\overset{(*)}{\leq}&\frac{\sqrt{2\log n}\theta_{\max}^2}{\theta_{\min}^2}\sqrt{2n^2(1-w)}\sqrt{(K^*)^2K'^2}\\
		=&\frac{2\theta_{\max}^2}{\theta_{\min}^2}nK^*K'\sqrt{\log n(1-w)},
	\end{align*}
	where by Hoeffding inequality $(*)$ holds with probability larger than $1-n^{-2}$. Thus, by \eqref{eqn_hoef1} again,
	\begin{align}
		&\mathbb{P}\left[\mathcal{III}\geq\frac{4\theta_{\max}^2}{\theta_{\min}^2}nK^*K'\sqrt{\log n(1-w)}\right]\nonumber\\
		\leq&n^{-2}+\sum_{k_1,k_2,l_1,l_2}\mathbb{P}\left[\Theta_{k_1,k_2,l_1,l_2}'\left|\hat{p}_{k_1,k_2,l_1,l_2}-B_{kl}\right|\geq\Theta_{k_1,k_2,l_1,l_2}'\sqrt{\frac{2\log n}{\Theta_{k_1,k_2,l_1,l_2}}}\right]\nonumber\\
		\leq&n^{-2}+2(K^*)^2K'^2n^{-4}\leq 3(K^*)^2n^{-2}.\label{eqn_probIII}
	\end{align}
	
	It remains to bound $\mathcal{II}$. Still by Hoeffding's inequality, we have conditioned on training entries,  
	\begin{align*}
		\mathbb{P}\left[\left.\left|\sum_{(i,j) \in \mathcal{T}_{k_1,k_2,l_1,l_2}}A_{ij}\hat{P}_{ij}^{(K')}-\mu_{k_1,k_2,l_1,l_2}\right|\geq\sqrt{2|\mathcal{T}_{k_1,k_2,l_1,l_2}|\log n}\right|\mathfrak{F}_{\mathcal{E}}\right]\leq&2\exp\left(-\frac{4|\mathcal{T}_{k_1,k_2,l_1,l_2}|\log n}{\sum_{(i,j) \in \mathcal{T}_{k_1,k_2,l_1,l_2}}\hat{P}^{(K')}_{ij}}\right)\\
		\leq&2n^{-4}.
	\end{align*}
	Therefore, by Cauchy-schwarz inequality again, we have
	\begin{align}
		&\mathbb{P}\left[\mathcal{II}\geq4nK^*K'\sqrt{\log n(1-w)}\right]\nonumber\\
		\leq&n^{-2}+\sum_{k_1,k_2,l_1,l_2}\mathbb{P}\left[\left|\sum_{(i,j) \in \mathcal{T}_{k_1,k_2,l_1,l_2}}A_{ij}\hat{P}_{ij}^{(K')}-\mu_{k_1,k_2,l_1,l_2}\right|\geq\sqrt{2|\mathcal{T}_{k_1,k_2,l_1,l_2}|\log n}\right]\nonumber\\
		\leq&n^{-2}+2(K^*)^2K'^2n^{-4}\leq 3(K^*)^2n^{-2}.\label{eqn_probII}
	\end{align}
	Combining \eqref{eqn_probI}, \eqref{eqn_probIII} and \eqref{eqn_probII}, we get the desired result.
\end{proof}

\begin{proof}[Proof of Theorem 2]
	Similarly to the proof of Theorem 1, we just need to prove the consistency conditioned on $\Omega$. We just need to prove that, 
	\begin{equation}
		\mathbb{P}\left[L_{K^*}(A,\mathcal{E}^c)>L_{K'}(A,\mathcal{E}^c)\right]\to0,\quad\forall\ K'\leq 2K^*\text{ and }K'\neq K^*, \label{dcbm_cons_bounded}
	\end{equation}
	and
	\begin{equation}
		\sum_{K'>2K^*}\mathbb{P}\left[L_{K^*}(A,\mathcal{E}^c)>L_{K'}(A,\mathcal{E}^c)\right]\to0. \label{dcbm_cons_unbounded}
	\end{equation}
	In Propositions \ref{prop_DCBMupper} -- \ref{prop_dcbmover1}, we proved that 
	\begin{equation}
		\ell_{K^*}(A, \mathcal{E}^c) - \ell_0(A, \mathcal{E}^c) =O_{\mathbb{P}}\left(\max\left\{\frac{n^{3/2}\rho_n}{w},\sqrt{\frac{n^3\log n\rho_n^3}{w}}\right\}(1-w)\right),\label{dcbm_bound1}
	\end{equation}
	for all $K'<K^*$,
	\begin{equation}
		\ell_{K'}(A, \mathcal{E}^c) - \ell_0(A, \mathcal{E}^c) =\Omega_{\mathbb{P}}\left(n^2\rho_n^2(1-w)\right),\label{dcbm_bound2}
	\end{equation}
	and for all $K^*<K'\leq 2K^*$,
	\begin{equation}
		\ell_{K'}(A, \mathcal{E}^c) - \ell_0(A, \mathcal{E}^c) \geq-O_{\mathbb{P}}\left(\sqrt{n^2\rho_n(1-w)}\right).\label{dcbm_bound3}
	\end{equation}
	
	We first prove \eqref{dcbm_cons_bounded}. When $K^*<K'\leq 2K^*$, notice that
	\begin{align*}
		&\mathbb{P}\left[L_{K^*}(A,\mathcal{E}^c)>L_{K'}(A,\mathcal{E}^c)\right]=\mathbb{P}[-\ell_{K',K^*}(A,\mathcal{E}^c)>(d_{K'}-d_{K^*})|\mathcal{E}^c|\lambda_n]\\
		\leq&\mathbb{P}\left[\left(\ell_{K^*}(A,\mathcal{E}^c)-\ell_{0}(A,\mathcal{E}^c)\right)-\left(\ell_{K'}(A,\mathcal{E}^c)-\ell_{0}(A,\mathcal{E}^c)\right)>1/4(K'(K'+3)-K^*(K^*+3))n(n-1)(1-w)\lambda_n\right]\to0
	\end{align*}
	by \eqref{dcbm_bound1}, \eqref{dcbm_bound3} and the assumption $\lambda_n=\omega_{\mathbb{P}}\left(\max\left\{\rho_n/wn^{1/2},\sqrt{\rho_n^3\log n/wn}\right\}\right)$.
	Similarly \eqref{dcbm_cons_bounded} holds when $K'<K^*$, by \eqref{dcbm_bound1}, \eqref{dcbm_bound2} and $\lambda_n=o_{\mathbb{P}}(\rho_n^2)$.
	
	Now we prove \eqref{dcbm_cons_unbounded}. Since the true $K^*$ is fixed, for any $\delta>0$, there exists an event $\Omega_0(\delta)$, such that with $n$ sufficiently large, this event has probability larger than $1-\delta$, and there exist constant $C_1(\delta)$ such that on this event, 
	\begin{equation}
		\ell_{K^*}(A, \mathcal{E}^c) - \ell_0(A, \mathcal{E}^c) \leq C_1(\delta)\max\left\{\frac{n^{3/2}\rho_n}{w},\sqrt{\frac{n^3\log n\rho_n^3}{w}}\right\}(1-w).\label{dcbm_bound4}
	\end{equation}
	Thus when $K'>2K^*$, we have
	\begin{align*}
		&\mathbb{P}\left[L_{K^*}(A,\mathcal{E}^c)>L_{K'}(A,\mathcal{E}^c)\right]\\
		\leq&\mathbb{P}\left[\left(\ell_{K^*}(A,\mathcal{E}^c)-\ell_{0}(A,\mathcal{E}^c)\right)-\left(\ell_{K'}(A,\mathcal{E}^c)-\ell_{0}(A,\mathcal{E}^c)\right)>1/4(K'(K'+3)-K^*(K^*+3))n(n-1)(1-w)\lambda_n\right]\\
		\leq&\mathbb{P}\left[\left(\ell_{K^*}(A,\mathcal{E}^c)-\ell_{0}(A,\mathcal{E}^c)\right)-\left(\ell_{K'}(A,\mathcal{E}^c)-\ell_{0}(A,\mathcal{E}^c)\right)>1/8K'^2n(n-1)(1-w)\lambda_n\right]\\
		\overset{(*)}{\leq}&\mathbb{P}\left[C_1(\delta)\max\left\{\frac{n^{3/2}\rho_n}{w},\sqrt{\frac{n^3\log n\rho_n^3}{w}}\right\}(1-w)+2(K^*)^2K'^2\log n+\frac{8\theta_{\max}^2}{\theta_{\min}^2}nK^*K'\sqrt{\log n(1-w)}\right.\\
		&\left.>1/8K'^2n(n-1)(1-w)\lambda_n\right]+\frac{8(K^*)^2}{n^2}\\
		\leq&\mathbb{P}\left[C_1(\delta)\max\left\{\frac{n^{3/2}\rho_n}{w},\sqrt{\frac{n^3\log n\rho_n^3}{w}}\right\}(1-w)>1/24(K^*)^2n(n-1)(1-w)\lambda_n\right]\\
		&+\mathbb{P}\left[2(K^*)^2K'^2\log n>1/24K'^2n(n-1)(1-w)\lambda_n\right]\\
		&+\mathbb{P}\left[\frac{8\theta_{\max}^2}{\theta_{\min}^2}nK^*K'\sqrt{\log n(1-w)}>1/24K^*K'n(n-1)(1-w)\lambda_n\right]+\frac{8(K^*)^2}{n^2},
	\end{align*}
	where (*) comes from \eqref{dcbm_bound4} and Proposition \ref{prop_dcbmover2}. Notice that for any $C>0$, we have $$\mathbb{P}\left[\lambda_n<\max\left\{\frac{\rho_n}{w\sqrt{n}},\sqrt{\frac{\rho_n^3\log n}{wn}}\right\}\right]\ll1/n,$$ then as $n\to\infty$, $$\mathbb{P}\left[C_1(\delta)\max\left\{\frac{n^{3/2}\rho_n}{w},\sqrt{\frac{n^3\log n\rho_n^3}{w}}\right\}(1-w)>1/24(K^*)^2n(n-1)(1-w)\lambda_n\right]\ll\frac{1}{n},$$
	thus
	$$\sum_{K'>2K}\mathbb{P}\left[C_1(\delta)\max\left\{\frac{n^{3/2}\rho_n}{w},\sqrt{\frac{n^3\log n\rho_n^3}{w}}\right\}(1-w)>1/24(K^*)^2n(n-1)(1-w)\lambda_n\right]=o(1).$$
	Similarly, since for any $C>0$, we have $\mathbb{P}\left[\lambda_n<C\sqrt{\frac{\log n}{n^2(1-w)}}\right]\ll1/n$, then as $n\to\infty$,
	$$\sum_{K'>2K}\mathbb{P}\left[2(K^*)^2K'^2\log n>1/24K'^2n(n-1)(1-w)\lambda_n\right]=o(1),$$
	and
	$$\sum_{K'>2K}\mathbb{P}\left[\frac{8\theta_{\max}^2}{\theta_{\min}^2}nK^*K'\sqrt{\log n(1-w)}>1/24K^*K'n(n-1)(1-w)\lambda_n\right]=o(1).$$
	Therefore, as $n\to\infty$,
	$$\sum_{K'>2K^*}\mathbb{P}\left[L_{K^*}(A,\mathcal{E}^c)>L_{K'}(A,\mathcal{E}^c)\right]\leq o(1)+o(1)+o(1)+\sum_{K'>2K^*}\frac{8(K^*)^2}{n^2}\leq o(1)+\frac{8(K^*)^2}{n}\to0,$$
	and \eqref{dcbm_cons_unbounded} is proved conditioned on $\Omega\cap\Omega_0(\delta)$. Therefore for all $\delta>0$, when n is sufficiently large,
	$$\mathbb{P}[\hat{K}=K^*]>1-\delta.$$
	Let $\delta\to0$, and thus Theorem 2 holds.
\end{proof}

\subsection{Proof of Theorem 3}

By Theorem 1, $\hat{K}$ obtained by Algorithm 2 is consistent, so we may condition on the event that $\hat{K}=K^*$. We just need to prove that when $\hat{K}=K^*$, Algorithm 4 can choose the right model with probability tending to 1.

Denote 
$$\ell_{1,2}(A,\mathcal{E}^c):=\ell_{1,K^*}(A,\mathcal{E}^c)-\ell_{2,K^*}(A,\mathcal{E}^c),$$
where 
$$\ell_{r,K^*}(A,\mathcal{E}^c):=\sum_{(i,j)\in\mathcal{E}^c}\left(A_{ij}-\hat{P}^{(r,K^*)}_{ij}\right)^2 \text{for }r\in\{1,2\}.$$

\begin{proposition}\label{propa3}
	If the true model is $\delta^{(1,K^*)}$, then $$|\ell_{1,2}(A,\mathcal{E}^c)|\leq O_{\mathbb{P}}\left(\frac{n\rho_n(1-w)}{w}+\frac{1-w}{w^2}\right).$$
	Therefore, under $n\rho_nw\to\infty$, we have $|\ell_{1,2}(A,\mathcal{E}^c)|\leq O_{\mathbb{P}}\left(\frac{n\rho_n(1-w)}{w}\right)$.
\end{proposition}

\begin{proof}[Proof of Proposition \ref{propa3}]
	Since the model $\delta^{(1,K^*)}$ is a special case of the SBM with $K^*$ communities $\delta^{(2,K^*)}$, the result of the upper bound of Step 1 in Propostion \ref{propa1} holds in this situation, that is,
	$$|\ell_{2,K^*}(A,\mathcal{E}^c)-\ell_0(A,\mathcal{E}^c)|\leq O_{\mathbb{P}}\left(\frac{n\rho_n(1-w)}{w}+\frac{1-w}{w^2}\right),$$
	where $\ell_0(A,\mathcal{E}^c)$ is the oracle loss.
	
	Now we give an upper bound for $\ell_{1,K^*}(A,\mathcal{E}^c)-\ell_{0}(A,\mathcal{E}^c)$. We use the same notation $\mathcal{Q}_{k_1,k_2,l_1,l_2}$, $\mathcal{U}_{k_1,k_2,l_1,l_2}$ and $\mathcal{T}_{k_1,k_2,l_1,l_2}$ as in the proof of Proposition \ref{propa1}. Define $\mathcal{Q}_{1,1}:=\{(i,j):c_i=c_j,\hat{c}_i=\hat{c}_j\}$, $\mathcal{Q}_{1,2}:=\{(i,j):c_i\neq c_j,\hat{c}_i=\hat{c}_j\}$, $\mathcal{Q}_{2,1}:=\{(i,j):c_i=c_j,\hat{c}_i\neq\hat{c}_j\}$ and $\mathcal{Q}_{2,2}:=\{(i,j):c_i\neq c_j,\hat{c}_i\neq\hat{c}_j\}$, and define $\mathcal{U}_{m,n}$, $\mathcal{T}_{m,n}$ for $m,n\in\{1,2\}$ similarly. Define $\mathcal{Q}_{\cdot,1}:=\mathcal{Q}_{1,1}\cup\mathcal{Q}_{2,1}$, and define the notations for $\mathcal{U}$ and $\mathcal{T}$ similarly. Since the affiliation model is a special case of the SBM, Lemma \ref{lem2} still holds in this situation, and thus \eqref{propa1UT} holds. Notice that 
	\begin{align*}
		\mathcal{U}_{\cdot,1}\Delta\mathcal{U}_{1,\cdot}&=
		\mathcal{U}_{1,2}\cup\mathcal{U}_{2,1}=\left(\bigcup_k(\mathcal{U}_{k,k,\cdot,\cdot}/\mathcal{U}_{k,k,k,k})\right)\bigcup\left(\bigcup_k(\mathcal{U}_{\cdot,\cdot,k,k}/\mathcal{U}_{k,k,k,k})\right)\\
		&= \bigcup_k\left((\mathcal{U}_{k,k,\cdot,\cdot}/\mathcal{U}_{k,k,k,k})\cup(\mathcal{U}_{\cdot,\cdot,k,k}/\mathcal{U}_{k,k,k,k})\right)=\bigcup_k\left(\mathcal{U}_{k,k,\cdot,\cdot}\Delta\mathcal{U}_{\cdot,\cdot,k,k}\right),
	\end{align*}
	thus by \eqref{propa1UT}, $|\mathcal{U}_{\cdot,1}\Delta\mathcal{U}_{1,\cdot}|=O_{\mathbb{P}}(n/\rho_n)$. Similarly, we have $\mathcal{U}_{\cdot,2}\Delta\mathcal{U}_{2,\cdot}=\bigcup_k\left(\mathcal{U}_{k,k,\cdot,\cdot}\Delta\mathcal{U}_{\cdot,\cdot,k,k}\right)$ and $|\mathcal{U}_{\cdot,2}\Delta\mathcal{U}_{2,\cdot}|=O_{\mathbb{P}}(n/\rho_n)$. A similar inequality for the test set also holds. By Bernstein's inequality, we have
	\begin{align*}
		|\hat{p}_{\mathrm{wc}}-p_{\mathrm{wc}}|=&|\frac{\sum_{\mathcal{U}_{1,\cdot}}A_{ij}}{|\mathcal{U}_{1,\cdot}|}-p_{\mathrm{wc}}|\nonumber\\
		\leq&\frac{|\mathcal{U}_{\cdot,1}|}{|\mathcal{U}_{1,\cdot}|}\left|\frac{\sum_{\mathcal{U}_{\cdot,1}}A_{ij}}{|\mathcal{U}_{\cdot,1}|}-p_{\mathrm{wc}}\right|+\left|1-\frac{|\mathcal{U}_{\cdot,1}|}{|\mathcal{U}_{1,\cdot}|}\right|p_{\mathrm{wc}}\nonumber\\
		&+\frac{|\mathcal{U}_{\cdot,1}\Delta \mathcal{U}_{1,\cdot}|}{|\mathcal{U}_{1,\cdot}|}\frac{\sum_{\mathcal{U}_{\cdot,1}\Delta \mathcal{U}_{1,\cdot}}A_{ij}}{|\mathcal{U}_{\cdot,1}\Delta \mathcal{U}_{1,\cdot}|}\nonumber\\
		\leq& O_{\mathbb{P}}\left(\sqrt{\frac{\rho_n}{n^2w}}\right)+O_{\mathbb{P}}\left(\frac{1}{n\rho_nw}\right)\rho_n+O_{\mathbb{P}}\left(\frac{1}{n\rho_nw}\right)\rho_n=O_{\mathbb{P}}\left(\frac{1}{nw}\right).
	\end{align*}
	Similarly, one can obtain that $|\hat{p}_{\mathrm{bc}}-p_{\mathrm{bc}}|=O_{\mathbb{P}}(1/nw)$, and thus following the procedure in the proof of Proposition \ref{propa1}, we can get 
	$$\ell_{1,K^*}(A,\mathcal{E}^c)-\ell_{0}(A,\mathcal{E}^c)=O_{\mathbb{P}}\left(\frac{n\rho_n(1-w)}{w}+\frac{1-w}{w^2}\right).$$
	
	Then
	\begin{align*}
		\ell_{1,2}(A,\mathcal{E}^c)=&\left(\ell_{1,K^*}(A,\mathcal{E}^c)-\ell_{0}(A,\mathcal{E}^c)\right)-\left(\ell_{2,K^*}(A,\mathcal{E}^c)-\ell_{0}(A,\mathcal{E}^c)\right)\\
		\leq&O_{\mathbb{P}}\left(\frac{n\rho_n(1-w)}{w}+\frac{1-w}{w^2}\right)+O_{\mathbb{P}}\left(\frac{n\rho_n(1-w)}{w}+\frac{1-w}{w^2}\right)=O_{\mathbb{P}}\left(\frac{n\rho_n(1-w)}{w}+\frac{1-w}{w^2}\right).
	\end{align*}
\end{proof}

\begin{proposition}\label{propa4}
	When model $\delta^{(2,K^*)}$ is true and model $\delta^{(1,K^*)}$ is false, i.e., the model is SBM with $K^*$ communities but not an affiliation model, we have $$\ell_{1,2}(A,\mathcal{E}^c)\geq\Omega_{\mathbb{P}}\left(n^2\rho_n^2(1-w)\right)-O_{\mathbb{P}}\left(\frac{n\rho_n(1-w)}{w}+\frac{1-w}{w^2}\right).$$
	Therefore as long as $n\rho_nw\rightarrow\infty$, we have $\ell_{1,2}(A,\mathcal{E}^c)=\Omega_{\mathbb{P}}\left(n^2\rho_n^2(1-w)\right)$.
\end{proposition}

\begin{proof}[Proof of Proposition \ref{propa4}]
	Define the following sets analogously to those in Proposition \ref{propa1}:
	$$\mathcal{T}_{1,l_1,l_2}=\{(i,j)\in\mathcal{E}^c:c_i=l_1,c_j=l_2,\hat{c}_i=\hat{c}_j\}$$
	$$\mathcal{T}_{2,l_1,l_2}=\{(i,j)\in\mathcal{E}^c:c_i=l_1,c_j=l_2,\hat{c}_i\neq\hat{c}_j\},$$
	where $c$ is the true label and $\hat{c}$ is the estimated label from the algorithm. By Step 1 of the proof of Proposition \ref{propa1}, we have 
	$$\ell_{2,K^*}(A,\mathcal{E}^c)-\ell_0(A,\mathcal{E}^c)\leq O_{\mathbb{P}}\left(\frac{n\rho_n(1-w)}{w}+\frac{1-w}{w^2}\right).$$
	Now, since model 1 is false, either there exists $i,j,k$ distinct such that $B_{ij}\neq B_{ik}$ or there exists $i\neq j$ such that $B_{ii}\neq B_{jj}$. We first consider the case that the former is true. 
	
	Without loss of generality, assume that $B_{12}\neq B_{13}$. By Lemma \ref{lem2}, we have that with probability tending to 1,
	$$|\mathcal{T}_{l_1,l_2,l_1,l_2}|\geq\frac{1-w}{2}|G_{l_1}\times G_{l_2}|-O_P\left(\frac{n}{\rho_nw}\right)\geq c(1-w)n^2$$
	by the assumption of balanced community structure. Therefore, for any $l_1\neq l_2$, we have $|T_{2,l_1,l_2}|\geq|T_{l_1l_2l_1l_2}|\geq c(1-w)n^2$. Now, we have
	\begin{align*}
		&\ell_{1,K^*}(A,\mathcal{E}^c)-\ell_0(A,\mathcal{E}^c)\\
		=&\sum_{k_1,k_2,l_1,l_2}\sum_{(i,j)\in T_{k_1,k_2,l_1,l_2}}[(A_{ij}-\hat{P}^{(1)}_{ij})^2-(A_{ij}-B_{l_1l_2})^2]\\
		=&\sum_{(i,j)\in T_{2,1,2}}[(A_{ij}-\hat{P}^{(1)}_{ij})^2-(A_{ij}-B_{12})^2]+\sum_{(i,j)\in T_{2,1,3}}[(A_{ij}-\hat{P}^{(1)}_{ij})^2-(A_{ij}-B_{13})^2]\\
		&+\sum_{other\,(k_1,k_2,l_1,l_2)}\sum_{(i,j)\in T_{k_1,k_2,l_1,l_2}}[(A_{ij}-\hat{M}^{(1)}_{ij})^2-(A_{ij}-B_{l_1l_2})^2].
	\end{align*}
	Notice for any index set $T$, $\sum_{(i,j)\in T}(A_{ij}-p)^2$ is minimized at $p=\frac{1}{|T|}\sum_{(i,j)\in T}A_{ij}$. Therefore, we get
	\begin{align*}
		&\ell_{1,K^*}(A,\mathcal{E}^c)-\ell_0(A,\mathcal{E}^c)\\
		\geq&\sum_{(i,j)\in T_{2,1,2}}[(A_{ij}-\hat{p})^2-(A_{ij}-B_{12})^2]+\sum_{(i,j)\in T_{2,1,3}}[(A_{ij}-\hat{p})^2-(A_{ij}-B_{13})^2]\\
		&+\sum_{other\,(k_1,k_2,l_1,l_2)}\sum_{(i,j)\in T_{k_1,k_2,l_1,l_2}}[(A_{ij}-\hat{p}_{k_1,k_2,l_1,l_2})^2-(A_{ij}-B_{l_1l_2})^2]\\
		:=&\mathcal{I}+\mathcal{II}+\mathcal{III},
	\end{align*}
	where $\hat{p}$ is the average of $A_{ij}$ over $T_{2,1,2}\cup T_{2,1,3}$ and $\hat{p}_{k_1,k_2,l_1,l_2}$ is the average of $A_{ij}$ over $T_{k_1,k_2,l_1,l_2}$. Define $f(p)=\sum_{T_{2,1,2}}(A_{ij}-p)^2$. Notice $\hat{p}=t\hat{p}_1+(1-t)\hat{p}_2$, where $\hat{p}_1=\hat{p}_{2,1,2}$ and $\hat{p}_2=\hat{p}_{2,1,3}$, with $t=\frac{|\mathcal{T}_{2,1,2}|}{|\mathcal{T}_{2,1,2}|+|\mathcal{T}_{2,1,3}|}.$ We have $|\mathcal{T}_{2,1,2}|\sim|\mathcal{T}_{2,1,3}|$, so $t$ and $1-t$ are of constant order. Denote correspondingly $p_1=B_{12}$, $p_2=B_{13}$. Then
	$$\mathcal{I}=f(t\hat{p}_1+(1-t)\hat{p}_2)-f(p_1)\geq|\mathcal{T}_{2,1,2}|(1-t)^2|\hat{p}_1-\hat{p}_2|^2+f(\hat{p}_1)-f(p_1).$$
	By the above argument, $|\mathcal{T}_{2,1,2}|,|\mathcal{T}_{2,1,3}|\geq cn^2(1-w)$, so by Bernstein's inequality, 
	$$|\hat{p}_1-\hat{p}_2|\geq|p_1-p_2|-|\hat{p}_1-p_1|-|\hat{p}_2-p_2|\geq|p_1-p_2|-O_{\mathbb{P}}\left(\sqrt{\frac{\rho_n}{n^2(1-w)}}\right)\geq c_1\rho_n.$$
	We still need a lower bound for $f(\hat{p}_1)-f(p_1)$. Notice here we have concentration bound $|\hat{p}_1-p_1|=O_{\mathbb{P}}\left(\sqrt{\frac{\rho_n}{n^2(1-w)}}\right)$ by Bernstein's inequality, we have
	\begin{align*}
		|f(\hat{p}_1)-f(p_1)|=&\sum_{(i,j)\in\mathcal{T}_{2,1,2}}\left[\left(A_{ij}-\hat{p}_1\right)^2-\left(A_{ij}-p_1\right)^2\right]\\
		\leq&2\sum_{(i,j)\in\mathcal{T}_{2,1,2}}\left|\hat{p}_1-p_1\right|\left(A_{ij}+p_1+\left|\hat{p}_1-p_1\right|\right)\\
		\leq&O_{\mathbb{P}}\left(\sqrt{\frac{\rho_n}{n^2(1-w)}}\right)n^2\rho_n(1-w)+O_{\mathbb{P}}\left(\frac{\rho_n}{n^2(1-w)}\right)n^2(1-w)\\
		=&O_{\mathbb{P}}\left(n\rho_n^{3/2}(1-w)^{1/2}\right).
	\end{align*}
	Thus we get
	$$\mathcal{I}=f(t\hat{p}_1+(1-t)\hat{p}_2)-f(p_1)=\Omega_{\mathbb{P}}\left(n^2\rho_n^2(1-w)\right).$$
	Similarly $\mathcal{II}=\Omega_{\mathbb{P}}\left(n^2\rho_n^2(1-w)\right)$. The remaining term $\mathcal{III}$ is negative and we have
	$$\mathcal{III}=-\sum_{other\,(k_1,k_2,l_1,l_2)}|\mathcal{T}_{k_1, k_2, l_1, l_2}|(\hat{p}_{k_1, k_2, l_1, l_2}-B_{l_1l_2})^2\geq-O_{\mathbb{P}}(\rho_n).$$
	Combining the bounds, the desired result follows.
	
	For the second case, i.e., there exists $i\neq j$ such that $B_{ii}\neq B_{jj}$, the proof follows analogously by comparing two within-community blocks, both of which are fitted by the common parameter
	$\widehat p_{wc}$.
\end{proof}

\begin{proof}[Proof of Theorem 3]
	Similarly to the proof of Theorem 1, we just need to prove the consistency conditioned on $\Omega$.
	We need to prove that
	\begin{equation}
		\mathbb{P}\left[L_{1,K^*}(A,\mathcal{E}^c)>L_{2,K^*}(A,\mathcal{E}^c)\right]\to0,\quad\text{when $\delta^{(1,K^*)}$ is true}, \label{theo2_cons1}
	\end{equation}
	and 
	\begin{equation}
		\mathbb{P}\left[L_{1,K^*}(A,\mathcal{E}^c)<L_{2,K^*}(A,\mathcal{E}^c)\right]\to0,\quad\text{when $\delta^{(2,K^*)}$ is true}. \label{theo2_cons2}
	\end{equation}
	When $\delta^{(1,K)}$ is true, notice that
	\begin{align*}
		\mathbb{P}\left[L_{1,K^*}(A,\mathcal{E}^c)>L_{2,K^*}(A,\mathcal{E}^c)\right]=&\mathbb{P}[\ell_{1,2}(A,\mathcal{E}^c)>(d_{2,K^*}-d_{1,K^*})|\mathcal{E}^c|\lambda_n]\\
		\leq&\mathbb{P}[\ell_{1,2}(A,\mathcal{E}^c)>1/2(d_{2,K^*}-d_{1,K^*})n(n-1)(1-w)\lambda_n]\to0
	\end{align*}
	by $d_{2,K^*}-d_{1,K^*}>0$ being of constant order, $\lambda_n=\omega_{\mathbb{P}}(\frac{\rho_n}{nw})$ and Proposition \ref{propa3}.
	Similarly \eqref{theo2_cons2} holds when $\delta^{(2,K^*)}$ is true, by $\lambda_n=o(\rho_n^2)$ and Proposition \ref{propa4}.
\end{proof}

\subsection{Proof of Theorem 4}
Still we can condition on $\hat{K}=K^*$ by the consistency result established in Theorem 2. Denote 
$$\ell_{1,2}(A,\mathcal{E}^c)=\ell_{1,K^*}(A,\mathcal{E}^c)-\ell_{2,K^*}(A,\mathcal{E}^c)$$
as in the proof of Theorem 2.
Using Proposition \ref{prop_DCBMupper}, a direct consequence is 
\begin{proposition}\label{propa6}
	If the true model is $\delta^{(1,K^*)}$, then 
	$$\ell_{1,2}(A,\mathcal{E}^c)\leq O_{\mathbb{P}}\left(\max\left\{\frac{n^{3/2}\rho_n}{w},\sqrt{\frac{n^3\log n\rho_n^3}{w}}\right\}(1-w)\right)$$
	as long as the assumptions for Theorem 4 hold.
\end{proposition}
\begin{proof}[Proof of Proposition \ref{propa6}]
	Notice $\ell_{1,2}(A,\mathcal{E}^c)=(\ell_{1,K^*}(A,\mathcal{E}^c)-\ell_0(A,\mathcal{E}^c))-(\ell_{2,K^*}(A,\mathcal{E}^c)-\ell_0(A,\mathcal{E}^c))$. By the proof of Proposition \ref{propa1}, we know that
	$$\ell_{1,K^*}(A,\mathcal{E}^c)-\ell_0(A,\mathcal{E}^c)\leq O_{\mathbb{P}}\left(\frac{n\rho_n(1-w)}{w}\right).$$
	
	Since SBM is a special case of DCBM with $\theta_i\equiv1$, Proposition \ref{prop_DCBMupper} holds in this framework. That is,
	$$\ell_{2,K^*}(A,\mathcal{E}^c)-\ell_0(A,\mathcal{E}^c)\geq -O_{\mathbb{P}}\left(\max\left\{\frac{n^{3/2}\rho_n}{w},\sqrt{\frac{n^3\log n\rho_n^3}{w}}\right\}(1-w)\right).$$
	Combining the two formulas we can get the desired result.
\end{proof}

\begin{proposition}\label{propa7}
	If the true model is $\delta^{(2,K^*)}$ with degree heterogeneity of order at least $a_{n,K^*}$, then $$\ell_{1,2}(A,\mathcal{E}^c)=\Omega_{\mathbb{P}}(n^2\rho_n^2a_{n,K^*}(1-w))$$ as long as the assumptions of Theorem 4 hold.
\end{proposition}
\begin{proof}[Proof of Proposition \ref{propa7}]
	Still by Proposition \ref{prop_DCBMupper}, we have
	\begin{equation}
		\ell_{2,K^*}(A,\mathcal{E}^c)-\ell_0(A,\mathcal{E}^c)\leq O_{\mathbb{P}}\left(\max\left\{\frac{n^{3/2}\rho_n}{w},\sqrt{\frac{n^3\log n\rho_n^3}{w}}\right\}(1-w)\right). \label{est(2)}
	\end{equation}
	We just need to derive a lower bound for $\ell_{1,K^*}(A,\mathcal{E}^c)-\ell_0(A,\mathcal{E}^c)$. We have
	$$\ell_{1,K^*}(A,\mathcal{E}^c)-\ell_0(A,\mathcal{E}^c)=\sum_{k_1,k_2,l_1,l_2}\sum_{(i,j)\in \mathcal{T}_{k_1,k_2,l_1,l_2}}[(A_{ij}-\hat{P}^{(1)}_{ij})^2-(A_{ij}-P_{ij})^2].$$
	By the definition of $\hat{P}^{(1)}$ in the algorithm, it is constant on $\mathcal{T}_{k_1,k_2,\cdot ,\cdot}$. Denote such a constant by $$\hat{p}_{k_1,k_2}:=\frac{\sum_{\mathcal{U}_{k_1,k_2,\cdot,\cdot}}A_{ij}}{|\mathcal{U}_{k_1,k_2,\cdot,\cdot}|},$$ 
	which depends only on the entries in the training set.
	
	For any vector $\mathbf{x}=(x_1,\cdots,x_n)^T$, function $f(\mathbf{x},p):=\sum_{i=1}^n(x_i-p)^2$ is minimized at $p=\bar{\mathbf{x}}=\sum_{i=1}^nx_i/n$. Furthermore, since $f$ is quadratic in $p$, we have 
	\begin{equation}
		f(\mathbf{x},p)=f(\mathbf{x},\bar{\mathbf{x}})+n(p-\bar{\mathbf{x}})^2. \label{qualoss}
	\end{equation}
	Therefore, by taking expectation over the test entries, we have
	\begin{align*}
		\mathbb{E}[\ell_{1,K^*}(A,\mathcal{E}^c)-\ell_0(A,\mathcal{E}^c)]=&\sum_{k_1,k_2}\sum_{(i,j)\in \mathcal{T}_{k_1,k_2,\cdot,\cdot}}\mathbb{E}\{(A_{ij}-\hat{p}_{k_1,k_2})^2-(A_{ij}-P_{ij})^2\}\\
		=&\sum_{k_1,k_2}\sum_{(i,j)\in \mathcal{T}_{k_1,k_2,\cdot,\cdot}}(\hat{p}_{k_1,k_2}-P_{ij})^2\\
		=&\sum_{k_1,k_2}\left[\sum_{(i,j)\in \mathcal{T}_{k_1,k_2,\cdot,\cdot}}\left(\frac{\sum_{\mathcal{T}_{k_1,k_2,\cdot,\cdot}}P_{i'j'}}{|\mathcal{T}_{k_1,k_2,\cdot,\cdot}|}-P_{ij}\right)^2\right.\\
		&\left.+|\mathcal{T}_{k_1,k_2,\cdot,\cdot}|\left(\hat{p}_{k_1,k_2}-\frac{\sum_{\mathcal{T}_{k_1,k_2,\cdot,\cdot}}P_{i'j'}}{|\mathcal{T}_{k_1,k_2,\cdot,\cdot}|}\right)^2\right]\\
		\geq&\sum_{k_1,k_2,l_1,l_2}\sum_{(i,j)\in \mathcal{T}_{k_1,k_2,l_1,l_2}}\left(\frac{\sum_{\mathcal{T}_{k_1,k_2,l_1,l_2}}P_{i'j'}}{|\mathcal{T}_{k_1,k_2,l_1,l_2}|}-P_{ij}\right)^2\\
		&+\sum_{k_1,k_2}|\mathcal{T}_{k_1,k_2,\cdot,\cdot}|\left(\hat{p}_{k_1,k_2}-\frac{\sum_{\mathcal{T}_{k_1,k_2,\cdot,\cdot}}P_{i'j'}}{|\mathcal{T}_{k_1,k_2,\cdot,\cdot}|}\right)^2\\
		=:&\mathcal{I}+\mathcal{II}.
	\end{align*}
	Since we use spherical spectral clustering in both procedures, by Lemma \ref{lem_ssc}, we have 
	$$|\mathcal{Q}_{k,l,\cdot,\cdot}\Delta \mathcal{Q}_{\cdot,\cdot,k,l}|=O_{\mathbb{P}}(n^\frac{3}{2}\rho_n^{-\frac{1}{2}}w^{-\frac{1}{2}}).$$
	Therefore, we have 
	\begin{equation}
		\frac{|\mathcal{U}_{k,l,\cdot,\cdot}\Delta \mathcal{U}_{\cdot,\cdot,k,l}|}{|\mathcal{U}_{k,l,\cdot,\cdot}|},\frac{|\mathcal{U}_{k,l,\cdot,\cdot}\Delta \mathcal{U}_{\cdot,\cdot,k,l}|}{|\mathcal{U}_{\cdot,\cdot,k,l}|},\frac{|\mathcal{T}_{k,l,\cdot,\cdot}\Delta \mathcal{T}_{\cdot,\cdot,k,l}|}{|\mathcal{T}_{k,l,\cdot,\cdot}|},\frac{|\mathcal{T}_{k,l,\cdot,\cdot}\Delta \mathcal{T}_{\cdot,\cdot,k,l}|}{|\mathcal{T}_{\cdot,\cdot,k,l}|}=O_{\mathbb{P}}\left(\sqrt{\frac{1}{n\rho_nw}}\right).  \label{UTcontrol}
	\end{equation}
	Thus we have
	\begin{align*}
		\mathcal{I}=&\sum_{k_1,k_2,l_1,l_2}\left[\sum_{(i,j)\in \mathcal{T}_{k_1,k_2,l_1,l_2}}\left(\frac{\sum_{\mathcal{T}_{\cdot,\cdot,l_1,l_2}}P_{i'j'}}{|\mathcal{T}_{\cdot,\cdot,l_1,l_2}|}-P_{ij}\right)^2\right.\\
		&\left.-|\mathcal{T}_{k_1,k_2,l_1,l_2}|\left(\frac{\sum_{\mathcal{T}_{\cdot,\cdot,l_1,l_2}}P_{i'j'}}{|\mathcal{T}_{\cdot,\cdot,l_1,l_2}|}-\frac{\sum_{\mathcal{T}_{k_1,k_2,l_1,l_2}}P_{i'j'}}{|\mathcal{T}_{k_1,k_2,l_1,l_2}|}\right)^2\right]\\
		=&\sum_{l_1,l_2}\sum_{(i,j)\in \mathcal{T}_{\cdot,\cdot,l_1,l_2}}\left(\frac{\sum_{\mathcal{T}_{\cdot,\cdot,l_1,l_2}}P_{i'j'}}{|\mathcal{T}_{\cdot,\cdot,l_1,l_2}|}-P_{ij}\right)^2-\sum_{l_1,l_2}|\mathcal{T}_{l_1,l_2,l_1,l_2}|\left(\frac{\sum_{\mathcal{T}_{\cdot,\cdot,l_1,l_2}}P_{ij}}{|\mathcal{T}_{\cdot,\cdot,l_1,l_2}|}-\frac{\sum_{\mathcal{T}_{l_1,l_2,l_1,l_2}}P_{ij}}{|\mathcal{T}_{l_1,l_2,l_1,l_2}|}\right)^2\\
		&-\sum_{(k_1,k_2)\neq(l_1,l_2)}|\mathcal{T}_{k_1,k_2,l_1,l_2}|\left(\frac{\sum_{\mathcal{T}_{\cdot,\cdot,l_1,l_2}}P_{ij}}{|\mathcal{T}_{\cdot,\cdot,l_1,l_2}|}-\frac{\sum_{\mathcal{T}_{k_1,k_2,l_1,l_2}}P_{ij}}{|\mathcal{T}_{k_1,k_2,l_1,l_2}|}\right)^2\\
		=:&\sum_{l_1,l_2}\sum_{(i,j)\in \mathcal{T}_{\cdot,\cdot,l_1,l_2}}\left(\frac{\sum_{\mathcal{T}_{\cdot,\cdot,l_1,l_2}}P_{i'j'}}{|\mathcal{T}_{\cdot,\cdot,l_1,l_2}|}-P_{ij}\right)^2-\mathcal{III}-\mathcal{IV}.
	\end{align*}
	Here $$\mathcal{III}=\sum_{l_1,l_2}|\mathcal{T}_{l_1,l_2,l_1,l_2}|\left(\frac{\sum_{\mathcal{T}_{\cdot,\cdot,l_1,l_2}/\mathcal{T}_{l_1,l_2,l_1,l_2}}P_{ij}}{|\mathcal{T}_{\cdot,\cdot,l_1,l_2}|}-\frac{|\mathcal{T}_{\cdot,\cdot,l_1,l_2}/\mathcal{T}_{l_1,l_2,l_1,l_2}|\sum_{\mathcal{T}_{l_1,l_2,l_1,l_2}}P_{ij}}{|\mathcal{T}_{\cdot,\cdot,l_1,l_2}||\mathcal{T}_{l_1,l_2,l_1,l_2}|}\right)^2.$$
	By \eqref{UTcontrol}, we have 
	$$\frac{\sum_{\mathcal{T}_{\cdot,\cdot,l_1,l_2}/\mathcal{T}_{l_1,l_2,l_1,l_2}}P_{ij}}{|\mathcal{T}_{\cdot,\cdot,l_1,l_2}|}=O_P\left(\rho_n\frac{|\mathcal{T}_{\cdot,\cdot,l_1,l_2}/\mathcal{T}_{l_1,l_2,l_1,l_2}|}{|\mathcal{T}_{\cdot,\cdot,l_1,l_2}|}\right)=O_{\mathbb{P}}\left(\sqrt{\frac{\rho_n}{nw}}\right),$$
	$$\frac{|\mathcal{T}_{\cdot,\cdot,l_1,l_2}/\mathcal{T}_{l_1,l_2,l_1,l_2}|\sum_{\mathcal{T}_{l_1,l_2,l_1,l_2}}P_{ij}}{|\mathcal{T}_{\cdot,\cdot,l_1,l_2}||\mathcal{T}_{l_1,l_2,l_1,l_2}|}=O_{\mathbb{P}}\left(\rho_n\frac{|\mathcal{T}_{\cdot,\cdot,l_1,l_2}/\mathcal{T}_{l_1,l_2,l_1,l_2}|}{|\mathcal{T}_{\cdot,\cdot,l_1,l_2}|}\right)=O_{\mathbb{P}}\left(\sqrt{\frac{\rho_n}{nw}}\right).$$
	Thus
	\begin{equation}
		\mathcal{III}=O_{\mathbb{P}}(\sum_{l_1,l_2}|\mathcal{T}_{l_1,l_2,l_1,l_2}|\frac{\rho_n}{nw})=O_{\mathbb{P}}(n\rho_n(1-w)/w). \label{IIIest}
	\end{equation}
	
	We also have by \eqref{UTcontrol},
	\begin{equation}
		\mathcal{IV}=O_{\mathbb{P}}(\sum_{(k_1,k_2)\neq(l_1,l_2)}|\mathcal{T}_{k_1,k_2,l_1,l_2}|\rho_n^2)=O_{\mathbb{P}}(\sum_{l_1,l_2}|\mathcal{T}_{\cdot,\cdot,l_1,l_2}/\mathcal{T}_{l_1,l_2,l_1,l_2}|\rho_n^2)=O_{\mathbb{P}}(n^\frac{3}{2}{\rho_n}^{\frac{3}{2}}w^{-\frac{1}{2}}(1-w)).\label{IVest}
	\end{equation}
	Since $n\rho_nw\to\infty$, combining \eqref{IIIest} and \eqref{IVest}, we have 
	\begin{equation}
		\mathcal{I}\geq\sum_{l_1,l_2}\sum_{(i,j)\in \mathcal{T}_{\cdot,\cdot,l_1,l_2}}\left(\frac{\sum_{\mathcal{T}_{\cdot,\cdot,l_1,l_2}}P_{i'j'}}{|\mathcal{T}_{\cdot,\cdot,l_1,l_2}|}-P_{ij}\right)^2-O_{\mathbb{P}}(n^\frac{3}{2}{\rho_n}^{\frac{3}{2}}w^{-\frac{1}{2}}(1-w)). \label{Iest}
	\end{equation}
	
	Now we control the term $\mathcal{II}$.
	\begin{align*}
		\mathcal{II}=&\sum_{k_1,k_2}|\mathcal{T}_{k_1,k_2,\cdot,\cdot}|\left(\frac{\sum_{\mathcal{U}_{k_1,k_2,\cdot,\cdot}}A_{ij}}{|\mathcal{U}_{k_1,k_2,\cdot,\cdot}|}-\frac{\sum_{\mathcal{T}_{k_1,k_2,\cdot,\cdot}}P_{ij}}{|\mathcal{T}_{k_1,k_2,\cdot,\cdot}|}\right)^2\\
		=&\sum_{k_1,k_2}|\mathcal{T}_{k_1,k_2,\cdot,\cdot}|\left[\left(\frac{\sum_{\mathcal{U}_{\cdot,\cdot,k_1,k_2}}A_{ij}}{|\mathcal{U}_{\cdot,\cdot,k_1,k_2}|}-\frac{\sum_{\mathcal{T}_{\cdot,\cdot,k_1,k_2}}P_{ij}}{|\mathcal{T}_{\cdot,\cdot,k_1,k_2}|}\right)+\left(\frac{\sum_{\mathcal{U}_{k_1,k_2,\cdot,\cdot}}A_{ij}}{|\mathcal{U}_{k_1,k_2,\cdot,\cdot}|}-\frac{\sum_{\mathcal{U}_{\cdot,\cdot,k_1,k_2}}A_{ij}}{|\mathcal{U}_{\cdot,\cdot,k_1,k_2}|}\right)\right.\\
		&+\left.\left(\frac{\sum_{\mathcal{T}_{\cdot,\cdot,k_1,k_2}}P_{ij}}{|\mathcal{T}_{\cdot,\cdot,k_1,k_2}|}-\frac{\sum_{\mathcal{T}_{k_1,k_2,\cdot,\cdot}}P_{ij}}{|\mathcal{T}_{k_1,k_2,\cdot,\cdot}|}\right)\right]^2.
	\end{align*}
	We have by \eqref{UTcontrol},
	\begin{align*}
		&\left|\frac{\sum_{\mathcal{U}_{k_1,k_2,\cdot,\cdot}}A_{ij}}{|\mathcal{U}_{k_1,k_2,\cdot,\cdot}|}-\frac{\sum_{\mathcal{U}_{\cdot,\cdot,k_1,k_2}}A_{ij}}{|\mathcal{U}_{\cdot,\cdot,k_1,k_2}|}\right|\\
		\leq&\frac{\sum_{\mathcal{U}_{k_1,k_2,\cdot,\cdot}/\mathcal{U}_{k_1,k_2,k_1,k_2}}A_{ij}}{|\mathcal{U}_{k_1,k_2,\cdot,\cdot}|}+\frac{\sum_{\mathcal{U}_{\cdot,\cdot,k_1,k_2}/\mathcal{U}_{k_1,k_2,k_1,k_2}}A_{ij}}{|\mathcal{U}_{\cdot,\cdot,k_1,k_2}|}\\
		&+\frac{|\mathcal{U}_{\cdot,\cdot,k_1,k_2}\Delta \mathcal{U}_{k_1,k_2,\cdot,\cdot}|\sum_{\mathcal{U}_{k_1,k_2,k_1,k_2}}A_{ij}}{|\mathcal{U}_{\cdot,\cdot,k_1,k_2}||\mathcal{U}_{k_1,k_2,\cdot,\cdot}|}\\
		=&O_{\mathbb{P}}\left(\rho_n\sqrt{\frac{1}{n\rho_nw}}\right)+O_{\mathbb{P}}\left(\rho_n\sqrt{\frac{1}{n\rho_nw}}\right)+O_{\mathbb{P}}\left(\rho_n\sqrt{\frac{1}{n\rho_nw}}\right)=O_{\mathbb{P}}\left(\sqrt{\frac{\rho_n}{nw}}\right).
	\end{align*}
	Similarly, we have
	$$\left|\frac{\sum_{\mathcal{T}_{\cdot,\cdot,k_1,k_2}}P_{ij}}{|\mathcal{T}_{\cdot,\cdot,k_1,k_2}|}-\frac{\sum_{\mathcal{T}_{k_1,k_2,\cdot,\cdot}}P_{ij}}{|\mathcal{T}_{k_1,k_2,\cdot,\cdot}|}\right|=O_{\mathbb{P}}\left(\sqrt{\frac{\rho_n}{nw}}\right).$$
	
	Therefore by \eqref{UTcontrol} again,
	\begin{align}
		\mathcal{II}\geq&\sum_{k_1,k_2}|\mathcal{T}_{k_1,k_2,\cdot,\cdot}|\left[\left(\frac{\sum_{\mathcal{U}_{\cdot,\cdot,k_1,k_2}}A_{ij}}{|\mathcal{U}_{\cdot,\cdot,k_1,k_2}|}-\frac{\sum_{\mathcal{T}_{\cdot,\cdot,k_1,k_2}}P_{ij}}{|\mathcal{T}_{\cdot,\cdot,k_1,k_2}|}\right)^2-O_{\mathbb{P}}({\rho_n}^{\frac{3}{2}}n^{-\frac{1}{2}}w^{-\frac{1}{2}})\right]\nonumber\\
		=&\sum_{k_1,k_2}|\mathcal{T}_{k_1,k_2,\cdot,\cdot}|\left(\frac{\sum_{\mathcal{U}_{\cdot,\cdot,k_1,k_2}}A_{ij}}{|\mathcal{U}_{\cdot,\cdot,k_1,k_2}|}-\frac{\sum_{\mathcal{T}_{\cdot,\cdot,k_1,k_2}}P_{ij}}{|\mathcal{T}_{\cdot,\cdot,k_1,k_2}|}\right)^2-O_{\mathbb{P}}(n^\frac{3}{2}{\rho_n}^{\frac{3}{2}}w^{-\frac{1}{2}}(1-w))\nonumber\\
		\geq&\sum_{k_1,k_2}(|\mathcal{T}_{\cdot,\cdot,k_1,k_2}|-|\mathcal{T}_{\cdot,\cdot,k_1,k_2}\Delta \mathcal{T}_{k_1,k_2,\cdot,\cdot}|)\left(\frac{\sum_{\mathcal{U}_{\cdot,\cdot,k_1,k_2}}A_{ij}}{|\mathcal{U}_{\cdot,\cdot,k_1,k_2}|}-\frac{\sum_{\mathcal{T}_{\cdot,\cdot,k_1,k_2}}P_{ij}}{|\mathcal{T}_{\cdot,\cdot,k_1,k_2}|}\right)^2\nonumber\\
		&-O_{\mathbb{P}}(n^\frac{3}{2}{\rho_n}^{\frac{3}{2}}w^{-\frac{1}{2}}(1-w))\label{IIest}\\
		=&\sum_{k_1,k_2}|\mathcal{T}_{\cdot,\cdot,k_1,k_2}|\left(\frac{\sum_{\mathcal{U}_{\cdot,\cdot,k_1,k_2}}A_{ij}}{|\mathcal{U}_{\cdot,\cdot,k_1,k_2}|}-\frac{\sum_{\mathcal{T}{\cdot,\cdot,k_1,k_2}}P_{ij}}{|\mathcal{T}_{\cdot,\cdot,k_1,k_2}|}\right)^2-O_{\mathbb{P}}(n^\frac{3}{2}{\rho_n}^{\frac{3}{2}}w^{-\frac{1}{2}}(1-w)).\nonumber
	\end{align}
	Combining \eqref{Iest} and \eqref{IIest}, we get
	\begin{align*}
		&\mathbb{E}[\ell_{1,K^*}(A,\mathcal{E}^c)-\ell_0(A,\mathcal{E}^c)]\\
		\geq&\sum_{k_1,k_2}\left[\sum_{(i,j)\in \mathcal{T}_{\cdot,\cdot,k_1,k_2}}\left(\frac{\sum_{\mathcal{T}_{\cdot,\cdot,k_1,k_2}}P_{i'j'}}{|\mathcal{T}_{\cdot,\cdot,k_1,k_2}|}-P_{ij}\right)^2+|\mathcal{T}_{\cdot,\cdot,k_1,k_2}|\left(\frac{\sum_{\mathcal{U}_{\cdot,\cdot,k_1,k_2}}A_{ij}}{|\mathcal{U}_{\cdot,\cdot,k_1,k_2}|}-\frac{\sum_{\mathcal{T}_{\cdot,\cdot,k_1,k_2}}P_{ij}}{|\mathcal{T}_{\cdot,\cdot,k_1,k_2}|}\right)^2\right]\\
		&-O_{\mathbb{P}}(n^\frac{3}{2}{\rho_n}^{\frac{3}{2}}w^{-\frac{1}{2}}(1-w))\\
		=:&\mathcal{V}-O_{\mathbb{P}}(n^\frac{3}{2}{\rho_n}^{\frac{3}{2}}w^{-\frac{1}{2}}(1-w)).
	\end{align*}
	We have
	\begin{align*}
		\mathcal{V}
		=&\sum_{k_1,k_2}\sum_{(i,j)\in \mathcal{T}_{\cdot,\cdot,k_1,k_2}}\left[\left(P_{ij}-\frac{\sum_{(i',j')\in \mathcal{T}_{\cdot,\cdot,k_1,k_2}}P_{i'j'}}{|\mathcal{T}_{\cdot,\cdot,k_1,k_2}|}\right)^2\right.\\
		&\left.+\left(\frac{\sum_{(i',j')\in \mathcal{U}_{\cdot,\cdot,k_1,k_2}}A_{i'j'}}{|\mathcal{U}_{\cdot,\cdot,k_1,k_2}|}-\frac{\sum_{(i',j')\in \mathcal{T}_{\cdot,\cdot,k_1,k_2}}P_{i'j'}}{|\mathcal{T}_{\cdot,\cdot,k_1,k_2}|}\right)^2\right]\\
		=&\sum_{k_1,k_2}\sum_{(i,j)\in Q_{\cdot,\cdot,k_1,k_2}}\left[\left(P_{ij}-\frac{\sum_{(i',j')\in \mathcal{T}_{\cdot,\cdot,k_1,k_2}}P_{i'j'}}{|\mathcal{T}_{\cdot,\cdot,k_1,k_2}|}\right)^2+\left(\frac{\sum_{(i',j')\in \mathcal{T}_{\cdot,\cdot,k_1,k_2}}P_{i'j'}}{|\mathcal{T}_{\cdot,\cdot,k_1,k_2}|}\right.\right.\\
		&-\left.\frac{\sum_{(i',j')\in \mathcal{U}_{\cdot,\cdot,k_1,k_2}}P_{i'j'}}{|\mathcal{U}_{\cdot,\cdot,k_1,k_2}|}\right)^2+\left(\frac{\sum_{(i',j')\in \mathcal{U}_{\cdot,\cdot,k_1,k_2}}A_{i'j'}}{|\mathcal{U}_{\cdot,\cdot,k_1,k_2}|}-\frac{\sum_{(i',j')\in \mathcal{T}_{\cdot,\cdot,k_1,k_2}}P_{i'j'}}{|\mathcal{T}_{\cdot,\cdot,k_1,k_2}|}\right)^2\\
		&\left.-\left(\frac{\sum_{(i',j')\in \mathcal{T}_{\cdot,\cdot,k_1,k_2}}P_{i'j'}}{|\mathcal{T}_{\cdot,\cdot,k_1,k_2}|}-\frac{\sum_{(i',j')\in \mathcal{U}_{\cdot,\cdot,k_1,k_2}}P_{i'j'}}{|\mathcal{U}_{\cdot,\cdot,k_1,k_2}|}\right)^2\right]\mathbbm{1}\{(i,j)\in\mathcal{E}^c\}\\
		=&\sum_{k_1,k_2}\sum_{(i,j)\in Q_{\cdot,\cdot,k_1,k_2}}\left[\left(P_{ij}-\frac{\sum_{(i',j')\in \mathcal{T}_{\cdot,\cdot,k_1,k_2}}P_{i'j'}}{|\mathcal{T}_{\cdot,\cdot,k_1,k_2}|}\right)^2+\left(\frac{\sum_{(i',j')\in \mathcal{T}_{\cdot,\cdot,k_1,k_2}}P_{i'j'}}{|\mathcal{T}_{\cdot,\cdot,k_1,k_2}|}\right.\right.\\
		&\left.\left.-\frac{\sum_{(i',j')\in \mathcal{U}_{\cdot,\cdot,k_1,k_2}}P_{i'j'}}{|\mathcal{U}_{\cdot,\cdot,k_1,k_2}|}\right)^2\right]\mathbbm{1}\{(i,j)\in\mathcal{E}^c\}+\sum_{k_1,k_2}|\mathcal{T}_{\cdot,\cdot,k_1,k_2}|\left(\frac{\sum_{(i,j)\in \mathcal{U}_{\cdot,\cdot,k_1,k_2}}A_{ij}}{|\mathcal{U}_{\cdot,\cdot,k_1,k_2}|}\right.\\
		&\left.-\frac{\sum_{(i,j)\in \mathcal{U}_{\cdot,\cdot,k_1,k_2}}P_{ij}}{|\mathcal{U}_{\cdot,\cdot,k_1,k_2}|}\right)\left(\frac{\sum_{(i,j)\in \mathcal{U}_{\cdot,\cdot,k_1,k_2}}A_{ij}}{|\mathcal{U}_{\cdot,\cdot,k_1,k_2}|}+\frac{\sum_{(i,j)\in \mathcal{U}_{\cdot,\cdot,k_1,k_2}}P_{ij}}{|\mathcal{U}_{\cdot,\cdot,k_1,k_2}|}-2\frac{\sum_{(i,j)\in \mathcal{T}_{\cdot,\cdot,k_1,k_2}}P_{ij}}{|\mathcal{T}_{\cdot,\cdot,k_1,k_2}|}\right)\\
		=&:\mathcal{VI}+\mathcal{VII}.
	\end{align*}
	For the second term, we have 
	\begin{align}
		\mathcal{VII}\geq&-\sum_{k_1,k_2}|\mathcal{T}_{\cdot,\cdot,k_1,k_2}|\left|\frac{\sum_{(i,j)\in \mathcal{U}_{\cdot,\cdot,k_1,k_2}}A_{ij}}{|\mathcal{U}_{\cdot,\cdot,k_1,k_2}|}-\frac{\sum_{(i,j)\in \mathcal{U}_{\cdot,\cdot,k_1,k_2}}P_{ij}}{|\mathcal{U}_{\cdot,\cdot,k_1,k_2}|}\right|\left|\frac{\sum_{(i,j)\in \mathcal{U}_{\cdot,\cdot,k_1,k_2}}A_{ij}}{|\mathcal{U}_{\cdot,\cdot,k_1,k_2}|}\right.\nonumber\\
		&\left.+\frac{\sum_{(i,j)\in \mathcal{U}_{\cdot,\cdot,k_1,k_2}}P_{ij}}{|\mathcal{U}_{\cdot,\cdot,k_1,k_2}|}-2\frac{\sum_{(i,j)\in \mathcal{T}_{\cdot,\cdot,k_1,k_2}}P_{ij}}{|\mathcal{T}_{\cdot,\cdot,k_1,k_2}|}\right|\nonumber\\
		\geq&-\sum_{k_1,k_2}|\mathcal{T}_{\cdot,\cdot,k_1,k_2}|O_{\mathbb{P}}\left(\sqrt{\frac{\rho_n}{|\mathcal{U}_{\cdot,\cdot,k_1,k_2}|}}\right)O_{\mathbb{P}}(\rho_n)=-O_{\mathbb{P}}\left(n{\rho_n}^{\frac{3}{2}}{w}^{-\frac{1}{2}}(1-w)\right).\label{VIIest}
	\end{align}
	For the first term, notice for $(i,j)\in\mathcal{Q}_{\cdot,\cdot,k_1,k_2}$, we have 
	\begin{align}
		&\left(P_{ij}-\frac{\sum_{(i',j')\in \mathcal{T}_{\cdot,\cdot,k_1,k_2}}P_{i'j'}}{|\mathcal{T}_{\cdot,\cdot,k_1,k_2}|}\right)^2+\left(\frac{\sum_{(i',j')\in \mathcal{T}_{\cdot,\cdot,k_1,k_2}}P_{i'j'}}{|\mathcal{T}_{\cdot,\cdot,k_1,k_2}|}-\frac{\sum_{(i',j')\in \mathcal{U}_{\cdot,\cdot,k_1,k_2}}P_{i'j'}}{|\mathcal{U}_{\cdot,\cdot,k_1,k_2}|}\right)^2\nonumber\\
		=&B_{k_1k_2}^2\left[\left(\theta_i\theta_j-\frac{\sum_{(i',j')\in \mathcal{T}_{\cdot,\cdot,k_1,k_2}}\theta_{i'}\theta_{j'}}{|\mathcal{T}_{\cdot,\cdot,k_1,k_2}|}\right)^2+\left(\frac{\sum_{({i'},{j'})\in \mathcal{T}_{\cdot,\cdot,k_1,k_2}}\theta_{i'}\theta_{j'}}{|\mathcal{T}_{\cdot,\cdot,k_1,k_2}|}-\frac{\sum_{({i'},{j'})\in \mathcal{U}_{\cdot,\cdot,k_1,k_2}}\theta_{i'}\theta_{j'}}{|\mathcal{U}_{\cdot,\cdot,k_1,k_2}|}\right)^2\right]\nonumber\\
		\overset{(*)}{\geq}&\frac{B_{k_1k_2}^2}{2}\left[\left(\theta_i\theta_j-\frac{\sum_{({i'},{j'})\in \mathcal{T}_{\cdot,\cdot,k_1,k_2}}\theta_{i'}\theta_{j'}}{|\mathcal{T}_{\cdot,\cdot,k_1,k_2}|}\right)\right.\nonumber\\
		&\left.-\frac{|U_{\cdot,\cdot,k_1,k_2}|}{|\mathcal{U}_{\cdot,\cdot,k_1,k_2}|+|\mathcal{T}_{\cdot,\cdot,k_1,k_2}|}\left(\frac{\sum_{({i'},{j'})\in \mathcal{U}_{\cdot,\cdot,k_1,k_2}}\theta_{i'}\theta_{j'}}{|\mathcal{U}_{\cdot,\cdot,k_1,k_2}|}-\frac{\sum_{({i'},{j'})\in \mathcal{T}_{\cdot,\cdot,k_1,k_2}}\theta_{i'}\theta_{j'}}{|\mathcal{T}_{\cdot,\cdot,k_1,k_2}|}\right)\right]^2\nonumber\\
		=&\frac{B_{k_1k_2}^2}{2}\left(\theta_i\theta_j-\frac{\sum_{({i'},{j'})\in \mathcal{Q}_{\cdot,\cdot,k_1,k_2}}\theta_{i'}\theta_{j'}}{|\mathcal{Q}_{\cdot,\cdot,k_1,k_2}|}\right)^2, \label{cauchy}
	\end{align}
	where $(*)$ is by $(x+y)^2\leq2(x^2+y^2)$. Therefore
	$$\mathcal{VI}\geq\sum_{k_1,k_2}\frac{B_{k_1k_2}^2}{2}\sum_{(i,j)\in Q_{\cdot,\cdot,k_1,k_2}}\left(\theta_i\theta_j-\frac{\sum_{({i'},{j'})\in \mathcal{Q}_{\cdot,\cdot,k_1,k_2}}\theta_{i'}\theta_{j'}}{|\mathcal{Q}_{\cdot,\cdot,k_1,k_2}|}\right)^2\mathbbm{1}\{(i,j)\in\mathcal{E}^c\}=:\mathcal{VIII}.$$
	We have 
	\begin{align*}
		\mathbb{E}(\mathcal{VIII})
		=&(1-w)\sum_{k_1,k_2}\frac{B_{k_1k_2}^2}{2}\sum_{(i,j)\in Q_{\cdot,\cdot,k_1,k_2}}\left(\theta_i\theta_j-\frac{\sum_{({i'},{j'})\in \mathcal{Q}_{\cdot,\cdot,k,l}}\theta_{i'}\theta_{j'}}{|\mathcal{Q}_{\cdot,\cdot,k,l}|}\right)^2\\
		=&\frac{1-w}{4}\sum_{k_1,k_2}\frac{B_{k_1k_2}^2}{|\mathcal{Q}_{\cdot,\cdot,k,l}|}\sum_{(i_1,j_1),(i_2,j_2)\in \mathcal{Q}_{\cdot,\cdot,k,l}}(\theta_{i_1}\theta_{j_1}-\theta_{i_2}\theta_{j_2})^2\\
		=&\Omega_{\mathbb{P}}\left(n^2\rho_n^2a_{n,K}(1-w)\right) 
	\end{align*}
	by Assumption 8. This order tends to $\infty$ as $n\to\infty$. We also notice that
	\begin{align*}
		\mathcal{VIII}=&\sum_{i,j}\frac{B_{c_ic_j}^2}{2}\left(\theta_i\theta_j-\frac{\sum_{({i'},{j'})\in \mathcal{Q}_{\cdot,\cdot,c_i,c_j}}\theta_{i'}\theta_{j'}}{|\mathcal{Q}_{\cdot,\cdot,c_i,c_j}|}\right)^2\mathbbm{1}\{(i,j)\in\mathcal{E}^c\}\\
		=&\sum_{i<j}B_{c_ic_j}^2\left(\theta_i\theta_j-\frac{\sum_{({i'},{j'})\in \mathcal{Q}_{\cdot,\cdot,c_i,c_j}}\theta_{i'}\theta_{j'}}{|\mathcal{Q}_{\cdot,\cdot,c_i,c_j}|}\right)^2\mathbbm{1}\{(i,j)\in\mathcal{E}^c\},
	\end{align*}
	being the sum of independent variables. Thus by $\min_{k,l}B_{0,kl}>c>0$ in Assumption 7 and Hoeffding's inequality, with probability tending to 1,
	\begin{equation}
		\mathcal{VI}\geq\mathcal{VIII}\geq\frac{1}{2}\mathbb{E}(\mathcal{VIII})=\Omega_{\mathbb{P}}\left(n^2\rho_n^2a_{n,K^*}(1-w)\right).\label{VIest}
	\end{equation}
	Combining \eqref{VIIest} and \eqref{VIest}, we have
	\begin{align*}
		\mathbb{E}[\ell_{1,K^*}(A,\mathcal{E}^c)-\ell_0(A,\mathcal{E}^c)]&\geq\Omega_{\mathbb{P}}(n^2\rho_n^2a_{n,K^*}(1-w))-O_{\mathbb{P}}(n^\frac{3}{2}{\rho_n}^{\frac{3}{2}}w^{-\frac{1}{2}}(1-w))\\
		&=\Omega_{\mathbb{P}}(n^2\rho_n^2a_{n,K^*}(1-w)).
	\end{align*}
	By Hoeffding's inequality, but this time only on the edge in the test set $\mathcal{E}^c$, we obtain with probability tending to 1,
	\begin{equation}
		\ell_{1,K^*}(A,\mathcal{E}^c)-\ell_0(A,\mathcal{E}^c)\geq\frac{1}{2}\mathbb{E}[\ell_{1,K^*}(A,\mathcal{E}^c)-\ell_0(A,\mathcal{E}^c)]=\Omega_{\mathbb{P}}(n^2\rho_n^2a_{n,K^*}(1-w)). \label{est(1)}
	\end{equation}
	Finally, combining \eqref{est(2)} and \eqref{est(1)}, we can get the desired result.
\end{proof}

\begin{proof}[Proof of Theorem 4]
	Similarly to the proof of Theorem 3, we just need to prove the consistency conditioned on $\Omega$ and $\hat{K}=K^*$.
	We need to prove that
	\begin{equation}
		\mathbb{P}\left[L_{1}(A,\mathcal{E}^c)>L_{2}(A,\mathcal{E}^c)\right]\to0,\quad\text{when $\delta^{(1,K^*)}$ is true}, \label{theo3_cons1}
	\end{equation}
	and 
	\begin{equation}
		\mathbb{P}\left[L_{1}(A,\mathcal{E}^c)<L_{2}(A,\mathcal{E}^c)\right]\to0,\quad\text{when $\delta^{(2,K^*)}$ is true and satisfies Assumption 8}. \label{theo3_cons2}
	\end{equation}
	When $\delta^{(1,K^*)}$ is true, notice that
	\begin{align*}
		\mathbb{P}\left[L_{1}(A,\mathcal{E}^c)>L_{2}(A,\mathcal{E}^c)\right]=&\mathbb{P}[\ell_{1,2}(A,\mathcal{E}^c)>(d_{2,K^*}-d_{1,K^*})|\mathcal{E}^c|\lambda_n]\\
		\leq&\mathbb{P}[\ell_{1,2}(A,\mathcal{E}^c)>1/2(d_{2,K^*}-d_{1,K^*})n(n-1)(1-w)\lambda_n]\to0
	\end{align*}
	by $d_{2,K^*}-d_{1,K^*}>0$ being of constant order, $\lambda_n=\omega_{\mathbb{P}}\left(\max\left\{{\rho_n}/{w\sqrt{n}},\sqrt{{\rho_n^3\log n}/{wn}}\right\}\right)$ and Proposition \ref{propa6}.
	Similarly \eqref{theo3_cons2} holds when $\delta^{(2,K^*)}$ is true, by $\lambda_n=o_{\mathbb{P}}(\rho_n^2a_{n,K^*})$ and Proposition \ref{propa7}.
\end{proof}

\subsection{Proof of Proposition 2}

First, we provide a quick illustration of how the NS algorithm fits on the partially observed adjacency matrix $Y$. \cite{zhang2017estimating} introduced the following distance between nodes:
$$\tilde{d}^2(i,i'):=\max_{k\neq i,i'}|\langle A_{i\cdot}-A_{i'\cdot},A_{k\cdot}\rangle|/n,$$
and defined the neighborhood $\mathcal{N}_i$ as follows,
$$\mathcal{N}_i:=\{i'\neq i:\tilde{d}(i,i')\leq q_i(h)\},$$
where $q_i(h)$ denotes the $h$-th sample quantile of the set $\{\tilde{d}(i,i'),i'\neq i\}$ with $h$ a tuning parameter. Thus, the general neighborhood smoothing estimator is defined by
\begin{equation}
	\tilde{P}_{ij}=\frac{\sum_{i'\in\mathcal{N}_i}A_{i'j}}{|\mathcal{N}_i|}, \label{Ptilde}
\end{equation}
and when the network is undirected (symmetric), a symmetric estimator $\hat{P}=(\tilde{P}+\tilde{P}^\top)/2$
is used as a replacement.

In our setting, in one split, after the training edge set $\mathcal{E}$ is generated, denote the corresponding partially observed adjacency matrix as $Y$. We modify the above process by replacing the whole adjacency matrix $A$ with $Y$ in $\tilde{d}$ and $\tilde{P}$, and scale the final symmetric estimator by $\hat{P}^{(2)}=\hat{P}/w$; here $w$ is the proportion of the training set.

The proof of Proposition 2 exactly follows the lines of the proof of Theorem 1 in \cite{zhang2017estimating} with some Bernstein inequalities changed. Therefore, we will just give a sketch of the proof, showing some of the key formulas that changed. Recall the requirement for the underlying graphon function in \cite{zhang2017estimating}:
\begin{definition}[Piecewise-Lipschitz graphon family]
	For any $\delta, L > 0$, let $\mathcal{F}_{\delta,L}$ denote a family of piecewise-Lipschitz graphon functions $f : [0,1]^2 \to [0,1]$ such that
	\begin{itemize}
		\item[(i)] there exists an integer $K \geq 1$ and a sequence $0 = x_0 < \dots < x_K = 1$ satisfying
		\[
		\min_{0 \leq s \leq K-1} (x_{s+1} - x_s) \geq \delta,
		\]
		\item[(ii)] both
		\[
		|f(u_1,v) - f(u_2,v)| \leq L |u_1 - u_2| \quad \text{and} \quad |f(u,v_1) - f(u,v_2)| \leq L |v_1 - v_2|
		\]
		hold for all $u, u_1, u_2 \in [x_s, x_{s+1}]$, $v, v_1, v_2 \in [x_t, x_{t+1}]$ \textit{and} $0 \leq s,t \leq K-1$.
	\end{itemize}
\end{definition}

\begin{asm}
	For $f_0$, there exist a global constant $L$ and $\delta=\delta(n)$ with $\lim_{n\to\infty}\delta/(n^{-1}\log n)^{1/2}\to\infty$, such that $f_0\in\mathcal{F}_{\delta,L}$. More precisely, let $0=x_0<x_1<\ldots<x_K=1$, $I_k:=[x_{k-1},x_k)$ for $1\leq k\leq K$, then $f_0$ is Lipschitz on each $I_k\times I_l$.
\end{asm}

\begin{asm}[Sparsity condition for neighborhood smoothing]\label{a2_NS}
	The sparsity parameter $\rho_n$ and the training proportion $w$ satisfy
	\[
	\frac{nw^2\rho_n^2}{\log n}\to\infty.
	\]
\end{asm}

Assumption~\ref{a2_NS} ensures that the training network is sufficiently
dense for the uniform concentration arguments required by the neighborhood
smoothing procedure. In particular, it guarantees that the deviation level
\[
\epsilon_n=\left(\frac{C\log n}{nw^2\rho_n^2}\right)^{1/2}
\]
used in the Bernstein bounds below satisfies $\epsilon_n=o(1)$.

This condition can be derived from Assumption 10. In fact, since $P_{ij}\leq \rho_n$, we have $b_{n,\mathcal{K}}\leq C\rho_n^2$ and thus the third term in the lower bound imposed in Assumption 10 for $b_{n,\mathcal{K}}$ implies
$$\rho_n^2\gtrsim b_{n,\mathcal{K}}\gg \sqrt{\frac{\rho_n^3\log n}{w\sqrt{n}}}\Longleftrightarrow\sqrt{n}w\rho_n \gg\log n,$$
which directly implies Assumption \ref{a2_NS}.

\begin{proof}[Proof Sketch of Proposition 2]
	Let $I(\xi)$ denote the $I_k$ that contains $\xi$. Denote $S_i(\Delta)=[\xi_i-\Delta,\xi_i+\Delta]\cap I(\xi_i)$. Since the split is independent of $A$, the partially observed matrix $Y$ has independent Bernoulli entries with
	$$\mathbb{P}(Y_{ij}=1)=w\rho_nf_0(\xi_i,\xi_j)=wP_{ij}.$$
	To prove the proposition, it suffices to show that with high probability,
	$$
	\frac{1}{n}\sum_j\left( \tilde{P}_{ij}-wP_{ij} \right)^2 \leq w\rho_n\left(\frac{\log n}{n}\right)^{1/2}
	$$
	holds for all $i$, where $\tilde{P}$ is \eqref{Ptilde} with $A$ substituted by $Y$. We have
	\begin{align*}
		\frac{1}{n}\sum_j\left( \tilde{P}_{ij}-wP_{ij} \right)^2&=\left[ \frac{\sum_{i'\in \mathcal{N}_i}\left\{( Y_{i'j} - wP_{i'j}) + (wP_{i'j}-wP_{ij})\right\}}{|\mathcal{N}_i|} \right]^2 \\
		&\leq  2\left\{\frac{\sum_{i'\in \mathcal{N}_i}(Y_{i'j}-wP_{i'j})}{|\mathcal{N}_i|}\right\}^2 +  2\left\{\frac{\sum_{i'\in \mathcal{N}_i}(wP_{i'j}-wP_{ij})}{|\mathcal{N}_i|}\right\}^2 = 2J_1(i,j) + 2J_2(i,j). 
	\end{align*}
	
	\begin{lemma}[Lemma 1 in \cite{zhang2017estimating}]\label{lem11}
		For arbitrary global constants $C_1$, $\tilde{C}_1>0$, define $\Delta_n=\left\{C_1+\left(\tilde{C}_1+4\right)^{1/2}\right\}(n^{-1}\log n)^{1/2}$.  For $n$ large enough so that 
		$\left\{(\tilde{C}_1+4)\log n/n\right\}^{1/2}\leq 1$ and $\Delta_n < \min_k|I_k|/2$,
		we have
		\begin{equation}
			\mathbb{P}\left\{ \min_i\frac{|\{i'\neq i:\xi_{i'}\in S_i(\Delta_n)\}|}{n-1}\geq C_1(n^{-1}\log n)^{1/2} \right\} \geq 1-2n^{-\tilde{C}_1/4} . 
		\end{equation}
	\end{lemma}
	
	Notice 
	$$\tilde{d}(i,i') = \max_{k\neq i,i'}\left|\langle Y_{i\cdot} - Y_{i'\cdot}, Y_{k\cdot} \rangle\right|\big/ n = \max_{k\neq i,i'}\left|(Y^2/n)_{ik} - (Y^2/n)_{i'k}\right|.$$
	
	\begin{lemma}[Modification of Lemma 2 in \cite{zhang2017estimating}]\label{lem12}
		Suppose we select the neighborhood $\mathcal{N}_i$ by thresholding at the lower $h$-th quantile of $\{\tilde{d}(i, k)\}_{k\neq i}$, where we set $h=C_0(n^{-1}\log n)^{1/2}$ with an arbitrary global constant $C_0$ satisfying $0<C_0\leq C_1$ for the $C_1$ from Lemma \ref{lem11}. Let $C_2, \tilde{C}_2>0$ be arbitrary global constants and assume $n\geq 6$ is large enough so that
		\begin{enumerate}[(i)]
			\item All conditions on $n$ in Lemma \ref{lem11} are satisfied;
			\item $\left\{(C_2+2)\log n/n\right\}^{1/2}\leq 1$;
			\item $C_1\left(n\log n\right)^{1/2}\geq 4$; and
			\item $4/n\leq w\rho_n\left\{\left(C_2+\tilde{C}_2+2\right)^{1/2} - \left(C_2+2\right)^{1/2} \right\}(n^{-1}\log n)^{1/2}$. \label{Ncondition::eliminate4divn}
		\end{enumerate}
		Then the neighborhood $\mathcal{N}_i$ has the following properties:
		\begin{enumerate}
			\item $|\mathcal{N}_i|\geq C_0\left(n\log n\right)^{1/2}$.
			\item With probability $1-2n^{-\tilde{C}_1/4}-2n^{-\tilde{C}_2/4}$, for all $i$ and $i'\in\mathcal{N}_i$, we have
			\begin{equation*}
				\|wP_{i'\cdot}-wP_{i\cdot}\|_2^2/n\leq w\rho_n\left[ 6L\left\{C_1+\left(\tilde{C}_2+4\right)^{1/2}\right\}^{1/2} + 8\left(C_2+\tilde{C}_2+2\right)^{1/2}  \right](n^{-1}\log n)^{1/2}.
			\end{equation*}
		\end{enumerate}
	\end{lemma}
	\begin{proof}[Proof Sketch of Lemma \ref{lem12}]
		By Bernstein's inequality, for any $0 < \epsilon \leq 1$  and $n\geq 3$ we have
		$$\mathbb{P}\left\{  \frac{|\sum_{k\neq i,j}(Y_{ik}Y_{kj} - w^2P_{ik}P_{kj})|}{n-2} \geq w^2\rho_n^2\epsilon  \right\} \leq 2\exp\left\{ -\frac{\frac{1}{2}(n-2)(w^2\rho_n^2\epsilon)^2}{w^2\rho_n^2+\frac{1}{3}w^2\rho_n^2\epsilon} \right\}\leq 2\exp\left(-\frac{1}{4}nw^2\rho_n^2\epsilon^2\right).$$
		Set $\epsilon = \left\{(C_2+2)\log n/nw^2\rho_n^2\right\}^{1/2}$ and take a union bound, we have
		$$\mathbb{P}\left\{  \max_{i,j:i\neq j}\frac{|\sum_{k\neq i,j}(Y_{ik}Y_{kj} - w^2P_{ik}P_{kj})|}{n-2} \geq  w\rho_n \left\{\frac{(C_2+2)\log n}{n}\right\}^{1/2}\right\} \leq 2n^{-\tilde{C}_2/4}. $$
		Following the lines in \cite{zhang2017estimating}, with probability $1-2n^{-\tilde{C}_2/4}$,
		$$\max_{i,j:i\neq j}\left|\left(Y^2/n\right)_{ij}-w^2\left(P^2/n\right)_{ij}\right|  \leq   w\rho_n\left\{\frac{(C_2+2)\log n}{n}\right\}^{1/2} + \frac{4}{n}\leq w\rho_n\left\{\frac{(C_2+\tilde{C}_2+2)\log n}{n}\right\}^{1/2}.$$
		We also have for all $i$ and any $\tilde{i}$ such that $\xi_{\tilde{i}}\in S_i(\Delta_n)$,
		$$\left|w^2\left(P^2/n\right)_{ik}-w^2\left(P^2/n\right)_{\tilde{i} k}\right| = w^2|\langle P_{i\cdot}, P_{k\cdot} \rangle  -  \langle P_{\tilde{i}\cdot}, P_{k\cdot} \rangle|/n\leq w^2\|P_{i\cdot} - P_{\tilde{i}\cdot}\|_2\|P_{k\cdot}\|_2/n\leq Lw^2\rho_n^2\Delta_n.$$
		Then for $\tilde{i}\in S_i(\Delta_n)$, with probability $1-2n^{-\tilde{C}_2/4}$,
		\begin{align}
			&\tilde{d}(i, \tilde{i}) =\max_{k\neq i, \tilde{i}}|(Y^2/n)_{ik} - (Y^2/n)_{\tilde{i} k}| \nonumber\\
			\leq& \max_{k\neq i, \tilde{i}}|w^2(P^2/n)_{ik} - w^2(P^2/n)_{\tilde{i} k}| + 2\max_{i, j: i\neq j}|(Y^2/n)_{ij} - w^2(P^2/n)_{ij}|\nonumber\\
			\leq&  Lw^2\rho_n^2\Delta_n+2w\rho_n\left\{\frac{(C_2+\tilde{C}_2+2)\log n}{n}\right\}^{1/2}.  \label{bounddtilda1}
		\end{align}
		By definition of $\mathcal{N}_i$ and our setting for $\Delta_n$ and $h$, combining \eqref{bounddtilda1} and Lemma \ref{lem11}, we have with probability $1-2n^{-\tilde{C}_1/4}-2n^{-\tilde{C}_2/4}$,
		$$\tilde{d}(i,i')\leq Lw^2\rho_n^2\Delta_n+2w\rho_n\left\{\frac{(C_2+\tilde{C}_2+2)\log n}{n}\right\}^{1/2}  $$
		holds for all $i$ and for all $i'\in\mathcal{N}_i$ simultaneously. Following the argument in \cite{zhang2017estimating}, we obtain the desired result.
	\end{proof}
	
	By Lemma \ref{lem12}, with probability $1-2n^{-\tilde{C}_1/4}-2n^{-\tilde{C}_2/4}$, 
	\begin{align}
		&\frac{1}{n}\sum_{j}J_2(i,j)=\frac{1}{n}\sum_j\left\{ \frac{\sum_{i'\in \mathcal{N}_i}(wP_{i'j}-wP_{ij})}{|\mathcal{N}_i|} \right\}^2\nonumber\\
		\leq& \frac{\sum_{i'\in \mathcal{N}_i} \sum_j (wP_{i'j}-wP_{ij})^2/n}{|\mathcal{N}_i|} = \frac{\sum_{i'\in \mathcal{N}_i} \|wP_{i'\cdot} - wP_{i\cdot}\|_2^2/n}{|\mathcal{N}_i|}\nonumber\\
		\leq& w\rho_n\left[ 6L\left\{C_1+\left(\tilde{C}_2+4\right)^{1/2}\right\}^{1/2} + 8\left(C_2+\tilde{C}_2+2\right)^{1/2}  \right](n^{-1}\log n)^{1/2}, 
		\label{boundJ2final}
	\end{align}
	
	For $n^{-1}\sum_jJ_1(i,j)$, we can decomposite it as 
	\begin{align*}
		&\frac{1}{n}\sum_j J_1(i, j) = \frac{1} 
		{n|\mathcal{N}_i|^2}\sum_j\left\{ \sum_{i'\in \mathcal{N}_i} (Y_{i'j} - wP_{i'j}) \right\}^2\nonumber\\
		= & \frac{1}{n|\mathcal{N}_i|^2}\sum_j\left\{ \sum_{i'\in 
			\mathcal{N}_i} (Y_{i'j} - wP_{i'j})^2 + \sum_{i'\in \mathcal{N}_i}\sum_{i''\neq i', i''\in\mathcal{N}_i}(Y_{i'j}-wP_{i'j})(Y_{i''j}-wP_{i''j}) \right\}. 
	\end{align*}
	By Bernstein's inequality, we have 
	\begin{equation}
		\sum_j (Y_{i'j} - wP_{i'j})^2/n = \|Y_{i'\cdot} - wP_{i'\cdot}\|_2^2/n \leq Cw\rho_n. \label{boundJ11}
	\end{equation}
	with probability tending to 1. Furthermore, 
	\begin{align}
		\frac{1}{n|\mathcal{N}_i|^2}&\sum_j \sum_{i'\in \mathcal{N}_i}\sum_{i''\neq i', i''\in\mathcal{N}_i}(Y_{i'j}-wP_{i'j})(Y_{i''j}-wP_{i''j})\nonumber\\
		\leq & \frac{1}{|\mathcal{N}_i|^2} \sum_{i',i''\in\mathcal{N}_i: i'\neq i''} \left\{  \frac{1}{n-2}\left|\sum_{j\neq i', i''}(Y_{i'j}-wP_{i'j})(Y_{i''j}-wP_{i''j})\right| + \frac{2}{n}  \right\} \label{boundJ12}
	\end{align}
	and by Bernstein's inequality again and union bound, we have
	\begin{equation}
		\mathbb{P}\left[ \max_{i_1, i_2, i_1\neq i_2}\frac{1}{n- 
			2}\left| \sum_{j\neq i_1, i_2} \left( Y_{i_1 j} - wP_{i_1 j} \right) \left( Y_{i_2 j} - wP_{i_2 j} \right) \right| \geq w\rho_n\left\{\frac{(C_3+8)\log n}{n}\right\}^{1/2} \right] \leq 2n^{-\tilde{C}_3/4},
		\label{boundtermJ1}
	\end{equation}
	with $C_3,\tilde{C}_3$ and n large enough such that $1/\{C_0\left(n\log n\right)^{1/2}\} + 2/n\leq \left\{ \left(C_3+\tilde{C}_3+8\right)^{1/2} - \left(C_3+8\right)^{1/2} \right\}(n^{-1}\log n)^{1/2}$.
	
	Combining \eqref{boundJ11}, \eqref{boundJ12} and \eqref{boundtermJ1}, we can get
	\begin{equation}
		\frac{1}{n}\sum_j J_1(i, j) \leq w\rho_n\left\{\frac{(C_3+\tilde{C}_3+8)\log n}{n}\right\}^{1/2}.\label{boundJ1final}
	\end{equation}
	Combining \eqref{boundJ1final} and \eqref{boundJ2final}, dividing by $w^2$ we can get the desired result. Here, symmetrization will not affect the final rate.
\end{proof}

\subsection{Proof of Theorem 5}
We first consider the case that the true model is SBM with $K^*$ communities. By the consistency result of Algorithm 2, we have $\mathbb{P}(\hat{K}=K^*)\to1$. As before, in this case, we can condition on the event that $\hat{K}=K^*$.
\begin{proposition}\label{propa8}
	If the true model is $\delta^{(1,K^*)}$, then we have
	$$\ell_{1,K^*}(A,\mathcal{E}^c)-\ell_2(A,\mathcal{E}^c)\leq n^2O_{\mathbb{P}}\left(\max\left\{\sqrt{a_{n,w}(1-w)\rho_n},a_{n,w}\right\}+\frac{\rho_n(1-w)}{nw}\right).$$
\end{proposition}
\begin{proof}[Proof of Proposition \ref{propa8}]
	We have $\ell_{1,K^*}(A,\mathcal{E}^c)-\ell_2(A,\mathcal{E}^c)=\left(\ell_{1,K^*}(A,\mathcal{E}^c)-\ell_0(A,\mathcal{E}^c)\right)-\left(\ell_{2}(A,\mathcal{E}^c)-\ell_0(A,\mathcal{E}^c)\right).$ 
	Exactly as in the proof of Proposition \ref{propa1}, we have
	$$\ell_{1,K^*}(A,\mathcal{E}^c)-\ell_0(A,\mathcal{E}^c)\leq O_{\mathbb{P
	}}\left(\frac{n\rho_n(1-w)}{w}\right).$$
	Following the lines of the Case 1 in the proof of Proposition 1, we have
	$$|\ell_{2}(A,\mathcal{E}^c)-\ell_0(A,\mathcal{E}^c)|\leq O_{\mathbb{P}}\left(\|\hat{P}^{(2)}-P\|_F\sqrt{\|P\|_{\infty}|\mathcal{E}^c|}\right)+\|\hat{P}^{(2)}-P\|_F^2.$$
	Therefore, by Assumption 9, we have
	$$\ell_{2}(A,\mathcal{E}^c)-\ell_{0}(A,\mathcal{E}^c)\geq -n^2O_{\mathbb{P}}\left(\sqrt{a_{n,w}(1-w)\rho_n}+a_{n,w}\right)=-n^2O_{\mathbb{P}}\left(\max\left\{\sqrt{a_{n,w}(1-w)\rho_n},a_{n,w}\right\}\right).$$
	Combining the above formulas, we can get the desired result.
\end{proof}
\begin{proof}[Proof of Theorem 5 (SBM case)]
	We just need to prove that
	\begin{equation}
		\mathbb{P}\left[L_{1,K^*}(A,\mathcal{E}^c)>L_{2}(A,\mathcal{E}^c)\right]\to0,\quad\text{when $\delta^{(1,K^*)}$ is true}. \label{theo4_cons1}
	\end{equation}
	Notice that
	\begin{align*}
		\mathbb{P}\left[L_{1,K^*}(A,\mathcal{E}^c)>L_{2}(A,\mathcal{E}^c)\right]=&\mathbb{P}[\ell_{1,K^*}(A,\mathcal{E}^c)-\ell_2(A,\mathcal{E}^c)>(d_{2}-d_{1,K})|\mathcal{E}^c|\lambda_n]\\
		\leq&\mathbb{P}[\ell_{1,K^*}(A,\mathcal{E}^c)-\ell_2(A,\mathcal{E}^c)>1/2(d_{2}-d_{1,K})n^2(1-w)\lambda_n]\to0
	\end{align*}
	by $d_2-d_{1,K}\sim \sqrt{n/\log n}$, $\lambda_n=\omega_{\mathbb{P}}\left({\max\left\{\sqrt{a_{n,w}(1-w)\rho_n},a_{n,w}\right\}}\bigg/{(1-w)\cdot \sqrt{n/\log n}}\right)$, $\lambda_n=\omega_{\mathbb{P}}(\rho_n/nw)$ and Proposition \ref{propa8}.
\end{proof}

Now we consider the case that the true model is the graphon model. Here the estimator $\hat{K}$ may take value in all candidates in $\mathcal{K}$, and thus we need to bound the loss carefully. Denote the estimated label as $\hat{c}$, and denote 
$$\mathcal{T}_{k,l}:=\{(i,j)\in\mathcal{E}^c:\hat{c}_i=k,\hat{c}_j=l\},$$
$$\mathcal{U}_{k,l}:=\{(i,j)\in\mathcal{E}:\hat{c}_i=k,\hat{c}_j=l\},$$
$$\mathcal{Q}_{k,l}:=\{(i,j):\hat{c}_i=k,\hat{c}_j=l\},$$
and for any possible label $c$, similarly denote
$$\mathcal{T}_{c,k,l}:=\{(i,j)\in\mathcal{E}^c:c_i=k,c_j=l\},$$
$$\mathcal{U}_{c,k,l}:=\{(i,j)\in\mathcal{E}:c_i=k,c_j=l\},$$
$$\mathcal{Q}_{c,k,l}:=\{(i,j):c_i=k,c_j=l\},$$
together with $\mathcal{G}_{c,k}:=\{i\in[n]:c_i=k\}.$
\begin{lemma}\label{lem13}
	Assume Assumption 11 holds, and $w^{-1}\max_{k\in\mathcal{K}}K<n^{1/4}/\log^{1/4} n$. With probability tending to 1, we have 
	\begin{equation}
		\min_{k,l}|\mathcal{U}_{k,l}|\geq\frac{\pi_0'^2n^2w}{2\max_{K\in\mathcal{K}}K^2},\quad \max_{k,l}|\mathcal{T}_{k,l}|\leq4\left((1-\pi_0')^2+\frac{\pi_0'^2}{\min_{K\in\mathcal{K}}K^2}\right)n^2(1-w).\label{graphonUTcon}
	\end{equation}
\end{lemma}
\begin{proof}
	Denote the set of all balanced labels that satisfy Assumption 11 with $k$ communities as $\mathcal{L}_k$. We have $$\min_{k,l}|\mathcal{U}_{k,l}|\geq\min_{c\in\mathcal{L}_{\hat{K}}}\min_{k,l}|\mathcal{U}_{c,k,l}|\geq\min_{k\in\mathcal{K}}\min_{c\in\mathcal{L}_k}\min_{l_1,l_2}|\mathcal{U}_{c,l_1,l_2}|.$$
	Therefore, we have
	\begin{align}
		\mathbb{P}\left[\min_{k,l}|\mathcal{U}_{k,l}|\leq\frac{\pi_0'^2n^2w}{2\max_{K\in\mathcal{K}}K^2}\right]&\leq\mathbb{P}\left[\min_{k\in\mathcal{K}}\min_{c\in\mathcal{L}_k}\min_{l_1,l_2}|\mathcal{U}_{c,l_1,l_2}|\leq\frac{\pi_0'^2n^2w}{2\max_{K\in\mathcal{K}}K^2}\right]\nonumber\\
		&\leq\sum_{k\in\mathcal{K}}\sum_{c\in\mathcal{L}_k}\sum_{l_1,l_2}\mathbb{P}\left[|\mathcal{U}_{c,l_1,l_2}|\leq\frac{\pi_0'^2n^2w}{2\max_{K\in\mathcal{K}}K^2}\right]\nonumber\\
		&\leq\sum_{k\in\mathcal{K}}\sum_{c\in\mathcal{L}_k}\sum_{l_1,l_2}\mathbb{P}\left[|\mathcal{U}_{c,l_1,l_2}|\leq\frac{\pi_0'^2n^2w}{2k^2}\right].\label{graphoncontrol}
	\end{align}
	For each given label $c$ with $k$ communities, we have by Hoeffding's inequality, when $l_1\neq l_2$
	\begin{align*}
		\mathbb{P}\left[|\mathcal{U}_{c,l_1,l_2}|\leq\frac{\pi_0'^2n^2w}{2k^2}\right]&=\mathbb{P}\left[|(\mathcal{G}_{c,l_1}\times\mathcal{G}_{c,l_2})\cap\mathcal{E}|\leq\frac{\pi_0'^2n^2w}{2k^2}\right]\\
		&\leq\mathbb{P}\left[|(\mathcal{G}_{c,l_1}\times\mathcal{G}_{c,l_2})\cap\mathcal{E}|\leq\frac{n_{c,l_1}n_{c,l2}w}{2}\right]\\
		&\leq\exp\left(-\frac{n_{c,l_1}n_{c,l2}w^2}{4}\right)\leq\exp\left(-\frac{\pi_0'^2n^2w^2}{4k^2}\right)\leq\exp\left(-\frac{\pi_0'^2n^2w^2}{4\max_{K\in\mathcal{K}}K^2}\right),
	\end{align*}
	and when $l_1=l_2$, a similar inequality holds.
	Plug in this bound into \eqref{graphoncontrol}, we have 
	\begin{align*}
		\mathbb{P}\left[\min_{k,l}|\mathcal{U}_{k,l}|\leq\frac{\pi_0'^2n^2w}{2\max_{K\in\mathcal{K}}K^2}\right]&\leq\sum_{k\in\mathcal{K}}\sum_{c\in\mathcal{L}_k}\sum_{l_1,l_2}\exp\left(-\frac{\pi_0'^2n^2w^2}{4\max_{K\in\mathcal{K}}K^2}\right)\\
		&\leq(\max_{k\in\mathcal{K}}K)^{n+3}\exp\left(-\frac{\pi_0'^2n^2w^2}{4\max_{K\in\mathcal{K}}K^2}\right)\\
		&=\exp\left((n+3)\log\max_{k\in\mathcal{K}}K-\frac{\pi_0'^2n^2w^2}{4\max_{K\in\mathcal{K}}K^2}\right).
	\end{align*}
	Thus when $w^{-1}\max_{k\in\mathcal{K}}K<n^{1/4}/\log^{1/4} n$, the above formula tends to 0. The other inequality holds similarly.
\end{proof}

\begin{lemma}\label{lem:uniform-block}
	Let $\mathcal L_k$ be the
	collection of balanced labelings with $k$ communities defined in the proof of Lemma \ref{lem13}. Denote $K_{\max}=\max_{K\in\mathcal{K}} K$. For $c\in\mathcal L_k$, define
	\[
	U_{c,r,s}:=\{(i,j)\in\mathcal E:c_i=r,\ c_j=s\}.
	\]
	On the high-probability event in Lemma~\ref{lem13},
	\[
	\max_{k\in\mathcal K}\max_{c\in\mathcal L_k}\max_{r,s\in[k]}\left|\frac{\sum_{(i,j)\in U_{c,r,s}}(A_{ij}-P_{ij})}{|U_{c,r,s}|}\right|=O_{\mathbb P}(r_{n,\mathcal K}),
	\]
	where
	\[
	r_{n,\mathcal K}:=K_{\max}\sqrt{\frac{\rho_n\log(K_{\max})}{nw}}+\frac{K_{\max}^2\log(K_{\max})}{nw}.
	\]
\end{lemma}

\begin{proof}
	Conditional on the sampling set $\mathcal E$, for every fixed labeling
	$c\in\mathcal L_k$ and every fixed pair $(r,s)$, Bernstein's inequality
	yields
	\[
	\mathbb P\left(\left|\sum_{(i,j)\in U_{c,r,s}}(A_{ij}-P_{ij})\right|\geq |U_{c,r,s}|t\,\middle|\,\mathcal E\right)\leq2\exp\left\{-\frac{|U_{c,r,s}|t^2}{2\rho_n+2t/3}\right\}.
	\]
	By Lemma \ref{lem13},
	\[
	|U_{c,r,s}|\geq\frac{\pi_0^{\prime 2}n^2w}{2K_{\max}^2}
	\]
	uniformly over all admissible $c,r,s$. Since the total number of
	labelings and block pairs is at most $K_{\max}^{n+3}$, the conclusion
	follows by taking a union bound and choosing $t$ to be a sufficiently
	large constant multiple of $r_{n,\mathcal K}$.
\end{proof}
Now in order to prove the consistency when the true model is graphon, we just need to condition on the event $E$ that \eqref{graphonUTcon} holds.
\begin{proposition}\label{propa9}
	If the true model is $\delta^{(2)}$ and at least $b_{n,\mathcal{K}}$ variable, then conditioned on $E$, we have
	$$\ell_{1,\hat{K}}(A,\mathcal{E}^c)-\ell_2(A,\mathcal{E}^c)\geq\Omega_{\mathbb{P}}((1-w)n^2b_{n,\mathcal{K}}).$$
\end{proposition}
\begin{proof}[Proof of Proposition \ref{propa9}]
	Since the true model is the graphon model, the formula 
	\begin{equation}
		\left|\ell_{2}(A,\mathcal{E}^c)-\ell_{0}(A,\mathcal{E}^c)\right|=n^2O_{\mathbb{P}}\left(\max\left\{\sqrt{a_{n,w}(1-w)\rho_n},a_{n,w}\right\}\right). \label{finaluppest}
	\end{equation}
	still holds. We first obtain a lower bound for
	$$\mathbb{E}[\ell_{1,\hat{K}}(A,\mathcal{E}^c)-\ell_{0}(A,\mathcal{E}^c)]=\sum_{(i,j)\in\mathcal{E}_2}(P_{ij}-\hat{P}^{(1,\hat{K})}_{ij})^2,$$
	where the expectation is taken over the test entries. We have
	\begin{align}
		&\mathbb{E}\{\ell_{1,\hat{K}}(A,\mathcal{E}^c)-\ell_{0}(A,\mathcal{E}^c)\}= \sum_{k,l}\sum_{(i,j)\in \mathcal{T}_{k,l}}(P_{ij}-\hat{P}^{(1,\hat{K})}_{ij})^2 \nonumber\\
		\overset{(*)}{=}&\sum_{k,l}\left[\sum_{(i,j)\in \mathcal{T}_{k,l}}\left(P_{ij}-\frac{\sum_{(i',j')\in \mathcal{T}_{k,l}}P_{i'j'}}{|\mathcal{T}_{k,l}|}\right)^2+|T_{k,l}|\left(\frac{\sum_{(i',j')\in \mathcal{U}_{k,l}}A_{i'j'}}{|\mathcal{U}_{k,l}|}-\frac{\sum_{(i',j')\in \mathcal{T}_{k,l}}P_{i'j'}}{|\mathcal{T}_{k,l}|}\right)^2\right],\label{eq4}
	\end{align}
	where the $(*)$ follows from \eqref{qualoss}. Expand this formula further, we obtain   
	\begin{align*}
		&\sum_{k,l}\left(\sum_{(i,j)\in \mathcal{T}_{k,l}}\left(P_{ij}-\frac{\sum_{(i',j')\in \mathcal{T}_{k,l}}P_{i'j'}}{|\mathcal{T}_{k,l}|}\right)^2+|\mathcal{T}_{k,l}|\left(\frac{\sum_{(i',j')\in \mathcal{U}_{k,l}}A_{i'j'}}{|\mathcal{U}_{k,l}|}-\frac{\sum_{(i',j')\in \mathcal{T}_{k,l}}P_{i'j'}}{|\mathcal{T}_{k,l}|}\right)^2\right)\\
		=&\sum_{k,l}\sum_{(i,j)\in \mathcal{T}_{k,l}}\left[\left(P_{ij}-\frac{\sum_{(i',j')\in \mathcal{T}_{k,l}}P_{i'j'}}{|\mathcal{T}_{k,l}|}\right)^2+\left(\frac{\sum_{(i',j')\in \mathcal{U}_{k,l}}A_{i'j'}}{|\mathcal{U}_{k,l}|}-\frac{\sum_{(i',j')\in \mathcal{T}_{k,l}}P_{i'j'}}{|\mathcal{T}_{k,l}|}\right)^2\right]\\
		=&\sum_{k,l}\sum_{(i,j)\in Q_{k,l}}\left[\left(P_{ij}-\frac{\sum_{(i',j')\in \mathcal{T}_{k,l}}P_{i'j'}}{|\mathcal{T}_{k,l}|}\right)^2+\left(\frac{\sum_{(i',j')\in \mathcal{T}_{k,l}}P_{i'j'}}{|\mathcal{T}_{k,l}|}-\frac{\sum_{(i',j')\in \mathcal{U}_{k,l}}P_{i'j'}}{|\mathcal{U}_{k,l}|}\right)^2+\left(\frac{\sum_{(i',j')\in \mathcal{U}_{k,l}}A_{i'j'}}{|\mathcal{U}_{k,l}|}\right.\right.\\
		&-\left.\left.\frac{\sum_{(i',j')\in \mathcal{T}_{k,l}}P_{i'j'}}{|\mathcal{T}_{k,l}|}\right)^2-\left(\frac{\sum_{(i',j')\in \mathcal{T}_{k,l}}P_{i'j'}}{|\mathcal{T}_{k,l}|}-\frac{\sum_{(i',j')\in \mathcal{U}_{k,l}}P_{i'j'}}{|\mathcal{U}_{k,l}|}\right)^2\right]\mathbbm{1}\{(i,j)\in\mathcal{E}^c\}\\ 
		=&\sum_{k,l}\sum_{(i,j)\in Q_{k,l}}\left[\left(P_{ij}-\frac{\sum_{(i',j')\in \mathcal{T}_{k,l}}P_{i'j'}}{|\mathcal{T}_{k,l}|}\right)^2+\left(\frac{\sum_{(i',j')\in \mathcal{T}_{k,l}}P_{i'j'}}{|\mathcal{T}_{k,l}|}-\frac{\sum_{(i',j')\in \mathcal{U}_{k,l}}P_{i'j'}}{|\mathcal{U}_{k,l}|}\right)^2\right]\mathbbm{1}\{(i,j)\in\mathcal{E}^c\}\\
		&+\sum_{k,l}|\mathcal{T}_{k,l}|\left(\frac{\sum_{(i,j)\in \mathcal{U}_{k,l}}A_{ij}}{|\mathcal{U}_{k,l}|}-\frac{\sum_{(i,j)\in \mathcal{U}_{k,l}}P_{ij}}{|\mathcal{U}_{k,l}|}\right)\left(\frac{\sum_{(i,j)\in \mathcal{U}_{k,l}}A_{ij}}{|\mathcal{U}_{k,l}|}+\frac{\sum_{(i,j)\in \mathcal{U}_{k,l}}P_{ij}}{|\mathcal{U}_{k,l}|}-2\frac{\sum_{(i,j)\in \mathcal{T}_{k,l}}P_{ij}}{|\mathcal{T}_{k,l}|}\right)\\
		=&:\mathcal{I}+\mathcal{II}.
	\end{align*}
	For the second term, we have
	\begin{align*}
		\mathcal{II}\geq&-\sum_{k,l}|\mathcal{T}_{k,l}|\left|\frac{\sum_{(i,j)\in \mathcal{U}_{k,l}}A_{ij}}{|\mathcal{U}_{k,l}|}-\frac{\sum_{(i,j)\in \mathcal{U}_{k,l}}P_{ij}}{|\mathcal{U}_{k,l}|}\right|\left|\frac{\sum_{(i,j)\in \mathcal{U}_{k,l}}A_{ij}}{|\mathcal{U}_{k,l}|}+\frac{\sum_{(i,j)\in \mathcal{U}_{k,l}}P_{ij}}{|\mathcal{U}_{k,l}|}-2\frac{\sum_{(i,j)\in \mathcal{T}_{k,l}}P_{ij}}{|\mathcal{T}_{k,l}|}\right|\nonumber\\
		\geq&-\sum_{k,l}|\mathcal{T}_{k,l}|\left|\frac{\sum_{(i,j)\in \mathcal{U}_{k,l}}A_{ij}}{|\mathcal{U}_{k,l}|}-\frac{\sum_{(i,j)\in \mathcal{U}_{k,l}}P_{ij}}{|\mathcal{U}_{k,l}|}\right|\left(\left|\frac{\sum_{(i,j)\in \mathcal{U}_{k,l}}A_{ij}}{|\mathcal{U}_{k,l}|}-\frac{\sum_{(i,j)\in \mathcal{U}_{k,l}}P_{ij}}{|\mathcal{U}_{k,l}|}\right|+2\frac{\sum_{(i,j)\in \mathcal{U}_{k,l}}P_{ij}}{|\mathcal{U}_{k,l}|}\right.\\
		&\left.+2\frac{\sum_{(i,j)\in \mathcal{T}_{k,l}}P_{ij}}{|\mathcal{T}_{k,l}|}\right).
	\end{align*}
	By Lemma \ref{lem:uniform-block}, the uniform concentration event applies
	to the data-dependent labeling $\widehat c$. Hence,
	\[
	\max_{k,l}\left|\frac{\sum_{(i,j)\in U_{k,l}}A_{ij}}{|U_{k,l}|}-\frac{\sum_{(i,j)\in U_{k,l}}P_{ij}}{|U_{k,l}|}\right|=O_{\mathbb P}(r_{n,\mathcal K}).
	\]
	Consequently,
	\[
	\mathcal{II}
	\geq
	-O_{\mathbb P}\left(
	n^2(1-w)
	\left\{
	\rho_n r_{n,\mathcal K}
	+
	r_{n,\mathcal K}^2
	\right\}
	\right).
	\]
	Under the condition $K_{\max}=o\left({n^{1/4}}/{\log^{1/4}n}\right)$
	and the sparsity condition for the neighborhood smoothing estimator (which is a corollary of Assumption 10),
	\[
	\rho_n r_{n,\mathcal K}
	+
	r_{n,\mathcal K}^2
	=
	o\left(
	\sqrt{\frac{\rho_n^3\log n}{w\sqrt n}}
	\right).
	\]
	Therefore, we have
	\begin{equation}
		\mathcal{II}\geq-O_{\mathbb{P}}\left(\frac{1-w}{\sqrt{w}}n^{7/4}\sqrt{\log n}\rho_n^{3/2}\right).\label{finalIIest}
	\end{equation}
	
	For the first term, similarly as \eqref{cauchy}, we have for $(i,j)\in\mathcal{Q}_{k,l}$,
	$$\left(P_{ij}-\frac{\sum_{(i',j')\in \mathcal{T}_{k,l}}P_{i'j'}}{|\mathcal{T}_{k,l}|}\right)^2+\left(\frac{\sum_{(i',j')\in \mathcal{T}_{k,l}}P_{i'j'}}{|\mathcal{T}_{k,l}|}-\frac{\sum_{(i',j')\in \mathcal{U}_{k,l}}P_{i'j'}}{|\mathcal{U}_{k,l}|}\right)^2\geq\left(P_{ij}-\frac{\sum_{(i',j')\in \mathcal{Q}_{k,l}}P_{i'j'}}{|\mathcal{Q}_{k,l}|}\right)^2.$$
	Therefore,
	$$\mathcal{I}\geq\sum_{k,l}\sum_{(i,j)\in \mathcal{Q}_{k,l}}\left(P_{ij}-\frac{\sum_{(i',j')\in \mathcal{Q}_{k,l}}P_{i'j'}}{|\mathcal{Q}_{k,l}|}\right)^2\mathbbm{1}\{(i,j)\in\mathcal{E}^c\}.$$
	We claim that under Assumption 10, 
	\begin{equation}
		\min_{k\in\mathcal{K}}\min_{c\in\mathcal{L}_k}\left\{\sum_{k,l}\sum_{(i,j)\in \mathcal{Q}_{c,k,l}}\left(P_{ij}-\frac{\sum_{(i',j')\in \mathcal{Q}_{c,k,l}}P_{i'j'}}{|\mathcal{Q}_{c,k,l}|}\right)^2\mathbbm{1}\{(i,j)\in\mathcal{E}^c\}\right\}=\Omega_{\mathbb{P}}((1-w)n^2b_{n,\mathcal{K}}), \label{corlowerest}
	\end{equation}
	where $Q_{c,k,l}$ denotes the $(k,l)$-th block under label $c$.
	
	For any given label $c$, denote $$a_{ij}:=\left(P_{ij}-\frac{\sum_{(i',j')\in\mathcal{Q}_{c,c_ic_j}}P_{i'j'}}{|\mathcal{Q}_{c,c_ic_j}|}\right)^2.$$ 
	Notice that $a_{ij}=a_{ji}$. Then by Hoeffding's inequality, we have
	$$\mathbb{P}\left[\sum_{i,j}a_{ij}\mathbbm{1}\{(i,j)\in\mathcal{E}^c\}\leq\frac{1-w}{2}\sum_{i,j}a_{ij}\right]\leq\exp\left(-\frac{(1-w)^2(\sum_{i,j}a_{ij})^2}{4\sum_{i,j}a_{ij}}\right)\leq\exp\left(-\frac{(1-w)^2n^2b_{n,\mathcal{K}}}{4}\right),$$
	since $\sum_{i,j}a_{ij}=\Omega(n^2b_{n,\mathcal{K}})$. This leads to 
	$$\mathbb{P}\left[\sum_{i,j}a_{ij}\mathbbm{1}\{(i,j)\in\mathcal{E}^c\}\leq\frac{1-w}{2}n^2b_{n,\mathcal{K}}\right]\leq\exp\left(-\frac{(1-w)^2(\sum_{i,j}a_{ij})^2}{4\sum_{i,j}a_{ij}}\right)\leq\exp\left(-\frac{(1-w)^2n^2b_{n,\mathcal{K}}}{4}\right).$$
	Therefore, by taking a union bound over all the possible $c$ and all possible $k$, we get
	\begin{align*}
		\mathbb{P}&\left[\min_{k\in\mathcal{K}}\min_{c\in\mathcal{L}_k}\left\{\sum_{k,l}\sum_{(i,j)\in \mathcal{Q}_{c,k,l}}\left(P_{ij}-\frac{\sum_{(i',j')\in \mathcal{Q}_{c,k,l}}P_{i'j'}}{|\mathcal{Q}_{c,k,l}|}\right)^2\mathbbm{1}\{(i,j)\in\mathcal{E}^c\}\right\}\leq\frac{1-w}{2}n^2b_{n,\mathcal{K}}\right]\\
		\leq& \left(\max_{k\in\mathcal{K}} K\right)^{n+1}\exp\left(-\frac{(1-w)^2n^2b_{n,\mathcal{K}}}{4}\right)=\exp\left((n+1) \max_{k\in\mathcal{K}}\log K-\frac{(1-w)^2n^2b_{n,\mathcal{K}}}{4}\right).
	\end{align*}
	By Assumption 10, $b_{n,\mathcal{K}}\gg\frac{\max_{K\in\mathcal{K}}\log K}{n(1-w)^2}$, and thus the probability above tends to 0 as $n\to\infty$. Thus \eqref{corlowerest} holds. It follows that
	\begin{equation}
		\mathcal{I}\geq\min_{k\in\mathcal{K}}\min_{c\in\mathcal{L}_k}\left\{\sum_{k,l}\sum_{(i,j)\in \mathcal{Q}_{c,k,l}}\left(P_{ij}-\frac{\sum_{(i',j')\in \mathcal{Q}_{c,k,l}}P_{i'j'}}{|\mathcal{Q}_{c,k,l}|}\right)^2\mathbbm{1}\{(i,j)\in\mathcal{E}^c\}\right\}=\Omega_{\mathbb{P}}((1-w)n^2b_{n,\mathcal{K}}). \label{finalIest}
	\end{equation}
	Combining \eqref{finalIIest} and \eqref{finalIest}, we have
	$$\mathbb{E}\{\ell_{1,\hat{K}}(A,\mathcal{E}^c)-\ell_{0}(A,\mathcal{E}^c)\}\geq\Omega_{\mathbb{P}}((1-w)n^2b_{n,\mathcal{K}})-O_{\mathbb{P}}\left(\frac{1-w}{\sqrt{w}}n^{7/4}\sqrt{\log n}\rho_n^{3/2}\right)=\Omega_{\mathbb{P}}((1-w)n^2b_{n,\mathcal{K}})$$
	by Assumption 10. Now apply Hoeffding's inequality on the test entries, we have with probability tending to 1,
	\begin{equation}
		\ell_{1,\hat{K}}(A,\mathcal{E}^c)-\ell_{0}(A,\mathcal{E}^c)\geq\frac{1}{2}\mathbb{E}\{\ell_{1,\hat{K}}(A,\mathcal{E}^c)-\ell_{0}(A,\mathcal{E}^c)\}=\Omega_{\mathbb{P}}((1-w)n(n-1)b_{n,\mathcal{K}}). \label{finalunderest}
	\end{equation}
	Combining \eqref{finaluppest} and \eqref{finalunderest}, with the order assumption between $a_{n,w}$ and $b_{n,\mathcal{K}}$ in Assumption 10, we have
	$$\ell_{1,\hat{K}}(A,\mathcal{E}^c)-\ell_2(A,\mathcal{E}^c)=(\ell_{1,\hat{K}}(A,\mathcal{E}^c)-\ell_0(A,\mathcal{E}^c))-(\ell_2(A,\mathcal{E}^c)-\ell_0(A,\mathcal{E}^c))=\Omega_{\mathbb{P}}((1-w)n^2b_{n,\mathcal{K}}).$$
	
\end{proof}

\begin{proof}[Proof of Theorem 5 (graphon case)]
	As before, in addition to the event $E$, we also condition on the event $\Omega$. We just need to prove 
	\begin{equation}
		\mathbb{P}\left[L_{1,\hat{K}}(A,\mathcal{E}^c)<L_{2}(A,\mathcal{E}^c)\right]\to0,  \label{theo4_cons2}
	\end{equation}
	when $\delta^{(2)}$ is true and satisfies Assumption 10. Notice that
	\begin{align*}
		\mathbb{P}\left[L_{1,\hat{K}}(A,\mathcal{E}^c)<L_{2}(A,\mathcal{E}^c)\right]=&\mathbb{P}[\ell_{1,\hat{K}}(A,\mathcal{E}^c)-\ell_2(A,\mathcal{E}^c)<(d_{2}-d_{1,K})|\mathcal{E}^c|\lambda_n]\\
		\leq&\mathbb{P}[\ell_{1,\hat{K}}(A,\mathcal{E}^c)-\ell_2(A,\mathcal{E}^c)<1/2(d_{2}-d_{1,K})n^2(1-w)\lambda_n]\to0
	\end{align*}
	by $d_2-d_{1,K}\sim \sqrt{n/\log n}$, $\lambda_n\cdot \sqrt{n/\log n}=o_{\mathbb{P}}(b_{n,\mathcal{K}})$ and Proposition \ref{propa9}.
\end{proof}

\begin{remark}
	The proofs of Theorem 3, 4 and 5 imply that, in our procedures, even without a penalty term, underfitting can be avoided by controlling the gap between the original prediction loss and the oracle loss defined by
	\begin{equation}
		\ell_0(A,\mathcal{E}^c):=\sum_{(i,j)\in\mathcal{E}^c}(A_{ij}-P_{ij})^2.\label{oracle}
	\end{equation}
	Therefore, in each procedure, we introduce a penalty term primarily to prevent overfitting.
\end{remark}

\begin{remark}
	In the proof of Proposition~\ref{propa9}, we obtained
	\[
	\mathcal{I}\geq\Omega_{\mathbb P}\bigl((1-w)n^2b_{n,\mathcal K}\bigr),\qquad\mathcal{II}\geq-O_{\mathbb P}\left(n^2(1-w)(\rho_nr_{n,\mathcal{K}}+r^2_{n,\mathcal{K}})\right),
	\]
	where 
	$$r_{n,\mathcal K}=K_{\max}\sqrt{\frac{\rho_n\log(K_{\max})}{nw}}+\frac{K_{\max}^2\log(K_{\max})}{nw}.$$
	Since $P_{ij}\leq\rho_n$, the blockwise approximation error satisfies $b_{n,\mathcal K}=O(\rho_n^2)$. Hence, when $w$ is of constant order and $K_{\max}=\Omega(n^{1/2})$, the first term of $r_{n,\mathcal{K}}$ is of order $\Omega(\sqrt{\rho_n\log n})$, and the available bound for the negative remainder is at least of order $n^2\sqrt{\log n}\rho_n^{3/2},$ whereas the largest possible benchmark order for the signal lower bound is $n^2\rho_n^2$. Their ratio is at least of order $\sqrt{{\log n}/{\rho_n}}\to\infty.$
	Consequently, the current argument can no longer guarantee that $\mathcal{I}+\mathcal{II}$ remains positive in this regime. This identifies $K_{\max}=\Omega(n^{1/2})$ as a technical barrier of the present proof, rather than an intrinsic impossibility for the proposed procedure.
\end{remark}

\begin{remark}
	The maximum allowable order for $\max_{K\in\mathcal{K}}K$ directly influences the third term for the formula
	$$b_{n,\mathcal{K}}\gg \max\left\{\sqrt{\frac{a_{n,w}\rho_n}{(1-w_n)}},\frac{a_{n,w}}{1-w_n},\sqrt{\frac{\rho_n^3\log n}{w_n\sqrt{n}}},\frac{\max_{K\in\mathcal{K}}\log K}{n(1-w_n)^2}\right\}$$
	stated in Assumption 10. In fact, a more general condition on $b_{n,\mathcal{K}}$ could be written as
	$$b_{n,\mathcal{K}}\gg \max\left\{\sqrt{\frac{a_{n,w}\rho_n}{(1-w_n)}},\frac{a_{n,w}}{1-w_n},\rho_nr_{n,\mathcal{K}}+r^2_{n,\mathcal{K}},\frac{\max_{K\in\mathcal{K}}\log K}{n(1-w_n)^2}\right\}.$$
	Specifically, for example, if $\max_{K \in \mathcal{K}} K = \Omega(n^{3/8})$, then this third term becomes dominant when the NS procedure is applied. In this regime, the tolerance for the sparsity parameter $\rho_n$ also decreases accordingly.
\end{remark}

\newpage

\section{Derivations for the Illustrative Examples}
\subsection{Derivation for Example 1}
Consider an SBM with $K$ communities, where $1\leq K< n$, characterized by a node membership vector $c$ and a block-wise connection probability matrix $B$. 
Let $\mathcal{G}_k=\{i\in[n]:c_i=k\}$ denote the set of nodes belonging to community $k$. Since $K<n$, at least one community contains at least two nodes. By relabeling the communities if necessary, we may assume without loss of generality that $|\mathcal{G}_K|\geq2$. 
Partition $\mathcal{G}_K$ into two nonempty, disjoint subsets $\mathcal{G}'_K$ and $\mathcal{G}'_{K+1}$ such that $\mathcal{G}_K=\mathcal{G}'_K\cup\mathcal{G}'_{K+1}$.
Define a new membership vector $c'$ corresponding to $K+1$ communities by $c'_i=c_i$ if $i\notin \mathcal{G}_K$, $c'_i=K$ if $i\in\mathcal{G}'_K$ and $c'_i=K+1$ if $i\in\mathcal{G}'_{K+1}$. Next, define a symmetric $(K+1)\times(K+1)$ matrix $B'$ by 

$$B'_{k_1k_2}=\begin{cases}
	B_{k_1k_2}, & k_1\leq k_2\in[K-1], \\
	B_{k_1K},& k_1\leq K\ \mathrm{and}\ k_2\geq K,\\
	B_{KK},& k_1=k_2=K+1.\\
\end{cases}$$
Under this construction, the probability matrix satisfies $P_{ij}=B'_{c'_ic'_j}(1-\delta_{ij})$, so the original probability matrix can also be represented by an SBM with $K+1$ communities. Hence, the class of $K$-community SBMs is nested within the class of $(K+1)$-community SBMs.

\subsection{Derivation for Example 2}

In Example 2, the inclusions $\mathcal{P}^{(1,K)}\subset \mathcal{P}^{(2,K)}\subset \mathcal{P}^{(3,K)}$ follow directly from the definitions. It therefore remains only to establish that $\mathcal{P}^{(3,K)}\subset \mathcal{P}^{(4)}$.

We show this inclusion at the level of admissible probability matrices, in accordance with Definition~1 of the main paper. Let $P\in\mathcal P^{(3,K)}$ be arbitrary. Then there exist a membership vector $c\in\mathcal C_K$, a symmetric probability matrix $B\in\mathcal B_K$, and a degree parameter vector $\theta\in\Theta_c$ such that
\[
P_{ij}=\theta_i\theta_j B_{c_i c_j}(1-\delta_{ij}),\qquad i,j\in[n].
\]
For each $k\in[K]$, let $\mathcal G_k=\{i\in[n]:c_i=k\}$, $n_k=|\mathcal G_k|$, and enumerate the nodes in $\mathcal G_k$ as
\[
\mathcal G_k=\{i_{k,1},\ldots,i_{k,n_k}\}.
\]
Set $\pi_k=n_k/n$, and define $a_0=0$ and $a_k=\sum_{\ell=1}^k\pi_\ell$ for $k=1,\ldots,K$. Let
\[
\mathcal J_k=[a_{k-1},a_k)
\]
denote the interval associated with community $k$, with the convention that the final interval $\mathcal J_K$ contains the endpoint 1. 
We further partition $\mathcal J_k$ into $n_k$ subintervals of equal length:
\[
\mathcal I_{k,r}=\left[a_{k-1}+\frac{(r-1)\pi_k}{n_k},\,a_{k-1}+\frac{r\pi_k}{n_k}\right),\qquad r=1,\ldots,n_k.
\]
For $x\in \mathcal I_{k,r}$, define
\[
\kappa(x)=k,\qquad \iota(x)=i_{k,r},\qquad \vartheta(x)=\theta_{i_{k,r}}.
\]
Now define a symmetric measurable function $f:[0,1]^2\to[0,1]$ by
\[
f(x,y)=
\begin{cases}
	\vartheta(x)\vartheta(y)B_{\kappa(x)\kappa(y)}, & \iota(x)\ne \iota(y),\\
	0, & \iota(x)=\iota(y).
\end{cases}
\]
The function $f$ is a graphon. The off-diagonal values are in $[0,1]$ because $P$ is an admissible DCBM probability matrix, and the values on the diagonal rectangles $\mathcal I_{k,r}\times \mathcal I_{k,r}$ are set to $0$, which is immaterial for recovering the hollow probability matrix.

For each node $i_{k,r}\in \mathcal G_k$, choose any latent position $\xi_{i_{k,r}}\in \mathcal I_{k,r}$. Then, for any distinct $i,j\in[n]$,
\[
f(\xi_i,\xi_j)=\theta_i\theta_j B_{c_i c_j}=P_{ij}.
\]
Moreover, since the diagonal entries of the probability matrix are zero,
\[
f(\xi_i,\xi_i)(1-\delta_{ii})=0=P_{ii}.
\]
Therefore,
\[
P_{ij}=f(\xi_i,\xi_j)(1-\delta_{ij}),\qquad i,j\in[n],
\]
which shows that $P\in\mathcal P^{(4)}$. Since $P\in\mathcal P^{(3,K)}$ was arbitrary, we conclude that
\[
\mathcal P^{(3,K)}\subset \mathcal P^{(4)}.
\]
Combining the preceding inclusions yields
\[
\mathcal P^{(1,K)}\subset \mathcal P^{(2,K)}\subset \mathcal P^{(3,K)}\subset \mathcal P^{(4)}.
\]
This nested structure highlights the flexibility of the graphon model in accommodating a broad range of network models.

The construction also provides a latent-variable interpretation of graphon models. Specifically, if $\xi\sim \operatorname{Unif}(0,1)$, then $\kappa(\xi)$ follows a multinomial distribution on $[K]$ with probabilities $(\pi_1,\ldots,\pi_K)$. 
Conditional on $\kappa(\xi)=k$, the degree factor $\vartheta(\xi)$ takes the values $\{\theta_i:i\in \mathcal G_k\}$ with equal probabilities $1/n_k$. Hence the pair $(\kappa(\xi),\vartheta(\xi))$ is an i.i.d. latent variable across nodes when $\xi_1,\ldots,\xi_n$ are i.i.d. uniform, but $\vartheta(\xi)$ need not be independent of $\kappa(\xi)$. In fact, the conditional distribution of $\vartheta(\xi)$ may vary across communities, and the DCBM identifiability constraint implies
\[
\mathbb E\{\vartheta(\xi)^2\mid \kappa(\xi)=k\}=\frac{1}{n_k}\sum_{i\in \mathcal G_k}\theta_i^2=1.
\]
Thus the DCBM can be viewed as a graphon model whose latent position encodes both community membership and degree heterogeneity.

\subsection{An Additional Example in Comparing SBMs and DCBMs}
As suggested by a reviewer, we provide a simple illustrative example to clarify the relationship between the degree-corrected stochastic block model (DCBM) and the stochastic block model (SBM). 
Consider a DCBM with K communities in which the degree parameters $\theta_i$ take two distinct values within each community. Such a model can be equivalently represented as an SBM with $2K$ communities, by further partitioning each original community according to the two degree levels.

In this setting, our method favors the more parsimonious representation, namely, a DCBM with $K$ communities, which will be explained below. This behavior follows from the hierarchical structure of our model-selection procedure. 
We first determine the number of communities within the DCBM class. Because the DCBM explicitly accommodates degree heterogeneity, the variation induced by the two-point distribution of the degree parameters can be captured without further splitting the original communities. Consequently, a DCBM with $K$ communities is preferred over a DCBM with $2K$ communities.
We then compare the SBM-$K$ and DCBM-$K$. At this stage, the DCBM with $K$ communities is favored over the SBM with $K$ communities because the latter cannot adequately capture the degree heterogeneity.

As a result, our proposed method tends to favor a model with fewer communities and a richer within-community structure, rather than over-partitioning the network to compensate for unmodeled degree heterogeneity. This behavior is also practically desirable, since substantial degree heterogeneity is commonly observed in real-world networks.

\newpage
\section{Pseudo-code for Algorithms}

\subsection{Algorithm 2}
\begin{table}[H]
	\centering\renewcommand\arraystretch{1}{
		\begin{tabular}{p{15cm}}
			\hline
			\textbf{Algorithm 2} Model selection within the SBM models\\
			\hline
			\textbf{Input:} an adjacency matrix $A$, the training proportion $w$, the candidate set of number of communities $\mathcal{K}$ (optional: default is $[n]$), the number of replications $S$, the model complexity parameter $d_{k}:k\in\mathcal{K}$ for each rank, and the penalty order $\lambda_n$.\\
			\textbf{Output:} the estimated number $\hat{K}$.\\
			\textbf{Step 1:} For $s=1,\ldots,S.$
			\begin{enumerate}[(a)]
				\item Randomly choose a subset of node pairs $\mathcal E_s$ with probability $w$ as the training set.
				\item For each $k$, run the rank-$k$ truncated singular value thresholding on the partially observed matrix $Y$ to obtain an estimated $\hat{A}^{(k)}_s$. 
				\item Run the spectral clustering algorithm on $\hat{A}^{(k)}_s$ to get the estimated label $\hat{c}^{(k)}_s$, and then use formula (2) in Section 3 to obtain a corresponding estimator $\hat{P}^{(k)}_s$.
				\item Evaluate the penalized loss of each $k$ by
				\[
				L_{k}(A,\mathcal{E}_s^c)=\frac{1}{|\mathcal E_s^c|}\sum_{(i,j)\in\mathcal{E}_s^c}(A_{ij}-\hat{P}^{(k)}_{s,ij})^2+d_{k}\lambda_n.
				\]
			\end{enumerate}
			\textbf{Step 2:} Determine the best model by
			\[
			\hat{K}=\arg\min_{k} S^{-1}\sum_{s=1}^S L_{k}(A,\mathcal{E}_s^c).
			\]\\
			\hline
	\end{tabular}}
\end{table}

\subsection{Algorithm 3}
\begin{table}[H]
	\centering\renewcommand\arraystretch{1}{
		\begin{tabular}{p{15cm}}
			\hline
			\textbf{Algorithm 3} Model selection within the DCBM models\\
			\hline
			\textbf{Input:} an adjacency matrix $A$, the training proportion $w$, the candidate set of number of communities $\mathcal{K}$ (optional: default is $[n]$), the number of replications $S$, the model complexity parameter $d_{k}:k\in\mathcal{K}$ for each rank, and the penalty order $\lambda_n$.\\
			\textbf{Output:} the estimated number $\hat{K}$.\\
			\textbf{Step 1:} For $s=1,\ldots,S.$
			\begin{enumerate}[(a)]
				\item Randomly choose a subset of node pairs $\mathcal E_s$ with probability $w$ as the training set.
				\item For each $k$, run the rank-$k$ truncated singular value thresholding on the partially observed matrix $Y$ to obtain an estimated $\hat{A}^{(k)}_s$. 
				\item Run the spherical spectral clustering algorithm on $\hat{A}^{(k)}_s$ to get the estimated label $\hat{c}^{(k)}_s$, and then use formula (4) in Section 3 to obtain a corresponding estimator $\hat{P}^{(k)}_s$.
				\item Evaluate the penalized loss of each $k$ by
				\[
				L_{k}(A,\mathcal{E}_s^c)=\frac{1}{|\mathcal E_s^c|}\sum_{(i,j)\in\mathcal{E}_s^c}(A_{ij}-\hat{P}^{(k)}_{s,ij})^2+d_{k}\lambda_n.
				\]
			\end{enumerate}
			\textbf{Step 2:} Determine the best model by
			\[
			\hat{K}=\arg\min_{k} S^{-1}\sum_{s=1}^S L_{k}(A,\mathcal{E}_s^c).
			\]\\
			\hline
	\end{tabular}}
\end{table}

\subsection{Algorithm 4}
\begin{table}[H]
	\centering\renewcommand\arraystretch{1}{
		\begin{tabular}{p{15cm}}
			\hline
			\textbf{Algorithm 4} Model selection between the SBM and affiliation models\\
			\hline
			\textbf{Input:} an adjacency matrix $A$, the training proportion $w$, the candidate set $\mathcal{K}$ (optional), the number of replications $S$, the model class complexity parameter $d_{1\cdot},d_{2\cdot}$ for two model classes (with $d_{1k}<d_{2k}$ for each $k$), and the penalty order $\lambda_n$.\\
			\textbf{Output:} the best model with $(\hat{m},\hat{K})$.\\
			\textbf{Step 1:} Apply Algorithm 2 to the adjacency matrix $A$ with the community number complexity parameter $d_{2k}$ and the penalty order $\lambda_n$ to obtain the estimated number of communities $\hat{K}$.\\
			\textbf{Step 2:} For $s=1,\ldots,S.$
			\begin{enumerate}[(a)]
				\setlength{\itemsep}{0.05cm}
				\item Randomly choose a subset of node pairs $\mathcal E_s$ with probability $w$ as the training set.
				\item Run the rank-$\hat{K}$ truncated singular value thresholding on the partially observed matrix $Y$ to obtain an estimated $\hat{A}_s$. Then run the spectral clustering algorithm on $\hat{A}_s$ to get the estimated label $\hat{c}_s$. 
				\item For model $\delta^{(m)}$, use the corresponding estimation procedures introduced above to obtain the corresponding estimator $\hat{P}^{(m)}_s$.
				\item Evaluate the penalized loss of each $m$ by
				\[
				L_{m}(A,\mathcal{E}_s^c)=\frac{1}{|\mathcal E_s^c|}\sum_{(i,j)\in\mathcal{E}_s^c}(A_{ij}-\hat{P}^{(m)}_{s,ij})^2+d_{m\hat{K}}\lambda_n.
				\]
				
			\end{enumerate}
			\textbf{Step 3:} Determine the best model by
			\[
			\hat{m}=\arg\min_{m} S^{-1}\sum_{s=1}^S L_{m}(A,\mathcal{E}_s^c),
			\]
			together with the corresponding $\hat{K}$ obtained in Step 1.\\
			\hline
	\end{tabular}}
\end{table}

\subsection{Algorithm 5}
\begin{table}[H]
	\centering\renewcommand\arraystretch{1}{
		\begin{tabular}{p{15cm}}
			\hline
			\textbf{Algorithm 5} Model selection between the SBM and DCBM models\\
			\hline
			\textbf{Input:} an adjacency matrix $A$, the training proportion $w$, the candidate set $\mathcal{K}$, the number of replications $S$, the model class complexity parameter $d_{1\cdot},d_{2\cdot}$ for two model classes (with $d_{1k}<d_{2k}$ for each $k$), and the penalty order $\lambda_n$.\\
			\textbf{Output:} the best model with $(\hat{m},\hat{K})$.\\
			\textbf{Step 1:} Apply Algorithm 3 to the adjacency matrix $A$ with the community number complexity parameter $d_{2k}$ and the penalty order $\lambda_n$ to obtain the estimated number of communities $\hat{K}$.\\
			\textbf{Step 2:} For $s=1,\ldots,S.$
			\begin{enumerate}[(a)]
				\item Randomly choose a subset of node pairs $\mathcal E_s$ with probability $w$ as the training set.
				\item Apply the rank-$\hat{K}$ truncated singular value thresholding to the partially observed matrix $Y$ to obtain an estimate $\hat{A}_s$. Perform spherical spectral clustering algorithm on $\hat{A}_s$ to obtain community labels $\hat{c}_s$. 
				\item For each model $m$, apply the corresponding estimation procedure to obtain the probability matrix estimator $\hat{P}^{(m)}_s$.
				\item Evaluate the penalized loss for each model $\delta^{(m,\hat K)}$ as:
				\[
				L_{m}(A,\mathcal{E}_s^c)=\frac{1}{|\mathcal E_s^c|}\sum_{(i,j)\in\mathcal{E}_s^c}(A_{ij}-\hat{P}^{(m)}_{s,ij})^2+d_{m\hat{K}}\lambda_n.
				\]
			\end{enumerate}
			\textbf{Step 3:} Determine the best model by
			\[
			\hat{m}=\arg\min_{m} S^{-1}\sum_{s=1}^S L_{m}(A,\mathcal{E}_s^c),
			\]
			while retaining the community number $\hat{K}$ estimated in Step 1.\\
			\hline
	\end{tabular}}
\end{table}

\subsection{Algorithm 6}
\begin{table}[H]
	\centering\renewcommand\arraystretch{1}{
		\begin{tabular}{p{15cm}}
			\hline
			\textbf{Algorithm 6} Model selection between the SBM and graphon models\\
			\hline
			\textbf{Input:} an adjacency matrix $A$, the training proportion $w$, the candidate set of number of communities $\mathcal{K}$ for SBM, a graphon estimation procedure, the number of replications $S$, the model complexity parameter $d_{1,k}:k\in\mathcal{K}$ for each rank $k$ in the SBM and $d_2$ for the graphon model, and the penalty order $\lambda_n$.\\
			\textbf{Output:} the best model $(\hat{m},\hat{k}_m)$.\\
			\textbf{Step 1:} Estimate the number of communities $\hat{K}$ using Algorithm 2 with candidate set $\mathcal{K}$ on adjacency matrix $A$.\\
			\textbf{Step 2:} For $s=1,\ldots,S.$
			\begin{enumerate}[(a)]
				\item Randomly choose a subset of node pairs $\mathcal E_s$ with probability $w$ as the training set.
				\item Use the same procedure described in Section 3 with specific rank chosen as $\hat{K}$ to obtain an estimator $\hat{P}^{(1)}_s$ for the SBM model. 
				\item Apply the chosen graphon estimation procedure to the adjusted (or unadjusted) $Y$ to obtain the corresponding estimator $\hat{P}^{(2)}_s$.
				\item Evaluate the penalized loss of each $m$ by
				\[
				L_{m}(A,\mathcal{E}_s^c)=\frac{1}{|\mathcal E_s^c|}\sum_{(i,j)\in\mathcal{E}_s^c}(A_{ij}-\hat{P}^{(m)}_{s,ij})^2+d_{m}'\lambda_n,
				\]
				where $d_2'=d_2$ and $d_1'=d_{1\hat{K}}$.
			\end{enumerate}
			\textbf{Step 3:} Determine the best model by
			\[
			\hat{m}=\arg\min_{m}\ S^{-1}\sum_{s=1}^S L_{m}(A,\mathcal{E}_s^c),
			\]
			while retaining the community number $\hat{K}$ estimated in Step 1 when $\hat{m}=1$.\\
			\hline
	\end{tabular}}
\end{table}

\newpage

\section{Detailed Discussion on the DCBM and the Graphon Models}
For the comparison between the graphon model and the DCBM, we consider the case in which $\theta$ is known, an assumption also adopted in 
\cite{hu2020corrected}, \cite{lei2016goodness} and \cite{gao2018community}. In this setting, we first apply a consistent procedure, such as BHMC, to estimate the number of communities $K$ under the DCBM. We then compare the selected DCBM with the graphon model using a procedure analogous to that described in Algorithm 6. The complete procedure is summarized in Algorithm 7.

\begin{table}[H]
	\centering\renewcommand\arraystretch{1}{
		\begin{tabular}{p{15cm}}
			\hline
			\textbf{Algorithm 7} Model selection between the DCBM and graphon models\\
			\hline
			\textbf{Input:} an adjacency matrix $A$, the training proportion $w$, the candidate set of number of communities $\mathcal{K}$ for SBM, a graphon estimation procedure, the number of replications $S$, the model complexity parameter $d_{1,k}:k\in\mathcal{K}$ for each rank $k$ in the DCBM and $d_2$ for the graphon model, and the penalty order $\lambda_n$.\\
			\textbf{Output:} the best model $(\hat{m},\hat{k}_m)$.\\
			\textbf{Step 1:} Estimate the number of communities $\hat{K}$ using a consistent procedure for community detection in Degree-Corrected Block Models (e.g., BHMC) with candidate set $\mathcal{K}$ on adjacency matrix $A$.\\
			\textbf{Step 2:} For $s=1,\ldots,S.$
			\begin{enumerate}[(a)]
				\item Randomly choose a subset of node pairs $\mathcal E_s$ with probability $w$ as the training set.
				\item Use the same procedure described in Section 4 with specific rank chosen as $\hat{K}$ to obtain an estimator $\hat{P}^{(1)}_s$ for the DCBM model. 
				\item Apply the chosen graphon estimation procedure to the adjusted (or unadjusted) $Y$ to obtain the corresponding estimator $\hat{P}^{(2)}_s$.
				\item Evaluate the penalized loss of each $m$ by
				\[
				L_{m}(A,\mathcal{E}_s^c)=\frac{1}{|\mathcal E_s^c|}\sum_{(i,j)\in\mathcal{E}_s^c}(A_{ij}-\hat{P}^{(m)}_{s,ij})^2+d_{m}'\lambda_n,
				\]
				where $d_2'=d_2$ and $d_1'=d_{1\hat{K}}$.
			\end{enumerate}
			\textbf{Step 3:} Determine the best model by
			\[
			\hat{m}=\arg\min_{m}\ S^{-1}\sum_{s=1}^S L_{m}(A,\mathcal{E}_s^c),
			\]
			while retaining the community number $\hat{K}$ estimated in Step 1 when $\hat{m}=1$.\\
			\hline
	\end{tabular}}
\end{table}

We introduce the following definition, which is analogous to Definition 5.

\begin{definitionD}\label{def 6}
	We say that an $n\times n$ probability matrix $P$, generated by a graphon $f=\rho_nf_0$, is at least $c_{n,\mathcal{K}}$-varying from the DCBM class if, for every $k\in\mathcal{K}$, every balanced labeling $c$ with $k$ communities, and every balanced degree-heterogeneity parameter vector $\theta$, the following condition holds:
	\begin{equation} \label{vardcbmgraphon}
		\frac{1}{n^2}\sum_{k_1,k_2\in[k]}\sum_{(i,j)\in \mathcal G_{k_1}\times\mathcal G_{k_2}}\left(P_{ij}-\theta_i\theta_j\frac{\sum_{(i',j')\in \mathcal G_{k_1}\times\mathcal G_{k_2}}\theta_{i'}\theta_{j'}P_{i'j'}}{\sum_{(i',j')\in \mathcal G_{k_1}\times\mathcal G_{k_2}}\theta_{i'}^2\theta_{j'}^2}\right)^2=\Omega(c_{n,\mathcal{K}}),
	\end{equation}
	where $\mathcal G_{k}$ denotes the $k$-th community under the labeling $c$.
\end{definitionD}

We further impose the following assumption, which is an analogue of Assumption 11.

\begin{asmD}\label{a11}
	The true model is either a DCBM satisfying Assumptions 4 and 6, or a graphon model whose true probability matrix is at least $c_{n,\mathcal{K}}$-varying from the DCBM class, where 
	$$
	c_{n,\mathcal{K}}\gg \max\left\{\sqrt{\frac{a_{n,w}\rho_n}{(1-w_n)}},\frac{a_{n,w}}{1-w_n},\sqrt{\frac{\rho_n^3\log n}{w_n\sqrt{n}}},\frac{\max_{K\in\mathcal{K}}\log K}{n(1-w_n)^2}\right\}.
	$$
\end{asmD}

Using arguments similar to those employed in the proof of Theorem 5, we obtain the following corollary. 
Since the proof is essentially the same as that for the SBM case, we omit the full details for brevity and provide only a proof sketch below.

\begin{cor}[DCBM versus the graphon model with known $\theta$]\label{them6}
	Assume that Assumptions 6--7, 9, 11 and \ref{a2_NS} hold. 
	Let $d_{1k}=k(k+3)/2$ and $d_{2}=3\sqrt{n/\log n}$ denote the model complexities of the DCBM and graphon model. Furthermore, we assume that $w_n^{-1}\max_{K\in\mathcal{K}} K=o(n^{1/4}/\log^{1/4} n)$. 
	Suppose further that $\lambda_n$ satisfies the following conditions:
	(1) $\lambda_n\cdot \sqrt{n/\log n}=o_{\mathbb{P}}(c_{n,\mathcal{K}})$ and (2) $\lambda_n=\omega_{\mathbb{P}}\left(\frac{\max\left\{\sqrt{a_{n,w}(1-w_n)\rho_n},a_{n,w}\right\}}{(1-w_n)\cdot \sqrt{n/\log n}}\right)$. Then, the true model $\delta^*$ can be chosen with probability tending to 1.
\end{cor}

\begin{proof}[Proof sketch of Corollary \ref{them6}]
	When the degree parameters are known, for any scalar block parameter $b$, we have
	$$
	(A_{ij}-\theta_i\theta_jb)^2=\theta_i^2\theta_j^2(\frac{A_{ij}}{\theta_i\theta_j}-b)^2.
	$$
	Here, Assumption~6 ensures that the degree parameters are bounded away from zero, so the above transformation is well defined.
	Consequently, for any fixed labeling, the DCBM block estimator is
	the weighted least-squares estimator
	$$
	\widehat B_{kl}=\frac{\sum_{(i,j)\in\mathcal U_{kl}}A_{ij}\theta_i\theta_j}{\sum_{(i,j)\in\mathcal U_{kl}}\theta_i^2\theta_j^2}.
	$$
	By Assumption~6, the weights satisfy
	$\theta_{\min}^4\leq\theta_i^2\theta_j^2\leq\theta_{\max}^4$.
	Therefore, the weighted block sizes and weighted quadratic errors are
	equivalent, up to multiplicative constants, to their unweighted counterparts.
	
	Suppose first that the true model is a DCBM. By the assumed consistency of the first-stage estimator, 
	\[ \mathbb P\left(\widehat K=K^*\right)\to 1. \]
	On the event $\{\widehat K=K^*\}$, the argument used for the SBM case in the proof
	of Theorem 5 applies directly to the weighted block estimator defined above. 
	
	Suppose next that the true model is a graphon model.  By Definition \ref{def 6}, for every \(K\in\mathcal K\), every admissible balanced labeling with \(K\) communities, and every admissible balanced degree-parameter vector, the DCBM approximation error satisfies
	$$\frac{1}{n^2}\sum_{k,l}\sum_{(i,j)\in \mathcal G_k\times \mathcal G_l}\left(P_{ij}-\theta_i\theta_j\frac{\sum_{(i',j')\in \mathcal G_k\times\mathcal  G_l}\theta_{i'}\theta_{j'}P_{i'j'}}{\sum_{(i',j')\in\mathcal G_k\times\mathcal G_l}\theta_{i'}^2\theta_{j'}^2}\right)^2=\Omega(c_{n,\mathcal K}).$$
	Thus, the separation condition holds uniformly over all admissible balanced labelings and degree-parameter vectors.
	Since the weights are uniformly bounded above and below, the uniform concentration argument used
	in the proof of Proposition \ref{propa9} continues to hold, up to multiplicative constants.
	It follows that
	$$
	\ell_{1,\widehat K}(A,\mathcal E^c)-\ell_2(A,\mathcal E^c)
	=
	\Omega_{\mathbb P}\left(
	n^2(1-w_n)c_{n,\mathcal K}
	\right).
	$$
	The upper-rate condition imposed on \(\lambda_n\) ensures that the penalty difference is asymptotically dominated by this separation. Consequently, the proposed procedure selects the graphon model with probability tending to one. This completes the proof.
\end{proof}

\newpage

\section{Additional Simulations}

All numerical experiments in the main article and in the supplementary material were conducted on the university high-performance computing servers, with all nodes running Linux CentOS 7.9. Due to differences in hardware architectures and system-level numerical libraries, minor variations in empirical results may occur when the experiments are reproduced on different computing platforms.

\subsection{Simulation Settings}

We first present the detailed simulation configurations.

\textbf{Simulation E.1: additional settings in comparing SBMs and DCBMs.} In Simulation 2, we consider a balanced setting when $K^*=3$ and an unbalanced structure when $K^*=5$. Here we consider a complementary setting, where we adopt an unbalanced setting when $K^*=3$ and a balanced configuration when $K^*=5$. Specifically, for $K^*=3$, we set $\Pi=\{1/6,1/3,1/2\}$ with $n\in\{300,600,1200,1800\}$, and 
for $K^*=5$ we consider 
$\Pi=\{1/5,\ldots,1/5\}$ with 
$n\in\{500,1000,1500\}$. Other settings such as the regularization parameter and the configuration for the block-wise connection probability matrix, are the same as in Simulation 2.

\textbf{Simulation E.2: general SBMs.} 
Since we have already compared our method with existing CV-based methods in Simulation~2 and Simulation~E.1 for simultaneous SBM--DCBM selection, we now focus on comparisons with information-criterion-based methods within a given model class. Specifically, we compare with the original BIC-type method \citep{airoldi2008mixed}, the composite likelihood BIC (CL-BIC) method \citep{saldana2017many} and the likelihood ratio BIC (LR-BIC) method \citep{wang2017likelihood}.

\begin{figure}[htbp]
	\centering
	\subfloat[Affiliation SBM model]{
		\includegraphics[width =6cm]{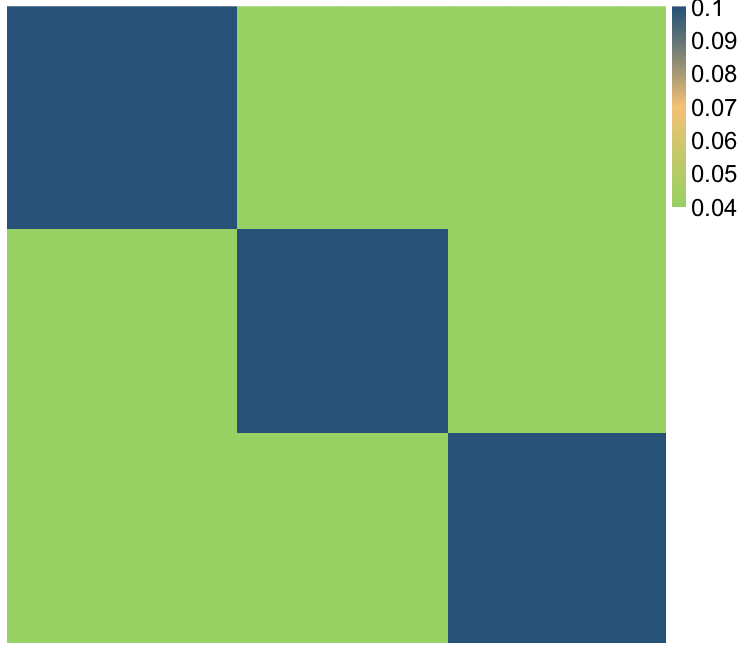}}
	\subfloat[General SBM model]{
		\includegraphics[width =6cm]{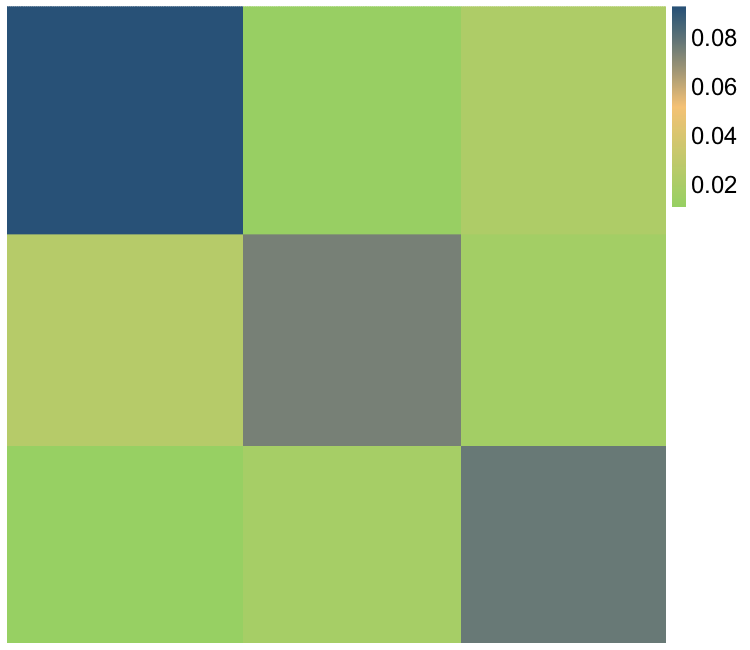}}\\
	\subfloat[DCBM model]{
		\includegraphics[width =6cm]{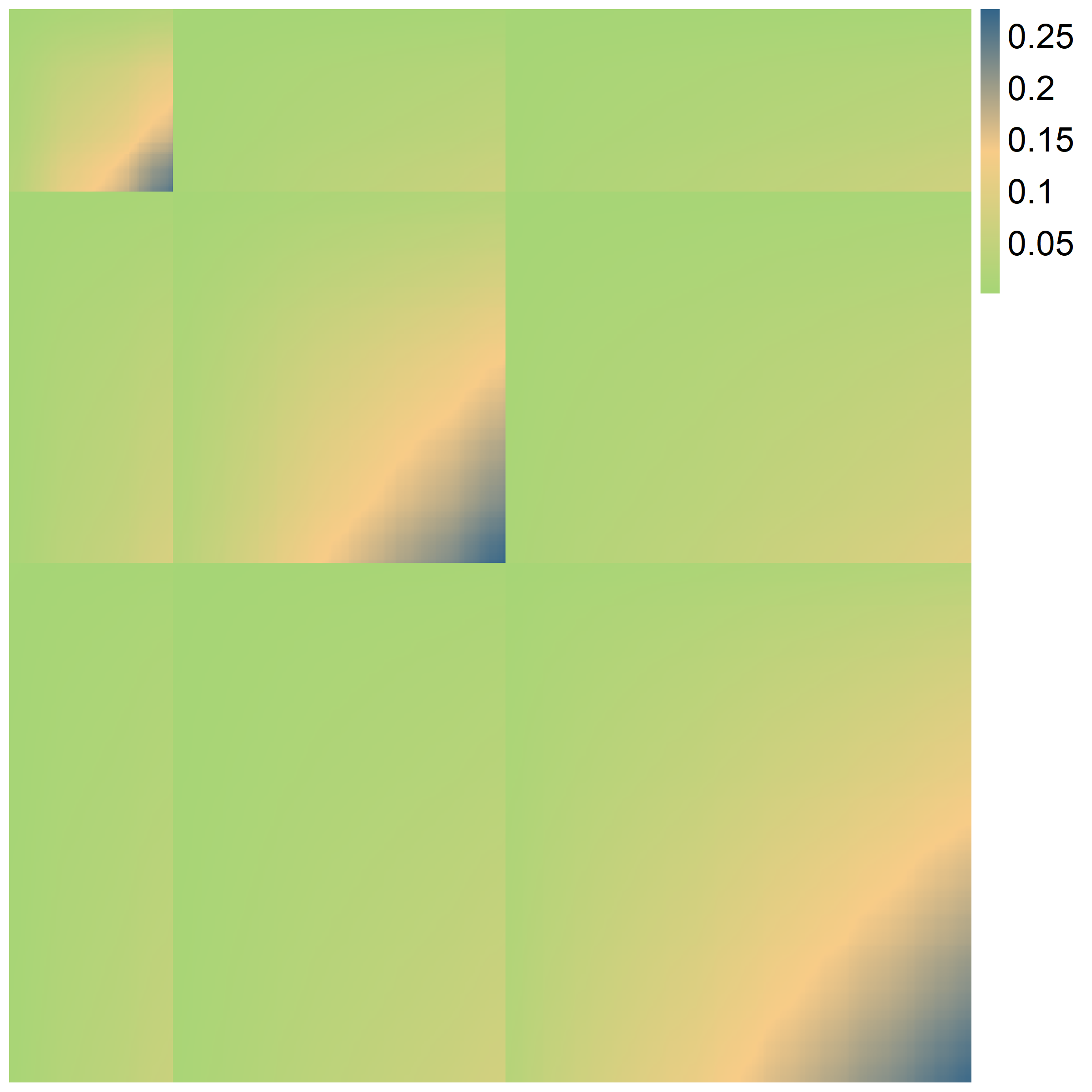}}
	\subfloat[NS graphon model]{
		\includegraphics[width =6cm]{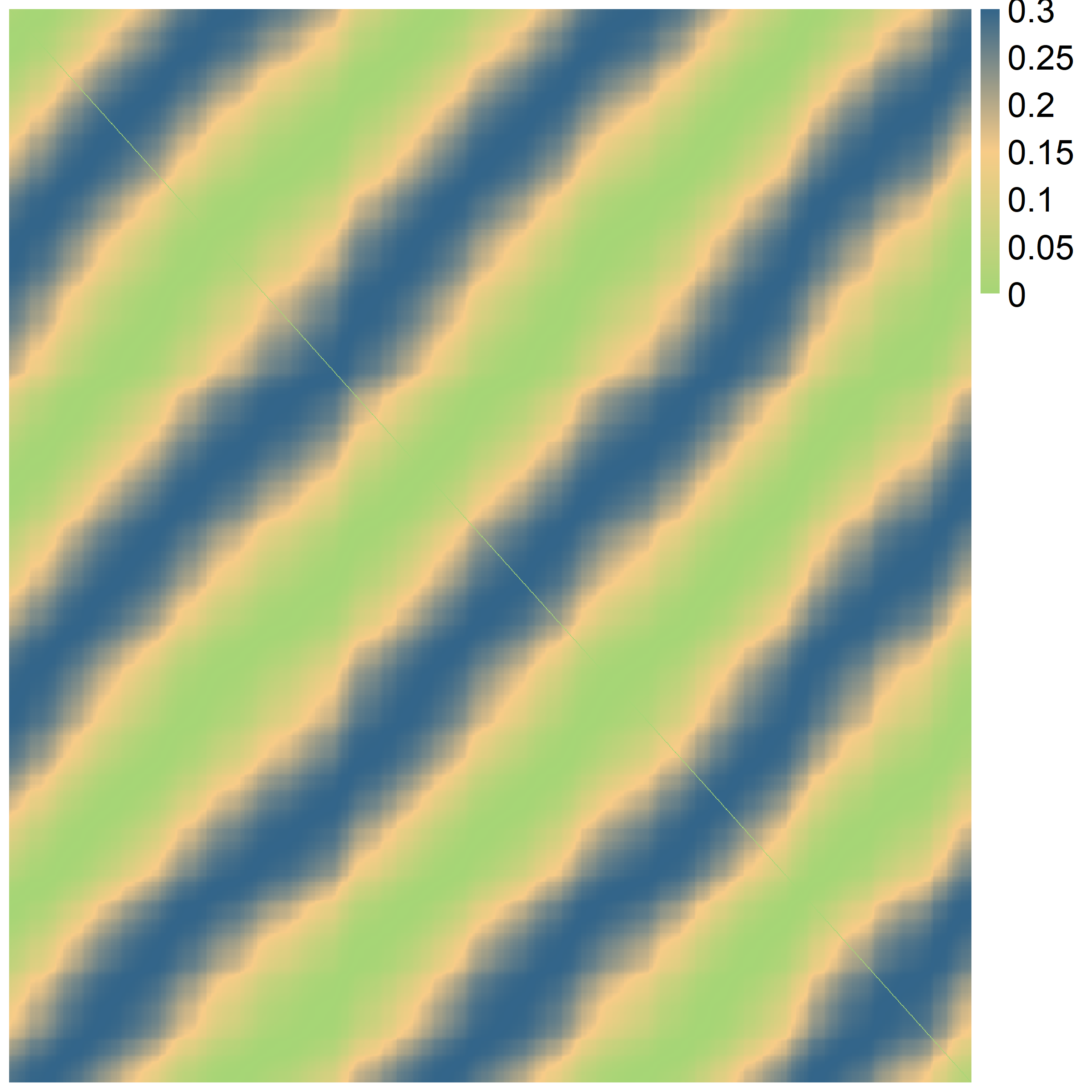}}\\
	\centering
	\caption{Heatmaps for different models in Simulation settings under $n=1000$.}\label{heatmap1}
\end{figure}

In this simulation, we set the community-wise edge probability matrix $B=rB_0$, where the diagonal entries of $B_0$ independently follow a uniform distribution $\mathcal U(0.6, 1)$, and off-diagonal entries are drawn independently from a uniform distribution $\mathcal U(0.1, 0.3)$. The membership vector $c$ is generated from a multinomial distribution $\mathcal M(n, \Pi)$ with equal community probabilities $\Pi=(1/K^*,\dots,1/K^*)$.
We consider several combinations of the community sizes and numbers. Specifically, under balanced community structure, given $r=0.1$, we set $n\in\{300,600,900\}$ under $K^*=3$ and $n\in\{500,1000,1500\}$ under $K^*=5$. Figure \ref{heatmap1}(b) depicts the probability matrix heatmap for a general SBM model under $K^*=3$ and $n=1000$.

The regularization parameter is set as $\lambda_n=0.001\hat{\rho}_n^2/\sqrt{\log n}$ as in the simulations in the main article, where we plug in an estimator $\sum_{ij} A_{ij}/n(n-1)$ for $\rho_n$. We also use the same adaptive search method to avoid exhaustive line search.

\textbf{Simulation E.3: general DCBMs.} In this simulation, we also compare our proposed PNN-CV method with the BIC-based methods, as introduced in Simulation E.2, with their DCBM variants. Here, for degree heterogeneous parameter $\theta$ and block-wise connection probability matrix $B$, we consider the same configuration as the DCBM part in Simulation 2. For the membership vector $c$, the network size and community sizes, we consider the same balanced setting as in Simulation E.2.

\textbf{Simulation E.4: General SBMs and DCBMs with known $K^*$.} 
To facilitate a direct comparison with testing-based procedures such as \citep{Bhadra2026unified}, which are developed under the assumption that the number of communities $K^*$ is known, we conduct additional simulations under this setting. This allows us to isolate and compare the performance of different approaches for model class selection between SBM and DCBM when $K^*$ is fixed. That is, here we assume the true number of communities $K^*$ is known as a ground truth, and compare our proposed PNN-CV with the $Q_1$ testing method \citep{Bhadra2026unified}.

Specifically, in this simulation, we use the same configuration for the SBM and the DCBM model (the block-wise connection probability matrix and the degree heterogeneity parameter) as in Simulation 2. We consider the balanced case, that is the membership vector $c$ is generated from a multinomial distribution $\mathcal M(n, \Pi)$ with equal community probabilities $\Pi=(1/K^*,\dots,1/K^*)$.
As in Simulation E.2, under balanced community structure, given $r=0.1$, we set $n\in\{300,600,900\}$ under $K^*=3$ and $n\in\{500,1000,1500\}$ under $K^*=5$. 

\textbf{Simulation E.5: DCBMs and graphon models.} In this simulation we select between DCBMs and graphon models, comparing PNN-CV, PNN-CVA, their unpenalized counterparts, and the hybrid method PNN-CV+BHMC where the BHMC method is first used to determine the possible block structure, along with its unpenalized counterpart. The regularization parameter configuration includes $\lambda_n=0.1\hat{\rho}_n^2/\sqrt{n}$, with the same plug-in estimator for $\rho_n$ as in Simulation E.1 and with network sizes $n\in\{300,600,900\}$. For neighborhood smoothing (NS) graphon estimation, the bandwidth parameter follows $h=\sqrt{\log n/n}$.
\begin{itemize}
	\item When the true model is the DCBM model, we set the model exactly the same as the unbalanced case for $K^*=3$ in Simulation E.1, except for fixing $r=0.3$. Representative connectivity pattern is shown in Figure \ref{heatmap1}(c) under $n=1000$.
	\item When the true model is the neighborhood smoothing graphon model, we set the probability matrix $P=rP_0$ with $P_{0,ij}=(f(u_i,u_j))$ and $f(u_i,u_j)=\sin[5\pi(u_i+u_j-1)+1]/2 +0.5$, where the $u_i$'s are independently drawn from uniform distribution $\mathcal U(0,1)$. We also fix $r=0.3$. Characteristic probability matrix structure is visualized in Figure~\ref{heatmap1}(d) under $n=1000$.
\end{itemize}

\textbf{Simulation E.6: affiliation SBMs with many communities.} This experiment illustrates the practical implications of allowing an unbounded number of communities, as highlighted in our theory. We consider an affiliation SBM model with 20 communities, similar to the setting in Simulation 1, with $\beta=0.2$ and $r=0.1$, and consider sample size $n\in\{2000,3000,4000\}$. We compare the proposed PNN-CV method, implemented with its default adaptive search strategy, to ECV, NCV, and NETCROP. For the competing methods, candidate models must be restricted to a finite set, and we consider two choices of the upper bound on the number of communities, namely the default choice $K_{\max}=15$ and a larger upper bound $K_{\max}=30$. All methods are implemented using their default tuning parameters unless otherwise specified.

\subsection{Simulation Results}
Table~\ref{simulationE1} reports complementary simulation results for SBM–DCBM selection under alternative configurations introduced in Simulation E.1, including unbalanced communities with $K^*=3$ and balanced communities with $K^*=5$. 
Across all settings, the proposed PNN-CV-based methods exhibit clear improvements with increasing sample size and achieve performance comparable to the strongest CV competitor (ECV), while consistently outperforming NCV and NETCROP. 
Moreover, the hybrid approach PNN-CV+BHMC demonstrates notably improved finite-sample performance in more challenging regimes, highlighting the flexibility of our framework in combining complementary strengths of different procedures.

\begin{table}[htbp]
	\centering\renewcommand\arraystretch{1}{\scriptsize
		\caption{Selection between SBM and DCBM. (complementary settings)
		}\label{simulationE1}
		\setlength{\tabcolsep}{4mm}{
			\begin{tabular}{ccccccccc}
				
				\hline
				\cline{1-9}
				&&\multicolumn{4}{c}{$K=3$}&\multicolumn{3}{c}{$K=5$}\\
				\cmidrule(lr){3-6}
				\cmidrule(lr){7-9}
				True model	 &  Method&$n=300$&$n=600$&$n=1200$&$n=1800$&$n=500$&$n=1000$&$n=1500$\\
				\hline
				
				SBM    & PNN-CV+BHMC    &0.17   &0.76   &0.99  &1.00  &0.23    &0.97   &1.00\\
				& PNN-CV	            &0.02   &0.13   &0.61  &0.81  &0.04    &0.87   &1.00\\
				& ECV	                &0.03   &0.13   &0.69  &0.88  &0.07    &0.87   &1.00\\
				& NCV                   &0.15   &0.20   &0.52  &0.68  &0.11    &0.75   &0.99\\
				& NETCROP               &0.00   &0.05   &0.33  &0.68  &0.01    &0.32   &0.25\\
				\hline
				
				DCBM   & PNN-CV+BHMC    &0.03   &0.56   &1.00  &1.00  &0.00    &0.92   &1.00\\
				& PNN-CV	            &0.00   &0.01   &0.40  &0.86  &0.00    &0.85   &1.00\\
				& ECV	                &0.03   &0.03   &0.58  &0.90  &0.03    &0.88   &1.00\\
				& NCV                   &0.00   &0.00   &0.14  &0.61  &0.00    &0.07   &0.80\\
				& NETCROP               &0.00   &0.01   &0.11  &0.74  &0.00    &0.14   &0.82\\
				\hline
				\cline{1-9}	
			\end{tabular}
	}}
\end{table}

Table \ref{simulationE2} reports the fraction of times over 100 replications where the correct community number is selected for different combinations of true community numbers and community sizes in Simulation E.2. 
Under sufficiently large sample sizes, our PNN-CV method achieves asymptotically equivalent performance to the LR-BIC method, with both exhibiting rapid convergence to the true community number as the sample size increases. In contrast, while CL-BIC and the classical BIC may show competitive performance in small sample regimes, their selection behavior does not stabilize as the network size grows. In particular, they tend to increasingly favor more complex models, indicating a clear tendency toward overfitting in large networks.

\begin{table}[htbp]
	\centering\renewcommand\arraystretch{1}{\scriptsize
		\caption{Selection in SBM framework.
		}\label{simulationE2}
		\setlength{\tabcolsep}{2.5mm}{
			\begin{tabular}{cccccccccccc}
				
				\hline
				\cline{1-12}
				\noalign{\vskip 2pt}
				Method&$(K_0,n)$&$\widehat K=2$&$\widehat K=3$ &$\widehat K=4$&others&  $(K_0,n)$&$\widehat K=3$ &$\widehat K=4$&$\widehat K=5$ &$\widehat K=6$&others  \\
				\hline
				
				PNN-CV &(3,300) &0.44   &0.50   &0.02   &0.04   & (5,500) &0.36   &0.36   &0.04   &0.01   &0.23   \\
				LR-BIC &        &0.32   &0.62   &0.00   &0.06   &         &0.24   &0.26   &0.24   &0.00   &0.23     \\
				CL-BIC &        &0.01   &0.97   &0.02   &0.00   &         &0.01   &0.52   &0.45   &0.02   &0.00     \\
				BIC    &        &0.00   &0.26   &0.43   &0.31   &         &0.00   &0.12   &0.49   &0.21   &0.18   \\
				
				\hline
				
				PNN-CV &(3,600) &0.01   &0.99   &0.00   &0.00   & (5,1000) &0.01   &0.12   &0.87   &0.00   &0.00   \\
				LR-BIC &        &0.01   &0.99   &0.00   &0.00   &         &0.00   &0.01   &0.99   &0.00   &0.00     \\
				CL-BIC &        &0.00   &0.93   &0.06   &0.01   &         &0.00   &0.01   &0.99   &0.00   &0.00     \\
				BIC    &        &0.00   &0.33   &0.21   &0.46   &         &0.00   &0.00   &0.70   &0.13   &0.17   \\

				\hline
				PNN-CV &(3,900) &0.00   &1.00   &0.00   &0.00   & (5,1500) &0.00   &0.00   &1.00   &0.00   &0.00   \\
				LR-BIC &        &0.00   &1.00   &0.00   &0.00   &         &0.00   &0.00   &1.00   &0.00   &0.00     \\
				CL-BIC &        &0.00   &0.66   &0.24   &0.10   &         &0.00   &0.00   &0.98   &0.01   &0.01     \\
				BIC    &        &0.00   &0.19   &0.25   &0.56   &         &0.00   &0.00   &0.32   &0.20   &0.48   \\
				\hline
				\cline{1-12}	
			\end{tabular}
	}}
\end{table}

Table \ref{simulationE3} reports the model selection performance under the degree-corrected block model. Here, CL-BIC and BIC occasionally produce undefined (NaN) values for DCBMs at small sample sizes which we will treat as a ``failure'' and count into the ``others'' category. In contrast to the SBM setting, CL-BIC exhibits noticeably improved behavior under DCBM and achieves near-consistent selection of the true number of communities as the network size increases, while both PNN-CV and LR-BIC still exhibit fast convergence to the true number of communities. Taken together with the SBM results, these findings suggest that our method and LR-BIC maintain robustness across various settings. CL-BIC and other BIC-based methods may benefit from degree correction in large networks, but its practical applicability remains limited in small samples due to numerical instability.

\begin{table}[htbp]
	\centering\renewcommand\arraystretch{1}{\scriptsize
		\caption{Selection in DCBM framework.
		}\label{simulationE3}
		\setlength{\tabcolsep}{2.5mm}{
			\begin{tabular}{cccccccccccc}
				
				\hline
				\cline{1-12}
				\noalign{\vskip 2pt}
				Method&$(K_0,n)$&$\widehat K=2$&$\widehat K=3$ &$\widehat K=4$&others&  $(K_0,n)$&$\widehat K=3$ &$\widehat K=4$&$\widehat K=5$ &$\widehat K=6$&others  \\
				\hline
				
				PNN-CV &(3,300) &0.09   &0.21   &0.00   &0.70   & (5,500) &0.00   &0.01   &0.00   &0.00   &0.99   \\
				LR-BIC &        &0.12   &0.18   &0.00   &0.70   &         &0.00   &0.00   &0.00   &0.00   &1.00     \\
				CL-BIC &        &0.11   &0.12   &0.00   &0.77   &         &0.21   &0.05   &0.00   &0.00   &0.74     \\
				BIC    &        &0.00   &0.02   &0.04   &0.94   &         &0.00   &0.05   &0.05   &0.09   &0.81   \\
				
				\hline
				
				PNN-CV &(3,600) &0.03   &0.96   &0.00   &0.01   & (5,1000) &0.00   &0.13   &0.85   &0.00   &0.02   \\
				LR-BIC &        &0.00   &1.00   &0.00   &0.00   &         &0.00   &0.05   &0.95   &0.00   &0.00     \\
				CL-BIC &        &0.00   &0.90   &0.00   &0.10   &         &0.00   &0.10   &0.86   &0.00   &0.04     \\
				BIC    &        &0.00   &0.62   &0.04   &0.34   &         &0.00   &0.00   &0.91   &0.00   &0.09   \\

				\hline
				PNN-CV &(3,900) &0.00   &1.00   &0.00   &0.00   & (5,1500) &0.00   &0.00   &1.00   &0.00   &0.00   \\
				LR-BIC &        &0.00   &1.00   &0.00   &0.00   &         &0.00   &0.00   &1.00   &0.00   &0.00     \\
				CL-BIC &        &0.00   &0.97   &0.00   &0.03   &         &0.00   &0.00   &0.98   &0.00   &0.02     \\
				BIC    &        &0.00   &0.71   &0.07   &0.22   &         &0.00   &0.00   &0.85   &0.01   &0.14   \\
				\hline
				\cline{1-12}	
			\end{tabular}
	}}
\end{table}

Table \ref{simulationE4} compares the performance of PNN-CV with the testing-based method $Q_1$ under the setting where the number of communities $K^*$ is assumed known, where the PNN-CV row reports the proportion of times that DCBM is selected, and the $Q_1$ row reports the proportion of times that the null hypothesis (that is, SBM is the true model) is rejected.
PNN-CV demonstrates fast convergence and high robustness across all scenarios: when the true model is SBM, it almost never misclassifies, and when the true model is DCBM, it quickly identifies the correct model as the network size increases.
In contrast, the testing-based method $Q_1$ controls the nominal level well under SBM (close to 0.05), but its power is limited under DCBM, especially for the more challenging $K=5$ case, where no clear convergence trend is observed.

\begin{table}[htbp]
	\centering\renewcommand\arraystretch{1}{\scriptsize
		\caption{Selection between SBM and DCBM under known $K^*$
		}\label{simulationE4}
		\setlength{\tabcolsep}{4mm}{
			\begin{tabular}{cccccccc}
				
				\hline
				\cline{1-8}
				&&\multicolumn{3}{c}{$K=3$}&\multicolumn{3}{c}{$K=5$}\\
				\cmidrule(lr){3-5}
				\cmidrule(lr){6-8}
				True model	 &  Method&$n=300$&$n=600$&$n=900$&$n=500$&$n=1000$&$n=1500$\\
				\hline
				
				SBM    & PNN-CV    &0.00   &0.00   &0.00  &0.00 &0.00    &0.00\\
				& $Q_1$	            &0.02   &0.00   &0.04  &0.00  &0.02    &0.06\\
				\hline
				
				DCBM   & PNN-CV    &0.61   &1.00   &1.00  &0.82  &1.00  &1.00\\
				& $Q_1$	            &0.23   &0.59   &0.99  &0.13    &0.08   &0.08\\
				\hline
				\cline{1-8}	
			\end{tabular}
	}}
\end{table}

Table~\ref{simulationE5} presents the proportion of times (out of 100 replications) that the correct model is selected across different network sizes in Simulation E.5. It is observed that for smaller sample sizes, the penalized versions of the hybrid method (PNN-CV+BHMC) outperform its unpenalized counterparts and the original PNN-CV based methods, highlighting the importance of incorporating the penalty term when the sample size is limited. Moreover, as the network size increases, all methods achieve near-perfect accuracy, corroborating the theoretical consistency results established in Section 5.

\begin{table}[htbp]
	\centering\renewcommand\arraystretch{1}{\scriptsize
		\caption{Selection between DCBM and graphon.
		}\label{simulationE5}
		\setlength{\tabcolsep}{1mm}{
			\begin{tabular}{ccccc}
				
				\hline
				\cline{1-5}
				True model	 &  Method&$n=300$&$n=600$&$n=900$\\
				\hline
				
				DCBM   & PNN-CV+BHMC   &0.68    &1.00    &1.00  \\
				& PNN-CV	       &0.42    &0.94    &0.99   \\
				& PNN-CVA	   &0.41    &0.94    &0.99    \\
				& PNN-CV0+BHMC  &0.58    &0.97    &1.00   \\
				& PNN-CV0	   &0.42    &0.94    &0.99   \\
				& PNN-CVA0	   &0.41    &0.94    &0.99  \\
				
				\hline
				
				NS	    & PNN-CV+BHMC   &0.99    &1.00    &1.00  \\
				& PNN-CV	    &0.99    &1.00    &1.00   \\
				& PNN-CVA	   &0.99    &1.00    &1.00    \\
				& PNN-CV0+BHMC  &1.00    &1.00    &1.00   \\
				& PNN-CV0	   &1.00    &1.00    &1.00   \\
				& PNN-CVA0	   &1.00    &1.00    &1.00 \\        
				\hline
				\cline{1-5}	
			\end{tabular}
	}}
\end{table}

Table~\ref{simulationE6} illustrates that the allowance of an unbounded number of communities, emphasized in our theoretical analysis, also has practical implications. Here ECV(15) denotes ECV with the default upper bound $K_{\max}=15$, while ECV(30) corresponds to an enlarged candidate set with $K_{\max}=30$; other competing methods are defined analogously.

When the true number of communities exceeds commonly used default bounds, methods that rely on a fixed candidate range may fail unless the upper bound is manually increased. While enlarging $K_{\max}$ can partially improve empirical performance, it raises two practical and theoretical issues: first, there is no principled guidance on how large $K_{\max}$ should be chosen, and excessively large bounds substantially increase computational cost; second, the resulting procedures deviate from the fixed-$K_{\max}$ regimes under which existing theoretical guarantees are derived. In contrast, PNN-CV employs an adaptive search strategy and does not require a pre-specified upper bound on $K$, remaining computationally stable and theoretically aligned while achieving strong empirical consistency in this setting.

\begin{table}[htbp]
	\centering\renewcommand\arraystretch{1}{\scriptsize
		\caption{Selection in affiliation SBM with many communities.
		}\label{simulationE6}
		\setlength{\tabcolsep}{1mm}{
			\begin{tabular}{cccc}
				
				\hline
				\cline{1-4}
				Method&$n=2000$&$n=3000$&$n=4000$\\
				\hline
				
				PNN-CV   &0.00    &0.40    &0.99  \\
				ECV(15)	     &0.00    &0.00    &0.00   \\
				NCV(15)	     &0.00    &0.00    &0.00    \\
				NETCROP(15)  &0.00    &0.00    &0.00   \\
				ECV(30)  &0.01   &0.41    &0.98 \\
				NCV(30)  &0.01   &0.21    &0.98 \\
				NETCROP(30)  &0.00    &0.00    &0.02\\
				\hline
				\cline{1-4}	
			\end{tabular}
	}}
\end{table}

To summarize, in the first scenario, PNN-CV demonstrates a clear consistency trend under the complementary settings for comparing SBMs and DCBMs, achieving top-tier performance among all CV methods, while its hybrid variant PNN-CV+BHMC outperforms all competitors. The second and the third simulations compare our proposed procedure with the likelihood-based method for selecting $K$ under different block structures, still offering competitive performance and plain consistency trend. The fourth setting demonstrates the strong performance of PNN-CV in selecting between SBMs and DCBMs when the number of communities is known, whereas the testing method $Q_1$ has relatively poor performance under rather complex scenarios. The fifth simulation illustrates that the penalized versions of PNN-CV+BHMC are especially effective in small-sample settings when distinguishing DCBMs from graphon models, while all methods perform well as the sample size increases. The final simulation emphasizes the importance of allowing an unbounded number of candidate communities, as proposed in our theory.

These results highlight the importance of incorporating structural penalties and demonstrate that our PNN-CV framework adapts well to a wide range of practical scenarios. Together, these supplementary simulations reinforce the robustness and flexibility of the PNN-CV framework across both well-specified and more complex generative scenarios.

\subsection{Sensitivity Analysis}

In this subsection, we examine the sensitivity of the proposed penalized CV procedure to the constant factor in the penalty term. 
We focus on robustness with respect to this constant while keeping the theoretical rate of $\lambda_n$ fixed, and consider three representative tasks: (i) selecting the number of communities under the DCBM, (ii) selecting between SBM and DCBM, and (iii) selecting between DCBM and graphon models.

\textbf{Simulation E.7: sensitivity in selecting DCBMs.} We first study the sensitivity of the penalty constant in the task of selecting the number of communities under the DCBM. We consider the grid $C\in(0.0005,0.001,0.002,0.005,0.01,0.02)$, which includes the recommended value 0.001. We adopt the same setting as Simulation E.3, that is, we use the same figuration for the degree heterogeneous parameter $\theta$ and block-wise connection probability matrix $B$ as the DCBM part of Simulation 2, and consider the balanced community structure as in Simulation E.2.

\textbf{Simulation E.8: sensitivity in selection between SBMs and DCBMs.} To focus on parametric model class selection, we fix the true number of communities $K^*$ and vary only the penalty constant, in order to eliminate the distractions induced by estimating $K$. We follow the same simulation setting for network structure as in Simulation E.4 and use the same grid for the constant factor as before.

\textbf{Simulation E.9: sensitivity in selection between DCBMs and graphon.} Now we focus on the comparison between parametric models and nonparametric models. Here, there are two constants that need to be determined, one is the constant $C_1$ in the complexity term $d_2=C_1 \sqrt{n/\log n}$, the other one is the constant $C_2$ in the penalty term $\lambda_n=C_2\hat{\rho}_n^2/\sqrt{n}$. We let the two constants take values in $C_1\in\{1,3,5\}$ and $C_2\in\{0.025,0.05,0.1,0.2,0.4\}$. The configuration for the network is the same as in Simulation E.5.

Table \ref{simulationE7} reports the sensitivity of PNN-CV in selecting the number of communities under DCBMs. The results show that, for both $K=3$ and $K=5$, the selection accuracy is stable across a broad range of penalty constants, particularly for moderate sample sizes and beyond. While overly large constants may slightly slow down convergence in finite samples, an effect that is more noticeable when K is larger, the correct number of communities is still consistently recovered as the sample size increases.

\begin{table}[htbp]
	\centering\renewcommand\arraystretch{1}{\scriptsize
		\caption{Sensitivity in selecting among DCBMs}\label{simulationE7}
		\setlength{\tabcolsep}{4mm}{
			\begin{tabular}{ccccccc}
				
				\hline
				\cline{1-7}
				&\multicolumn{3}{c}{$K=3$}&\multicolumn{3}{c}{$K=5$}\\
				\cmidrule(lr){2-4}
				\cmidrule(lr){5-7} Constant&$n=300$&$n=600$&$n=900$&$n=500$&$n=1000$&$n=1500$\\
				\hline
				
				0.0005    &0.21   &0.96   &1.00  &0.00 &0.85    &1.00\\
				0.001	  &0.21   &0.96   &1.00  &0.00 &0.85    &1.00\\
				0.002	  &0.20   &0.95   &1.00  &0.00 &0.85    &1.00\\
				0.005	  &0.20   &0.93   &1.00  &0.00 &0.82    &1.00\\
				0.01	  &0.19   &0.91   &0.98  &0.00 &0.71    &1.00\\
				0.02	  &0.14   &0.86   &0.95  &0.00 &0.44    &0.99\\
				\hline
				\cline{1-7}	
			\end{tabular}
	}}
\end{table}

Table \ref{simulationE8} presents the sensitivity results for selecting between SBMs and DCBMs given $K$. When the true model is SBM, PNN-CV achieves perfect selection across all considered constants and sample sizes, indicating that the penalty does not induce underfitting toward more complex models. When the true model is DCBM, small-sample performance exhibits mild dependence on the penalty constant, but this effect diminishes rapidly as $n$ increases, and consistency is preserved across all tuning choices.

\begin{table}[htbp]
	\centering\renewcommand\arraystretch{1}{\scriptsize
		\caption{Sensitivity in selecting between SBMs and DCBMs}\label{simulationE8}
		\setlength{\tabcolsep}{4mm}{
			\begin{tabular}{cccccccc}
				
				\hline
				\cline{1-8}
				&&\multicolumn{3}{c}{$K=3$}&\multicolumn{3}{c}{$K=5$}\\
				\cmidrule(lr){3-5}
				\cmidrule(lr){6-8} 
				True Model&Constant&$n=300$&$n=600$&$n=900$&$n=500$&$n=1000$&$n=1500$\\
				\hline
				
				SBM &0.0005    &1.00   &1.00   &1.00  &1.00 &1.00    &1.00\\
				&0.001	  &1.00   &1.00   &1.00  &1.00 &1.00    &1.00\\
				&0.002	  &1.00   &1.00   &1.00  &1.00 &1.00    &1.00\\
				&0.005	  &1.00   &1.00   &1.00  &1.00 &1.00    &1.00\\
				&0.01	  &1.00   &1.00   &1.00  &1.00 &1.00    &1.00\\
				&0.02	  &1.00   &1.00   &1.00  &1.00 &1.00    &1.00\\
				\hline
				DCBM &0.0005    &0.61   &1.00   &1.00  &0.82 &1.00    &1.00\\
				&0.001	  &0.61   &1.00   &1.00  &0.82 &1.00    &1.00\\
				&0.002	  &0.61   &1.00   &1.00  &0.80 &1.00    &1.00\\
				&0.005	  &0.61   &1.00   &1.00  &0.74 &1.00    &1.00\\
				&0.01	  &0.60   &1.00   &1.00  &0.69 &1.00    &1.00\\
				&0.02	  &0.52   &1.00   &1.00  &0.56 &1.00    &1.00\\
				\hline
				\cline{1-8}	
			\end{tabular}
	}}
\end{table}

Table~\ref{simulationE9} illustrates the sensitivity of the proposed procedure to the multiplicative constants in the complexity term ($C_1$) and the penalty term ($C_2$).
We observe that for a broad and practically relevant range of constants, the method exhibits stable and improving performance as the sample size increases.
When both $C_1$ and $C_2$ are chosen excessively large, the penalty may systematically favor parametric models and suppress the graphon alternative even at moderate sample sizes, reflecting an over-penalization regime.
This behavior is consistent with the role of penalization and highlights the importance of selecting $\lambda_n$ within the theoretically prescribed rate range.

\begin{table}[htbp]
	\centering\renewcommand\arraystretch{1}{\scriptsize
		\caption{Sensitivity in selecting between DCBMs and graphon models}\label{simulationE9}
		\setlength{\tabcolsep}{4mm}{
			\begin{tabular}{cccccccc}				
				\hline
				\cline{1-8}
				&&\multicolumn{3}{c}{DCBM-3}&\multicolumn{3}{c}{graphon}\\
				\cmidrule(lr){3-5}
				\cmidrule(lr){6-8} 
				$C_1$&$C_2$&$n=300$&$n=600$&$n=900$&$n=300$&$n=600$&$n=900$\\
				\hline
				
				1&0.025    &0.59   &0.97   &1.00  &1.00 &1.00    &1.00\\
				&0.05	  &0.59   &0.97   &1.00  &1.00 &1.00    &1.00\\
				&0.1	  &0.58   &0.97   &1.00  &1.00 &1.00    &1.00\\
				&0.2	  &0.56   &0.97   &1.00  &1.00 &1.00    &1.00\\
				&0.4	  &0.51   &0.97   &1.00  &1.00 &1.00    &1.00\\
				\hline
				3&0.025    &0.62   &0.99   &1.00  &1.00 &1.00    &1.00\\
				&0.05	  &0.68   &1.00   &1.00  &1.00 &1.00    &1.00\\
				&0.1	  &0.90   &1.00   &1.00  &0.99 &1.00    &1.00\\
				&0.2	  &1.00   &1.00   &1.00  &0.06 &1.00    &1.00\\
				&0.4	  &1.00   &1.00   &1.00  &0.00 &0.00    &0.00\\
				\hline
				5&0.025    &0.69   &1.00   &1.00  &1.00 &1.00    &1.00\\
				&0.05	  &0.93   &1.00   &1.00  &0.98 &1.00    &1.00\\
				&0.1	  &1.00   &1.00   &1.00  &0.04 &1.00    &1.00\\
				&0.2	  &1.00   &1.00   &1.00  &0.00 &0.00    &0.00\\
				&0.4	  &1.00   &1.00   &1.00  &0.00 &0.00    &0.00\\
				\cline{1-8}	
			\end{tabular}
	}}
\end{table}

Taken together, the results in Tables \ref{simulationE7}--\ref{simulationE9} demonstrate that PNN-CV is robust to the choice of tuning constants in practice. The method admits a broad stable region of penalty values, and its asymptotic consistency is not sensitive to precise calibration. This supports the practical applicability of the proposed framework and alleviates concerns about an additional layer of tuning.

\newpage
\renewcommand{\thefigure}{F.\arabic{figure}}
\renewcommand{\thetable}{F.\arabic{table}}
\setcounter{figure}{0}
\setcounter{table}{0}
\section{Additional Real Data}
\subsection{Additional Detail and Illustration for ``Political Books'' Dataset}

As introduced in Section 7, following the approach in \cite{zhan202224} to avoid betting on one particular data splitting ratio in CV, we vary the data splitting ratio $w$ over a wide range, representing different numbers of folds in cross-validation, and examine the output of our method under these different settings. Recall that we set the number of replications in each voting process to 20. 

For comparison, we also apply the ECV method to this dataset, selecting among SBM and DCBM models with no more than 15 communities, with the frequency results shown in Table \ref{polbooks2}.

\begin{table}[htbp]
	\centering\renewcommand\arraystretch{1}{\scriptsize
		\caption{PNN-CV on Political Books Data.
		}\label{polbooks}
		\setlength{\tabcolsep}{4mm}{
			\begin{tabular}{cccccccc}
				\hline
				\cline{1-8}
				Known ground truth    &   Model  & 2-fold & 3-fold & 5-fold & 8-fold & 10-fold & 15-fold\\
				\hline
				Yes    &   DCBM-3  &  1.00  &  1.00  &  1.00  &  1.00  &  1.00  &  1.00\\
				& SBM-3 &0.00 &0.00 &0.00 &0.00 &0.00 &0.00\\
				\hline
				No    &   DCBM-4  &  1.00  &  1.00  &  1.00  &  1.00  &  1.00  &  1.00\\
				& others &0.00 &0.00 &0.00 &0.00 &0.00 &0.00\\
				\hline
				\cline{1-8}
			\end{tabular}
		}
	}
\end{table}

\begin{table}[htbp]
	\centering\renewcommand\arraystretch{1}{\scriptsize
		\caption{ECV on Political Books Data.
		}\label{polbooks2}
		\setlength{\tabcolsep}{4mm}{
			\begin{tabular}{ccccccc}
				\hline
				\cline{1-7}
				Model  & 2-fold & 3-fold & 5-fold & 8-fold & 10-fold & 15-fold\\
				\hline
				DCBM-2  &  1.00 &  1.00  &  0.45  &  0.10  &  0.00  &  0.00\\
				DCBM-8  &  0.00 &  0.00  &  0.10  &  0.25  &  0.30  &  0.50\\
				DCBM-9  &  0.00 &  0.00  &  0.10  &  0.30  &  0.25  &  0.15\\
				DCBM-10 &  0.00 &  0.00  &  0.05  &  0.05  &  0.10  &  0.10\\
				Other DCBMs&0.00&  0.00  &  0.20  &  0.25  &  0.25  &  0.15\\
				SBMs    &  0.00 &  0.00  &  0.10  &  0.05  &  0.10  &  0.10\\
				\hline
				\cline{1-7}
			\end{tabular}
		}
	}
\end{table}

As shown in Table \ref{polbooks2}, the model selection results of the ECV method display notable variability across different data splitting ratios. In particular, with small numbers of folds (e.g., 2-fold and 3-fold), the selected model is consistently DCBM-2, while under larger fold settings (e.g., 10-fold and 15-fold), models with significantly more communities such as DCBM-8 or DCBM-9 are frequently selected. This suggests that the ECV method may be sensitive to the fold choice and may tend to favor models with a larger number of communities, especially when more data are used in the training step. While such behavior may reflect efforts to improve model fit, it also highlights potential instability and the lack of explicit complexity control in model selection.

In contrast, our proposed PNN-CV+BHMC framework exhibits consistent and stable behavior across a wide range of folds. As demonstrated in Table \ref{polbooks}, when incorporating the ground truth that $K=3$, the DCBM with three communities is consistently selected across all splitting ratios, strongly indicating that DCBM fits the data better than SBM. However, when no ground truth is assumed and taken advantage of, our method selects DCBM with four communities as the optimal model under all settings.

\begin{figure}[htbp]
	\centering
	\includegraphics[width =10cm]{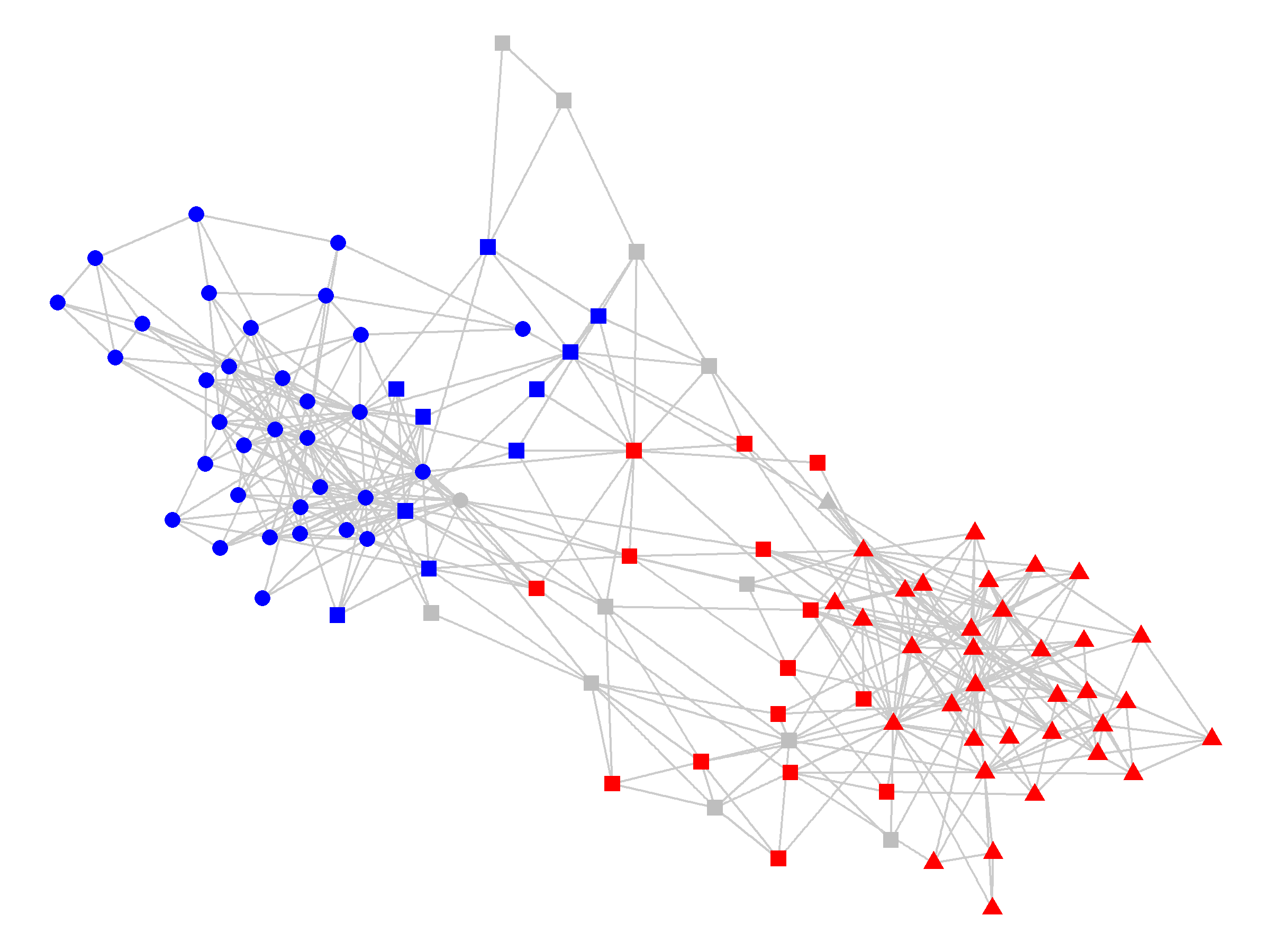}
	\caption{Estimated community structure for SBM-3}\label{polbooks_visualize_add}
\end{figure}

In addition to the illustration for DCBMs in Section 7, we also provide the corresponding visualization under SBM-3 in Figure \ref{polbooks_visualize_add}.

\subsection{Additional Real Data Applications}

We consider four additional benchmark network datasets of varying sizes, ranging from small networks with fewer than 100 nodes to large-scale networks with several thousand nodes.
For all datasets, we apply both PNN-CV and its hybrid variant PNN-CV+BHMC.
For small and moderate-sized networks, we focus on selecting among parametric block models (AM/SBM/DCBM), while for larger networks (more than one thousand) we additionally include graphon models, which are more appropriate in the large-sample regime. For the first three real datasets we use voting with 5 replications, and for the last one we use a single split due to its large size.

These real data examples also highlight an aspect that has been less explored in the
literature. While most existing methods focus on comparisons within parametric block
models, our framework allows comparisons between parametric block models and graphon
models in a unified manner, and in large-scale networks the latter are often selected, suggesting that simple block-structured assumptions may be slightly inadequate in capturing the underlying connectivity patterns in large complex networks.

We first report our result for two relatively small datasets.

\paragraph{Dolphin Network}
The dolphins network was compiled by \cite{lusseau2003dolphin},contains 62 nodes and 159 undirected edges and has subsequently been analyzed in the literature such as in \cite{hwang2024estimation}. Both PNN-CV and PNN-CV+BHMC select degree-corrected block models, reflecting pronounced degree heterogeneity in the data. The hybrid method (PNN-CV+BHMC) selects DCBM-2, which aligns well with the known biological labels, while PNN-CV alone selects a slightly larger model (DCBM-4), indicating mild over-partitioning in very small samples.

\paragraph{NCAA Football Network}
The NCAA football dataset was proposed in \cite{newman2004community} and has become a benchmark real data example. Both the hybrid method and PNN-CV method choose the stochastic block model, which is reasonable since its minimum degree is 7 and maximum degree is 12, indicating limited degree heterogeneity. Specifically, the hybrid method selects SBM with 10 communities and the PNN-CV method picks SBM with 11 communities, which are both between 5 (the number of conferences) and the reported estimate $\hat{K}=12$ in \cite{newman2004community}.

Now we consider two large datasets, where graphon models are included due to their flexibility in capturing complex connectivity patterns.

\paragraph{Weblogs Network}
The weblogs (or, political weblogs) network, collected and analyzed by \cite{adamic2005political}, contains 1,222 political blogs, with an edge indicating a hyperlink between a pair of blogs. The nodes are labeled as being either
liberal or conservative. When graphon models are included, both procedures select a graphon model, indicating substantial deviation from block-structured assumptions. This can be further verified by the goodness-of-fit statistic reported for this dataset by \cite{Jin2025goodness}, where they claimed that DCBM suffers from mild lack-of-fit on this network. If restricted to block models, PNN-CV+BHMC selects DCBM-8, while PNN-CV alone selects a more parsimonious DCBM-2, which aligns more closely with the known political labels.

\paragraph{``fb-CMU-Carnegie49'' network}
This is a subset of a social network compiled by \cite{rossi2015networkrepository} composed of 6,637 individuals and 249,967 friendships on Facebook. Both procedures favor graphon models, suggesting that neither SBM nor DCBM adequately captures the underlying structure. When restricted to block models, PNN-CV+BHMC selects a large DCBM, while PNN-CV favors a more parsimonious alternative. Neither result aligns exactly with the three available labels, indicating that the observed network structure may not be well explained by simple community-based models.

Overall, these results reinforce the empirical patterns observed in the simulations: PNN-CV+BHMC tends to provide stable selections in small-sample regimes, while PNN-CV alone often yields more interpretable and parsimonious models in larger networks, and graphon models are naturally favored when block structures are insufficient.

\bibliographystyle{asa}
\bibliography{ref}

\end{document}